\documentclass[aps,twocolumn]{revtex4}
\usepackage{graphicx}
\usepackage{amsmath}
\usepackage{color}
\newcommand{\be}{\begin{equation}}
\newcommand{\ee}{\end{equation}}
\newcommand{\bea}{\begin{eqnarray}}
\newcommand{\eea}{\end{eqnarray}}
\newcommand{\bs}{\begin{split}}
\newcommand{\bse}{\begin{subequations}}
\newcommand{\ese}{\end{subequations}}

\begin{document}
 
\title{Influence of uniaxial single-ion anisotropy on the magnetic and thermal properties of Heisenberg antiferromagnets within unified molecular field theory}
\author {David C.\ Johnston} 
\affiliation {Ames Laboratory and Department of Physics and Astronomy, Iowa State University, Ames, Iowa 50011}

\date{\today}

\begin{abstract}

The influence of uniaxial single-ion anisotropy $-DS_z^2$ on the magnetic and thermal properties of Heisenberg antiferromagnets (AFMs) is investigated.  The uniaxial anisotropy is treated exactly and the Heisenberg interactions are treated within unified molecular field theory (MFT) [Phys.~Rev.~B {\bf 91}, 064427 (1915)], where thermodynamic variables are expressed in terms of directly measurable parameters.  The properties of collinear AFMs with ordering along the $z$~axis ($D>0$) in applied fields~$H_z=0$ are calculated versus~$D$ and temperature~$T$, including the ordered moment $\mu$, the N\'eel temperature $T_{\rm N}$, the magnetic entropy, internal energy, heat capacity and the anisotropic magnetic susceptibilities $\chi_\parallel$ and~$\chi_\perp$ in the paramagnetic (PM) and AFM states.  The high-field average magnetization per spin~$\mu_z(H_z,D,T)$ is found, and the critical field $H_{\rm c}(D,T)$ is derived at which the second-order AFM to PM phase transition occurs.  The magnetic properties of the spin-flop (SF) phase are calculated, including the zero-field properties $T_{\rm N}(D)$ and $\mu(D,T)$.  The high-field $\mu_z(H_z,D,T)$ is determined, together with the associated spin-flop field $H_{\rm SF}(D,T)$ at which a second-order SF to~PM phase transition occurs.  The free energies of the AFM, SF and PM phases are derived from which $H_z-T$ phase diagrams are constructed.  For $f_J =-1$ and~$-0.75$, where $f_J = \theta_{{\rm p}J}/T_{{\rm N}J}$ and $\theta_{{\rm p}J}$ and $T_{{\rm N}J}$ are the Weiss temperature in the Curie-Weiss law and the N\'eel temperature due to exchange interactions alone, respectively, phase diagrams in the $H_z-T$ plane similar to previous results are obtained.  However, for $f_J=0$ we find a topologically different phase diagram where a spin-flop bubble with PM and AFM boundaries occurs at finite $H_z$ and~$T$\@.  Also calculated are properties arising from a perpendicular magnetic field, including  the perpendicular susceptibility $\chi_\perp(D,T)$, the associated effective torque at low fields arising from the $-DS_z^2$ term in the Hamiltonian, the high-field perpendicular magnetization $\mu_\perp$ and the perpendicular critical field $H_{\rm c\perp}$ at which the second-order AFM to PM phase transition occurs.  In addition to the above results for $D>0$, the $T_{\rm N}(D)$ and ordered moment $\mu(T,D)$ for collinear AFM ordering along the $x$~axis with $D<0$ are determined.  In order to compare the  properties of the above spin systems with those of noninteracting systems  with $-DS_z^2$ uniaxial anisotropy with either sign of~$D$, an Appendix is included in which results for the thermal and magnetic properties of such noninteracting spin systems are provided.

\end{abstract}

%\pacs {75.30.Cr, 75.10.Jm, 75.40.Cx, 75.50.Ee}

\maketitle

\section{Introduction}

The presence of anisotropy in a spin system that otherwise has isotropic Heisenberg exchange interactions can significantly affect the thermal and magnetic properties of the system.  The origin of the anisotropy can take various forms \cite{Nagamiya1955, Kanamori1963, Darby1973}.  The ubiquitous magnetic dipole interaction between spins is well known.  A comprehensive study of the resulting anisotropic properties of spin systems with Heisenberg interactions within molecular field theory (MFT) recently appeared \cite{Johnston2016}.  Another potential source of anisotropy is anisotropy in the exchange interactions in spin space, leading, e.g., to the XY, Ising and intermediate XXZ models.  The anisotropy in the magnetic susceptibility~$\chi$ of noninteraction spin systems arising from single-ion magnetocrystalline anisotropy is also well known \cite{VanVleck1932, Carlin1986}, although a comprehensive study of the magnetic and thermal behaviors of these systems is lacking.

A MFT study of the influence of single-ion anisotropy on~$\chi$ of Heisenberg spin systems was carried out in 1951 \cite{Yosida1951} using the same MFT as for calculations in 1941 of the anisotropic $\chi$ below the antiferromagnetic (AFM) ordering temperature~$T_{{\rm N}J}$ for Heisenberg spin interactions \cite{VanVleck1941}.  These MFT predictions are highly constrained by the requirement that in the absence of the uniaxial anisotropy, the ratio $f_J = \theta_{{\rm p}J}/T_{{\rm N}J}$ of the Weiss temperature $\theta_{{\rm p}J}$ in the high-temperature Curie-Weiss law and $T_{{\rm N}J}$ is equal to~$-1$, which is rarely if ever observed in practice.  Here we distinguish between the Weiss temperature $\theta_{\rm p}$ and N\'eel temperature $T_{\rm N}$ obtained in the presence of both uniaxial anistropy and Heisenberg interactions from the above designations $\theta_{{\rm p}J}$ and $T_{{\rm N}J}$ resulting from exchange interactions alone.  Spin-wave theory has been applied to systems with single-ion anisotropy and Heisenberg interactions and the theory predicts that the anisotropy gives rise to energy gaps in the spin-wave spectra \cite{Keffer1966} in addition to modifying the spin wave branches.  Spin-wave calculations have also been useful in predicting the $\chi$ and magnetic heat capacity~$C_{\rm mag}$ of AFMs at temperatures~$T$ below their~$T_{\rm N}$ \cite{Keffer1966, Itoh1972}.  The influence of uniaxial single-ion anisotropy on $T_{\rm N}$ of Heisenberg spin systems was studied  using Green function techniques, and was found for spins with spin angular momentum quantum number $S=1$ on a simple-cubic lattice to be significantly stronger than inferred from MFT for small anisotropy parameters \cite{Lines1967}. Subsequent Green function treatments for $S=1$ showed that MFT accurately predicts $T_{\rm N}$ for large values of the single-ion anisotropy  \cite{Devlin1971, Tanaka1973}.

In this paper we greatlyly extend previous work by carrying out a comprehensive investigation of the influence of uniaxial single-ion $DS_z^2$ anisotropy on the thermal and magnetic properties of local-moment Heisenberg AFMs.  The anisotropy is treated exactly and the Heisenberg interactions by MFT\@.  We obtain expressions for arbitrary values of $f_J$ and for both positive and negative anisotropy parameters~$D$ of arbitrary magnitude.  Many plots of the properties are provided including phase diagrams in the field-temperature plane.  We confirm that the presence of ferromagnetic interactions in addition to the required AFM ones can result in first-order AFM to paramagnetic (PM) phase transitions for fields aligned along the AFM easy axis with $D>0$ \cite{Vilfan1986}. We also calculate the magnetic properties of systems with $D< 0$ where in-plane AFM ordering occurs. 

The unified MFT used in our calculations to treat the Heisenber interactions was recently presented for local-moment AFMs containing identical crystallographically-equivalent spins with Heisenberg interactions that does not use the concept of magnetic sublattices \cite{Johnston2012,Johnston2015,Johnston2015b}.  Instead, the magnetic and thermal properties are calculated simply from the interactions of a representative spin with its neighbors.  Another significant advantage of this MFT is that it is formulated in terms of physically measurable quantities.  These include the spin $S$ of the local moment, $f_J$, $T_{\rm N}$, $\chi(T_{\rm N})$ and $\theta_{\rm p}$ in the Curie-Weiss law.

The  Curie-Weiss law in the PM state at temperatures $T \geq T_{\rm N}$ is written for a representative spin as
\bse
\label{Eqs:CWLaw}
\be
\chi = \frac{C_1}{T-\theta_{\rm p}},
\ee
where
\be
C_1 = \frac{g^2S(S+1)\mu_{\rm B}^2}{3k_{\rm B}}
\label{C1}
\ee
\ese
is the single-spin Curie constant, $g$ is the spectroscopic splitting factor ($g$ factor), $\mu_{\rm B}$ is the Bohr magneton and $k_{\rm B}$ is Boltzmann's constant.   For simplicity it is assumed in this paper that the $g$-factor is isotropic as appropriate for $s$-state magnetic ions for which $g\approx 2$.  For moments that are aligned along a principal axis~$\alpha$, $g$ can be replaced by a variable $g_\alpha$ in the respective equations, where $g_\alpha$ is obtained theoretically and/or from experimental measurements.

The Hamiltonian associated with a representative spin~$i$ is taken to be 
\be
{\cal H} = {\bf S}\cdot \sum_j J_{ij}{\bf S}_j + g\mu_{\rm B}{\bf S}\cdot {\bf H} - DS_z^2,
\label{Eq:Ham1}
\ee
where the first term is the sum of the Heisenberg exchange interactions between spin~$i$ with spin operator {\bf S} and its neighbors ${\bf S}_j$ with which it interacts with strength $J_{ij}$, a positive (negative) $J_{ij}$ corresponds to AFM (ferromagnetic FM) interactions, and {\bf S} is in units of $\hbar$ where $\hbar$ is Planck's constant divided by $2\pi$.  The second term in Eq.~(\ref{Eq:Ham1}) is the Zeeman interaction $-\vec{\mu}_i\cdot{\bf H}$ of the magnetic moment operator $\vec{\mu}_i$ with the applied field~{\bf H}, where this operator is written in terms of~{\bf S} as
\be
\vec{\mu}_i = -g\mu_{\rm B}{\bf S}
\label{Eq:muiS}
\ee
and the negative sign originates from the negative charge on the electron which is usually taken to be a plus sign in the literature.  The third term in Hamiltonian~(\ref{Eq:Ham1}) is the uniaxial single-ion anisotropy with respect to the uniaxial $z$~axis.  The negative sign preceding this term is conventional and results in collinear AFM ordering along the $z$-axis for $D>0$.  The present paper is devoted to studying the influence of this term on the thermal and magnetic properties of Heisenberg spin systems.

The theory needed for the calculations of the thermal and magnetic properties with the Heisenberg interactions treated by the unified MFT is given in Sec.~\ref{MFTBckgrnd}.  This section includes the general expression for the exchange field expressed in terms of the MFT variables in Refs.~\cite{Johnston2012, Johnston2015}, the magnetic moment operators needed to calculate the thermal-average moments, expressions for the N\'eel and Weiss temperatures due to Heisenberg exchange interactions by themselves, treatment of the special case of two-sublattice AFM structures, the definitions of the dimensionless magnetic susceptibilities, the expressions used to calculate the magnetic entropy, internal energy, Helmholtz free energy and heat capacity within the context of MFT, and the second-order perturbation theory for both integer and half-integer spins that is used to provide formulas for the perpendicular susceptibilities of various spin configurations.  The parallel susceptibility $\chi_\parallel$ is defined as the magnetic susceptibility parallel to the easy axis of a collinear AFM taken to be the $z$-axis for $D>0$, and the perpendicular susceptibility $\chi_\perp$ measured with the applied field perpdicular to the easy axis, taken to be the $x$~axis.

The remainder of the paper presents applications of the theory in Sec.~\ref{MFTBckgrnd} to the influences of the quantum uniaxial anisotropy on the thermal and magnetic properties of various Heisenberg spin configurations within the unified MFT, mostly for $D>0$.  Many plots of the predicted properties versus $T$ and/or~{\bf H} are provided.  The $\chi_\parallel(D,T)$ and $\chi_\perp(D,T)$ behaviors are obtained for the paramagnetic (PM) state in Sec.~\ref{Sec:ChiPM} for both integer and half-integer spins, where second-order perturbation theory is used to derive $\chi_\perp(D,T)$.  The ordered moment in $H=0$ versus temperature, the N\'eel temperature versus~$D$ and the thermal properties of collinear AFMs with $D>0$ are studied versus $T$ and~$D$ in Sec.~\ref{Sec:CollAFMDgtr0}.

The properties of collinear AFMs with $D>0$ in parallel fields are obtained in Sec.~\ref{Sec:CollinearAFMDH}, including calculations of $\chi_\parallel(D,T)$ and the parallel magnetization in high fields, together with the associated critical fields ($H_{\rm c}$) for transitions from the AFM to the PM state versus~$T$\@.  The staggered magnetization (the AFM order parameter) versus $H_z$ and~$D>0$ is  also obtained.

Section~\ref{Sec:SFPhase} is devoted to a study of the spin-flop (SF) phase with $D>0$, where the ordered moments are flopped over from the collinear AFM phase along the $z$~axis into two sublattices that make equal angles with the $z$~axis.  In this section the zero-field N\'eel temperature and ordered moment of the SF phase versus $T$ and~$D$ are calculated, and the magnetization versus high applied $H_z$ field determined.  From the latter calculation the spin-flop field $H_{\rm SF}(D,T)$ for the second-order transition from the SF to the PM phase is found.

In Sec.~\ref{Sec:PhaseDiagram} the free energies of the AFM and SF phases versus $T$ and~$H_z$ are calculated for representative spin $S=1$ and $D=0.5k_{\rm B}T_{{\rm N}J}$.  From a comparison of their free energies, the first-order AFM to SF transition line in the $T-H_z$ plane is found.  Then together with the previous calculations of $H_{\rm c}(D,T)$ of the AFM phase and $H_{\rm SF}(D,T)$ of the SF phase, exemplary $H_z-T$ phase diagrams are constructed for $S=1$ and $D=0.5k_{\rm B}T_{{\rm N}J}$ with $f_J=-1,\ -0.75$ and~0. The phase diagrams for $f_J>-1$ correspond to the introduction of ferromagnetic exchange interactions between the spins.  For $f_J = -1$ and~$-0.75$ we obtain phase diagrams of the well-known type.  However, for $f_J=0$ we find a topological change in the phase diagram where the spin-flop phase appears as a bubble in the $H_z-T$ plane at finite $H_z$ and~$T$\@.

In Sec.~\ref{Sec:HPerp} the effects of fields~$H_x$ applied perpendicular to the easy axis of a collinear AFM with $D > 0$ are discussed.  Here we calculate $\chi_\perp(D,T)$ using the second-order perturbation theory in Sec.~\ref{MFTBckgrnd}.  Expressions for the Weiss temperature in the Curie-Weiss law~(\ref{Eqs:CWLaw}), the effective torque and the anisotropy constant~$K_1$ associated with the uniaxial anisotropy at low fields are also obtained. The latter expression agrees with a previous result at $T=0$ obtained using a different approach \cite{Kanamori1962}.  We also determine the $T$~dependence of~$K_1$.  The high-field perpendicular magnetization is then calculated and the critical field $H_{\rm c\perp}(D,T)$ for the second-order transition from the canted AFM state to the PM state determined.  In contrast to most previous MFT treatments of $\mu_\perp$ versus $H_\perp$ (e.g., \cite{Johnston2015}), we find that both the ordered moment and $\mu_\perp/H_\perp$ at a given $T$ in the AFM state depend on $H_\perp$ when $D>0$.  In Sec.~\ref{Sec:In-PlaneOrdDless0} collinear AFM ordering along the transverse $x$~axis with $D < 0$ is discussed, where the N\'eel temperature and ordered moment in the AFM state versus~$D$ and~$T$ in $H=0$ are calculated.

A brief summary of the results of this paper is given in Sec.~\ref{Sec:Summary}.  In order to compare these results with those for noninteracting spin systems as done in the main text, the thermal and magnetic properties of spin systems with no spin interactions but with axial single-ion anisotropy including plots of these properties versus~$T$ and/or~$H$ are described in the Appendix.

%c\clearpage

\section{\label{MFTBckgrnd} Theory}

The expressions in this section involving the unified MFT are either quoted from or derived from those in Refs.~\cite{Johnston2012, Johnston2015}.

\subsection{Exchange Field and Hamiltonian}

The basis states of the Hilbert space used for the Hamiltonian eigenfunctions in this paper for spin~$S$ are $|S,S_z\rangle$, with $z$~components of the spin angular momentum $S_z \equiv m_S = -S,\ -S+1,\ \ldots,\ S$.  Since the expectation value $\langle S_z^2\rangle = 1/4$ for the two values $m_S = \pm1/2$ of the spin magnetic quantum number for $S = 1/2$, the $DS_z^2$ single-ion anisotropy term in Eq.~(\ref{Eq:Ham1}) is a constant and hence produces no anisotropy for spins $S=1/2$. 

Within MFT, one approximates the exchange interactions~$J_{ij}$ of a given spin~$i$ with its neighbors $j$ in Eq.~(\ref{Eq:Ham1}) by an effective molecular (or exchange) field 
\bse
\be
{\bf H}_{{\rm exch}\,i} = -\frac{1}{g^2\mu_{\rm B}^2}\sum_j J_{ij}\vec{\mu}_j,
\label{HexchiVec}
\ee
where $\vec{\mu}_j$ is the thermal-average moment of spin~$j$. A moment $\vec{\mu}$  can arise from exchange interactions,  an applied field or both.  We will therefore ofter refer to such thermal-average moments as simply ``ordered moments''.  The exchange field is treated as if it were an applied field.  The component of the exchange field parallel to moment $\vec{\mu}_i$ is
\be
H_{{\rm exch}\,i} = \hat{\mu}_i\cdot {\bf H}_{{\rm exch}\,i} = \frac{1}{g^2\mu_{\rm B}^2}\sum_j J_{ij}\mu_j\cos\alpha_{ji},
\ee
\ese
where $\alpha_{ji}$ is the angle between $\vec{\mu}_j$ and $\vec{\mu}_i$ in the ordered and/or field-induced state.  In {\bf H} = 0, due to their crystallographic equivalence all ordered moments have the same magnitude defined as $\mu_0$, in which case $\alpha_{ji}\equiv\phi_{ji}$.  The $\phi_{ji}$ are given by the assumed magnetic structure in either the AFM or PM state.

Using Eqs.~(\ref{Eq:Ham1}) and (\ref{Eq:muiS}), within MFT the Hamiltonian associated with a representative spin including the {\bf H}, ${\bf H}_{\rm exch}$ and $DS_z^2$ terms is
\bse
\be
{\cal H} = -\vec{\mu}_i\cdot {\bf B}_i - DS_z^2 = g\mu_{\rm B}{\bf S}\cdot {\bf B}_i -DS_z^2,
\label{Eq:Hamds2bi0}
\ee
where
\be
{\bf B}_i = {\bf H}_{{\rm exch}i} + {\bf H}
\label{Eq:BDef}
\ee
\ese
is the local magnetic induction at the position of spin~$i$.  The {\bf B} and {\bf H} are normalized here  according to
\be
{\bf b} \equiv \frac{g\mu_{\rm B}{\bf B}}{k_{\rm B}T_{{\rm N}J}},\qquad {\bf h} \equiv \frac{g\mu_{\rm B}{\bf H}}{k_{\rm B}T_{{\rm N}J}},
\label{Eq:bhDef}
\ee
where $T_{{\rm N}J}$ is the N\'eel temperature for an assumed magnetic structure in $H=0$ that would occur due to the exchange interactions alone as derived in Sec.~\ref{Sec:TNJqJ} below. In terms of these reduced variables, one has
\be
{\bf b}_i = {\bf h}_{{\rm exch}i} + {\bf h}.
\label{Eq:bDef}
\ee   
All energies are also normalized by $k_{\rm B}T_{{\rm N}J}$, so the reduced Hamiltonian obtained from Eq.~(\ref{Eq:Hamds2bi0}) is
\be
\frac{\cal H}{k_{\rm B}T_{{\rm N}J}} = {\bf S}\cdot {\bf b}_i -dS_z^2,
\label{Eq:Hamds2bi}
\ee
where the reduced anisotropy constant~$d$ is
\be
d \equiv \frac{D}{k_{\rm B}T_{{\rm N}J}}.
\label{Eq:dDef}
\ee

The $2S+1$ reduced energy eigenvalues of the Hamiltonian~(\ref{Eq:Hamds2bi}) for a given spin~$S$ are denoted as 
\be
\epsilon_n = \frac{E_n}{k_{\rm B}T_{{\rm N}J}} \qquad (n = 1,\ 2,\ \ldots,\ 2S+1),
\label{Eq:EpsDef}
\ee
where 
\be
\epsilon_n  = \epsilon_n(h_\alpha,d, S)
\ee
and $\alpha = x$ or~$z$ here.  Within MFT, the final expressions for the energy eigenvalues are in general temperature dependent due to the temperature dependence of the ordered and/or field-induced moments contained in them that are solved for as described for different cases in subsequent sections.

\subsection{Magnetic Moment Operators and Thermal-Average Components of the Magnetic Moment}

In this paper, we consider ordered moments lying either along the $z$~axis as in collinear magnetic ordering along this axis, or in the $x-z$ plane as when a perpendicular field $H_x$ is applied to a collinear AFM structure that is aligned along the $z$ axis in $H=0$.  The $x-z$ plane ordered-moment alignment also applies to the spin-flop phase where in zero field the ordered moments are aligned along the $x$~axis, and tilt towards the $z$~axis in the presence of a field $H_z$ along the $z$~axis.  For collinear moment alignments along the $z$~axis, the exchange field ${\bf H}_{{\rm exch}i}$ seen by a representative spin~$i$ is also oriented along the $z$~axis, whereas for both the spin-flop phase and the AFM phase with an easy $z$~axis in a perpendicular ${\bf H} = H_x\hat{\bf i}$,  ${\bf H}_{{\rm exch}i}$ has components along both the $x$ and~$z$ axes in general.

In general, the eigenvalues of Hamiltonian~(\ref{Eq:Hamds2bi0}) thus contain both $x$ and $z$ components $\mu_{ix}$ and $\mu_{iz}$ of the central ordered moment~$\vec{\mu}_i$ which must both be solved for.  We therefore define magnetic moment operators $\mu_{nx}^{\rm op}$ and~$\mu_{nz}^{\rm op}$ in terms of the energy eigenvalues~$E_n$ of Hamiltonian~(\ref{Eq:Hamds2bi0}) as
\be
\mu_{n\alpha}^{\rm op} = -\frac{\partial E_n}{\partial B_\alpha} \qquad(\alpha=x,z),
\label{Eq:muxmuzDef}
\ee
where $B_x$ and $B_z$ are the $x$ and~$z$ components of the magnetic induction~{\bf B} in Eq.~(\ref{Eq:BDef}), respectively.  It is convenient to define dimensionless reduced magnetic moments
\bse
\label{Eqs:mubarDef}
\be
\bar{\mu} = \frac{\mu}{\mu_{\rm sat}},
\label{Eq:barmuDef}
\ee
where the saturation moment $\mu_{\rm sat}$ is
\be
\mu_{\rm sat} = gS\mu_{\rm B}.
\ee
\ese
In terms of the reduced variables in Eqs.~(\ref{Eq:bhDef}), (\ref{Eq:EpsDef}) and~(\ref{Eqs:mubarDef}), the magnetic moment operators~(\ref{Eq:muxmuzDef}) become
\be
\bar{\mu}_{n\alpha}^{\rm op} = -\frac{1}{S}\frac{\partial \epsilon_n}{\partial b_\alpha}.
\label{Eq:barmuOp}
\ee

The thermal-average values $\bar{\mu}_\alpha$ are calculated self-consistently from the conventional expression
\bse
\be
\bar{\mu}_\alpha = \frac{1}{Z_S}\sum_{n=1}^{2S+1}\bar{\mu}_\alpha^{\rm op}e^{-\epsilon_n/t}=-\frac{1}{SZ_S}\sum_{n=1}^{2S+1}\frac{\partial\epsilon_n}{\partial b_\alpha}e^{-\epsilon_n/t},
\label{Eq:barmualpha}
\ee
where the reduced temperature~$t$ is 
\be
t = \frac{T}{T_{{\rm N}J}}
\label{Eq:tDef}
\ee
and the partition function is
\be
Z_S = \sum_{n=1}^{2S+1}e^{-\epsilon_n/t}.
\ee
\ese
If both $\bar{\mu}_x$ and $\bar{\mu}_z$ are nonzero, then Eq.~(\ref{Eq:barmualpha}) becomes two simultaneous equations in these two variables from which the solutions to both $\bar{\mu}_x$ and $\bar{\mu}_z$ are obtained.  If all moments and fields are aligned along the $z$~axis, then $\epsilon_n\to\epsilon(m_S)$ and the above sums over~$n$ become sums over the spin magnetic quantum number $m_S = -S$ to~$S$ in integer increments. 

\subsection{\label{Sec:TNJqJ} N\'eel and Weiss Temperatures from Exchange Interactions Only}

The AFM transition temperature $T_{{\rm N}J}$ in $H=0$ and the Weiss temperature $\theta_{{\rm p}J}$ due to exchange interactions between spins of the same magnitude are given by
\bse
\label{Eqs:TNqpJ}
\bea
T_{{\rm N}J} &=& -\frac{S(S+1)}{3k_{\rm B}}\sum_j J_{ij}\cos\phi_{ji},\label{Eq:TNJGen}\\*
\theta_{{\rm p}J} &=& -\frac{S(S+1)}{3k_{\rm B}}\sum_j J_{ij}\label{Eq:ThetapGen},
\eea
\ese
where the sums are over all neighbors~$j$ of a given central spin~$i$ and the subscript $J$ on the left sides signifies that these quantities arise from exchange interactions only, and $\phi_{ji}$ is the angle between moments $j$ and~$i$ in the AFM structure at $T<T_{{\rm N}J}$.  The exchange field component in the direction of representative ordered moment $\vec{\mu}_i$ in $H=0$ is
\be
H_{{\rm exch}0} =\frac{T_{{\rm N}J}}{C_1}\mu_0 = \frac{3k_{\rm B}T_{{\rm N}J}}{g\mu_{\rm B}(S+1)}\bar{\mu}_0.
\label{Eq:Hexch1}
\ee
where the index $i$ has been dropped because the exchange field is the same for each spin since they are assumed to be identical and crystallographically equivalent and the subscript 0 in $\bar{\mu}_0 \equiv \bar{\mu}_i$ means that it is a zero-field property. The dimensionless reduced fields $h$ and $b$ associated with the field $H$ and $B$ are defined as
\be
h = \frac{g\mu_{\rm B}H}{k_{\rm B}T_{{\rm N}J}}, \qquad b = \frac{g\mu_{\rm B}B}{k_{\rm B}T_{{\rm N}J}}.
\label{Eq:hDef}
\ee
Thus Eqs.~(\ref{Eq:Hexch1}) and~(\ref{Eq:hDef}) give the magnitude of the reduced exchange field in the direction of each of the ordered moments in any AFM state with $H=0$ as
\be
h_{\rm exch0} = \frac{3\bar{\mu}_0}{S+1} \qquad {\rm (AFM~state},~H=0).
\label{Eq:hexch0def}
\ee

\subsection{Two-Sublattice Collinear AFM Structures}

Magnetic structures are studied later consisting of equal numbers of spins on two sublattices where  all moments $\vec{\mu}_i$ having the same magnitude and direction are on the same~(s) sublattice and the equal number of other moments $\vec{\mu}_j$ with a different magnitude and direction are on the different (d) sublattice.  

For the special case of a collinear AFM in $H=0$ where the moments on the two sublattices s and~d have the same magnitude but are antiparallel in direction,  Eqs.~(\ref{Eqs:TNqpJ}) give
\bse
\label{Eqs:TNJthetap}
\bea
T_{{\rm N}J} &=& -\frac{S(S+1)}{3k_{\rm B}} \bigg({\sum_j}^{\rm s}J_{ij} - {\sum_j}^{\rm d}J_{ij}\bigg),\label{Eq:TNJzAxis}\\*
\theta_{{\rm p}J} &=& -\frac{S(S+1)}{3k_{\rm B}} \bigg({\sum_j}^{\rm s}J_{ij} + {\sum_j}^{\rm d}J_{ij}\bigg).
\eea
\ese
Solving Eqs.~(\ref{Eqs:TNJthetap}) for the two sums gives
\bse
\label{Eqs:sumssumd}
\bea
{\sum_j}^{\rm s}J_{ij} &=& -\frac{3k_{\rm B}T_{{\rm N}J}(1+f_J)}{2S(S+1)},\\*
{\sum_j}^{\rm d}J_{ij} &=& \frac{3k_{\rm B}T_{{\rm N}J}(1-f_J)}{2S(S+1)},
\eea
where we used the definition
\be
f_J = \frac{\theta_{{\rm p}J}}{T_{{\rm N}J}}.
\label{Eq:fJDef}
\ee
\ese
Equations~(\ref{Eqs:sumssumd}) allow replacement of the respective sums wherever they occur by the more physically relevant parameters $T_{{\rm N}J}$ and~$\theta_{{\rm p}J}$.  One has $-\infty<f_J < 1$ for AFMs and $f_J = 1$ for FMs.  

From Eq.~(\ref{HexchiVec}), the exchange field seen by central moment $\vec{\mu}_i$ in $H=0$ in a two-sublattice AFM is given in general by
\bse
\label{Eqs:Hexchi2Sub}
\be
{\bf H}_{{\rm exch}i} = -\frac{1}{g^2\mu_{\rm B}^2}\bigg(\vec{\mu}_i{\sum_j}^{\rm s} J_{ij} + \vec{\mu}_j{\sum_j}^{\rm d} J_{ij}\bigg).
\label{Eq:VecHexchi}
\ee
Then Eqs.~(\ref{Eqs:sumssumd}) give
\be
{\bf H}_{{\rm exch}i} = \frac{3k_{\rm B}T_{{\rm N}J}}{2g^2\mu_{\rm B}^2S(S+1)}\left[\vec{\mu}_i(1+f_J) - \vec{\mu}_j(1-f_J)\right].
\label{Eq:HexchHGen}
\ee
Using Eqs.~(\ref{Eq:hDef}), the reduced exchange field seen by $\vec{\mu}_i$ is obtained from Eq.~(\ref{Eq:HexchHGen}) as
\bea
{\bf h}_{{\rm exch}i} &=& \frac{3}{2g\mu_{\rm B}S(S+1)}\left[\vec{\mu}_i(1+f_J) - \vec{\mu}_j(1-f_J)\right]\nonumber\label{Eq:RedHexchGen}\\*
&=& \frac{3}{2(S+1)}\left[\vec{\bar{\mu}}_i(1+f_J) - \vec{\bar{\mu}}_j(1-f_J)\right]\label{Eq:RedHexchGen2}
\eea

For collinear AFM ordering along a principal axis in $H=0$, one has $\vec{\bar{\mu}}_j=-\vec{\bar{\mu}}_i$, yielding Eq.~(\ref{Eq:hexch0def}), whereas in the paramagnetic (PM) state with $\vec{\bar{\mu}}_j=\vec{\bar{\mu}}_i$, Eq.~(\ref{Eq:RedHexchGen}) yields
\be
{\bf h}_{{\rm exch}i} = \frac{3f_J\vec{\bar{\mu}}_i}{S+1} \qquad {\rm (PM~state)}.
\label{Eq:RedHexchPM}
\ee
\ese
This may be compared with Eq.~(\ref{Eq:hexch0def}) where the factor $f_J$ does not appear.

\subsection{\label{Eq:MagSusc} Magnetic Susceptibilities}

As noted above, we define $\mu_\alpha$ as the thermal-average moment per spin induced by an applied field $H_\alpha$ and/or exchange field $H_{\rm exch\alpha}$ in the $\alpha$ principal-axis direction ($\alpha = z,\ x$ in this paper).  The magnetic susceptibility per spin $\chi_\alpha$ for the $\alpha$ direction is rigorously defined for nonferromagnetic materials  as 
\be
\chi_\alpha = \lim_{H_\alpha\to0}\frac{\mu_\alpha}{H_\alpha}.
\ee
For calculations with an infinitesimal $H_\alpha$ applied to a PM or to an AFM-ordered spin system such as in the perturbation-theory calculations outlined in Sec.~\ref{Sec:PerpPertThy} below, one has
\be
\chi_\alpha = \frac{\mu_\alpha}{H_\alpha}.
\ee

We define dimensionless reduced susceptibilities $\bar{\chi}_\alpha$ as
\be
\bar{\chi}_\alpha \equiv \frac{\chi_\alpha T_{{\rm N}J}}{C_1} = \bigg(\frac{3}{S+1}\bigg)\frac{\bar{\mu}_\alpha}{h_\alpha},
\label{Eq:ChiBarDef}
\ee
where $C_1$ is the single-spin Curie constant in Eq.~(\ref{C1}) and $T_{{\rm N}J}$ is given in Eq.~(\ref{Eq:TNJGen}).  The second equality is in terms of the more convenient reduced parameters $h_\alpha$ and~$\bar{\mu}_\alpha$ defined as in Eqs.~(\ref{Eq:bhDef}) and~(\ref{Eq:barmuDef}), respectively.

\subsection{Magnetic Entropy, Internal Energy, Helmholtz Free Energy and Heat Capacity}

As noted above, when an exchange field is present the eigenenergies of the reduced MFT Hamiltonian~(\ref{Eq:Hamds2bi}) are temperature dependent once the temperature-dependent ordered and/or induced moment $\vec{\mu}$ values are determined as described for various situations later.  Therefore the standard  statistical-mechanical expression $S_{\rm mag} = -\partial F_{\rm mag}/\partial T$ to derive the magnetic entropy $S_{\rm mag}(T)$ from the magnetic Helmholtz free energy $F_{\rm mag}(T)$ gives incorrect results.  However, $S_{\rm mag}$, $F_{\rm mag}$ and the magnetic internal energy $U_{\rm mag}$ are state functions and can therefore be correctly calculated directly once the temperature dependence of the ordered moments is calculated.  Then the magnetic heat capacity $C_{\rm mag}(T)$ can be derived from them.

After the exhange interactions between a representative spin~$i$ and its neighbors are taken into account by approximating them by an effective exchange field within MFT and $\vec{\mu}_i(t)$ is determined, the system can be considered to consist of noninteracting spins.  Then $S_{\rm mag}(t)$ per spin for fixed $d$ and~$S$ can be calculated from the Boltzmann expression
\bse
\label{Eqs:SmagCmag}
\bea
\frac{S_{\rm mag}(t)}{k_{\rm B}} &=& -\sum_{n=1}^{2S+1} P_n(t)\ln P_n(t),\label{Eq:SmagP}\\*
P_n(t) &=& \frac{1}{Z_S(t)} e^{-\epsilon_n(t)/t},\\*
Z_S(t) &=& \sum_{n=1}^{2S+1} e^{-\epsilon_n(t)/t},
\eea
\ese
where $\epsilon_n(t)$ are the reduced eigenenergies of the reduced Hamiltonian~(\ref{Eq:Hamds2bi}) and $P_n(t)$ is the probability that a spin is in eigenstate~$n$ at reduced temperature~$t$.  The reduced magnetic internal energy $u_{\rm mag}$ per spin is obtained from
\be
u_{\rm mag}(t) \equiv \frac{U_{\rm mag}(t)}{k_{\rm B}T_{{\rm N}J}} =  \frac{1}{Z_S(t)}\sum_{n=1}^{2S+1}\epsilon_ne^{-\epsilon_n(t)/t}.
\label{Eq:umag(t)}
\ee

Once numerical values of $S_{\rm mag}(t)$ or $u_{\rm mag}(t)$ are calculated, the reduced magnetic heat capacity per mole of spins can be obtained from either
\bse
\be
\frac{C_{\rm mag}(t)}{R} = t \frac{d[S_{\rm mag}(t)/R]}{dt},
\label{Eq:Cmag(t)1}
\ee
or 
\be
\frac{C_{\rm mag}(t)}{R} = \frac{du_{\rm mag}(t)}{dt}
\ee
\ese
where $R$ is the molar gas constant. The reduced Helmholtz free energy per spin $f_{\rm mag}(t)$ is obtained from the above single-spin results from either
\bse
\label{Eqs:fmagCalc}
\be
f_{\rm mag}(t) \equiv \frac{F_{\rm mag}(t)}{k_{\rm B}T_{{\rm N}J}} = -t \ln Z_S(t)
\label{Eq:fmagFromZ}
\ee
or
\be
f_{\rm mag}(t)  = u_{\rm mag}(t) - t [S_{\rm mag}(t)/k_{\rm B}].
\label{Eq:fmag(t)}
\ee
\ese

\subsection{\label{Sec:PerpPertThy} Generic Perturbation Theory for an Infinitesimal Perpendicular Magnetization}

The parallel axis is assumed here to be the $z$~axis and the perpendicular axis is taken to be the $x$~axis.  We consider a generic magnetic induction $B_x$ seen by a representative spin that can be comprised of either an exchange field or an applied field or both and $B_z$ which can arise from exchange interactions.  All spins respond identically to $B_x$ because they are identical and crystallographically equivalent by assumption.  The Hamiltonian associated with a representative spin is
\bse
\label{Eqs:GenPert}
\be
{\cal H} = -\vec{\mu}\cdot {\bf B} -DS_z^2 = g\mu_{\rm B}(B_xS_x+B_zS_z) - D S_z^2.
\ee
The unperturbed and perturbed parts of the Hamiltonian ${\cal H} = {\cal H}_0 + {\cal H}^\prime$ are respectively
\bea
{\cal H}_0 &=& g\mu_{\rm B}B_zS_z- D S_{z}^2,\label{Eq:H0Perp}\\*
{\cal H}^\prime &=& g\mu_{\rm B}B_xS_x = \frac{g\mu_{B}B_x}{2} (S_+ + S_-),\label{Eq:HprimePert}
\eea
\ese
where $S_+$ and $S_-$ are raising and lowering operators on the $z$~components of the basis states $|S,S_z\rangle$, which we abbreviate as $|S_z\rangle$ for an assumed value of the spin~$S$\@. The unperturbed eigenenergies obtained from Eq.~(\ref{Eq:H0Perp}) are
\be
E_0(m_S) = g\mu_{\rm B}B_zm_S - Dm_S^2,
\label{Eq:E0mSPerp}
\ee
where $m_S$ is the spin magnetic quantum number.  In order to apply the theory given in the following to a specific case, one must first derive the Hamiltonian per spin for that case and from that obtain the expressions for $B_x$ and/or~$B_z$ in Eqs.~(\ref{Eqs:GenPert}).

The perturbation theory for integer and half-integer spins to second order is different in general, because for half-integer spins the matrix elements $\langle\pm\frac{1}{2}|{\cal H}^\prime|\mp\frac{1}{2}\rangle$ are nonzero but the unperturbed eigenenergies of the $|\frac{1}{2}\rangle$ and~$|-\frac{1}{2}\rangle$ states are the same if $B_z$ in Eq.~(\ref{Eq:E0mSPerp}) is zero; hence these two states associated with half-integer spins must then be treated by degenerate perturbation theory.  On the other hand, if $B_z>0$, integer and half-integer spins can be treated using the same formulas.  In the following two sections we discuss the perturbation theory for these two cases separately.  The generic theory presented here in the context of MFT applies both to noninteracting spins and to spins interacting by arbitrary sets of Heisenberg exchange interactions. 

\subsubsection{\label{Sec:GenPertThyIntSpns} Integer Spins with $B_z\geq0$ and Half-Integer Spins with $B_z>0$}

The nonzero matrix elements of ${\cal H}^\prime$ are
\be
\langle m_S\pm1|{\cal H}^\prime |m_S\rangle = \frac{g\mu_{\rm B}B_x}{2}\sqrt{S(S+1)-m_S(m_S\pm1)},\nonumber
\ee
which are zero if $m_S=\pm S$, respectively.  Hence the first-order corrections to the eigenenergies are zero.  The eigenenergies of ${\cal H}^\prime$ at second order in $B_x$ are
\bse
\label{Eqs:E2KPerp}
\bea
E_2(m_S) &=& - \frac{g^2\mu_{\rm B}^2B_x^2}{2}K(m_S),\label{Eq:EnergySzs}\\*
K(m_S) &=& \frac{1}{2}\bigg[\frac{S(S+1)-m_S(m_S+1)}{g\mu_{\rm B}B_z -D(2m_S+1)}\label{Eq:KDef}\\*
&& \hspace{0.2in}  -\ \frac{S(S+1)-m_S(m_S-1)}{g\mu_{\rm B}B_z -D(2m_S-1)}\bigg].\nonumber
\eea
\ese
The magnetic moment operators $\mu_x^{\rm op}(m_S)$ associated with these eigenenergies are obtained using Eq.~(\ref{Eq:muxmuzDef}) as
\be
\mu_x^{\rm op}(m_S) = -\frac{\partial E_2(m_S)}{\partial B_x} = g^2\mu_{\rm B}^2B_xK(m_S).
\label{Eq:mux}
\ee
Since these $\mu_x^{\rm op}(m_S)$ operators are proportional to $B_x$, the associated moments are all induced by this field.

Weighting the magnetic moments according to the Boltzmann distribution yields the thermal-average $\mu_x$ to first order in $B_x$ as
\bse
\bea
\mu_x &=& \frac{1}{Z_S}\sum_{m_S=-S}^S \mu_x^{\rm op}(m_S) e^{-E(m_S)/k_{\rm B}T}\\*
&=& \frac{g^2\mu_{\rm B}^2B_x}{Z_S}\sum_{m_S=-S}^S K(m_S) e^{-E_0(m_S)/k_{\rm B}T},\nonumber\\*
Z_S &=&\sum_{m_S=-S}^S e^{-E_0(m_S)/k_{\rm B}T},\label{Eq:ZSx0}
\eea
where $E_0(m_S)$ is given in Eq.~(\ref{Eq:E0mSPerp}). This is more compactly written as
\bea
\mu_x &=& g^2\mu_{\rm B}^2B_xF_{x1},\label{Eq:muxPerpInt}\\*
F_{x1} &=& \frac{1}{Z_S}\sum_{m_S=-S}^S K(m_S) e^{-E_0(m_S)/k_{\rm B}T}.\nonumber
\eea
\ese

In terms of the reduced variables introduced in Sec.~\ref{MFTBckgrnd} that are more appropriate and useful when Heisenberg exchange interactions are present, Eqs.~(\ref{Eq:E0mSPerp}), (\ref{Eq:KDef}) and~(\ref{Eq:muxPerpInt}) become
\bse
\label{Eqs:muPerpIntSpins}
\bea
\epsilon_0(m_S) &=& b_zm_S - dm_S^2, \label{Eq:E0mSPerpRed}\\*
K(m_S) &=& \frac{1}{2}\bigg[\frac{S(S+1)-m_S(m_S+1)}{b_z -d(2m_S+1)}\hspace{0.5in}\label{Eq:KPerpInt}\\*
&& \hspace{0.2in}  -\ \frac{S(S+1)-m_S(m_S-1)}{b_z -d(2m_S-1)}\bigg],\nonumber\\*
\bar{\mu}_x &=& b_xF_{x1},\label{Eq:muxPerpIntRed}\\*
F_{x1} &=& \frac{1}{SZ_S}\sum_{m_S=-S}^S K(m_S)\, e^{-\epsilon_0(m_S)/t},\label{Eq:Fx1}\\*
Z_S &=& \sum_{m_S=-S}^S e^{-\epsilon_0(m_S)/t}.
\eea
If $b_z=0$, $K(m_S)$ in Eq.~(\ref{Eq:KPerpInt}) simplifies to
\bea
K(m_S) &=& \frac{S(S+1) + m_S^2}{d(4m_S^2-1)} \label{Eq:KPerpInt2}\\*
&&\hspace{-0.7in}(b_z=0,~{\rm integer~spins~only).}\nonumber
\eea
\ese
The definitions of the above variables are summarized as
\bea
\epsilon_0 &=& \frac{E_0}{{k_{\rm B}T_{{\rm N}J}}}\quad b_\alpha = \frac{g\mu_{\rm B}B_\alpha}{k_{\rm B}T_{{\rm N}J}},\quad d = \frac{D}{k_{\rm B}T_{{\rm N}J}},\hspace{0.3in} \label{ParDefs}\\*
 \bar{\mu}_x &=& \frac{\mu_x}{\mu_{\rm sat}},\quad \mu_{\rm sat} = gS\mu_{\rm B},\quad t = \frac{T}{T_{{\rm N}J}}.\nonumber
\eea

\subsubsection{Half-Integer Spins with $B_z=0$}

For half-integer spins $S = 3/2,\ 5/2,\ \ldots$ with $B_z=0$, we first diagonalize the $m_S = \pm1/2$ subspace with respect to ${\cal H}^\prime$ in Eq.~(\ref{Eq:HprimePert}), which yields the symmetric~(+) and antisymmetric~($-$) eigenfunctions
\be
|\pm\rangle = \frac{1}{\sqrt{2}}\left[\big|1/2\big\rangle \pm \big|-1/2\big\rangle\right].
\label{Eq:pmFcns}
\ee
The nonzero matrix elements involving these $|\pm\rangle$ states are
\bea
\langle \pm|S_z^2|\pm\rangle &=& \frac{1}{4},\\*
\langle \pm|{\cal H}^\prime|\pm\rangle &=& \pm\frac{g\mu_{\rm B}B_x}{2}\sqrt{S(S+1)+1/4},\nonumber\\*
\big\langle 3/2\big|{\cal H}^\prime\big|\pm\big\rangle &=& \frac{g\mu_{\rm B}B_x}{2\sqrt{2}}\sqrt{S(S+1)-3/4},\nonumber\\*
\big\langle -3/2\Big|{\cal H}^\prime\Big|\pm\Big\rangle &=&\pm \frac{g\mu_{\rm B}B_x}{2\sqrt{2}}\sqrt{S(S+1)-3/4}.\nonumber
\eea
where the first and third sets of matrix elements are twofold degenerate.  The eigenenergies of the $|\pm\rangle$ states to second order in $B_x$ are
\bea
E(\pm) &=& -\frac{D}{4} \pm\frac{g\mu_{\rm B}B_x}{2}\sqrt{S(S+1)+1/4}\label{Eq:Epm2}\\*
&&+\ \frac{g^2\mu_{\rm B}^2B_x^2}{8}\bigg[\frac{S(S+1)-3/4}{D}\bigg].\nonumber
\eea

The magnetic moment operators for these states are
\bea
\mu_x^{\rm op}(\pm) &=& -\frac{\partial E(\pm)}{\partial B_x} \nonumber\\*
&=& \mp\frac{g\mu_{\rm B}}{2}\sqrt{S(S+1)+1/4}\label{Eq:Mupm}\\*
&&  -\ \frac{g^2\mu_{\rm B}^2B_x}{4}\bigg[\frac{S(S+1)-3/4}{D}\bigg].
\nonumber
\eea
The first term corresponds to a permanent magnetic moment and the second to a magnetic moment induced by~$B_x$.  The thermal-average moments $\mu_x(\pm)$ of the $|\pm\rangle$ states to first order in $B_x$ are
\bea
 \mu_x(\pm) &=& \frac{1}{Z_S}\left[\mu_x^{\rm op}(+)e^{-E(+)/k_{\rm B}T} + \mu_x^{\rm op}(-)e^{-E(-)/k_{\rm B}T}\right]\nonumber\\*
&=& g^2\mu_{\rm B}^2B_x F_{x2},\label{Eq:muxpmAve}\\*
F_{x2} &=& \frac{e^{D/4k_{\rm B}T}}{2Z_S}\Bigg[\frac{S(S+1)+1/4}{k_{\rm B}T} -\ \frac{S(S+1)-3/4}{D}\Bigg],\nonumber
\eea
where the partition function~$Z_S$ is again given by Eq.~(\ref{Eq:ZSx0}).

The contributions of the remaining $m_S=\pm3/2,\ \pm5/2,\ \ldots,\ \pm S$ states to $\mu_x$ are the same as those for integer spins, given by Eq.~(\ref{Eq:muxPerpInt}) as 
\bea
\mu_x(m_S\geq3/2) \equiv g^2\mu_{\rm B}^2B_xF_{x3},\hspace{0.6in}\label{Eq:muxPerpHalfInt2}\\*
F_{x3} = \frac{2}{Z_S}\sum_{m_S=3/2}^S K(m_S) e^{-E_0(m_S)/k_{\rm B}T},\nonumber
\eea
where $K(m_S)$ is given in Eq.~(\ref{Eq:KDef}) and $E_0(m_S)$ in Eq.~(\ref{Eq:E0mSPerp}). Adding the two contributions~(\ref{Eq:muxpmAve}) and~(\ref{Eq:muxPerpHalfInt2}) gives the total thermal-average $x$-axis magnetic moment of representative spin~$i$ as
\bea
\mu_x &=&  g^2\mu_{\rm B}^2B_x F_{x4},\nonumber\\*
F_{x4} &=& F_{x2} + F_{x3}. \label{Eq:muxBxHalfInt}
\eea

When Heisenberg exchange interactions are present, the above results in Eqs.~(\ref{Eq:muxpmAve})--(\ref{Eq:muxBxHalfInt}) for half-integer spins are better expressed in terms of reduced variables as
\bse
\label{Eqs:muxbarFx4}
\bea
\bar{\mu}_x &=& b_x F_{x4},\label{Eq:muxpmAveRed}\\*
F_{x4} &=& F_{x2} + F_{x3},\\*
F_{x2} &=& \frac{e^{d/4t}}{2SZ_S}\bigg[\frac{S(S+1)+1/4}{t} -\ \frac{S(S+1)-3/4}{d}\bigg],\nonumber\\*
&&\\*
F_{x3} &=& \frac{2}{SZ_S}\sum_{m_S=3/2}^S K(m_S) e^{-\epsilon_0(m_S)/t},
\eea
\ese
where $\epsilon_0(m_S)$ and~$K(m_S)$ are given in Eqs.~(\ref{Eq:E0mSPerpRed}) and~(\ref{Eq:KPerpInt}), respectively, and the variable definitions are summarized in Eqs.~(\ref{ParDefs}).

%\clearpage

\section{\label{Sec:ChiPM} Magnetic Susceptibility in the Paramagnetic State with $D>0$}

In the PM state the moments induced by a field in a principal axis direction are parallel to each other and to the applied field.  The exchange field is also oriented in this direction.

\subsection{Parallel Susceptibility}

Here we consider the case $D>0$ with an infinitesimal field aligned along the uniaxial parallel $z$-axis direction.  According to Eqs.~(\ref{Eq:bDef}) and~(\ref{Eq:RedHexchPM}), the reduced magnetic induction seen by a each spin is given by
\be
b_z = \frac{3f_J\bar{\mu}_z}{S+1} + h_z.
\label{Eq:bzPM}
\ee
The reduced Hamiltonian~(\ref{Eq:Hamds2bi}) for each spin is diagonal with reduced energy eigenvalues
\be
\epsilon(m_S) = \left(\frac{3f_J\bar{\mu}_z}{S+1} + h_z\right)m_S - dm_S^2.
\label{Eq:EPMi}
\ee
The operator $\bar{\mu}_{z}^{\rm op}$ is given by Eqs.~(\ref{Eq:barmuOp}), (\ref{Eq:bzPM}) and~(\ref{Eq:EPMi}) as
\be
\bar{\mu}_{z}^{\rm op} = -\frac{1}{S}\frac{\partial \epsilon(m_S)}{\partial b_z} = -\frac{m_S}{S}.
\label{Eq:muzop}
\ee
The reduced thermal-average $\bar{\mu}_{z}$ is then obtained from Eq.~(\ref{Eq:barmualpha}) as
\bse
\label{Eqs:barmuiz}
\bea
\bar{\mu}_{z} &=& - \frac{1}{SZ_S}\sum_{m_S=-S}^S m_S e^{-\epsilon(m_S)/t},\\*
Z_S &=& \sum_{m_S=-S}^S e^{-\epsilon(m_S)/t}.
\eea
\ese

Equations~(\ref{Eq:EPMi})--(\ref{Eqs:barmuiz}) are valid for arbitrary values of $h_z>0$, $d$ and~$f_J<1$, but here we only consider infinitesimal $h_z$ and $\bar{\mu}_z$.  Using Eqs.~(\ref{Eq:bzPM}) and~(\ref{Eq:EPMi}), expanding Eqs.~(\ref{Eqs:barmuiz}) to first order in $h_z$ and $\bar{\mu}_z$ and then  solving for $\bar{\mu}_z$ gives
\bse
\label{Eqs:muzPar}
\bea
\bar{\mu}_z &=& \frac{\left(\frac{S+1}{3}\right)h_z}{\frac{(S+1)t}{3F_z} - f_J},\label{Eq:barmuzSoln}\\*
F_z(d,t) &=& \frac{1}{SZ_S} \sum_{m_S=-S}^S m_S^2 e^{dm_S^2/t},\label{FzDef}\\*
Z_S &=& \sum_{m_S=-S}^S e^{dm_S^2/t}.
\eea
\ese
The reduced parallel susceptibility is obtained from Eqs.~(\ref{Eq:ChiBarDef}) and~(\ref{Eq:barmuzSoln}) as
\be
\bar{\chi}_\parallel \equiv \frac{\chi_z T_{{\rm N}J}}{C_1} = \frac{1}{\frac{(S+1)t}{3F_z} - f_J} \qquad{\rm (PM~state)}.
\label{Eq:ChiParPM}
\ee
In the limit of high~$t$, one obtains a Curie law with $\bar{\chi}_\parallel = 1/t$, irrespective of $d$,~$S$ and~$f_J$.

Converting Eq.~(\ref{Eq:barmuzSoln}) to unreduced variables gives 
\be
\chi_\parallel \equiv \frac{\mu_z}{H_z} = \frac{C_1}{\frac{T}{F(d,t)}-\theta_{{\rm p}J}},
\label{Eq:chi}
\ee
where $C_1$ is the single-spin Curie constant in Eq.~(\ref{C1}).  If $d=0$ one obtains 
\be
\chi_\parallel = \frac{C_1}{T - \theta_{{\rm p}J}},
\ee
which is the Curie-Weiss law for Heisenberg exchange interactions with no uniaxial anisotropy as required.  At high temperatures, Eq.~(\ref{Eq:chi}) yields the Curie-Weiss law
\bse
\label{Eqs:qpqpqqD}
\bea
\chi_\parallel &=& \frac{C_1}{T - \theta_{{\rm p}\parallel}} \qquad{\rm (PM~state)},\\*
\theta_{{\rm p}\parallel} &=& \theta_{{\rm p}J} + \theta_{{\rm p}D\parallel},\\*
\theta_{{\rm p}D\parallel} &=& \left(\frac{D}{k_{\rm B}}\right)\frac{(2S-1)(2S+3)}{15}.\label{Eq:thetapD}
\eea
\ese
The expression for $\theta_{{\rm p}D\parallel}$ arising from the single-ion anisotropy is identical to that found in the Appendix in the absence of exchange interactions.  Thus the Weiss temperatures from the exchange and single-ion anisotropies are additive.  This is also found to be the case for magnetic dipole interactions combined with exchange interactions \cite{Johnston2016}.  Equation~(\ref{Eq:thetapD}) yields $\theta_{{\rm p}D\parallel} = 0$ if $S=1/2$ as required.

Because the $\chi$ anisotropy tensor in the PM state arising from single-ion anisotropy is traceless, one can immediately give the expression for the Weiss temperature associated with $\chi_\perp$ that is measured along an axis perpendicular to the parallel easy~($z$) axis of a uniaxial collinear AFM\@.  From Eq.~(\ref{Eq:thetapD}) one obtains
\be
\theta_{{\rm p}D\perp} = -\frac{\theta_{{\rm p}D\parallel}}{2} = -\left(\frac{D}{k_{\rm B}}\right)\frac{(2S-1)(2S+3)}{30}.
\label{Eq:thetaPerp}
\ee
This is confirmed by explicit calculations of the PM $\chi_\perp(T)$ in the following section.

\subsection{Perpendicular Susceptibility}

According to Eqs.~(\ref{Eq:bDef}) and~(\ref{Eq:RedHexchPM}), the reduced magnetic induction seen by each spin is in the $x$~direction and contains both exchange field and applied field parts, given by
\be
b_x = \frac{3f_J\bar{\mu}_x}{S+1} + h_x,
\label{Eq:bxPM}
\ee
where $\bar{\mu}_x$ is the reduced thermal-average moment in the $x$~direction.  

\subsubsection{Integer Spins}

To solve for $\chi_\perp$ we use Eqs.~(\ref{Eqs:muPerpIntSpins}) and set $b_z=0$.  The expressions in Eqs.~(\ref{Eqs:muPerpIntSpins}) appropriate to the present case are
\bse
\label{Eqs:muPerpIntSpins2}
\bea
\epsilon_0(m_S) &=& - dm_S^2, \label{Eq:E0mSPerpRed2}\\*
K(m_S) &=& \frac{S(S+1)+m_S^2}{d(4m_S^2-1)},\label{Eq:KmSDef}\\*
\bar{\mu}_x &=& b_xF_{x1},\label{Eq:muxPerpIntRed2}\\*
F_{x1} &=& \frac{1}{SZ_S}\sum_{m_S=-S}^S K(m_S)\, e^{dm_S^2/t},\label{Eq:Fx12}\\*
Z_S &=& \sum_{m_S=-S}^S e^{dm_S^2/t}.\\*
&& {\rm (integer~spins)}\nonumber.
\eea
\ese

The reduced $x$-axis moment per spin $\bar{\mu}_x$ is obtained from Eqs.~(\ref{Eq:bxPM}) and~(\ref{Eq:muxPerpIntRed2}) as
\be
\bar{\mu}_x = b_xF_{x1} = \left(\frac{3f_J\bar{\mu}_x}{S+1} + h_x\right)F_{x1}.
\ee
Solving for $\bar{\mu}_x$ gives
\be
\bar{\mu}_x = \frac{(S+1)h_x/3}{\frac{S+1}{3F_{x1}} - f_J}.
\label{barmuxPMPerp}
\ee
Using Eqs.~(\ref{Eq:ChiBarDef}) and~(\ref{barmuxPMPerp}), the normalized perpendicular susceptibility is obtained as
\be
\bar{\chi}_\perp \equiv \frac{\chi_\perp T_{{\rm N}J}}{C_1} = \frac{1}{\frac{S+1}{3F_{x1}} - f_J}.
\label{Eq:ChiPerpPM}
\ee
In the limit of low temperatures, we obtain
\be
\bar{\chi}_\perp(t\to0) = \left[\frac{d(S+1)(2S-1)}{3} - f_J\right]^{-1}.
\label{Eq:barchiPerp0}
\ee
whereas in the limit of high temperatures a Curie Law is obtained, $\bar{\chi}_\perp=1/t$.  Carrying out a Taylor series expansion of Eq.~(\ref{Eq:ChiPerpPM}) to second order in $1/t$ yields a Curie-Weiss law~(\ref{Eqs:CWLaw}) with Weiss temperature
\be
\theta_{{\rm p}\parallel} = \theta_{{\rm p}J} + \theta_{{\rm p}D\perp},
\label{Eq:thetaPMPerp}
\ee
with $\theta_{{\rm p}D\perp}$ the same as previously inferred in Eq.~(\ref{Eq:thetaPerp}).  

\subsubsection{Half-Integer Spins}

Here we use Eqs.~(\ref{Eqs:muxbarFx4}) since $b_z=0$. Utilizing Eq.~(\ref{Eq:bxPM}) for~$b_x$, Eqs.~(\ref{Eqs:muxbarFx4}) yield
\be
\bar{\mu}_x = \frac{(S+1)h_x/3}{\frac{S+1}{3F_{x4}} - f_J}.
\label{barmuxPMPerpHI}
\ee
Then Eqs.~(\ref{Eq:ChiBarDef}) and~(\ref{barmuxPMPerpHI}) give
\be
\bar{\chi}_\perp = \frac{1}{\frac{S+1}{3F_{x4}} - f_J} \quad{\rm (half~integer~spins)}.
\label{Eq:ChiPerpPMHI}
\ee
At high temperatures $\chi_\perp$ follows the same Curie-Weiss law as integer spins do.  For $t\to0$ one also obtains 
the same expression~(\ref{Eq:barchiPerp0}) as for integer spins.

\begin{figure}
\includegraphics [width=3.3in]{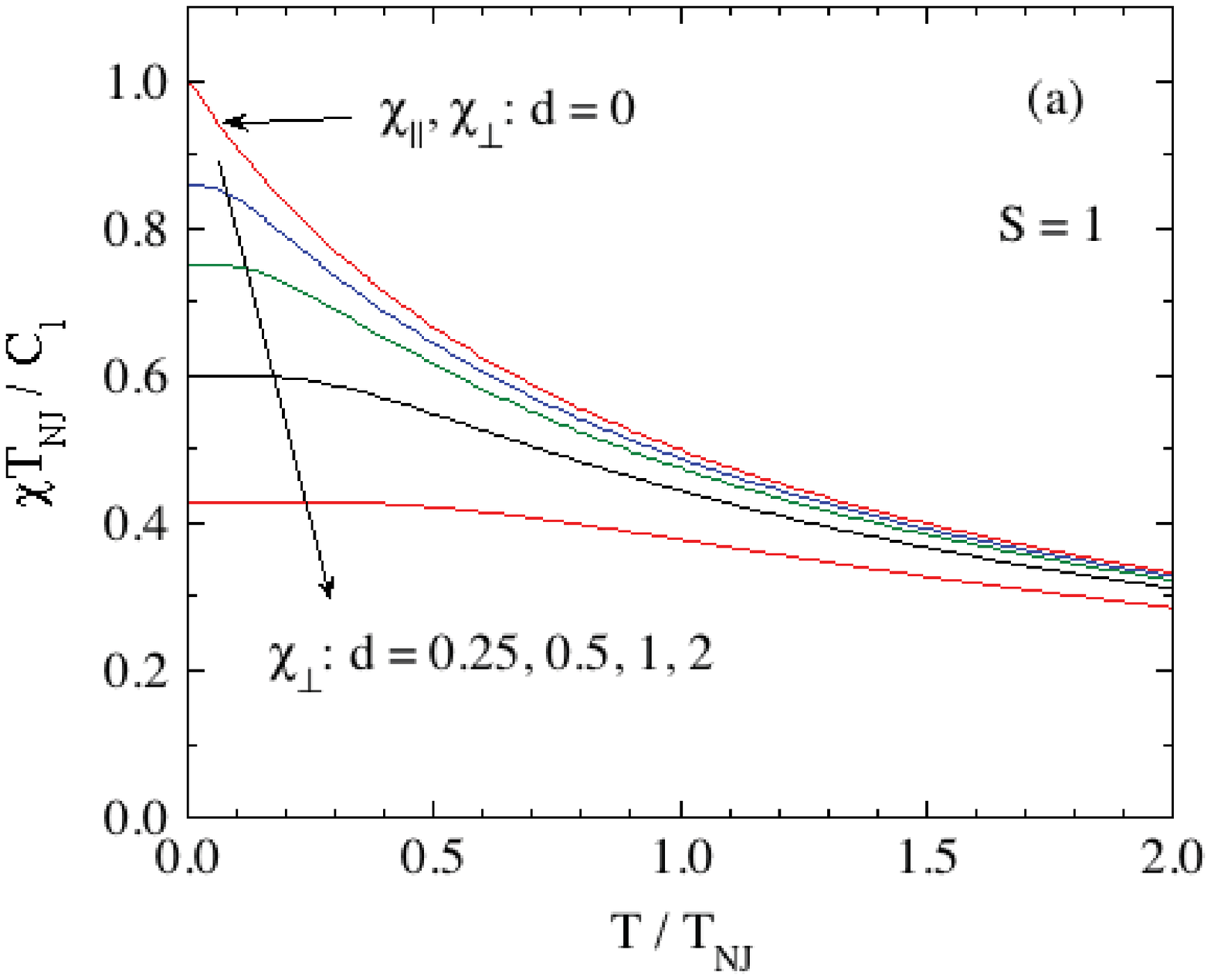}
\includegraphics [width=3.3in]{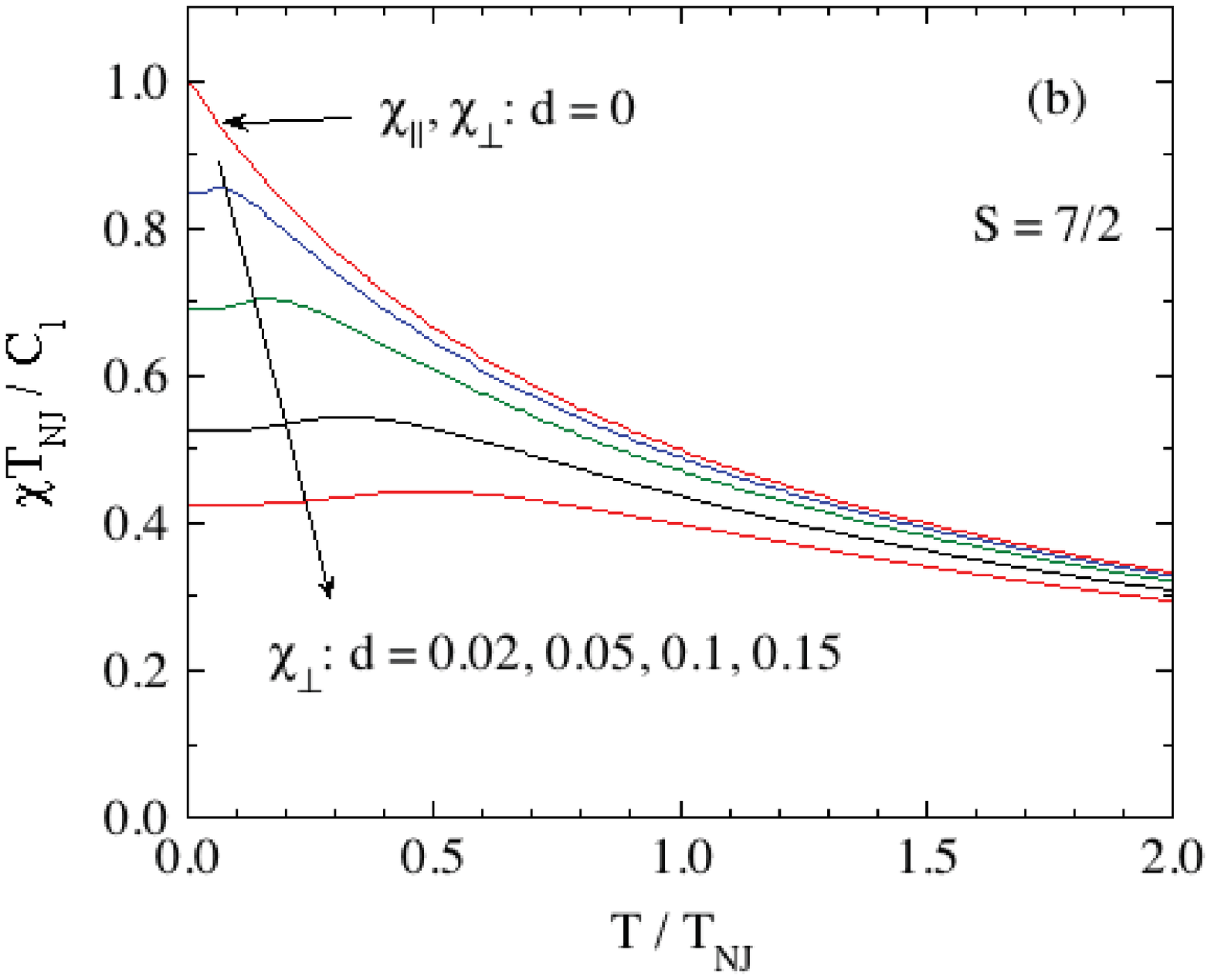}
\caption{(Color online) Reduced parallel and perpendicular paramagnetic susceptibilities $\bar{\chi}_\parallel$ with $d=0$ and $\bar{\chi}_\perp$ for the listed values of reduced anisotropy constants $d=D/k_{\rm B}T_{{\rm N}J}$ for spins (a)~$S=1$ obtained from Eq.~(\ref{Eq:ChiPerpPM}) and (b)~$S=7/2$ obtained from Eq.~(\ref{Eq:ChiPerpPMHI}).}
\label{Fig:ChiParPerpPMfJm1}
\end{figure}

Shown in Fig.~\ref{Fig:ChiParPerpPMfJm1} are the reduced parallel susceptibility $\bar{\chi}_\parallel$ for $d=0$ and the reduced perpendicular susceptibility $\bar{\chi}_\perp$ versus reduced temperature~$t$ for the listed values of $d$ for spins~$S=1$ and~7/2 obtained using Eqs.~(\ref{Eq:ChiPerpPM}) and~(\ref{Eq:ChiPerpPMHI}).  The value $\bar{\chi}_\parallel(t=0)=1$ is the same for all~$d$ and~$S$.  We thus find that $\bar{\chi}_\parallel(t)$ is not very sensitive to the value of~$d$ (not shown), whereas $\bar{\chi}_\perp(t)$ is quite sensitive to it as seen in Fig.~\ref{Fig:ChiParPerpPMfJm1}.  One also sees that the $\bar{\chi}_\perp$ curves for $S = 7/2$ in Fig.~\ref{Fig:ChiParPerpPMfJm1}(b) are far more sensitive to $d$ than are those for the much smaller spin~$S=1$ in Fig.~\ref{Fig:ChiParPerpPMfJm1}(a).  The regions in Fig.~\ref{Fig:ChiParPerpPMfJm1} at $t\lesssim1$ are not observed in practice because they are preempted by AFM ordering that occurs at $t\gtrsim1$ for $d\geq0$ as discussed in Sec.~\ref{Sec:TN}.

\section{\label{Sec:CollAFMDgtr0} Collinear $z$-Axis AFM Ordering with $D>0$ and $H=0$}

When the anisotropy constant $D>0$, $z$-axis AFM collinear ordering is favored over collinear or coplanar AFM ordering in the $xy$~plane.  When the ordered moment $\vec{\mu}_i$ and {\bf H} and/or ${\bf H}_{\rm exch}$ are all aligned along the $z$~axis, the Hamiltonian is diagonal in the basis vectors $|S,S_z\rangle$.  When $h=0$ as assumed in this section the reduced Hamiltonian~(\ref{Eq:Hamds2bi}) for representative spin~$i$ is
\be
\frac{\cal H}{k_{\rm B}T_{{\rm N}J}} = b_{zi}S_z -dS_z^2.
\label{Eq:HamdsAFMz}
\ee
According to Eq.~(\ref{Eq:hexch0def}) one has 
\be
b_{iz} = h_{\rm exch0} = \frac{3\bar{\mu}_0}{S+1},
\label{Eq:bizAFM}
\ee
where we assume that the representative moment~$i$ is directed in the $+z$ direction and hence $\bar{\mu}_0 = \bar{\mu}_{iz}$. The reduced eigenenergies obtained from Eq.~(\ref{Eq:HamdsAFMz}) are thus
\be
\epsilon(m_S) = \frac{3\bar{\mu}_0}{S+1}m_S -dm_S^2.
\label{Eq:Eamds2bi}
\ee

\subsection{\label{Sec:AFMmu0} Ordered Moment}

The reduced magnetic moment operator $\bar{\mu}_{z}^{\rm op}$ is obtained using Eqs.~(\ref{Eq:barmuOp}), (\ref{Eq:bizAFM}) and~(\ref{Eq:Eamds2bi}), which give the same expression as for the PM state in Eq.~(\ref{Eq:muzop}).  Using Eqs.~(\ref{Eq:barmualpha}) and~(\ref{Eq:muzop}), the reduced thermal-average $z$-component $\bar{\mu}_{iz}\equiv \bar{\mu}_0$ of moment $\vec{\mu}_i$ is then obtained from
\bse
\label{Eqs:mu0barzColAFM}
\be
\bar{\mu}_0 = -\frac{1}{SZ_S}\sum_{m_S = -S}^S m_S e^{dm_S^2/t}e^{-m_Sy}, \label{Eq:barmuizExch}
\ee
where the partition function is
\be
Z_S = \sum_{m_S = -S}^S e^{dm_S^2/t}e^{-m_Sy},
\ee
the variable $y$ is
\be
y \equiv y_0 = \frac{3\bar{\mu}_0}{(S+1)t},
\label{Eq:yDefAFM}
\ee
\ese
and the reduced temperature~$t$ is defined in Eq.~(\ref{Eq:tDef}).  We define the function
\be
G_S(y) = -\frac{1}{SZ_S}\sum_{m_S = -S}^S m_S e^{dm_S^2/t}e^{-m_Sy}
\label{Eqs:GSDef}
\ee
so Eq.~(\ref{Eq:barmuizExch}) becomes
\be
\bar{\mu}_0 = G_S(y_0),
\label{Eq:barmuFromGS}
\ee
which is analogous to $\bar{\mu}_0 = B_S(y_0)$ for noninteracting spins with $d=0$ where $B_S(y)$ is the Brillouin function and $y=g\mu_{\rm B}H/k_{\rm B}T$.

From Eq.~(\ref{Eqs:GSDef}) one obtains
\bea
{G_S}^\prime(y) &\equiv& \frac{dG_S(y)}{dy} \label{Eq:GSPrime}\\*
&=& \frac{1}{SZ_S} \left[\sum_{m_S=-S}^S m_S^2e^{dm_S^2/t}e^{-m_Sy}\right] - SG_S^2(y),\nonumber
\eea
which we will need later. For $y\ll1$, a Taylor series expansion of $G_S(y)$ in Eq.~(\ref{Eqs:GSDef}) to first order in~$y$ gives
\be
G_S(y) = yF_z(d,t)\qquad(y\ll1),
\label{Eq:GS2}
\ee
where $F_z(d,t)$ is defined in Eqs.~(\ref{Eqs:muzPar}).

\begin{figure}
\includegraphics [width=3.3in]{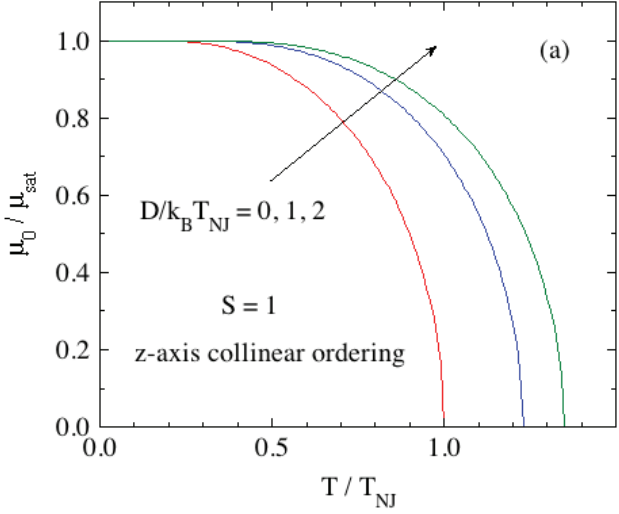}
\includegraphics [width=3.3in]{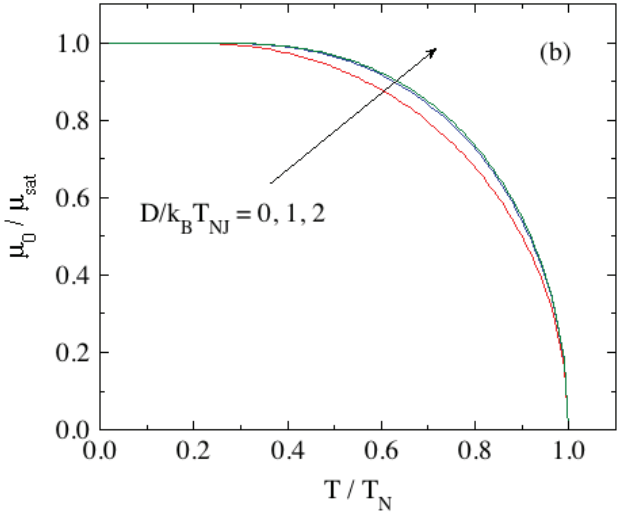}
\caption{(Color online) Reduced ordered moment $\bar{\mu}_0 = \mu_0/\mu_{\rm sat}$ versus reduced temperatures (a) $t = T/T_{{\rm N}J}$ and (b) $T/T_{\rm N}$ for $z$-axis collinear ordering in $h_z=0$ with spins $S=1$ and reduced anisotropy constants $d=D/k_{\rm B}T_{{\rm N}J} = 0$, 1 and~2 obtained by solving Eq.~(\ref{Eq:barmuFromGS}).}
\label{Fig:mu0S1}
\end{figure}

\begin{figure}
\includegraphics [width=3.3in]{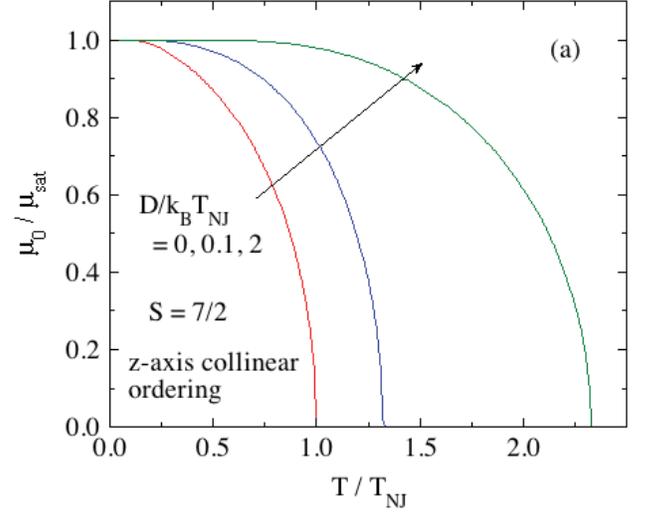}
\includegraphics [width=3.3in]{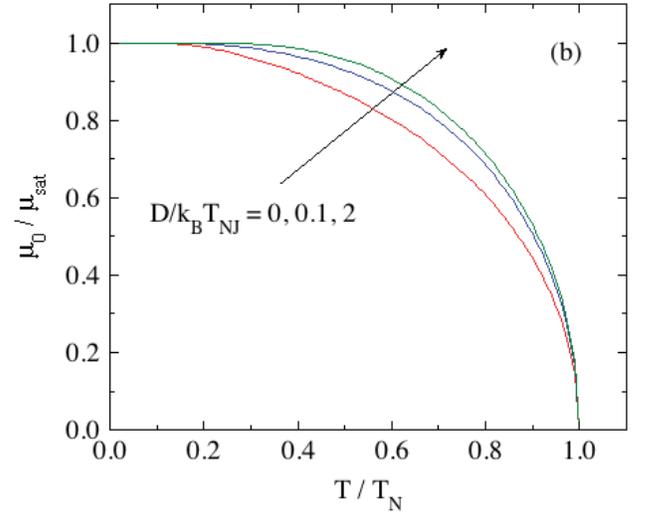}
\caption{(Color online) Same as Fig.~\ref{Fig:mu0S1} except that here $S = 7/2$ and $d=0$, 0.1 and~2.}
\label{Fig:mu0S72}
\end{figure}

Shown in Figs.~\ref{Fig:mu0S1}(a) and~\ref{Fig:mu0S1}(b) are plots of $\bar{\mu}_0$ versus $t = T/T_{{\rm N}J}$ and versus $T/T_{\rm N}$, respectively, for $S = 1$ and $d= 0$, 1 and~2, that were obtained by solving Eq.~(\ref{Eq:barmuFromGS}) using the {\tt FindRoot} utility of {\tt Mathematica}.  A similar variation in the curves with increasing~$D$ for $S=1$ as in Fig.~\ref{Fig:mu0S1}(b) computed using MFT was previously reported \cite{Cooper1960}.  Corresponding plots for $S=7/2$ with $d= 0$, 0.1 and~2 are shown in Figs.~\ref{Fig:mu0S72}(a) and~\ref{Fig:mu0S72}(b).  The N\'eel temperature $T_{{\rm N}J}$ arising from exchange interactions alone is given by Eq.~(\ref{Eq:TNJzAxis}) and the $T_{\rm N}$ including the influence of uniaxial anisotropy is calculated in the next section.  From Figs.~\ref{Fig:mu0S1}(a) and~\ref{Fig:mu0S72}(a) one sees that the $T_{\rm N}$ values (at which $\bar{\mu}_0\to0$) are strongly affected by $d>0$.  From Figs.~\ref{Fig:mu0S1}(b) and~\ref{Fig:mu0S72}(b), the shapes of the curves are also seen to be significantly affected upon varying~$d$.  The low-$t$ limits of $\bar{\mu}_0$ in Figs.~\ref{Fig:mu0S1} and~\ref{Fig:mu0S72} are unity.  Green function calculations for $S=1$ yield $\bar{\mu}_0(d\to0) = 0.92$ and indicate that this quantity increases with increasing~$d$ \cite{Devlin1971}.

\subsection{\label{Sec:TN} N\'eel Temperature}

\begin{figure}
\includegraphics [width=3.3in]{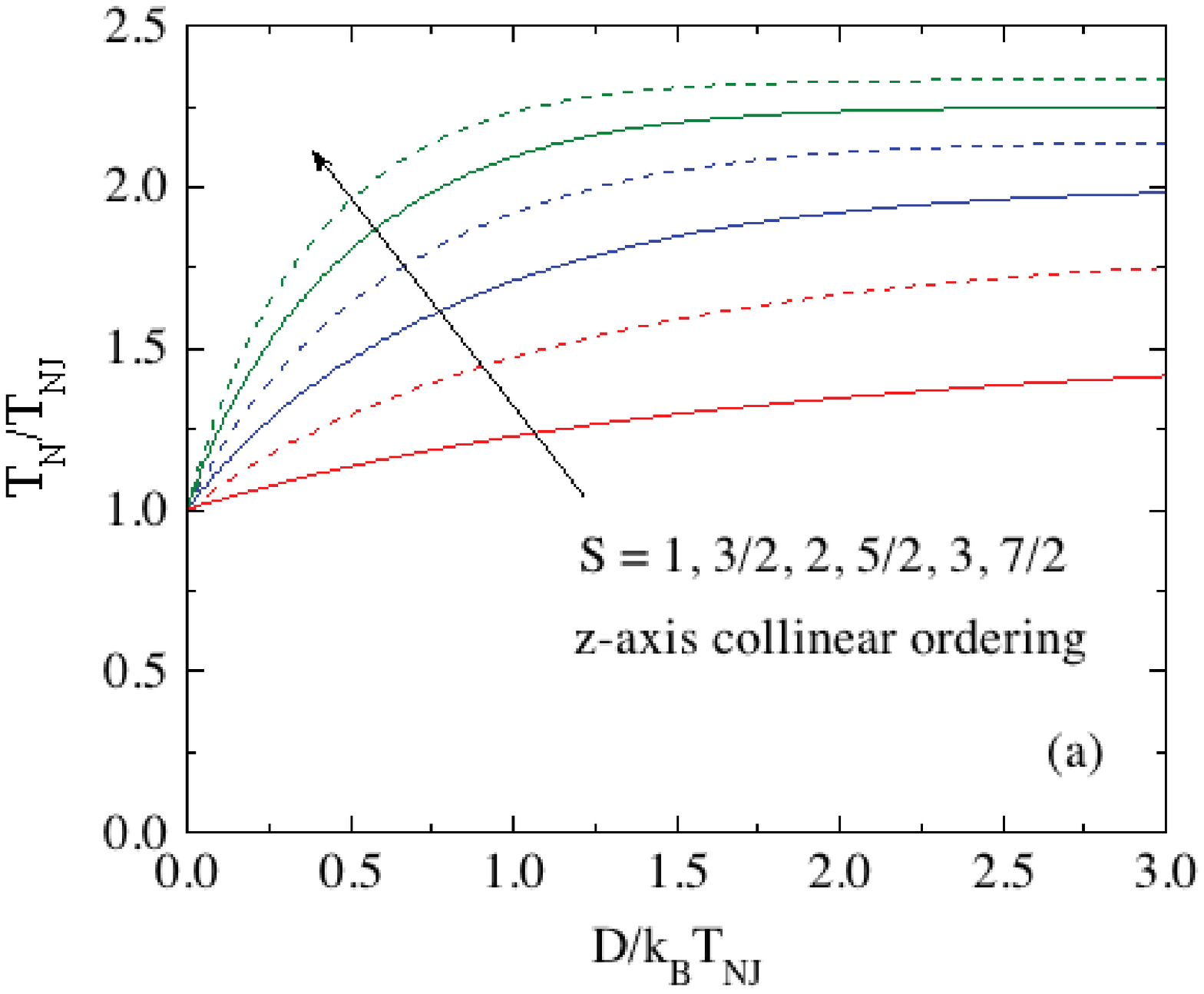}\vspace{-0.05in}
\includegraphics [width=3.3in]{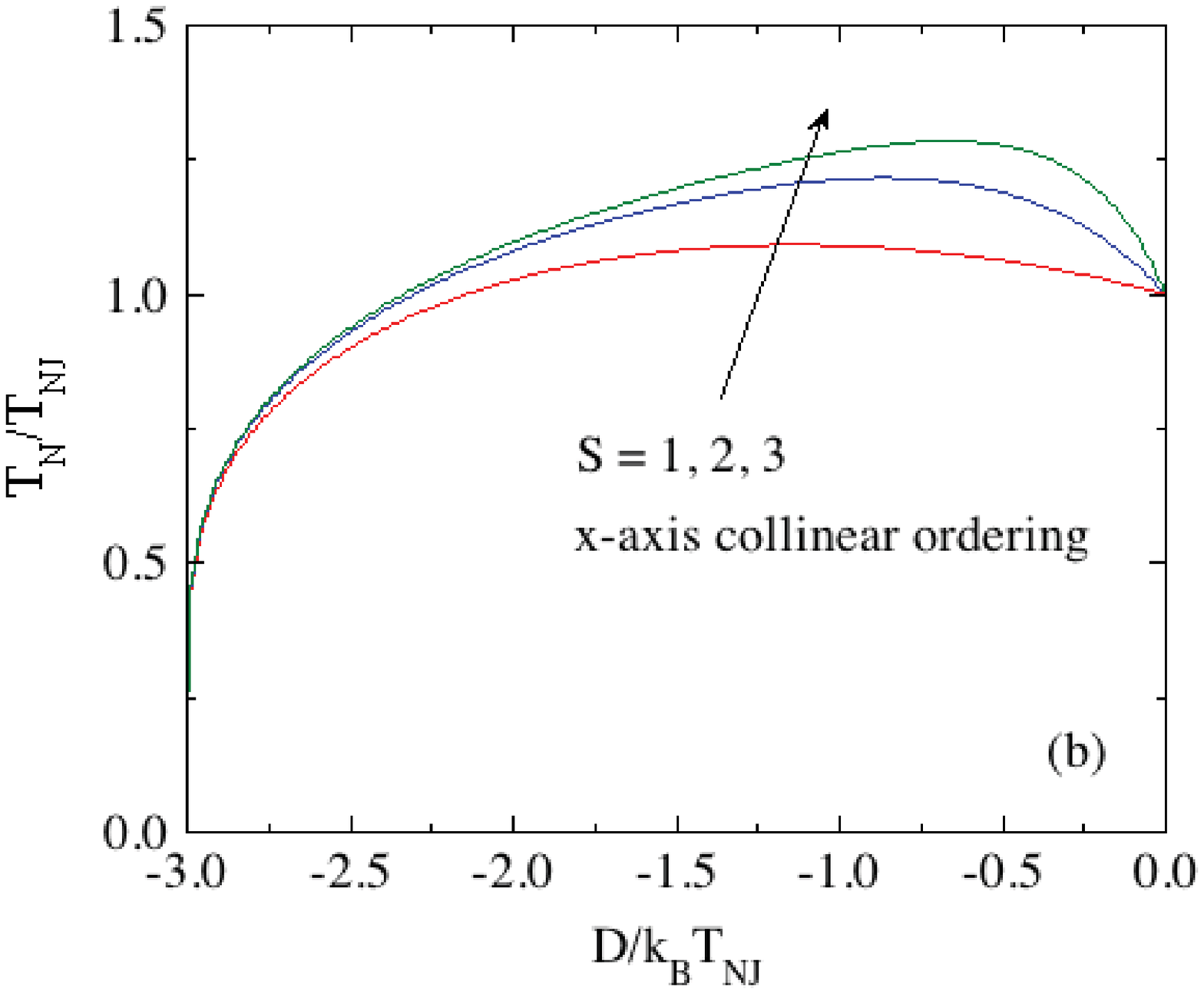}\vspace{-0.05in}
\includegraphics [width=3.3in]{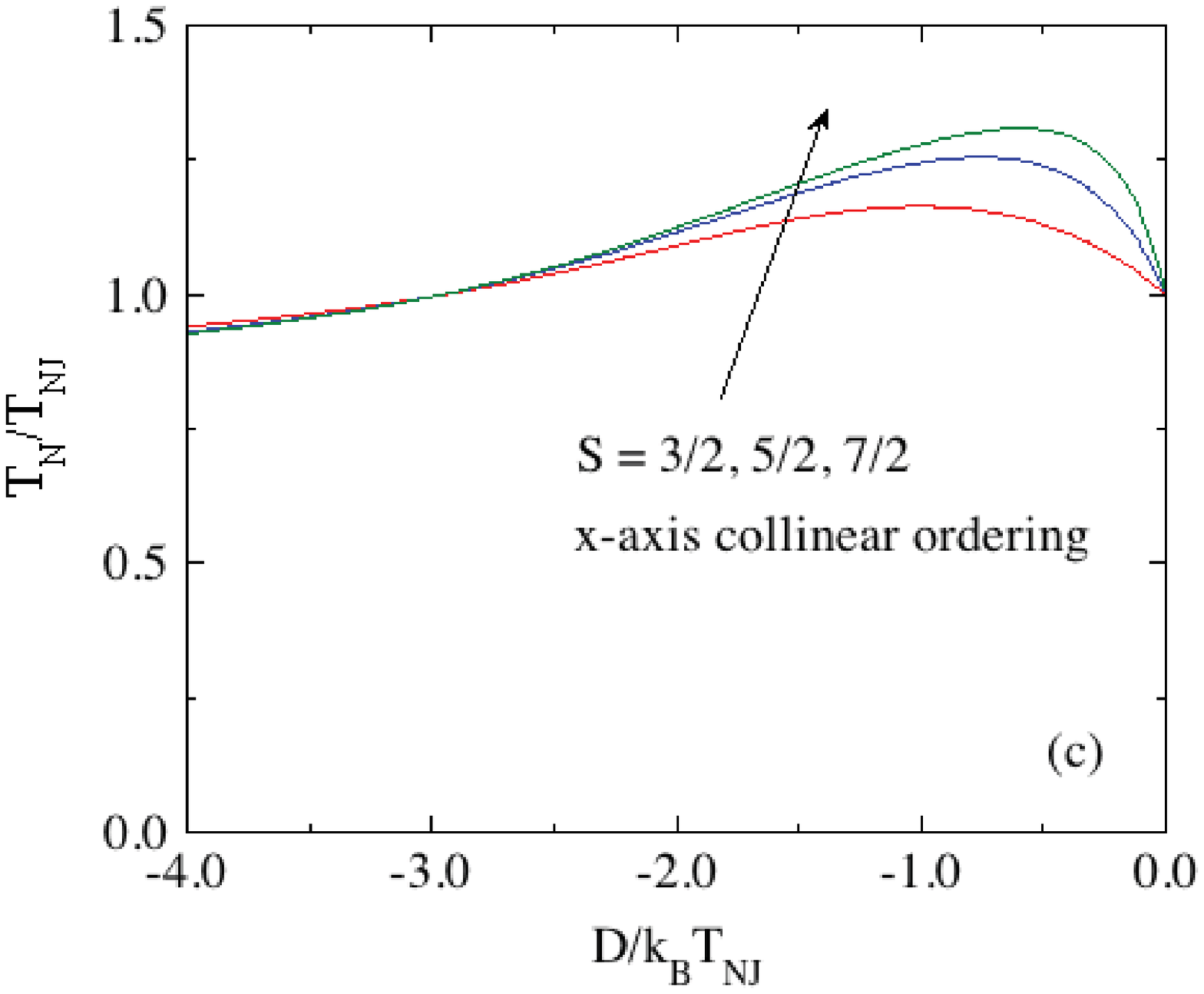}\vspace{-0.1in}
\caption{(Color online) Reduced AFM ordering temperature $t_{\rm N} = T_{\rm N}/T_{{\rm N}J}$ versus reduced anisotropy parameter $d = D/k_{\rm B}T_{{\rm N}J}$ for collinear ordering (a) along the $z$-axis calculated using Eq.~(\ref{Eq:tNz}) with $d\geq0$ and transverse $x$~axis ordering for (b) integer spins and (c) half-integer spins calculated using Eqs.~(\ref{Eqs:TNx}) below with $d\leq0$ for the spin~$S$ values listed. In~(b), AFM ordering does not occur for $d\leq3$.  $z$-axis ordering is favored for $d>0$ and $x$-axis ordering for $d<0$.}
\label{Fig:tNVSd}
\end{figure}

As $t$ approaches unity from below ($T\to T_{\rm N}^-$) one has $y_0\ll1$ in Eq.~(\ref{Eq:yDefAFM}) because $\bar{\mu}_0$ becomes infinitesimally small.  Then setting 
\be
t = t_{\rm N}\equiv \frac{T_{\rm N}}{T_{{\rm N}J}},
\label{Eq:tNDef}
\ee
Eqs.~(\ref{Eq:yDefAFM}), (\ref{Eq:barmuFromGS}) and~(\ref{Eq:GS2}) give
\be
\bar{\mu}_0 = \frac{\bar{\mu}_0}{t_{\rm N}}F_z(d,t_{\rm N}).
\ee
One solution is that the ordered moment~$\bar{\mu}_0$ is zero, which corresponds to $T\geq T_{\rm N}$.  Just below $T_{\rm N}$, $\mu_0>0$ and one can divide it out.  Then one has an expression from which $t_{\rm N}(d)$ can be calculated, i.e., 
\be
t_{\rm N} = F_z(d,t_{\rm N}),
\label{Eq:tNz}
\ee
where $F_z(d,t)$ is defined in Eqs.~(\ref{Eqs:muzPar}).  This is consistent with and is a generalization of Eq.~(A.4) in Ref.~\cite{Kanamori1962} to include arbitrary exchange interactions between arbitrary neighbors of a given spin, to the extent that these interactions give a classical $z$-axis collinear AFM structure as the ground-state magnetic structure.  One can express $t = T/T_{{\rm N}J}$ in terms of $T/T_{\rm N}$ according to
\be
\frac{T}{T_{\rm N}} = \frac{T}{T_{{\rm N}J}}\,\frac{T_{{\rm N}J}}{T_{\rm N}} = \frac{t}{t_{\rm N}},
\ee
and using Eq.~(\ref{Eq:tNz}) thereby plot quantities versus $T/T_{\rm N}$ instead of $t = T/T_{{\rm N}J}$ if desired as done above in Figs.~\ref{Fig:mu0S1}(b) and~\ref{Fig:mu0S72}(b).

In general, Eq.~(\ref{Eq:tNz}) must be solved numerically.  However, for $d\ll1$, one obtains
\bse
\be
t_{\rm N} = 1 + \frac{d}{15}(2S-1)(2S+3)\quad (d>0,\ d\ll1,\ {\rm all}\ S).
\label{tNdGTR0dll1}
\ee
Using the above definitions $t_{\rm N} = T_{\rm N}/T_{{\rm N}J}$ and $d = D/T_{{\rm N}J}$, Eq.~(\ref{tNdGTR0dll1}) gives
\be
T_{\rm N} = T_{{\rm N}J} + \frac{D}{15k_{\rm B}}(2S-1)(2S+3)\quad (d>0,\ d\ll1,\ {\rm all}\ S).
\label{Eq:TNTNJ}
\ee
\ese
A comparison of Eqs.~(\ref{Eqs:qpqpqqD}) and~(\ref{Eq:TNTNJ}) shows that for $d\ll1$, the N\'eel temperature and Weiss temperature increase by the same amount for a given $d$ and~$S$\@.  For $S=1/2$, there is no influence of the anisotropy on the N\'eel temperature (i.e., $T_{\rm N} = T_{{\rm N}J}$, independent of~$d$), as required.  For $d=0$ one obtains $T_{\rm N} = T_{{\rm N}J}$ as also required.

The variations of $t_{\rm N}$ versus (positive) $d$ for $S=1$ to $S=7/2$ obtained using Eq.~(\ref{Eq:tNz}) are shown in Fig.~\ref{Fig:tNVSd}(a).  One sees that the uniaxial anisotropy enhances $t_{\rm N}$ above the value $t_{\rm N}=1$ in the absence of the anisotropy.  However, increasing $d$ indefinitely does not increase $t_{\rm N}$ indefinitely.  In the limit of large $d$ only the $m_S = \pm S$ terms in the sums in Eqs.~(\ref{Eqs:muzPar}) survive, yielding from Eq.~(\ref{Eq:tNz}) the maximum $t_{\rm N}$ for a given~$S$ given by
\be
t_{\rm N}^{\rm max}(S) = \frac{3S}{S+1}.
\label{Eq:tNzMAX}
\ee

Figures~\ref{Fig:tNVSd}(b) and~\ref{Fig:tNVSd}(c) show the variations in the ordering temperatures for integer and half-integer spins, respectively, versus~$d$ for $x$-axis ordering with $d<0$ as derived and discussed later in Sec.~\ref{Sec:In-PlaneOrdDless0}.  For large $|d|$, one sees a qualitative difference between $t_{\rm N}(d)$ for integer and half-integer spins which arises from the nonmagnetic and magnetic nature of the ground states of these spin systems for negative~$d$, respectively.

\subsection{Magnetic Entropy, Internal Energy, Helmholtz Free Energy and Heat Capacity in $H=0$}

\begin{figure}
\includegraphics [width=3.3in]{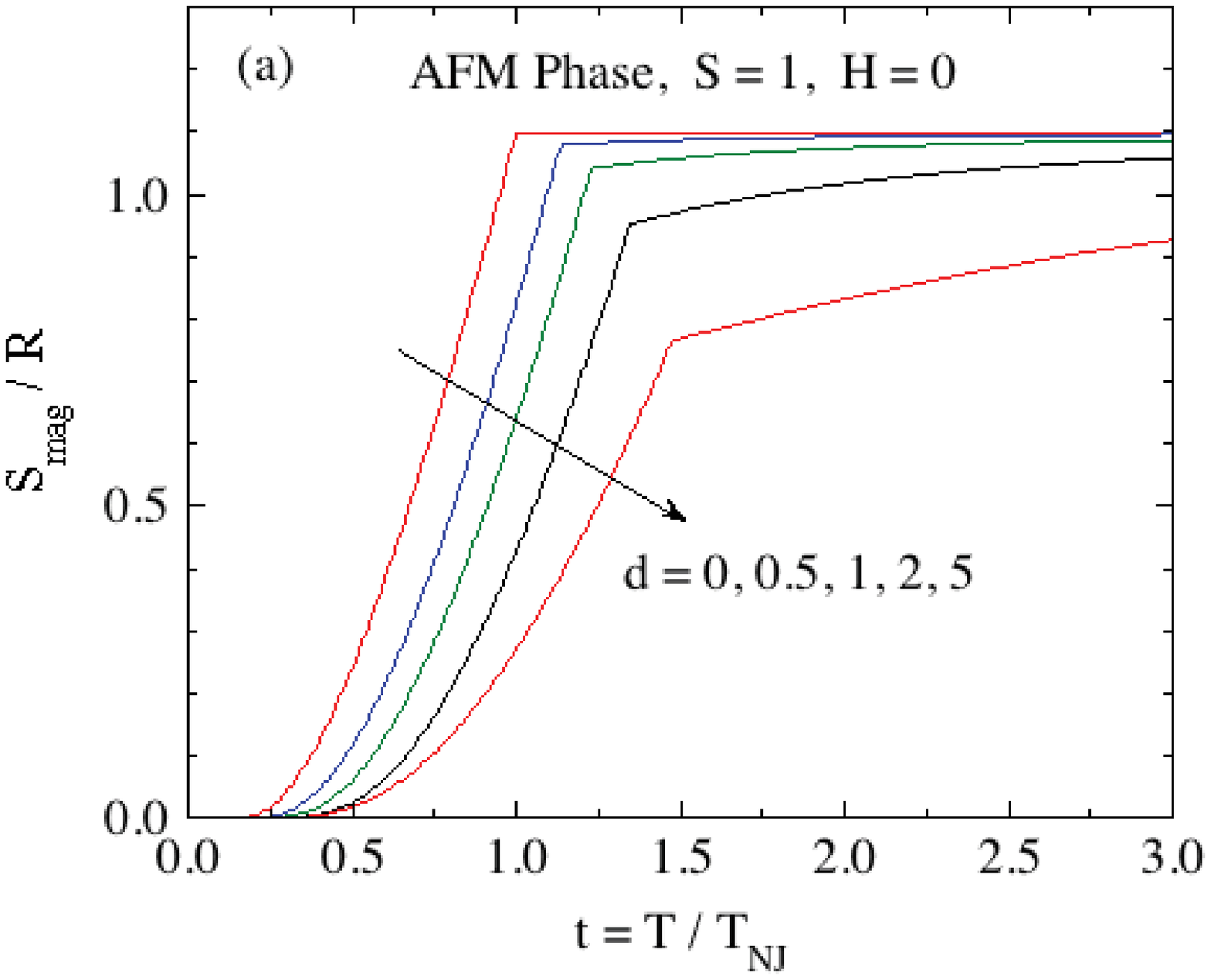}
\includegraphics [width=3.3in]{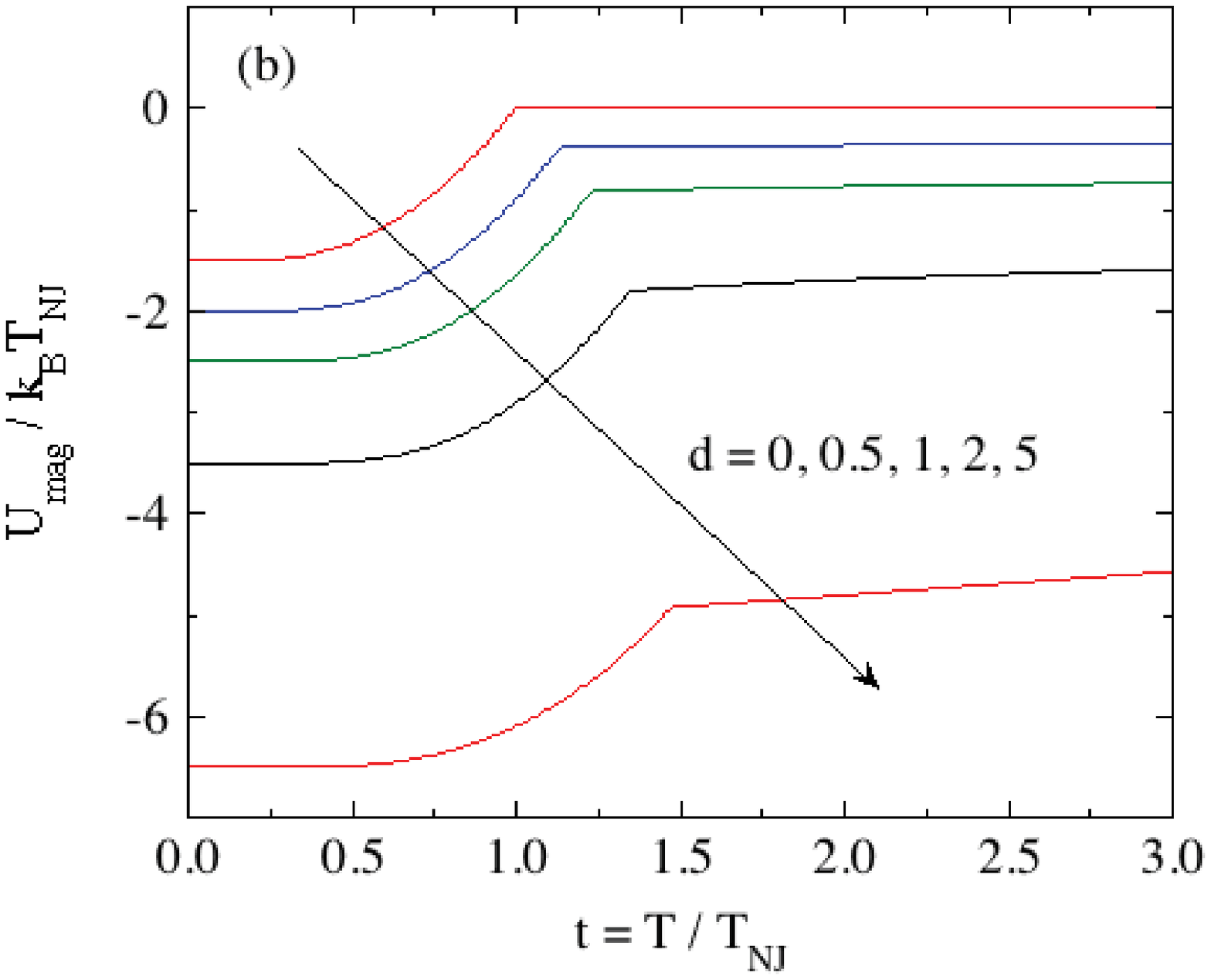}
\includegraphics [width=3.3in]{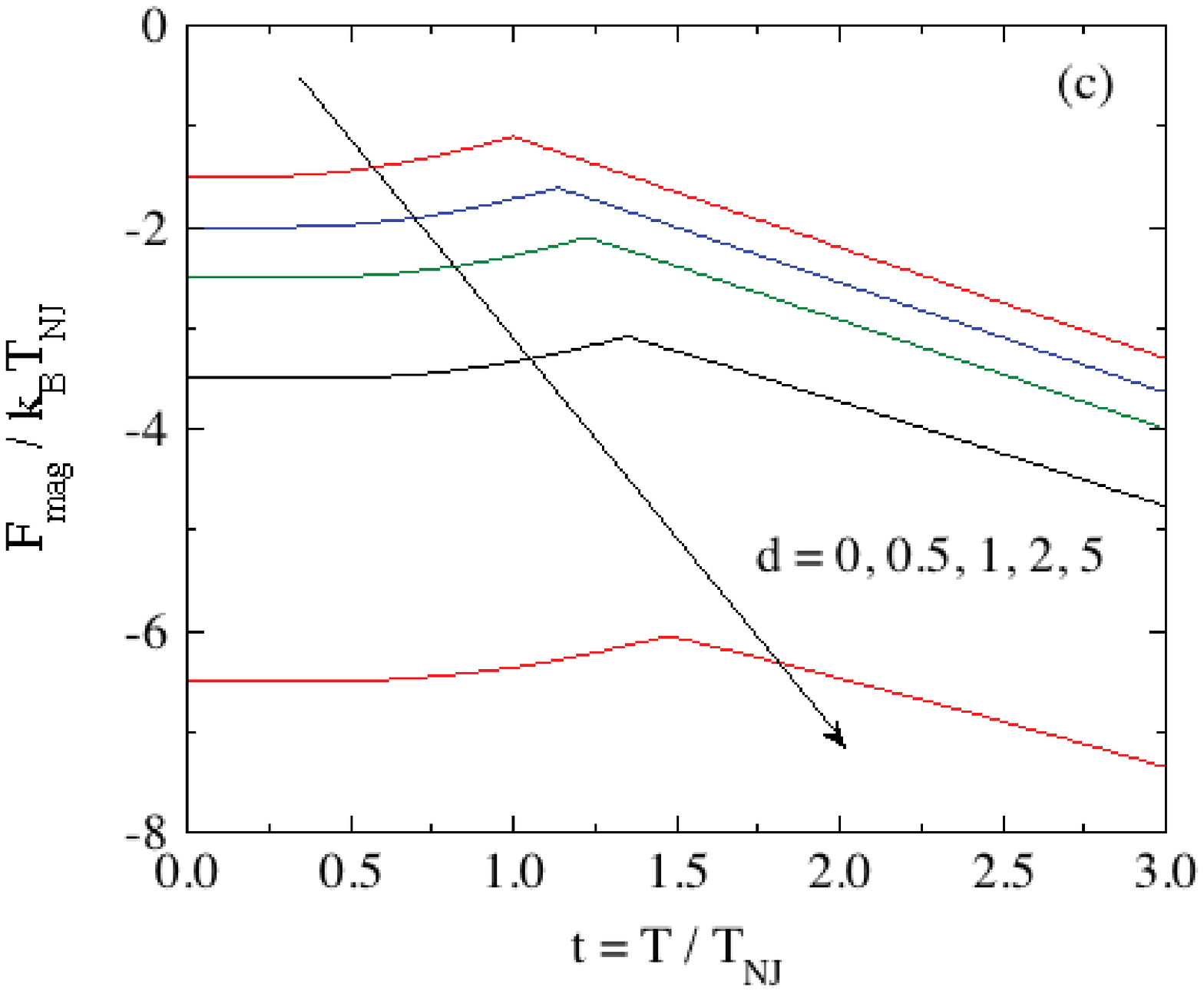}
\caption{(Color online) Reduced magnetic (a)~molar entropy $S_{\rm mag}/R$, (b) internal energy per spin $u_{\rm mag} = U_{\rm mag}/k_{\rm B}T_{{\rm N}J}$ and (c)~free energy per spin $f_{\rm mag} = F_{\rm mag}/k_{\rm B}T_{{\rm N}J}$ versus reduced temperature~$t$ in the collinear antiferromagnetic phase aligned along the $z$~axis for spins $S = 1$ with the listed values of the reduced anisotropy parameter~$d=D/k_{\rm B}T_{{\rm N}J}$, obtained by solving Eqs.~(\ref{Eqs:SmagCmag}).}
\label{Fig:SmagAFMH0S1d0to5}
\end{figure}

\begin{figure}
\includegraphics [width=3.3in]{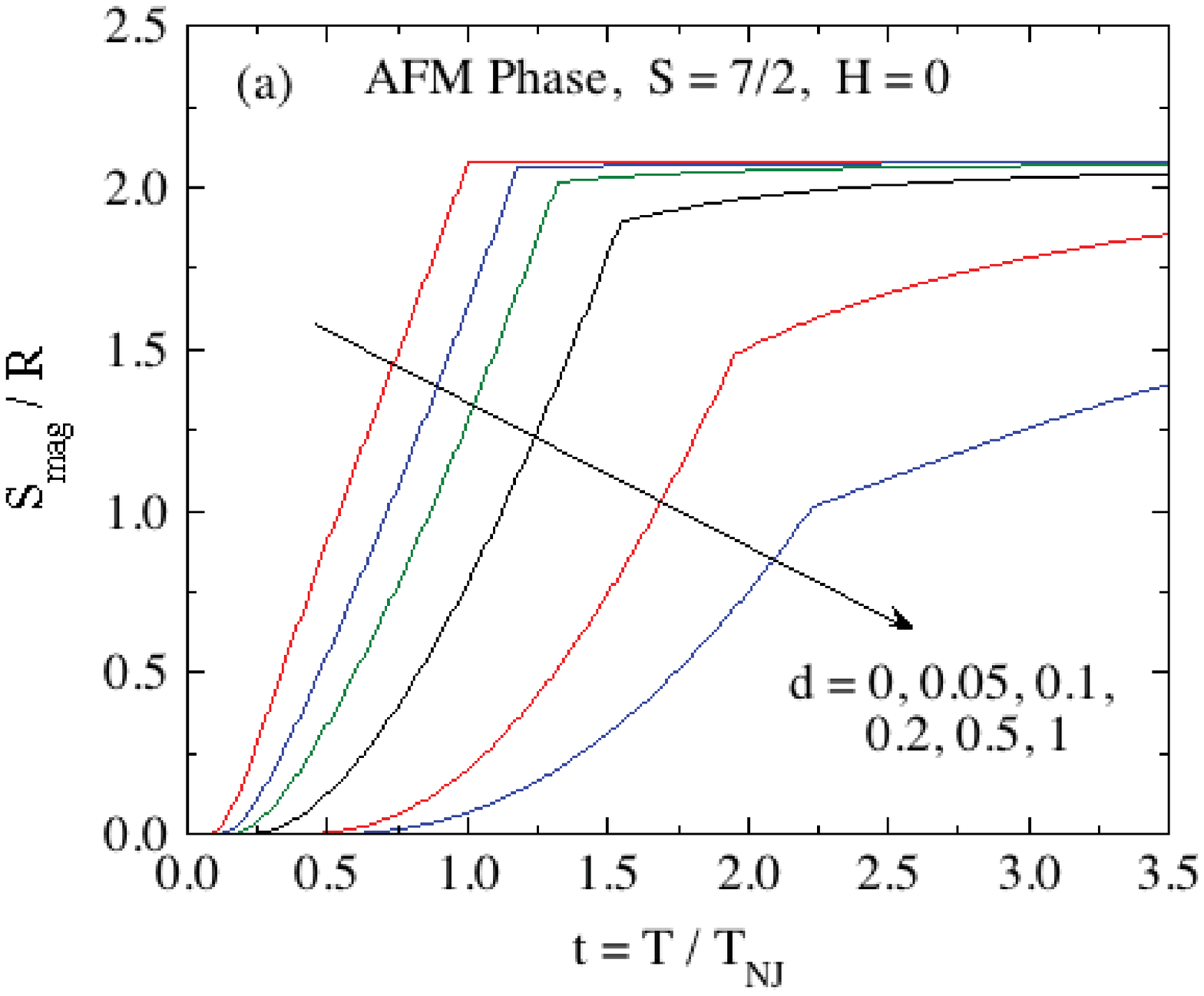}
\includegraphics [width=3.3in]{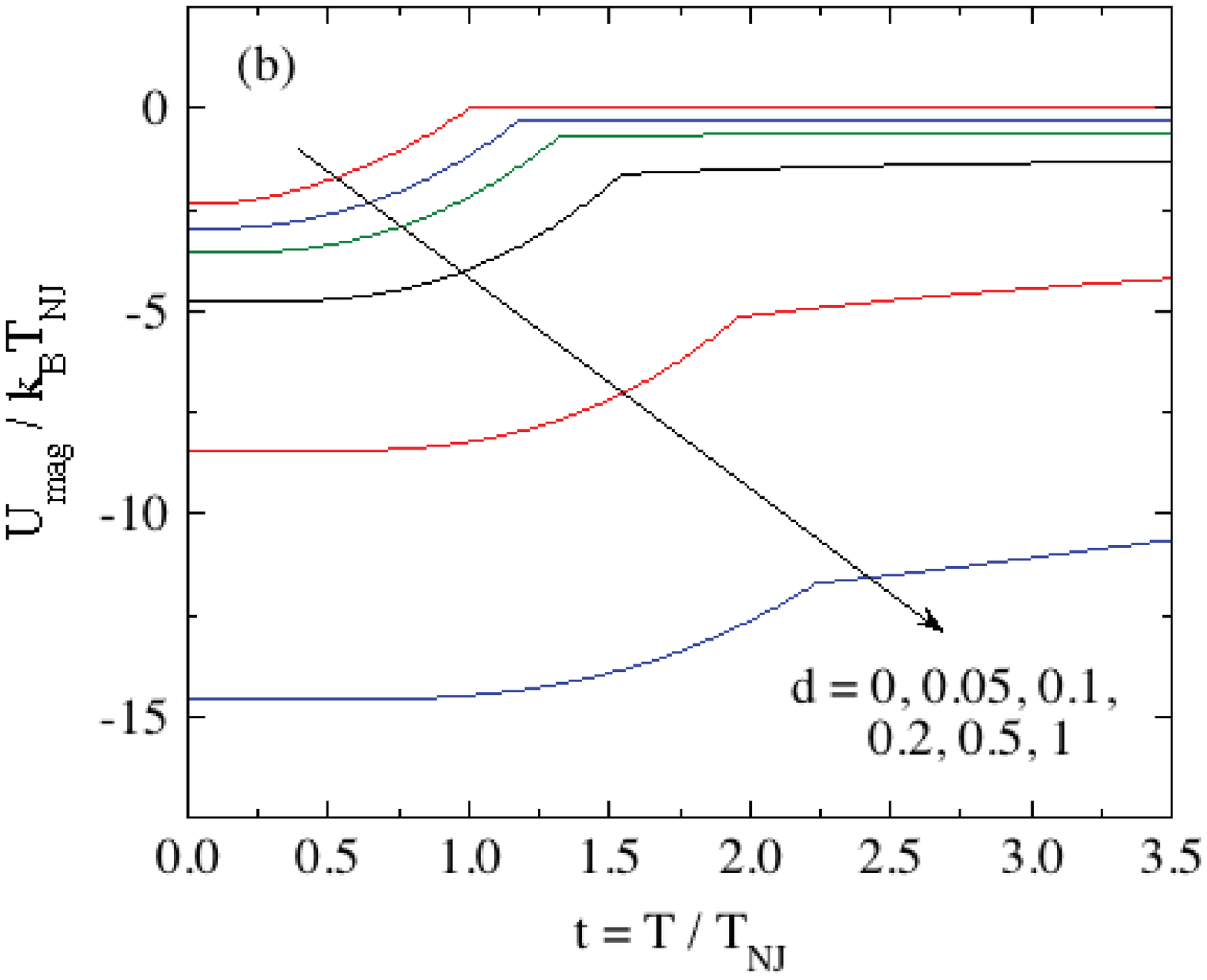}
\includegraphics [width=3.3in]{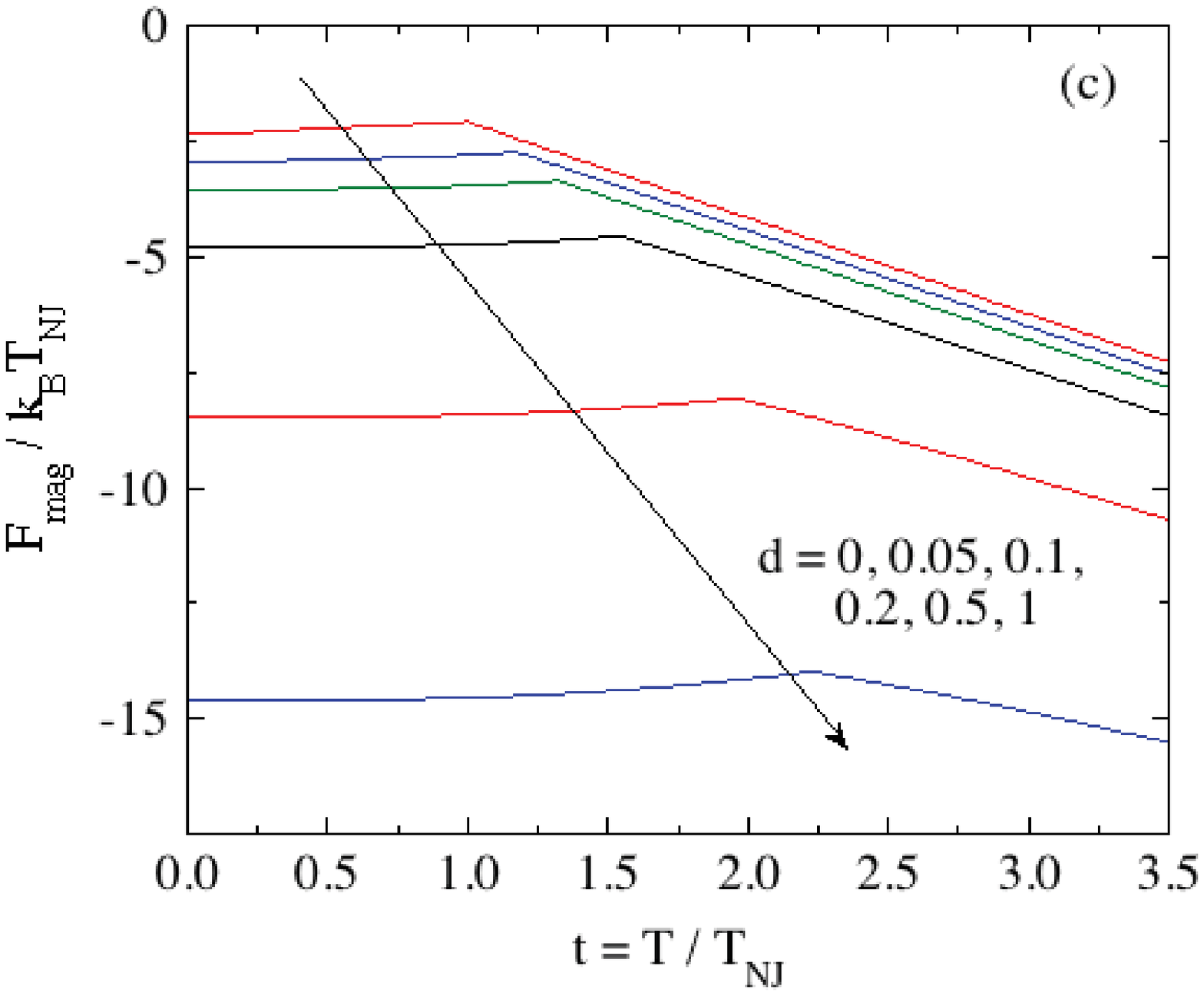}
\caption{(Color online) Same as Fig.~\ref{Fig:SmagAFMH0S1d0to5} except that $S = 7/2$ and $d=0$, 0.05, 0.1, 0.2, 0.5 and~1.}
\label{Fig:SmagAFMH0S72d0to1}
\end{figure}

The eigenenergies for collinear ordering along the $z$ axis are given above in Eq.~(\ref{Eq:Eamds2bi}), where $\bar{\mu}_0(t)$ is determined by solving Eq.~(\ref{Eq:barmuFromGS}).  Then the magnetic entropy $S_{\rm mag}$ versus~$t$ is obtained using Eqs.~(\ref{Eqs:SmagCmag}), where here the sums over eigenstates are sums over $m_S$.  The reduced internal energy $u_{\rm mag}$ and free energy $f_{\rm mag}(t)$ are determined using Eqs.~(\ref{Eq:umag(t)}) and~(\ref{Eq:fmagFromZ}), respectively.   Shown in Figs.~\ref{Fig:SmagAFMH0S1d0to5} and~\ref{Fig:SmagAFMH0S72d0to1} are plots of the zero-field molar $S_{\rm mag}/R$, single-spin $u_{\rm mag}$ and single-spin $f_{\rm mag}$ versus reduced temperature~$t$ for spins $S=1$ and $S=7/2$, respectively.  The cusp in each plot occurs at the respective reduced N\'eel temperature~$t_{\rm N}$.  Except for $d=0$ for which $t_{\rm N}=1$, the entropy continues to increase above $t_{\rm N}$ due to the uniaxial-anisotropy-induced zero-field splittings of the energy levels.

\begin{figure}
\includegraphics [width=3.3in]{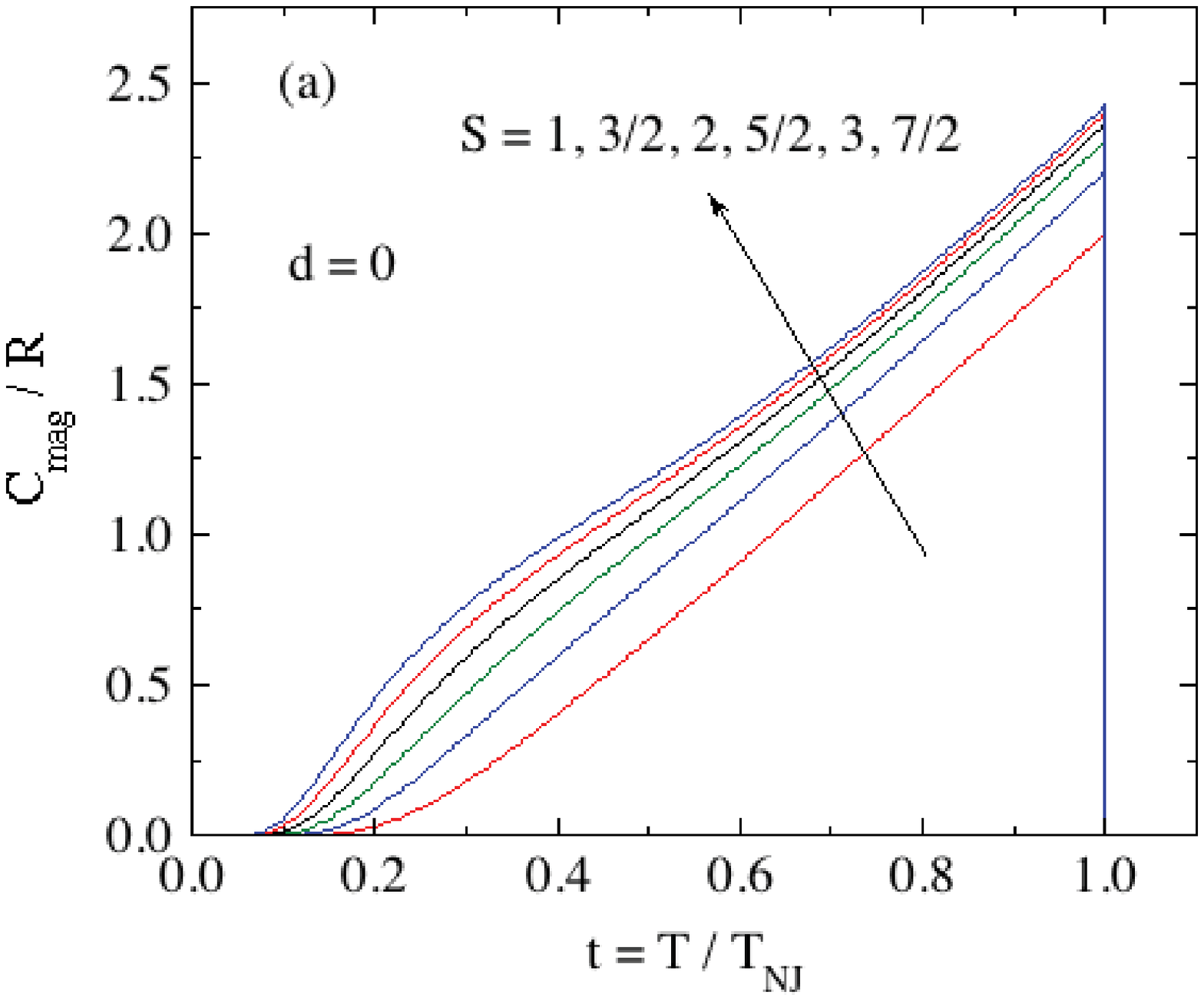}
\includegraphics [width=3.3in]{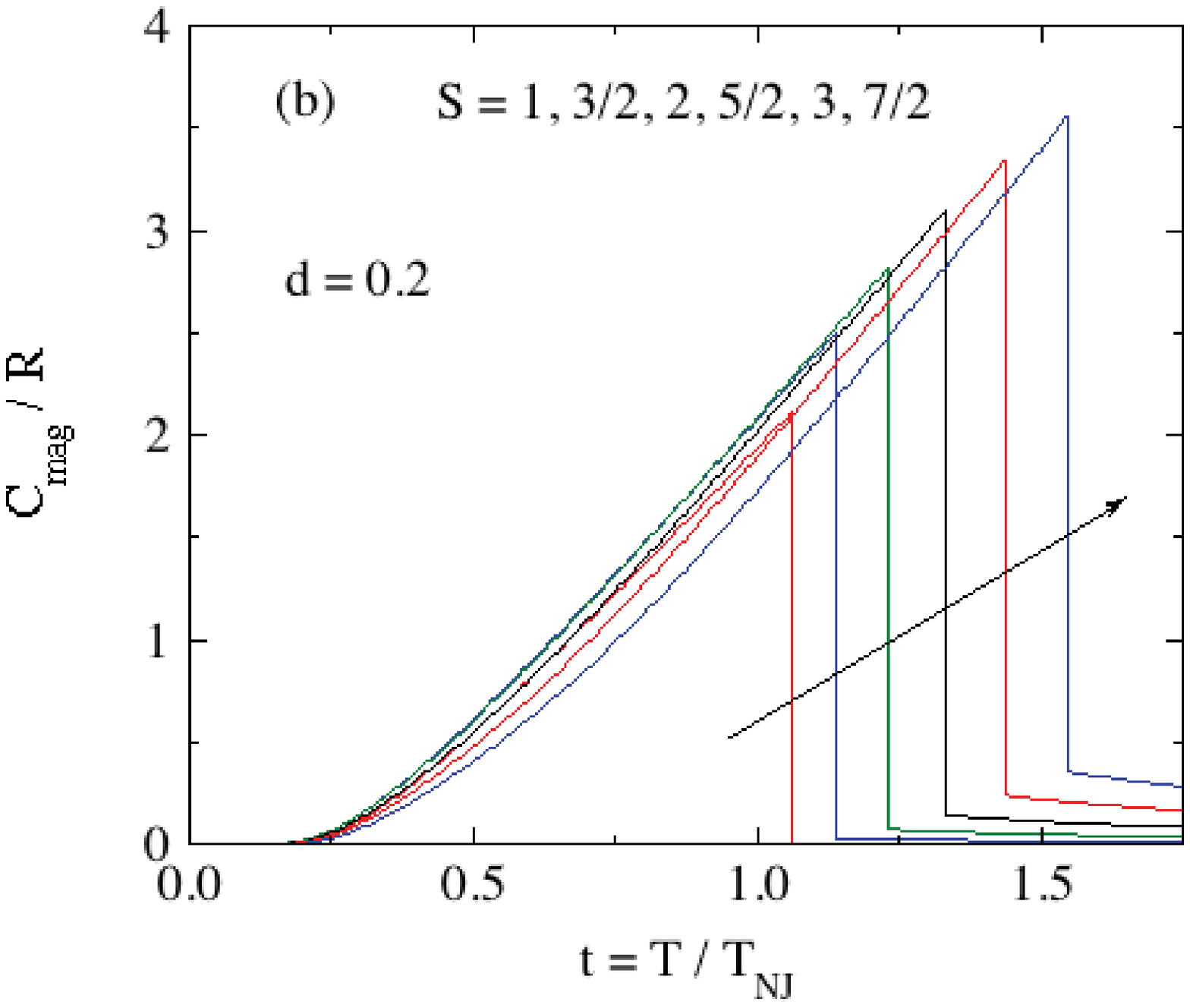}
\includegraphics [width=3.3in]{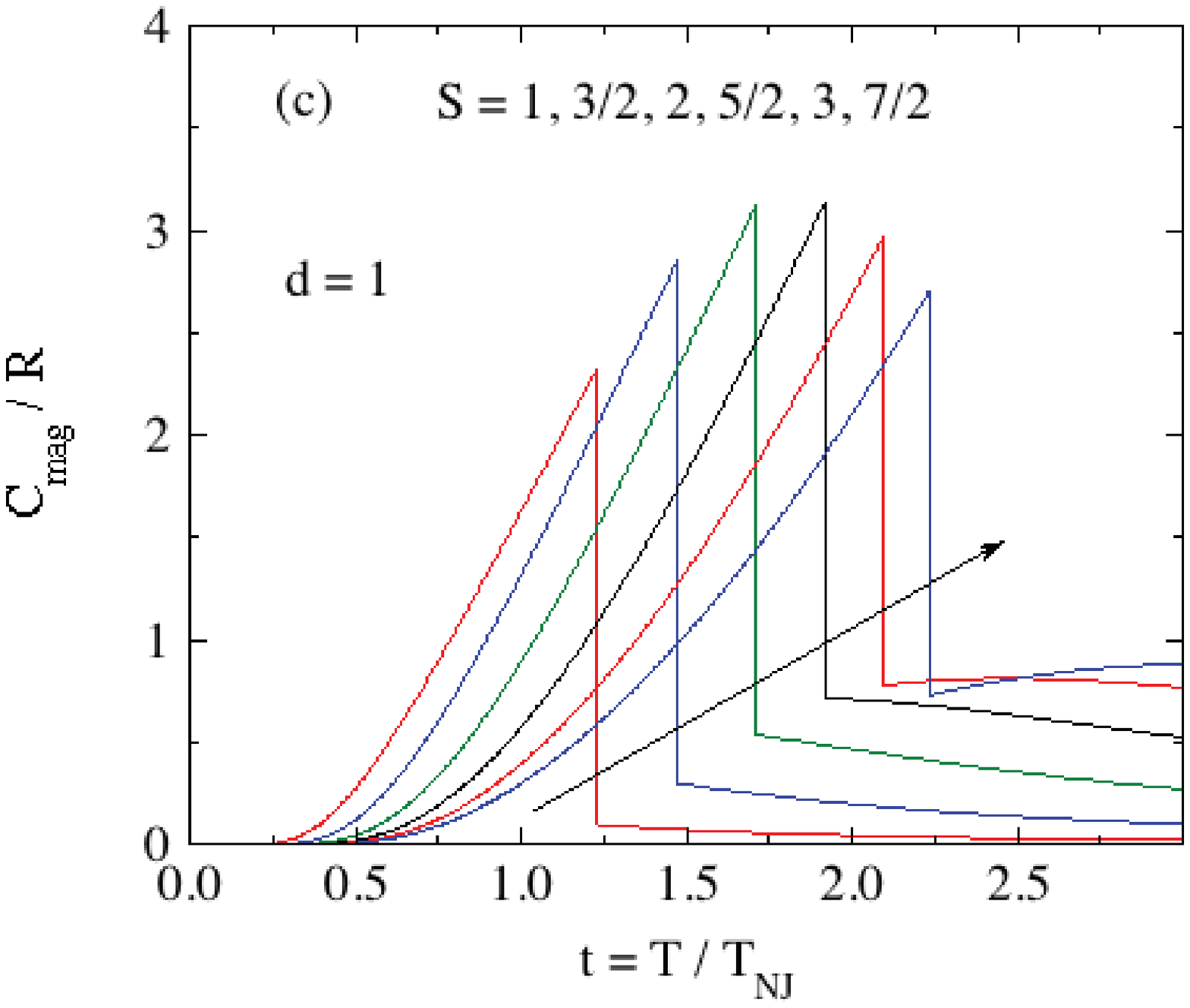}
\caption{(Color online) Molar magnetic heat capacity $C_{\rm mag}/R$ versus reduced temperature~$t$ in $H=0$ for spins $S=1$ to~7/2 with reduced anisotropy constants $d=D/k_{\rm B}T_{{\rm N}J}$ of (a)~0, (b)~0.2 and (c)~1, calculated using Eq.~(\ref{Eq:Cmag(t)1}).}
\label{Fig:CmagS1toS72d0}
\end{figure}

The molar $C_{\rm mag}(t)$ behaviors for $H = 0$ and spins $S=1$ to~7/2 obtained using Eq.~(\ref{Eq:Cmag(t)1}) are plotted for $d=0$, 0.2 and~1 in Figs.~\ref{Fig:CmagS1toS72d0}(a), \ref{Fig:CmagS1toS72d0}(b) and~\ref{Fig:CmagS1toS72d0}(c), respectively. With increasing~$d$, the hump in $C_{\rm mag}(t)$ at $t\sim 1/4$ for the larger $S$ values is progressively suppressed.  The corresponding loss of entropy is compensated by an increase of $C_{\rm mag}(t)$ at $t \lesssim t_{\rm N}$ for small~$d$. For the largest $d$~value shown, $d=1$, one sees that a significant amount of the entropy is present above $t_{\rm N}$ due to the presence of a Schottky anomaly as seen for noninteracting spins in Figs.~\ref{Fig:CmagS1d1m1Iso}(a) and~\ref{Fig:CmagS72d1m1Iso}(a) in the Appendix for $S=1$ and $S=7/2$, respectively.   From Fig.~\ref{Fig:CmagS1toS72d0}, the relative contribution above $t_{\rm N}$ of the Schottky anomaly increases with increasing~$d$ and~$S$\@.

\begin{figure}
\includegraphics [width=3.3in]{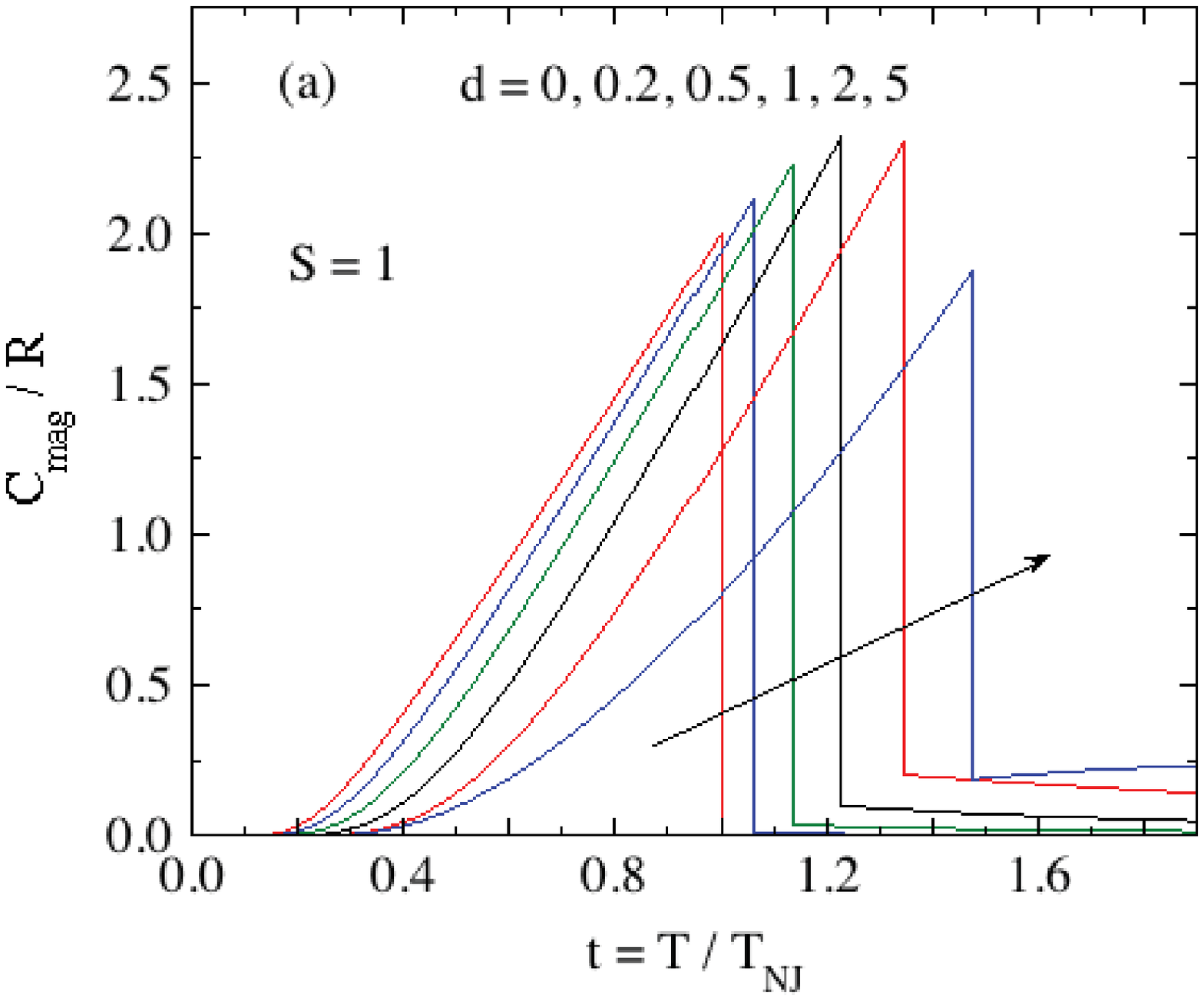}
\includegraphics [width=3.3in]{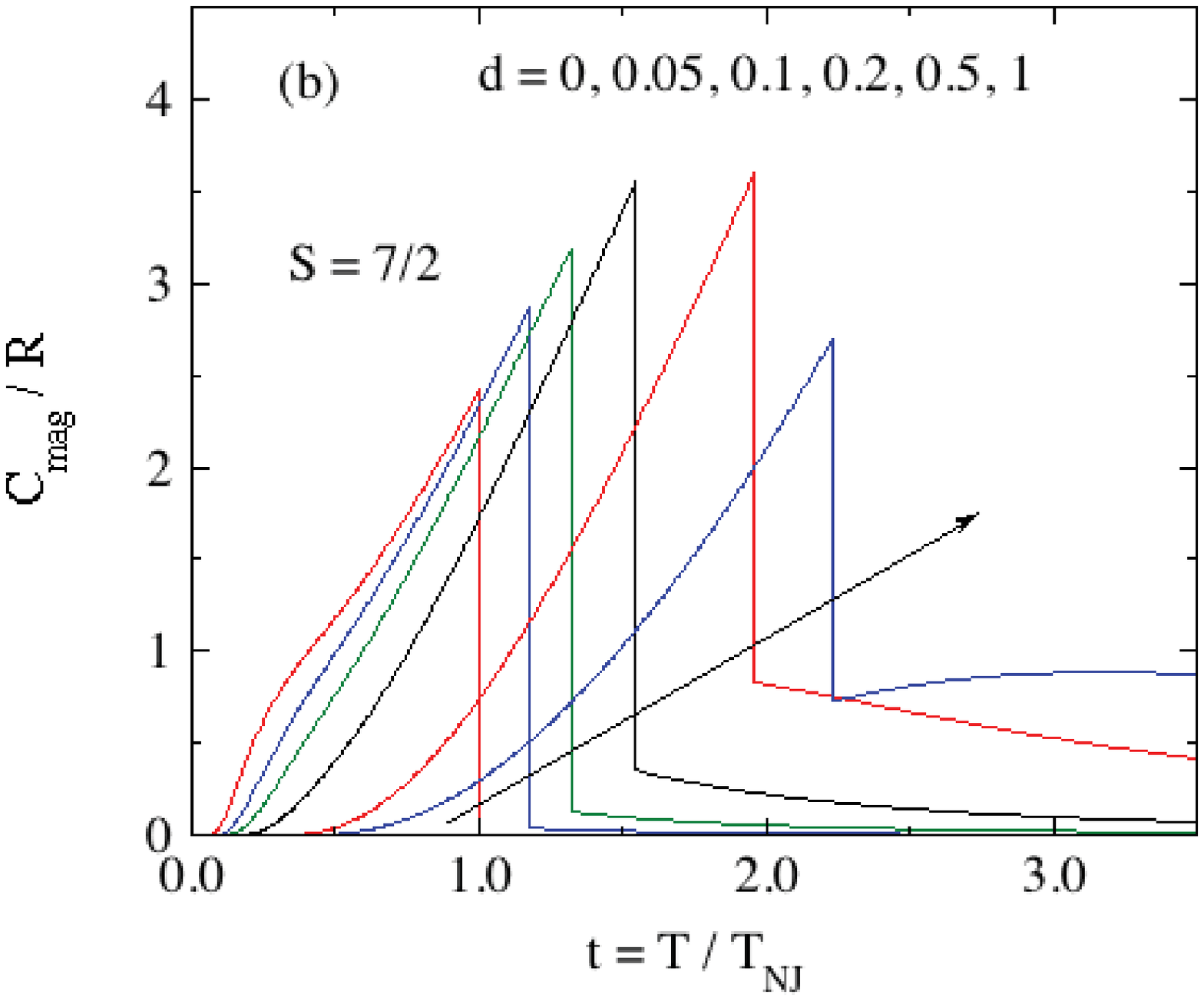}
\caption{(Color online) Molar magnetic heat capacity $C_{\rm mag}/R$ versus reduced temperature~$t$ for the listed reduced anisotropy parameters~$d$ and spins (a)~$S=1$ and (b)~$S=7/2$.}
\label{Fig:CmagS1d0To100}
\end{figure}

The dependences of $C_{\rm mag}$ on $t$ for variable $d$ and fixed $S=1$ and $S = 7/2$ are shown in Figs.~\ref{Fig:CmagS1d0To100}(a) and~\ref{Fig:CmagS1d0To100}(b), respectively.  Here one sees a strong increase in the influence of a given $d$ on $C_{\rm mag}(t)$ with increasing $S$ due to the Schottky anomaly contributions.  Indeed, for $d=5$ with $S=1$ and $d=1$ for $S=7/2$, the maxima of the Schottky anomalies are observed at $t>t_{\rm N}$.  Also, due to the increasing influence of $d$ on $C_{\rm mag}$ at $t\gtrsim t_{\rm N}$, the heat capacity jump at $t_{\rm N}$ first shows an increase with increasing~$d$, but then shows a decrease at the larger $d$~values for each~$S$ because the proportion of magnetic entropy in the Schottky anomaly above~$t_{\rm N}$ progressively increases with increasing~$d$.

\section{\label{Sec:CollinearAFMDH} Magnetic Fields Applied along the Uniaxial Easy Axis of Collinear Antiferromagnets}

\subsection{Magnetic Susceptibility}

Here we must distinguish the two sublattices in the collinear AFM state with $z$-axis alignment because they have different magnitudes in a finite applied field $H_z$.  The ordered moments on the same~(s) sublattice have the same value as a representative central spin $\vec{\mu}_i$ on that sublattice which is assumed to point in the $+z$ direction.  The moments on the second different~(d) sublattice $\vec{\mu}_j$ are pointed antiparallel to $\vec{\mu}_i$ in the $-z$ direction.  When a small field $dH_z$ is applied in the $+z$ direction, in general the magnitude $\mu_i$ of $\vec{\mu}_i$ increases slightly and that of $\vec{\mu}_j$ decreases by the same amount, so that
\be
d\vec{\mu}_j = d\vec{\mu}_i.
\label{Eq:dvecmuj}
\ee
When the spins are aligned along the $z$~axis, the differential of the exchange field  seen by $\vec{\mu}_i$ is given by Eq.~(\ref{Eq:HexchHGen}) as
\be
d{\bf H}_{{\rm exch}\,i} = \frac{3k_{\rm B}\theta_{{\rm p}J}}{g^2\mu_{\rm B}^2S(S+1)}d\vec{\mu}_i,
\label{Eq:dHexchi2}
\ee
where we used Eqs.~(\ref{Eq:fJDef}) and~(\ref{Eq:dvecmuj}).  Taking the $z$~components of the vectors and introducing the reduced $z$-axis moment definition
\be
\bar{\mu}_{iz} \equiv \frac{\mu_{iz}}{\mu_{\rm sat}} = \frac{\mu_{iz}}{gS\mu_{\rm B}}
\ee
as in Eqs.~(\ref{Eqs:mubarDef}), Eq.~(\ref{Eq:dHexchi2}) gives
\be
dH_{{\rm exch}\,iz} = \frac{3k_{\rm B}\theta_{{\rm p}J}}{g\mu_{\rm B}(S+1)}d\bar{\mu}_{iz},
\label{Eq:dHexchiz0}
\ee
which in reduced form is
\be
dh_{{\rm exch}\,iz} = \frac{3f_J}{S+1}d\bar{\mu}_{iz},
\label{Eq:dHexchiz}
\ee
where the reduced field $h_z$ and the parameter $f_J$ are defined in generic Eq.~(\ref{Eq:hDef}) and in  Eq.~(\ref{Eq:fJDef}), respectively. 

In the present case, Eq.~(\ref{Eq:barmuFromGS}) becomes
\be
\bar{\mu}_i = G_S(y)
\label{Eq:GSy2Def}
\ee
which is used to solve for $\bar{\mu}_i$, where 
\be
y = \frac{h_z}{t} + \frac{h_{{\rm exch}iz}}{t}
\label{Eq:y0AFM}
\ee
and the reduced temperature~$t$ is defined in Eq.~(\ref{Eq:tDef}).  Using Eq.~(\ref{Eq:dHexchiz}) and~(\ref{Eq:y0AFM}) one obtains
\be
dy = \frac{dh_z}{t} + \frac{3f_J}{(S+1)t}d\bar{\mu}_{iz}.\label{Eq:dyAFM}
\ee

Expanding Eq.~(\ref{Eq:GSy2Def}) in a Taylor series to first order in $d\bar{\mu}_{iz}$ gives
\be
d\bar{\mu}_{iz} = {G_S}^\prime(y_0)dy,
\label{Eq:dbarmuiz}
\ee
where ${G_S}^\prime(y)$ is given in Eq.~(\ref{Eq:GSPrime}) and $y_0$ in Eq.~(\ref{Eq:yDefAFM}).  Inserting Eq.~(\ref{Eq:dyAFM}) into~(\ref{Eq:dbarmuiz}) and solving for $d\bar{\mu}_{iz}$ yields
\be
d\bar{\mu}_{iz} = \frac{dh_z(S+1)/3}{\frac{(S+1)t}{3{G_S}^\prime(y_0)} - f_J }.
\label{Eq:dbarmuiz2}
\ee
Using Eq.~(\ref{Eq:ChiBarDef}) and~(\ref{Eq:dbarmuiz2}) one obtains the reduced parallel susceptibility~$\bar{\chi}_\parallel$ as
\bse
\label{Eqs:chiParallelz}
\bea
\bar{\chi}_\parallel(t) &\equiv& \frac{\chi_z(t)T_{{\rm N}J}}{C_1} = \frac{1}{\tau^\ast(t) - f_J},
\eea
where
\be
\tau^\ast(t) = \frac{(S+1)t}{3{G_S}^\prime(y_0)},\qquad y_0 = \frac{3\bar{\mu}_0}{(S+1)t},
\ee
\ese
and $\bar{\mu}_0(t)$ is calculated using Eq.~(\ref{Eq:barmuFromGS}).

Equations~(\ref{Eqs:chiParallelz}) are analogous to those for collinear AFM ordering from Heisenberg interactions in the absence of uniaxial anisotropy where here ${G_S}^\prime(y_0)$ replaces the derivative of the Brillouin function~${B_S}^\prime(y_0)$ in that case \cite{Johnston2012, Johnston2015}.  As in Refs.~\cite{Johnston2012, Johnston2015} for $d=0$, we find here for nonzero~$d$
\be
\tau^\ast(T = T_{\rm N}) = 1,
\ee
where $T_{\rm N}$ is the N\'eel temperature including both exchange interactions and single-ion anisotropy.  Then Eqs.~(\ref{Eqs:chiParallelz}) and the definition~(\ref{Eq:tNDef}) for $t_{\rm N}$ give
\bse
\label{Eqs:barchiparz}
\be
\bar{\chi}_\parallel(t = t_{\rm N}) =  \frac{1}{1-f_J}
\label{Eq:chibarz}
\ee
and
\be
\frac{\chi_\parallel(t)}{\chi_\parallel(t = t_{\rm N})} = \frac{1-f_J}{\tau^\ast(t) - f_J}.
\label{Eq:chionchiTN}
\ee
\ese

\begin{figure}
\includegraphics [width=3.3in]{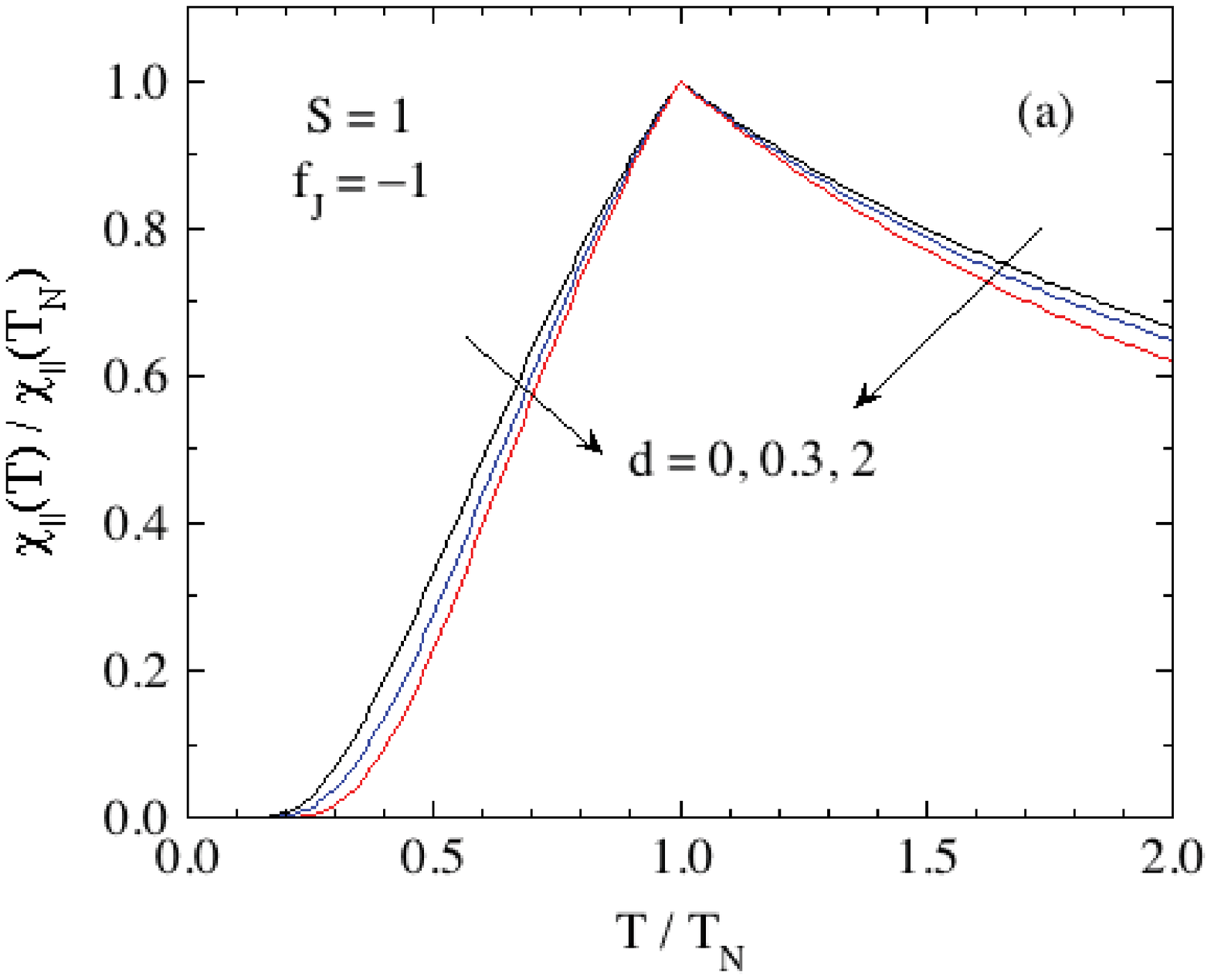}
\includegraphics [width=3.3in]{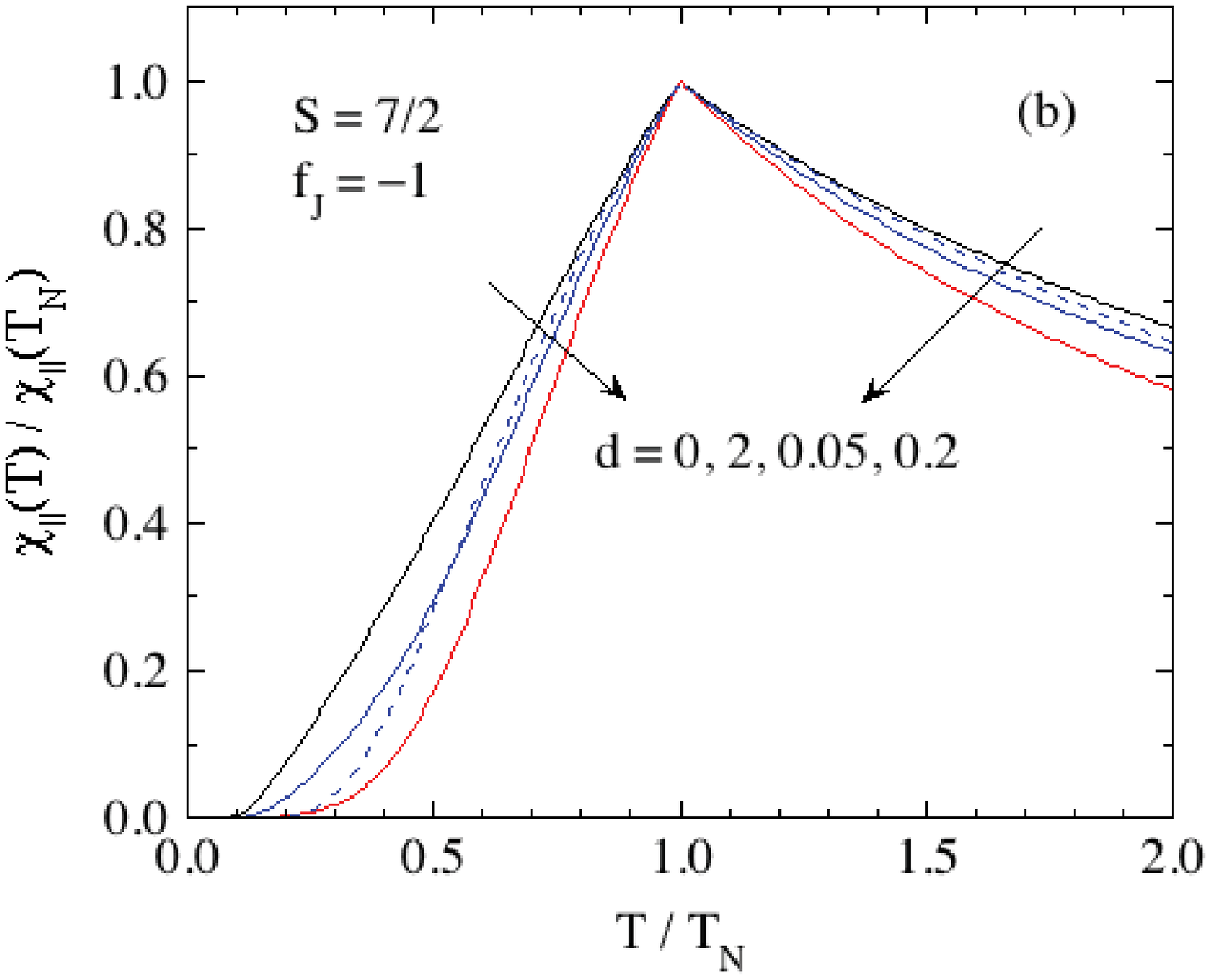}
\caption{(Color online) Normalized parallel susceptibility $\chi_\parallel(T)/\chi_\parallel(T_{\rm N})$ versus $T/T_{\rm N}$ obtained using Eq.~(\ref{Eq:chionchiTN}) for the parameter $f_J = \theta_{{\rm p}J}/T_{{\rm N}J}= -1$, the listed reduced anisotropy parameters $d=D/k_{\rm B}T_{{\rm N}J}$ and spins (a)~$S=1$ and (b)~$S=7/2$.}
\label{Fig:ChizbarS1fJm1d0to2}
\end{figure}

\begin{figure}
\includegraphics [width=3.3in]{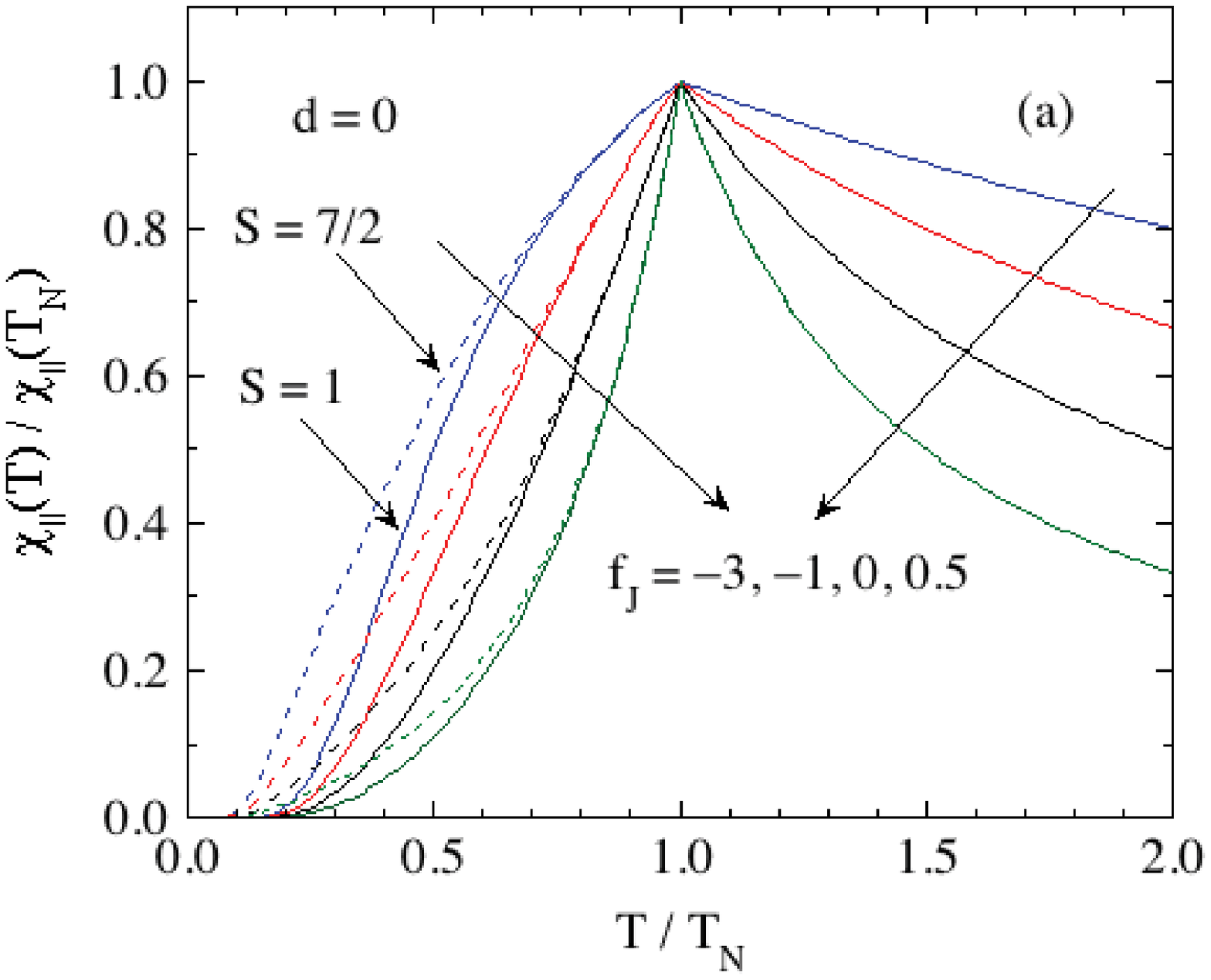}
\includegraphics [width=3.3in]{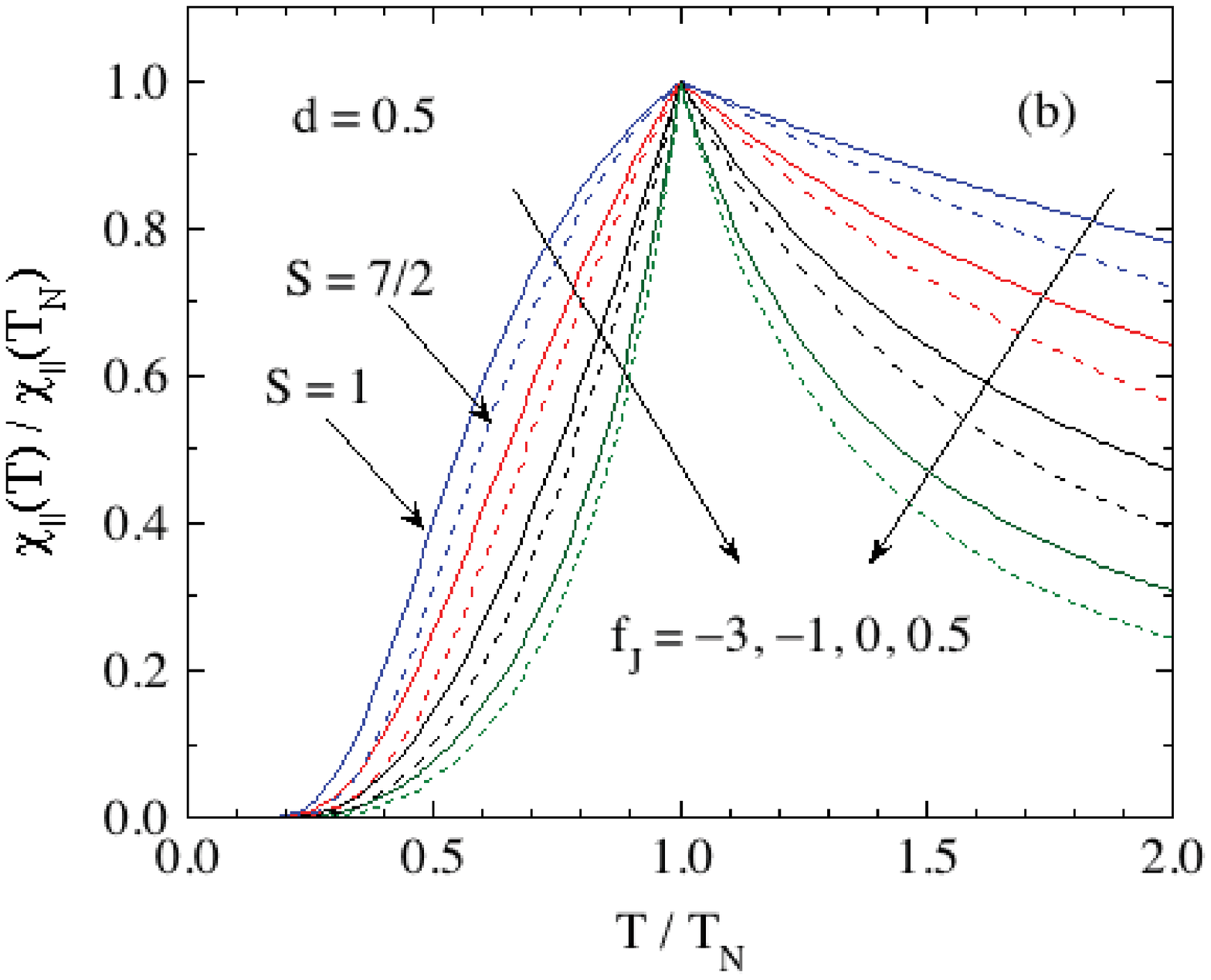}
\caption{(Color online) Normalized parallel susceptibility $\chi_\parallel(T)/\chi_\parallel(T_{\rm N})$ versus $T/T_{\rm N}$ obtained using Eq.~(\ref{Eq:chionchiTN}) for the listed values of $f_J$ and spins $S=1$ (solid curves) and $S=7/2$ (dashed curves) for (a)~$d=0$ and (b)~$d=0.5$.}
\label{Fig:ChibarVsTonTNS1S72d0}
\end{figure}

Shown in Fig.~\ref{Fig:ChizbarS1fJm1d0to2} are plots of the normalized parallel susceptibility~$\bar{\chi}_\parallel$ for spins~1 and~7/2 versus $T/T_{\rm N}$ (not versus $t=T/T_{{\rm N}J}$) for the listed values of~$d$.  One sees that these data are more strongly influenced by changes in~$d$ for $S=7/2$ compared with similar changes for $S=1$.  Figure~\ref{Fig:ChibarVsTonTNS1S72d0} shows how  $\chi_\parallel(T)/\chi_\parallel(T_{\rm N})$ versus~$T/T_{\rm N}$ depends on $f_J=\theta_{{\rm p}J}/T_{{\rm N}J}$ for $d=0$ and~$d=1/2$.  These two figures show that $\chi_\parallel(T)/\chi_\parallel(T_{\rm N})$ versus~$T/T_{\rm N}$ depends rather strongly for $T<T_{\rm N}$ on~$f_J$ compared with the dependences on $S$ and~$d$.

\subsection{\label{Eq:HiHCollAFM} Magnetization in a High Parallel Field}

In a finite $H_z$ applied along the easy $z$ collinear AFM ordering axis, one must again define two sublattices 1 and~2 because the magnitudes of the  ordered moments are not in general the same on the two sublattices.  In $H=0$, sublattice 1 is defined to have $\mu_{1z}>0$ and sublattice 2 then has $\mu_{2z}<0$ with equal moment magnitudes.  

Using Eq.~(\ref{Eq:HexchHGen}), the reduced exchange fields seen by spins on sublattices~1 and~2 are respectively
\bea
h_{{\rm exch}1z} &=& \frac{3}{2(S+1)}\left[\bar{\mu}_{1z}(1+f_J) - \bar{\mu}_{2z}(1-f_J)\right],\nonumber\\*
\label{Eqs:Hexchiz12}\\*
h_{{\rm exch}2z} &=& \frac{3}{2(S+1)}\left[-\bar{\mu}_{1z}(1-f_J) + \bar{\mu}_{2z}(1+f_J)\right].\nonumber
\eea
Thus there are now two simultaneous equations of the form of Eq.~(\ref{Eq:GSy2Def}), i.e., 
\bse
\label{Eqs:mu1zmu2z}
\be
\bar{\mu}_{1z} = G_S(y_1),\qquad \bar{\mu}_{2z} = G_S(y_2),
\ee
where
\bea
y_1 &=& \frac{h_z}{t} + \frac{3}{2(S+1)t}\left[\bar{\mu}_{1z}(1+f_J) - \bar{\mu}_{2z}(1-f_J)\right],\nonumber\\*
\label{Eqs:y1y2}\\*
y_2 &=& \frac{h_z}{t} + \frac{3}{2(S+1)t}\left[-\bar{\mu}_{1z}(1-f_J) + \bar{\mu}_{2z}(1+f_J)\right].\nonumber
\eea
\ese

\begin{figure}
\includegraphics [width=3.3in]{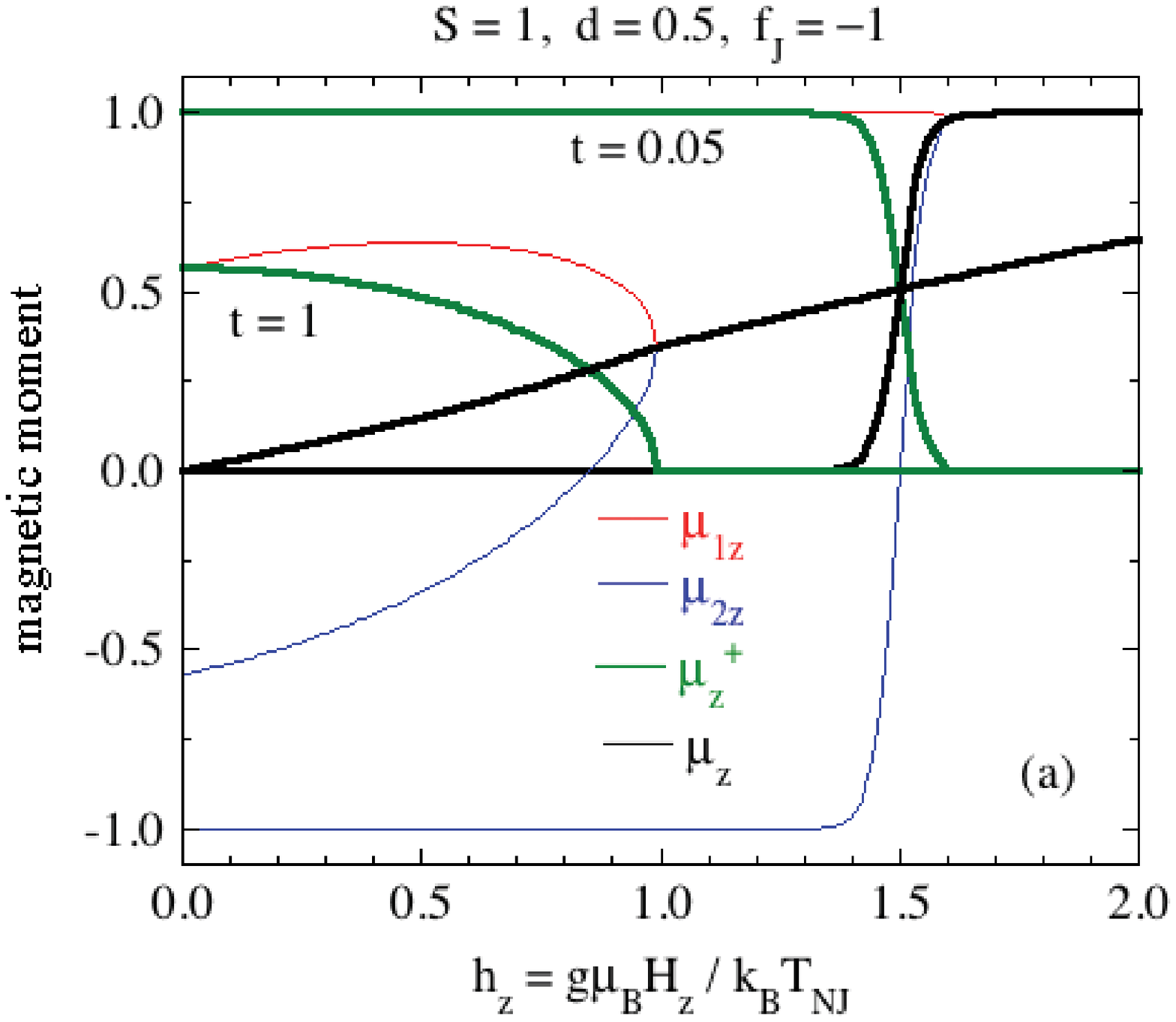}
\includegraphics [width=3.3in]{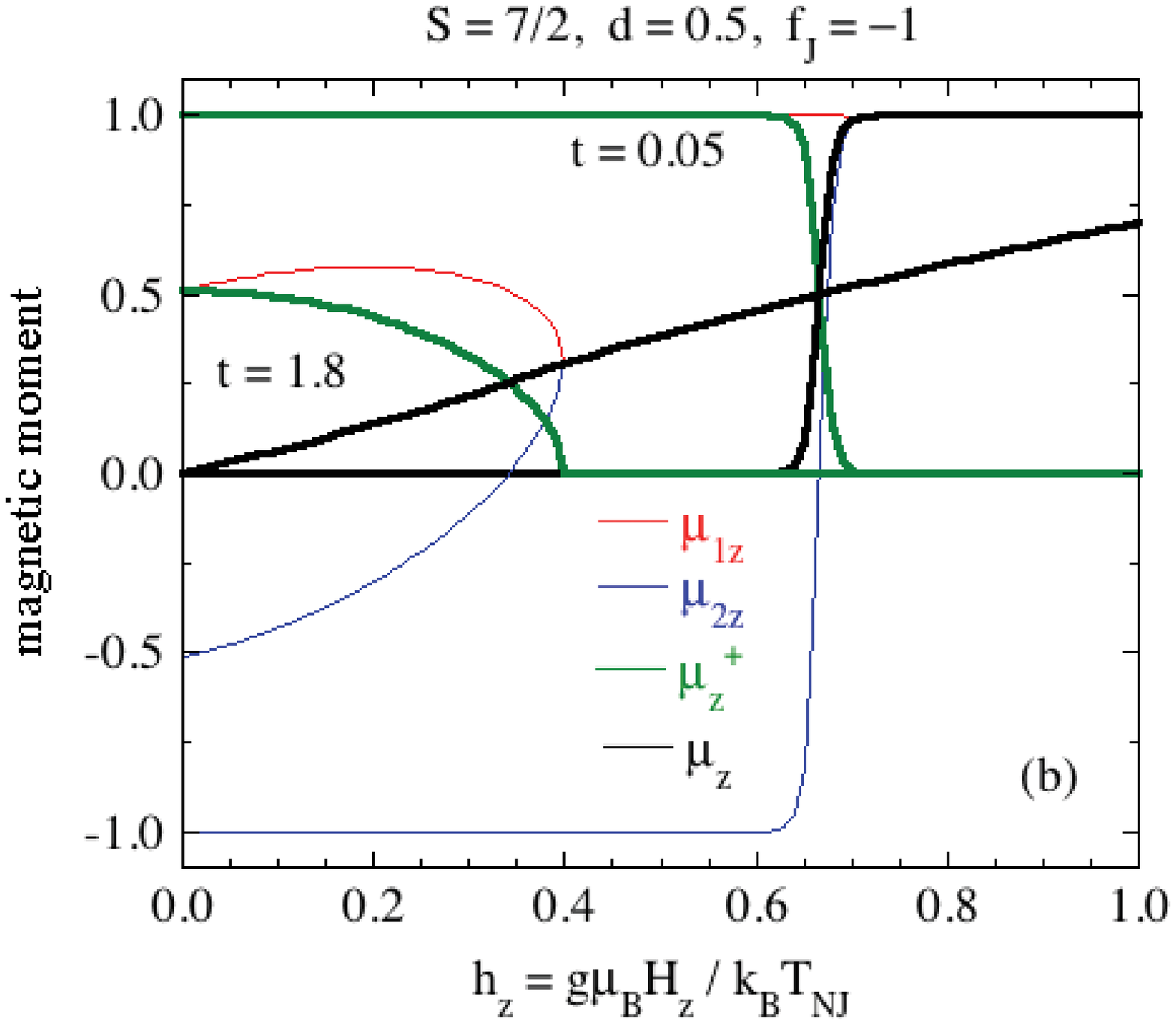}
\caption{(Color online) Reduced $z$-axis magnetic moments $\bar{\mu}_{1z}$ and $\bar{\mu}_{2z}$ versus reduced magnetic field $h_z = g\mu_{\rm B}H_z/k_{\rm B}T_{{\rm N}J}$ for $z$-axis collinear AFM ordering for anisotropy parameter $d = D/k_{\rm B}T_{{\rm N}J} = 0$ for spins (a)~$S = 1$ and (b)~$S=7/2$.  Also plotted versus $h_z$ are the reduced staggered moment $\bar{\mu}_z^\dagger = (\bar{\mu}_{1z}-\bar{\mu}_{2z})/2$ (the AFM order parameter) and the average moment $\bar{\mu}_z = (\bar{\mu}_{1z}+\bar{\mu}_{2z})/2$.  Note the different scales on the abscissas in (a) and~(b).}
\label{Fig:HzAFMS1fJm1d50mu}
\end{figure}

By numerically solving these two simultaneous equations, one obtains $\bar{\mu}_{1z}$ and $\bar{\mu}_{2z}$ as functions of $t$, $h_z$, $f_J$ and~$d$.  We solved Eqs.~(\ref{Eqs:mu1zmu2z}) iteratively.  Setting the initial value $\bar{\mu}_{1z}\sim1$, $\bar{\mu}_{2z}$ was calculated.  Then taking this value of $\bar{\mu}_{2z}$, $\bar{\mu}_{1z}$ was calculated.  This cycle was iterated until the difference between each of $\bar{\mu}_{1z}$ and $\bar{\mu}_{2z}$ and their respective subsequent interations was within $10^{-10}$. 

We find that if $f_J=-1$, which coincides with Van Vleck's value when calculating $\chi_\parallel(t)$ in the AFM state with $J_{ij}=J$ and only nearest-neighbor interactions on a bipartite spin lattice \cite{VanVleck1941}, then solutions to $\bar{\mu}_{1z}$ and $\bar{\mu}_{2z}$ have no first-order transitions versus $h_z$ at fixed~$t$, irrespective of the positive value of~$d$.  According to Eqs.~(\ref{Eqs:y1y2}), the criterion that $f_J=-1$ for second-order transitions is equivalent to requiring that $y_1$ is only a function of $\bar{\mu}_{2z}$ and conversely that $y_2$ is only a function of $\bar{\mu}_{1z}$.  Shown in Figs.~\ref{Fig:HzAFMS1fJm1d50mu}(a) and~\ref{Fig:HzAFMS1fJm1d50mu}(b) are plots for $S=1$ and $S=7/2$, respectively, of the field dependences with $d=0.5$ of $\bar{\mu}_{1z}$, $\bar{\mu}_{2z}$, the staggered ordered moment $\bar{\mu}^\dagger = (\bar{\mu}_{1z}-\bar{\mu}_{2z})/2$ which is the AFM order parameter, and the average $\bar{\mu}_z =  (\bar{\mu}_{1z}+\bar{\mu}_{2z})/2$ which is the quantity obtained from uniform magnetization measurements along the $z$~axis.  For $T\to0$, one sees from Fig.~\ref{Fig:HzAFMS1fJm1d50mu} that $\bar{\mu}_{1z} = 1$, $\bar{\mu}_{2z} = -1$, $\bar{\mu}^\dagger =1$ and $\bar{\mu}_z = 0$, all as expected.  For the two representative temperatures shown for each spin, $\bar{\mu}_{1z}>0$ for all $h_z$, whereas $\bar{\mu}_{2z}$ continuously increases with increasing field from its initial value of~$-1$ to become positive, eventually meeting up with $\bar{\mu}_{1z}$ at the reduced critical field $h_{\rm c}$ which is the second-order transition field from the AFM state to the PM state.  With increasing $t$ the transition from the AFM state to the PM state with increasing $h_z$ becomes less and less visible in plots of $\bar{\mu}_z$ versus~$h_z$.  

\begin{figure}
\includegraphics [width=3.in]{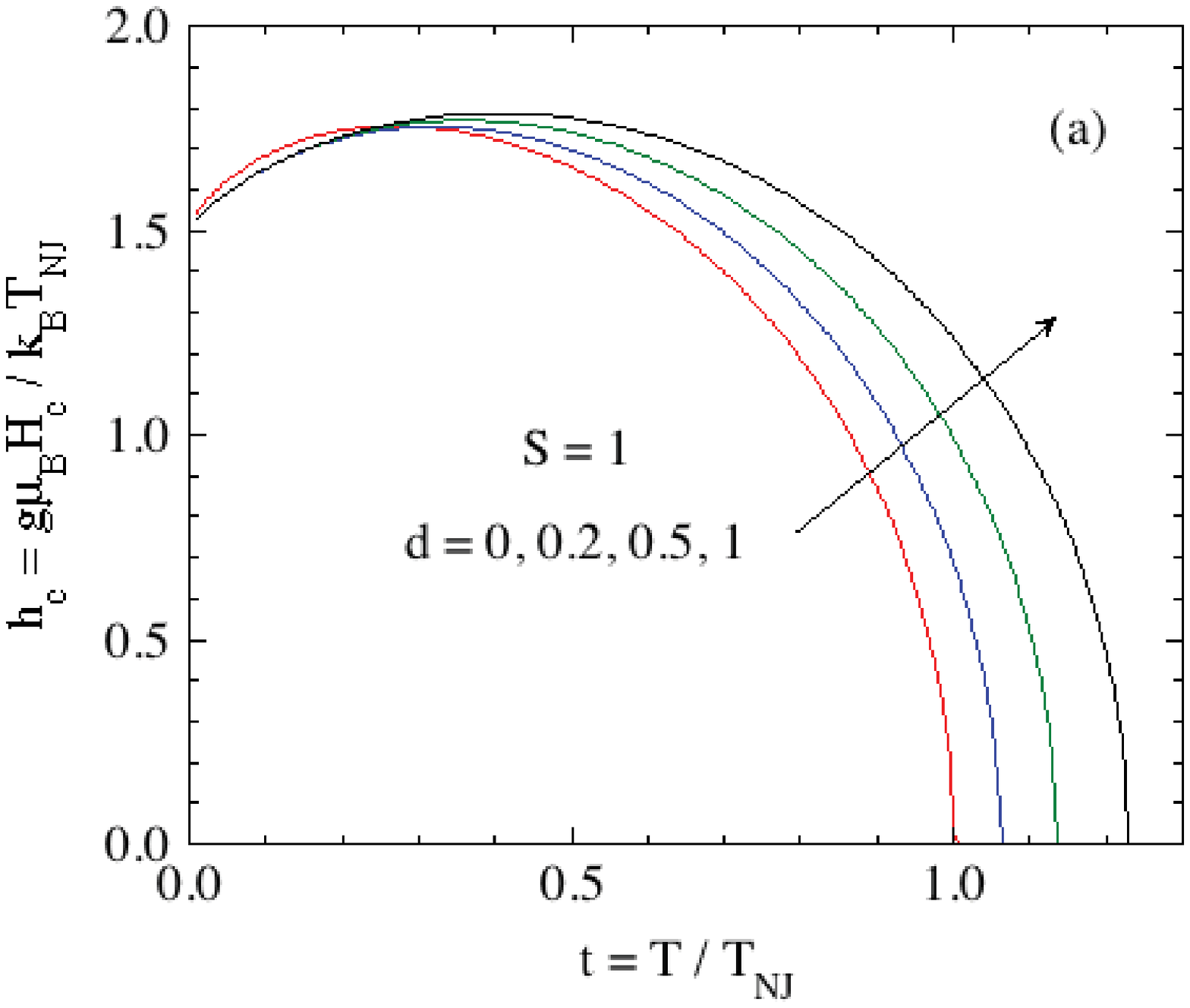}
\includegraphics [width=3.in]{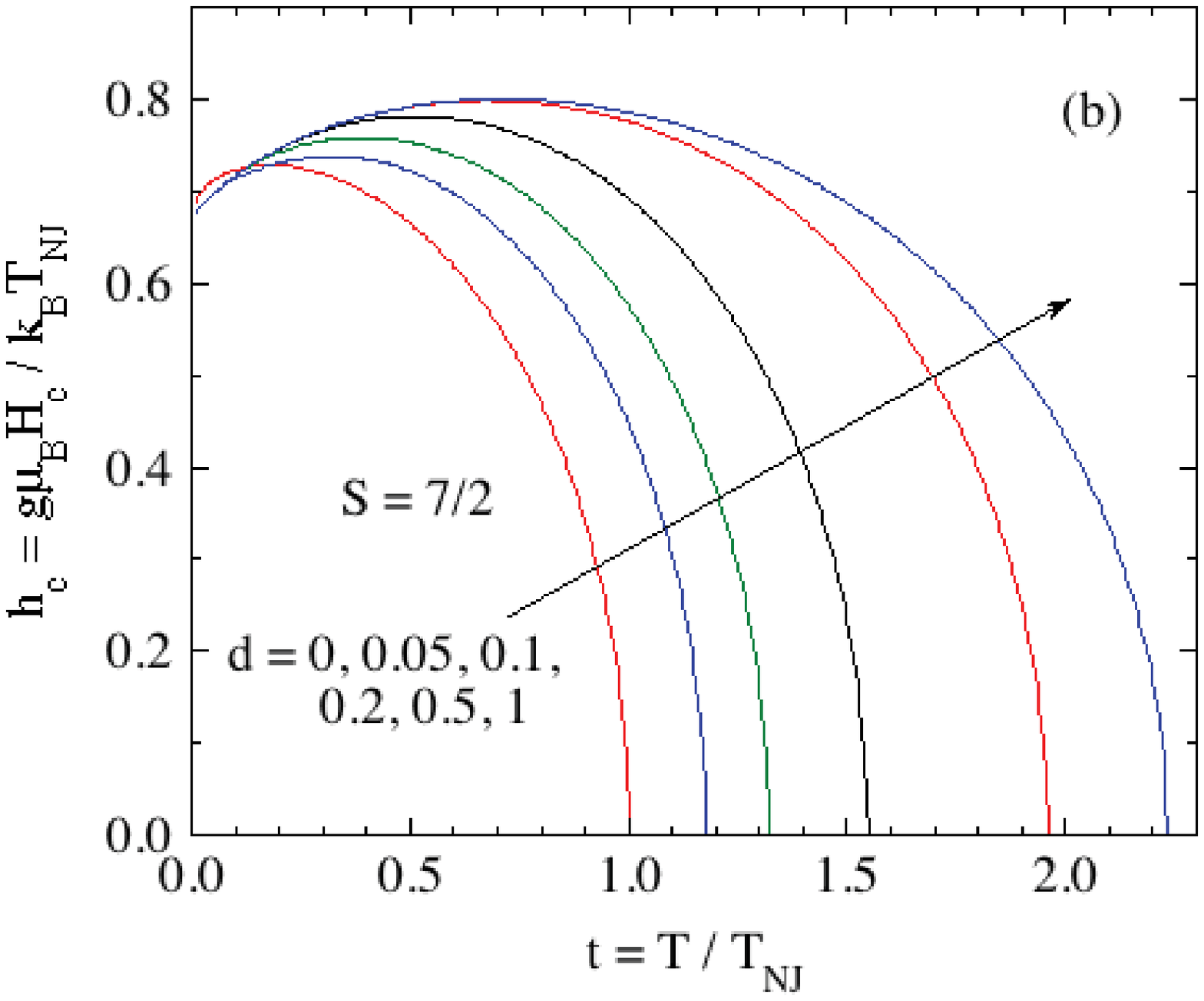}
\caption{(Color online) Reduced $z$-axis critical fields $h_{\rm c}$ for $z$-axis collinear AFM ordering versus reduced temperature $t$ for the listed values of the anisotropy parameter $d = D/k_{\rm B}T_{{\rm N}J}$ for spins (a)~$S = 1$ and (b)~$S=7/2$.}
\label{Fig:HcAFMS1fJm1d0To1}
\end{figure}

Plots of $h_{\rm c}$ versus~$t$ for several values of~$d$ for spins $S=1$ and $S=7/2$ are shown in Figs.~\ref{Fig:HcAFMS1fJm1d0To1}(a) and~\ref{Fig:HcAFMS1fJm1d0To1}(b), respectively.  The data for each spin show that $h_{\rm c}$ increases with increasing $t$ from a spin-dependent finite $h_{\rm c}(t=0)$ to a broad maximum at a temperature that increases with increasing~$d$.  The curves in Fig.~\ref{Fig:HcAFMS1fJm1d0To1} form the boundary between the low-field AFM and the high-field and/or high-temperature PM phases in the $H_z-T$ plane for a given $d$ value.  With increasing~$d$, for $h_z=0$ the system remains in the AFM state to increasing temperatures $t=T/T_{{\rm N}J}$ because $t_{\rm N}$ increases with increasing~$d$ as shown above in Fig.~\ref{Fig:tNVSd}(a). These observations do not take into account the competition with the spin-flop phase discussed in Secs.~\ref{Sec:SFPhase} and~\ref{Sec:PhaseDiagram} below.

\begin{figure}
\includegraphics [width=3.3in,viewport= 18 10 480 405,clip]{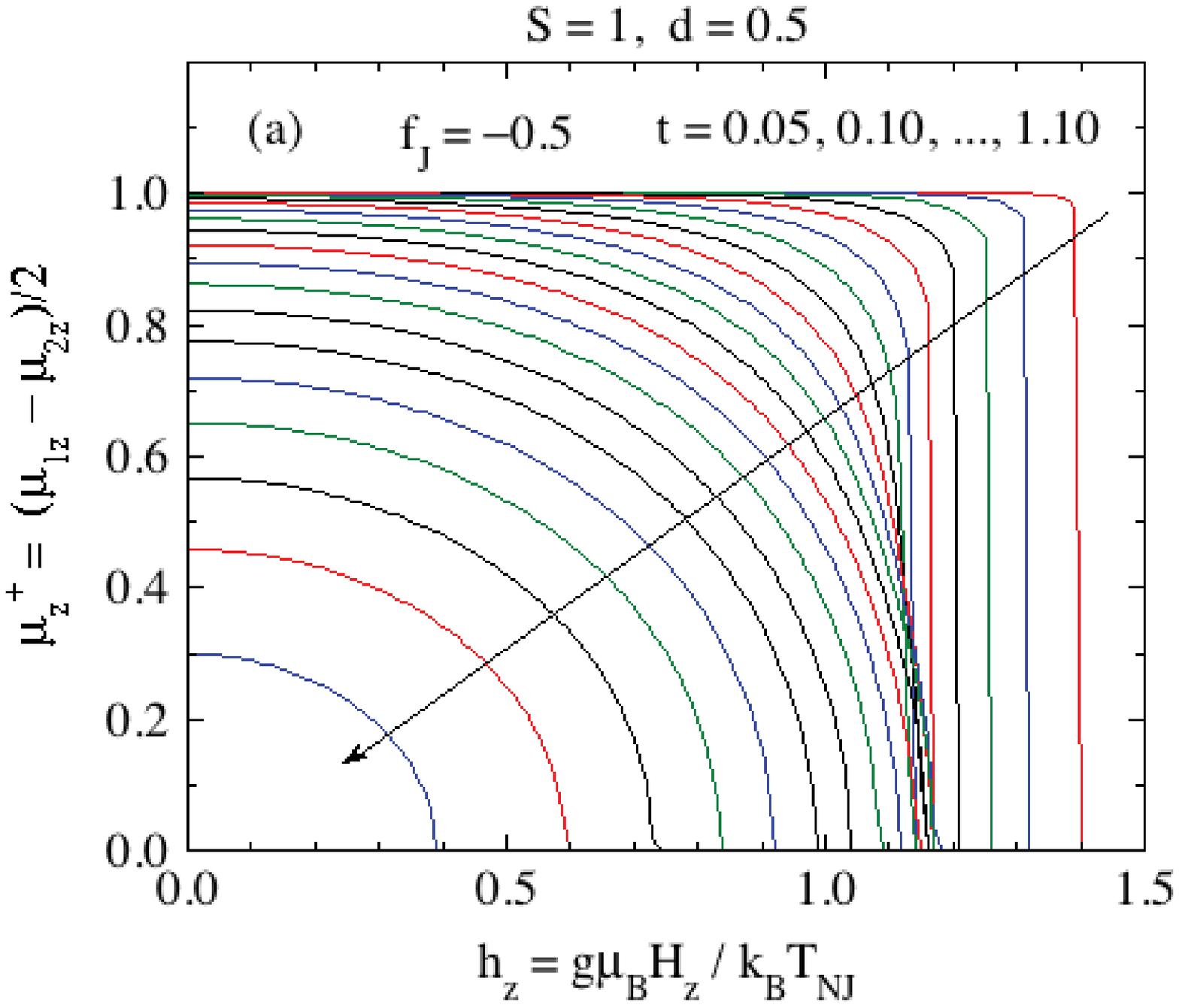}
\includegraphics [width=3.3in,viewport= 18 10 480 405,clip]{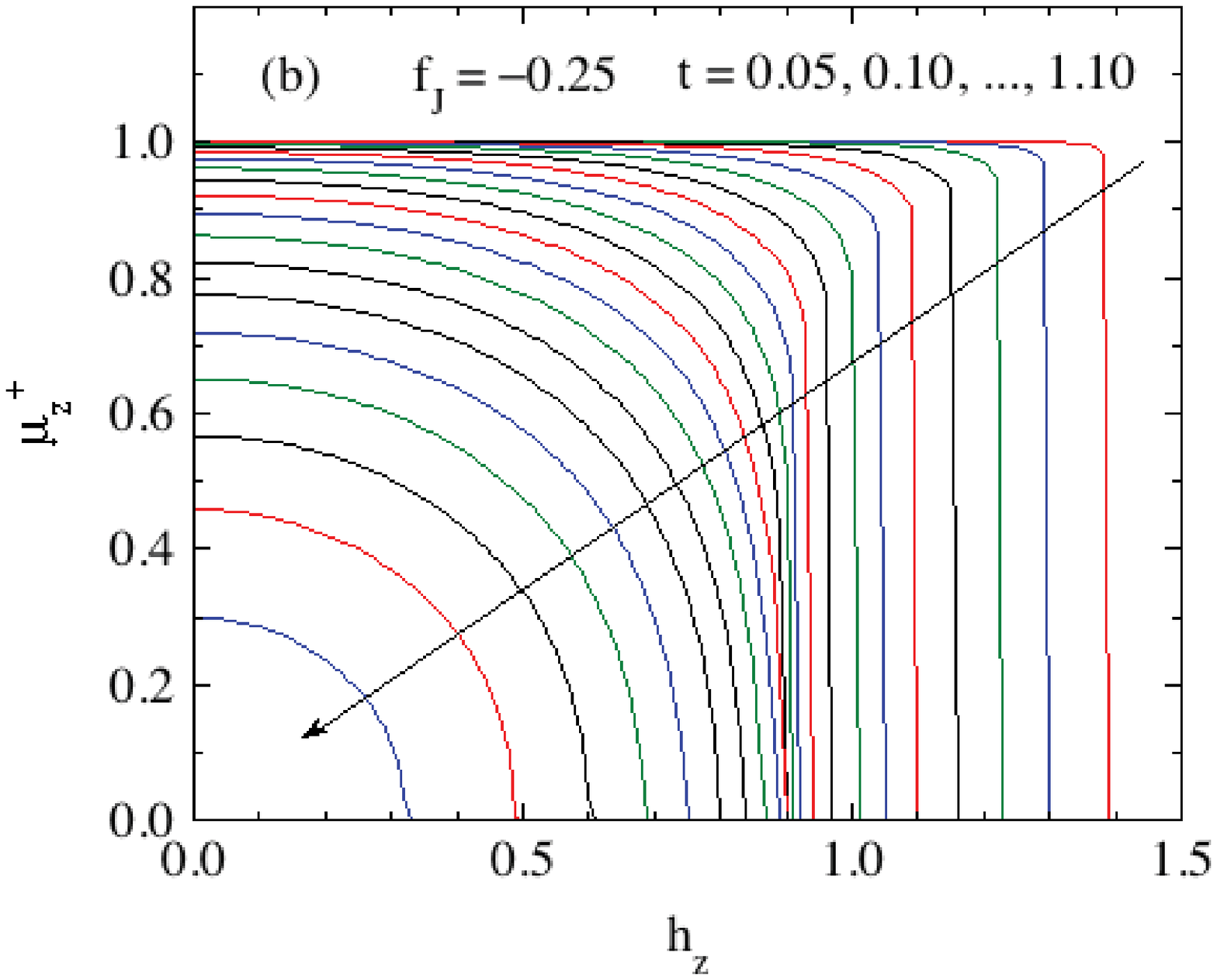}
\includegraphics [width=3.3in,viewport= 18 10 480 405,clip]{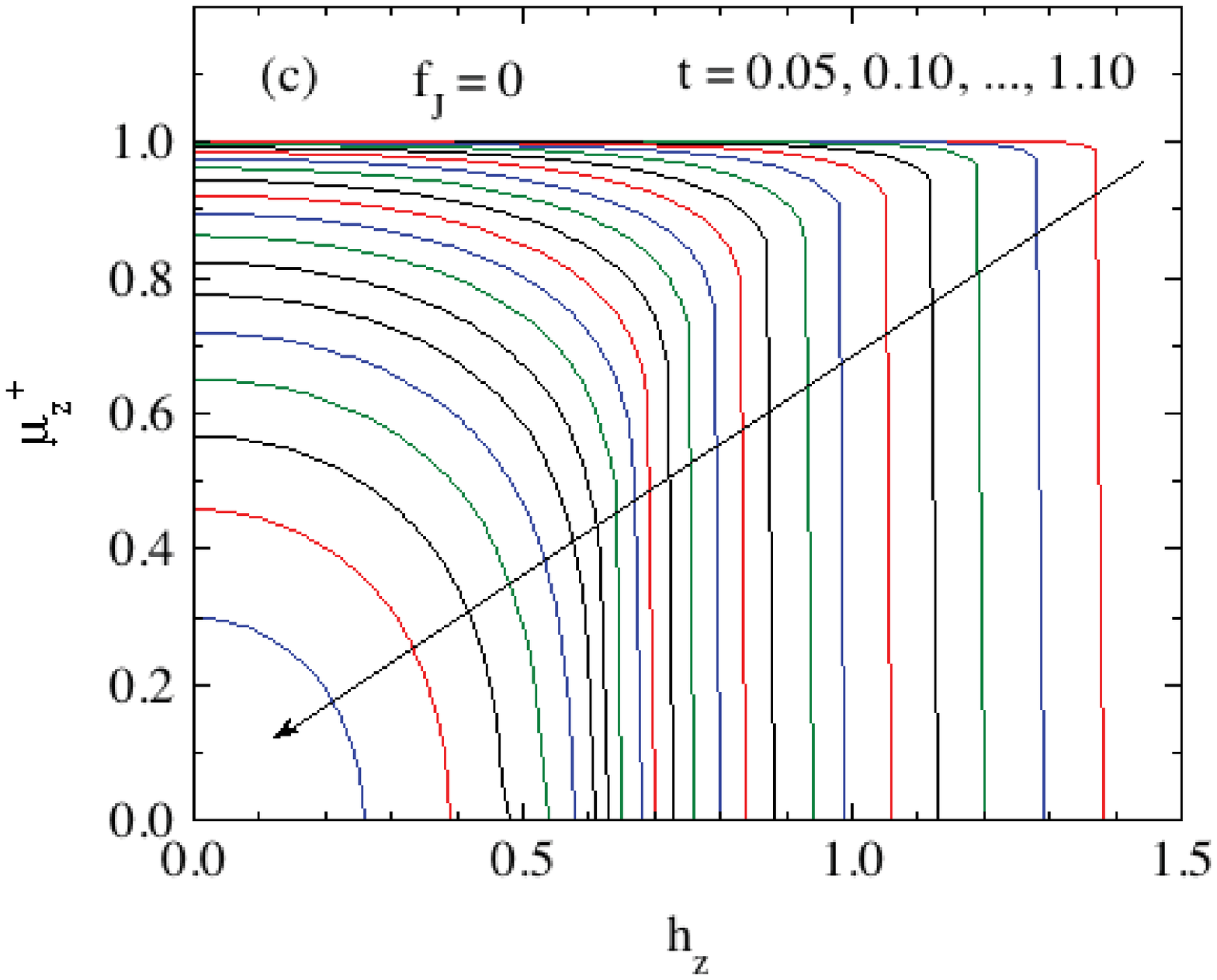}
\caption{(Color online) Staggered $z$-axis moment $\mu_z^\dagger$ (AFM order parameter) versus reduced field~$h_z$ for the listed values of reduced temperature~$t$ for spins $S=1$ with reduced anisotropy parameter~$d=0.5$ and parameter $f_J = \theta_{{\rm p}J}/T_{{\rm N}J}$ given by (a)~$f_J=-0.5$, (b)~$f_J=-0.5$, and (c)~$f_J=-0.5$.}
\label{Fig:muzVsHzAFMS1d50fJ}
\end{figure}

When $f_J = \theta_{{\rm p}J}/T_{{\rm N}J}$ is in the range $-1 < f_J < 1$ where the value $f_J=1$ corresponds to a ferromagnet, plots such as shown in Fig.~\ref{Fig:HzAFMS1fJm1d50mu} for $f_J=-1$ show first-order transitions versus field.  Such $f_J$ values result from one or more ferromagnetic Heisenberg interactions $J_{ij}$ between the central spin~$i$ and its neighbors~$j$ in addition to the AFM interactions necessary to yield collinear AFM ordering.  Shown in Fig.~\ref{Fig:muzVsHzAFMS1d50fJ} are plots of the staggered moment~$\mu_z^\dagger$ versus~reduced field~$h_z$ at various reduced temperatures~$t$ for spin~$S=1$ with reduced anisotropy parameter~$d=0.5$ and $f_J= -0.5,\ -0.25$ and~0.  One sees that as $f_J$ increases algebraically above $-1$, first-order transitions occur for an increasing range of temperature.  

\begin{figure}[t]
\includegraphics [width=3.3in]{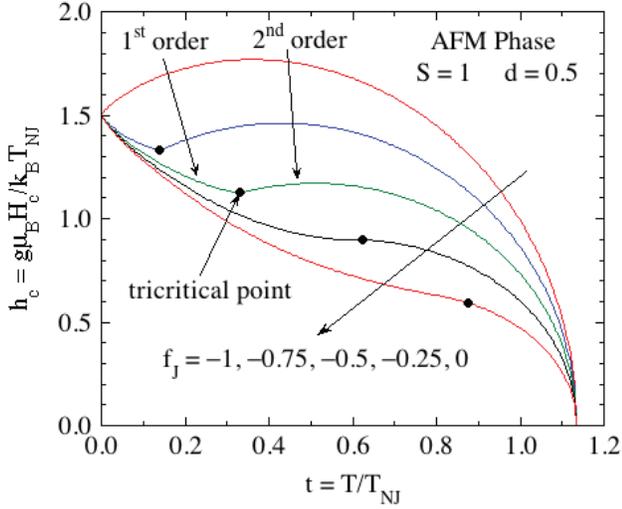}
\caption{(Color online) Reduced critical field~$h_{\rm c}$ separating the antiferromagnetic phase from the paramagnetic phase versus reduced temperature~$t$ for spin $S=1$ with reduced anisotropy parameter $d=0.5$ for several values of the parameter $f_{\rm J} = \theta_{{\rm p}J}/T_{{\rm N}J}$ as shown.  For $f_J=-1$ the critical field curve correponds to second-order transitions only on crossing the curve and is duplicated from Fig.~\ref{Fig:HcAFMS1fJm1d0To1}(a) for $d=0.5$.  For $f_J > -1$ the transition is second-order at high temperatures and first order at low temperatures, where the two regions are separated by a tricritical point for each such $f_J$ as shown by the filled black circles.}
\label{Fig:HcAFMS1d50fJ0T0m1}
\end{figure}

The reduced critical field~$h_{\rm c}$ representing the transition from the AFM to the PM phase is plotted versus reduced temperature~$t$ for $S=1$, $d=0.5$ and five $f_J$ values in the range $-1 \leq f_J \leq 0$ in Fig.~\ref{Fig:HcAFMS1d50fJ0T0m1}.  The first- and second-order regions of each transition curve with $f_J = -0.75,\ -0.5,\ -0.25$ and~0 are separated by a tricritical point as shown.  As discussed above, the curve for $f_J=-1$ represents  second-order transitions only.  The tricritical point is seen to move to higher termperatures with increasing values of $f_J$.

%\clearpage

\section{\label{Sec:SFPhase}  Magnetic Fields Applied along the Uniaxial Easy Axis:\hspace{2in} The Spin-Flop Phase}

\begin{figure}[t]
\includegraphics [width=2.25in]{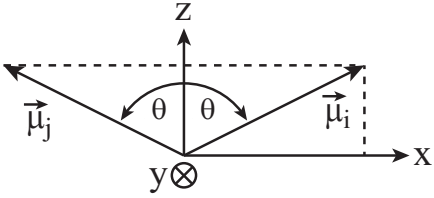}
\caption{(Color online) Geometry of two representative ordered moments~$\vec{\mu}_i$ and~$\vec{\mu}_j$ on the two sublattices in the spin-flop phase.  Both moments make equal angles~$\theta$ with respect to the $z$~axis along which the applied field~{\bf H} = $H_z\hat{\bf k}$ is aligned and have equal magnitudes~$\mu$ at a given $\theta$.}
\label{Fig:mui_muj}
\end{figure}

At sufficiently large $H_z$, the ordered moments in the collinear AFM phase aligned along the $z$~axis can flop to an approximately perpendicular orientation, resulting in a canted AFM phase with a lower free energy and a net moment along the $+z$ direction as shown in Fig.~\ref{Fig:mui_muj}.  Here we assume that the spin-flop (SF) phase is coplanar, where the ordered moments on the two sublattices are aligned within the $xz$~plane, each at an angle $\theta$ with the $z$~axis.

\subsection{Hamiltonian}

From Fig.~\ref{Fig:mui_muj}, the ordered moments on the two sublattices are described by
\bse
\label{Eqs:muimujki}
\bea
\vec{\mu}_i &=& \mu\left[\sin(\theta)\hat{\bf i} + \cos(\theta)\hat{\bf k} \right],\label{Eq:mui}\\*
\vec{\mu}_j &=& \mu\left[- \sin(\theta)\hat{\bf i} + \cos(\theta)\hat{\bf k} \right].
\eea
\ese
Substituting Eqs.~(\ref{Eq:mui}) into the general two-sublattice expression~(\ref{Eq:HexchHGen}) gives the exchange field seen by $\vec{\mu}_i$ as
\be
{\bf H}_{{\rm exch}i} = \frac{3 k_{\rm B}T_{{\rm N}J}\bar{\mu}}{g\mu_{\rm B}(S+1)}\left[\sin(\theta)\hat{\bf i}+ f_J\cos(\theta)\hat{\bf k} \right],
\label{Eq:HexchiSF3}
\ee
where the definition of $\bar{\mu}$ is given in Eqs.~(\ref{Eqs:mubarDef}).  Using
\be
\bar{\mu}_x = \bar{\mu}\sin\theta,\qquad	\bar{\mu}_z = \bar{\mu}\cos\theta,
\ee
Eq.~(\ref{Eq:HexchiSF3}) becomes
\be
{\bf H}_{{\rm exch}i} = \frac{3 k_{\rm B}T_{{\rm N}J}}{g\mu_{\rm B}(S+1)}\left(\bar{\mu}_x\hat{\bf i}+ f_J\bar{\mu}_z\hat{\bf k} \right).
\label{Eq:HexchiSF4}
\ee

Since the magnetic moment operator is $\vec{\mu}_i = -g\mu_{\rm B}{\bf S}$ where ${\bf S}$ is the spin operator for spin~$i$, the part of the Hamiltonian associated with spin~$i$ interacting with ${\bf H}_{{\rm exch}i}$ in Eq.~(\ref{Eq:HexchiSF4}) is
\bea
{\cal H}_{{\rm exch}i} &=& -\vec{\mu}_i\cdot {\bf H}_{{\rm exch}i} = g\mu_{\rm B}{\bf S}\cdot {\bf H}_{{\rm exch}i}\\*
 &=& \frac{3 k_{\rm B}T_{{\rm N}J}}{S+1} (\bar{\mu}_xS_x + f_J\bar{\mu}_zS_z).\nonumber
\eea
Using the dimensionless reduced parameters in Eqs.~(\ref{Eq:dDef}) and~(\ref{Eq:hDef}), the normalized Hamiltonian for spin~$i$ in the SF phase including the exchange field, the single-ion anisotropy and the applied field is
\bea
\frac{{\cal H}}{ k_{\rm B}T_{{\rm N}J} } &=& \frac{3\bar{\mu}_x }{S+1} S_x+ \bigg(\frac{3f_J\bar{\mu}_z}{S+1} + h_z\bigg)S_z - dS_z^2 \nonumber\\*
&& = b_xS_x + b_zS_z - dS_z^2. \label{Eqs:HSF0}
\eea
Given $S,\ f_J$ and~$d$, in general there are two unknowns $\bar{\mu}_x(t)$ and~$\bar{\mu}_z(t)$ to solve for at each~$t$ and~$h_z$.  The PM state at high~$h_z$ corresponds to $\bar{\mu}_x = 0$.  In that high-field regime, the energy eigenvalues of Hamiltonian~(\ref{Eqs:HSF0}) are identical to those already given in Eq.~(\ref{Eq:EPMi}) for the PM state.

\subsection{N\'eel Temperature in $H=0$}

Here we use the second-order perturbation theory described generically in Sec.~\ref{Sec:PerpPertThy} to calculate the reduced N\'eel temperature $t_{\rm N}$ for continuous (second-order) transitions of the SF phase versus~$d$ in $h_z=0$.  For $h_z = \bar{\mu}_z = 0$ for which $\theta=90^\circ$ in Fig.~\ref{Fig:mui_muj}, the reduced Hamiltonian~(\ref{Eqs:HSF0}) for the SF phase can be separated into unperturbed~${\cal H}_0$ and perturbed parts~${\cal H}^\prime$ as
\bse
\label{Eqs:epsSF}
\bea
\frac{{\cal H}}{ k_{\rm B}T_{{\rm N}J} } &=& \frac{{\cal H}_0}{ k_{\rm B}T_{{\rm N}J} } + \frac{{\cal H}^\prime}{ k_{\rm B}T_{{\rm N}J} }, \label{Eq:HamilSFH0}\\*
\frac{{\cal H}_0}{ k_{\rm B}T_{{\rm N}J} } &=& - dS_z^2, \nonumber\\*
\frac{{\cal H}^\prime}{ k_{\rm B}T_{{\rm N}J} } &=& b_x S_x,\nonumber
\eea
where 
\be
b_x =  \frac{3\bar{\mu}_0}{S+1}
\ee
\ese
is the reduced exchange field for AFM ordering in Eq.~(\ref{Eq:hexch0def}), assumed here to be infinitesimal.  Also $\bar{\mu}_0 \equiv \bar{\mu}_{0x}$ for the central moment~$\vec{\mu}_i$ under consideration that points in the $+x$ direction.

For $t\to t_{\rm N}^-$, $\bar{\mu}_0$ becomes infinitesimally small, as assumed in the present  perturbation theory treatment, and hence one can set $t=t_{\rm N}\equiv T_{\rm N}/T_{{\rm N}J}$ in this limit.  To first order in $\bar{\mu}_0$, for integer spins Eqs.~(\ref{Eqs:muPerpIntSpins}) yield the expression from which $t_{\rm N}$ can be numerically solved for, given by
\bse
\label{Eqs:tNSF}
\bea
&&\hspace{-0.6in} 1 = \frac{3}{dS(S+1) Z_S}\sum_{m_S=-S}^S\left[\frac{S(S+1)+m_S^2}{4m_S^2-1}\right] e^{dm_S^2/t_{\rm N}},\nonumber\\
Z_S &=& \sum_{m_S=-S}^S e^{dm_S^2/t_{\rm N}} \qquad ({\rm integer\ spins}),\label{Eq:tNSFH0}
\eea
where a multiplicative factor of $\bar{\mu}_0$ on both sides of the top equation has been divided out.  Using Eqs.~(\ref{Eqs:muxbarFx4}), $t_{\rm N}$ can be calculated for half-integer spins by solving for it in the expression
\bea
1 &=& \frac{3}{S(S+1)Z_S}\Bigg\{\frac{2}{d}\sum_{m_S = 3/2}^S \left[\frac{S(S+1)+m_S^2}{4m_S^2-1}\right]e^{dm_S^2/t_{\rm N}}\nonumber\\*
&&+\ \frac{e^{d/4t_{\rm N}}}{2}\left[\frac{S(S+1)+1/4}{t_{\rm N}}- \frac{S(S+1)-3/4}{d}\right]\bigg\}\nonumber\\*
&&\hspace{0.7in} ({\rm half\ integer\ spins}).
\eea
\ese
For numerical calculations of~$t_{\rm N}$ we used the {\tt FindRoot} utility of {\tt Mathematica}.

\begin{figure}
\includegraphics [width=3.3in]{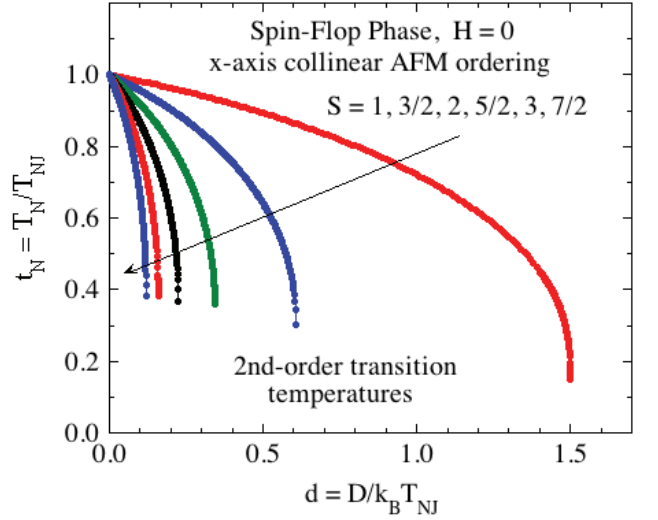}
\caption{(Color online) Reduced transition temperature~$t_{\rm N}$ of the spin-flop phase versus reduced anisotropy parameter~$d$ calculated from Eqs.~(\ref{Eqs:tNSF}) for the listed spin values.  These data give the $t_{\rm N}$ and $d$~ranges for second-order transitions of $\bar{\mu}_0$ versus temperature.  The missing part of each curve gives the $t_{\rm N}$ range for first-order transitions [see Fig.~\ref{Fig:mu0barSFS1_vs_t_d}(a) for $S=1$].}
\label{Fig:tN_SF_vs_d_S1.2.3}
\end{figure}

One sees from Eqs.~(\ref{Eqs:tNSF}) that $t_{\rm N}$ of the SF phase in $H=0$ only depends on $S$ and $d$ and not on $f_J$.  From its derivation, the $t_{\rm N}$ obtained from Eqs.~(\ref{Eqs:tNSF}) is for continuous (second-order) transitions only.  Plots of $t_{\rm N}$ versus~$d$ for $S = 1$ to~7/2 in 1/2 increments obtained using Eqs.~(\ref{Eqs:tNSF}) are shown in Fig.~\ref{Fig:tN_SF_vs_d_S1.2.3}.  All data sets have the correct limit $t_{\rm N}(d\to0) = 1$.  One also sees that second-order transitions only occur for $d$ values below an $S$-dependent maximum value to which a minimum $t_{\rm N}$ corresponds.  This feature is reflected in plots of $\bar{\mu}_0(t)$ in Fig.~\ref{Fig:mu0barSFS1_vs_t_d}(a) below which show first-order transitions versus~$t$ for $S=1$ with $d\geq3/2$ (cf.~Fig.~\ref{Fig:tN_SF_vs_d_S1.2.3}).  One also sees that with $d>0$, $t_{\rm N}$ is suppressed with respect to the value for $d=0$.  This is opposite to the behavior for AFM ordering along the $z$~axis, for which $d>0$ increases the N\'eel temperature.  Related to this feature, the stable phase for $H=0$ is shown later to be the AFM phase for all $t$; i.e., the SF phase is unstable at all temperatures in $H=0$ as would have been anticipated.
  
\subsection{Ordered Moment versus Temperature in Zero Field}

For $h_z=\bar{\mu}_z=0$ the reduced Hamiltonian for the SF phase is again given by Eq.~(\ref{Eq:HamilSFH0}), but where here $\bar{\mu}_0$ is not assumed to be small so perturbation theory cannot be used to calculate it.  The $2S+1$ eigenenergies of the nondiagonal Hamiltonian are labeled $\epsilon_n$. Using Eq.~(\ref{Eq:barmuOp}), the magnetic moment operator is given by
\be
\bar{\mu}_{0n}^{\rm op} = -\frac{1}{S}\frac{\partial \epsilon_n}{\partial h_{\rm exch0}} = -\left(\frac{S+1}{3S}\right)\frac{\partial\epsilon_n}{\partial \bar{\mu}_0}.
\ee
The thermal-average $\bar{\mu}_0(t)$ is obtained by solving the self-consistency equation
\bse
\label{Eqs:mubarH0SF}
\bea
\bar{\mu}_0 &=& -\frac{S+1}{3SZ_S}\sum_{n=1}^{2S+1}\frac{\partial\epsilon_n}{\partial \bar{\mu}_0}e^{-\epsilon_n/t},\\*
Z_S &=& \sum_{n=1}^{2S+1} e^{-\epsilon_n/t},
\eea
\ese
where $\bar{\mu}_0$ on the right sides of these equations is contained in the each of the $2S+1$ expressions for $\epsilon_n$.  Equations~(\ref{Eqs:mubarH0SF}) are valid for both integer and half-integer spins.

\begin{figure}
\includegraphics [width=3.3in]{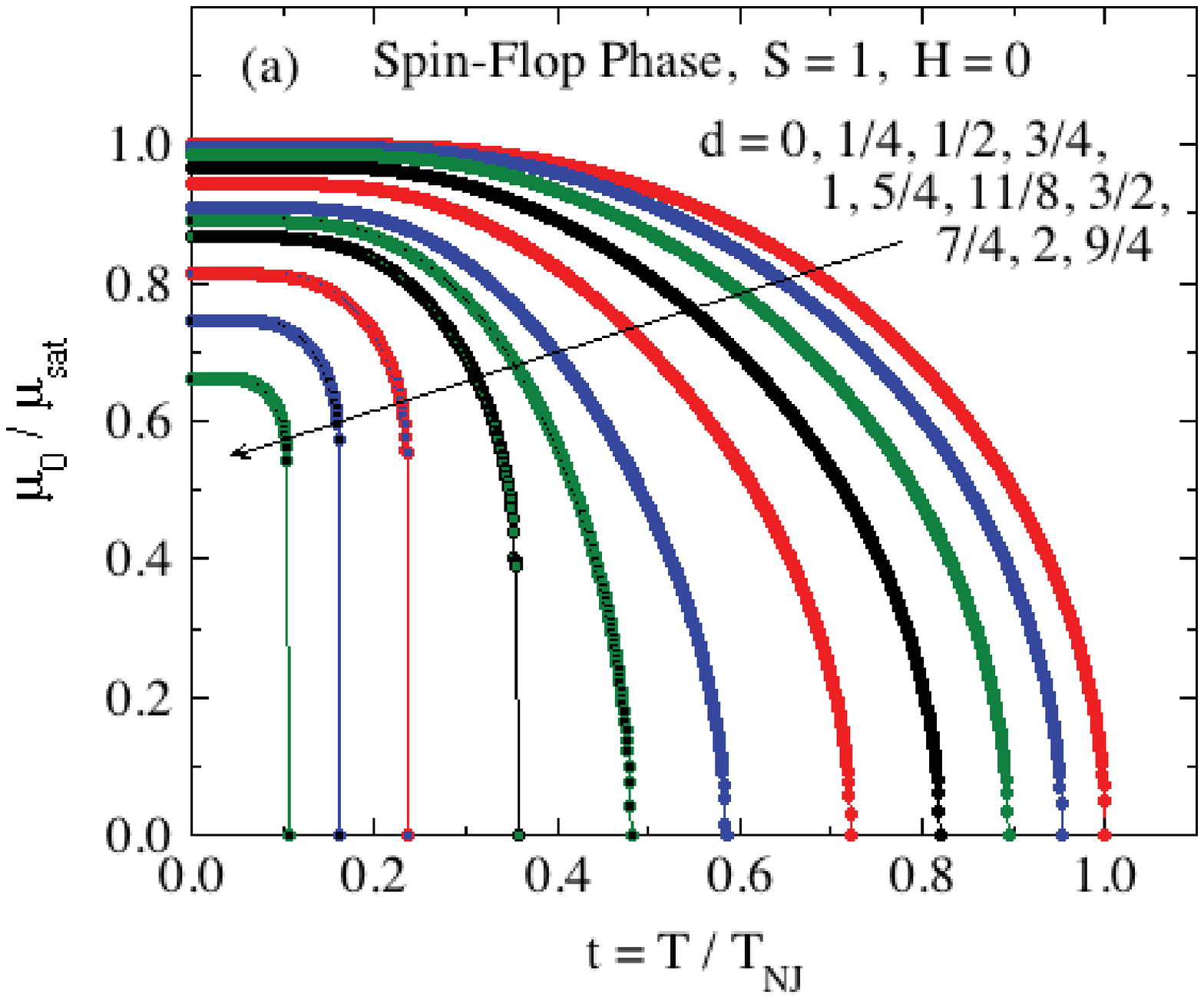}
\includegraphics [width=3.3in]{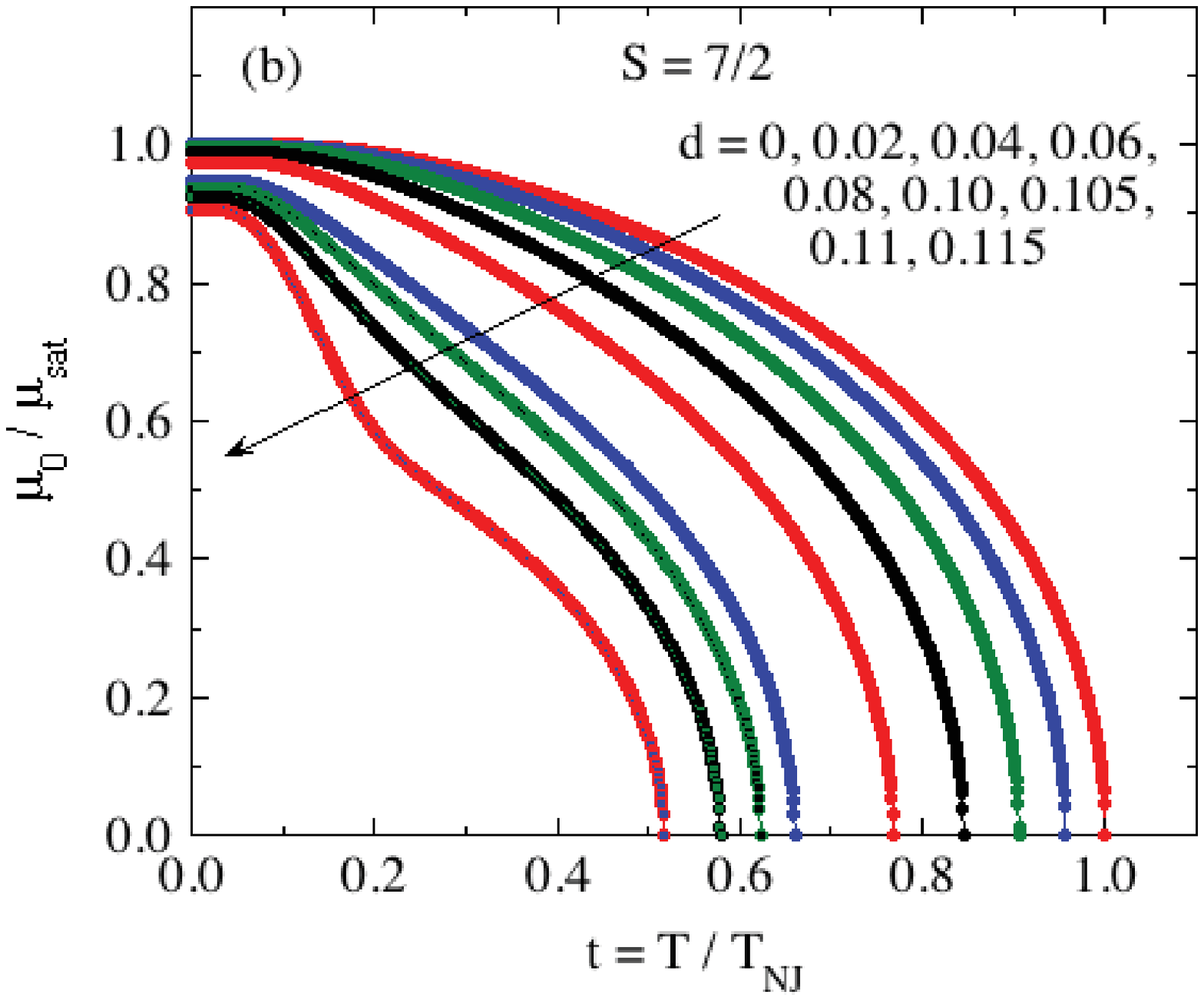}
\caption{(Color online) Reduced ordered moment $\bar{\mu}_0$ of the spin-flop phase in zero field for (a)~$S=1$ and (b)~$S=7/2$ for the listed values of reduced anisotropy parameter~$d$, calculated from Eqs.~(\ref{Eqs:mubarH0SF}).  Several transitions in~(a) are seen to be first order for sufficiently large~$d$, consistent with Fig.~\ref{Fig:tN_SF_vs_d_S1.2.3}. The $d$ values for $S = 7/2$ in~(b) are small enough that all transitions shown are second order (cf.~Fig.~\ref{Fig:tN_SF_vs_d_S1.2.3}).}
\label{Fig:mu0barSFS1_vs_t_d}
\end{figure}

Shown in Fig.~\ref{Fig:mu0barSFS1_vs_t_d} are plots of $\bar{\mu}_0$ versus reduced temperature~$t$ for $S=1$ and $S=7/2$ and several values of reduced anisotropy parameter~$d$ as listed.  For $S=1$, plots with $d\geq3/2$ are included for which no second-order transition exists for which $\bar{\mu}_0$ goes continuously to zero at the N\'eel temperature according to Fig.~\ref{Fig:tN_SF_vs_d_S1.2.3}.  Thus for these values of~$d$ the transitions are first order.  Furthermore, for $d>0$, the ordered moment at $t=0$ is less than unity.  This occurs because the ground state energy level has negative curvature (see Fig.~\ref{Fig:EofHxS1S72d1} in the Appendix), and because the exchange field at $t=0$ is finite.

\subsection{\label{Sec:MofHzSF} High-Field Magnetization}

Using the full reduced spin Hamiltonian~(\ref{Eqs:HSF0}) and the magnetic moment operators 
\bse
\bea
\bar{\mu}_x^{\rm op} &=& -\frac{1}{S}\frac{\partial\epsilon_n}{\partial b_x} = -\frac{S+1}{3S}\frac{\partial\epsilon_n}{\partial \bar{\mu}_x},\\*
\bar{\mu}_z^{\rm op} &=& -\frac{1}{S}\frac{\partial\epsilon_n}{\partial b_z}\Big|_{b_z = 3f_J\bar{\mu}_z/(S+1)+h_z},
\eea
\ese
the thermal-average values of~$\bar{\mu}_x$ and~$\bar{\mu}_z$ are calculated for each $t$ and $h_z$ by solving the two simultaneous equations
\bse
\label{Eqs:muxmuzSF}
\bea
\bar{\mu}_x &=& -\frac{1}{SZ_S}\sum_{n=1}^{2S+1}\frac{\partial\epsilon_n}{\partial b_x}e^{-\epsilon_n/t} \label{muxEqn}\\*
&=& -\frac{S+1}{3SZ_S}\sum_{n=1}^{2S+1}\frac{\partial\epsilon_n}{\partial \bar{\mu}_x}e^{-\epsilon_n/t},\nonumber\\*
\bar{\mu}_z &=& -\frac{1}{SZ_S}\sum_{n=1}^{2S+1}\frac{\partial\epsilon_n}{\partial b_z}\Big|_{b_z = h_z + 3\frac{f_J\bar{\mu}_z}{S+1}}e^{-\epsilon_n/t},\label{muzEqn}\\*
Z_S &=& \sum_{n=1}^{2S+1}e^{-\epsilon_n/t}.\nonumber
\eea
\ese
These two equations for $\bar{\mu}_x$ and~$\bar{\mu}_z$ were solved iteratively for given values of $S,\ f_J,\ d,\ t$ and $h_z$.  First a starting value of $\bar{\mu}_x\sim 1$ was inserted into Eq.~(\ref{muzEqn}), and $\bar{\mu}_z$ solved for.  This value of $\bar{\mu}_z$ was inserted into Eq.~(\ref{muxEqn}) and $\bar{\mu}_x$ solved for.  This procedure was iterated until the difference in each variable in subsequent iterations was less than $10^{-10}$.

\begin{figure}
\includegraphics [width=3.3in]{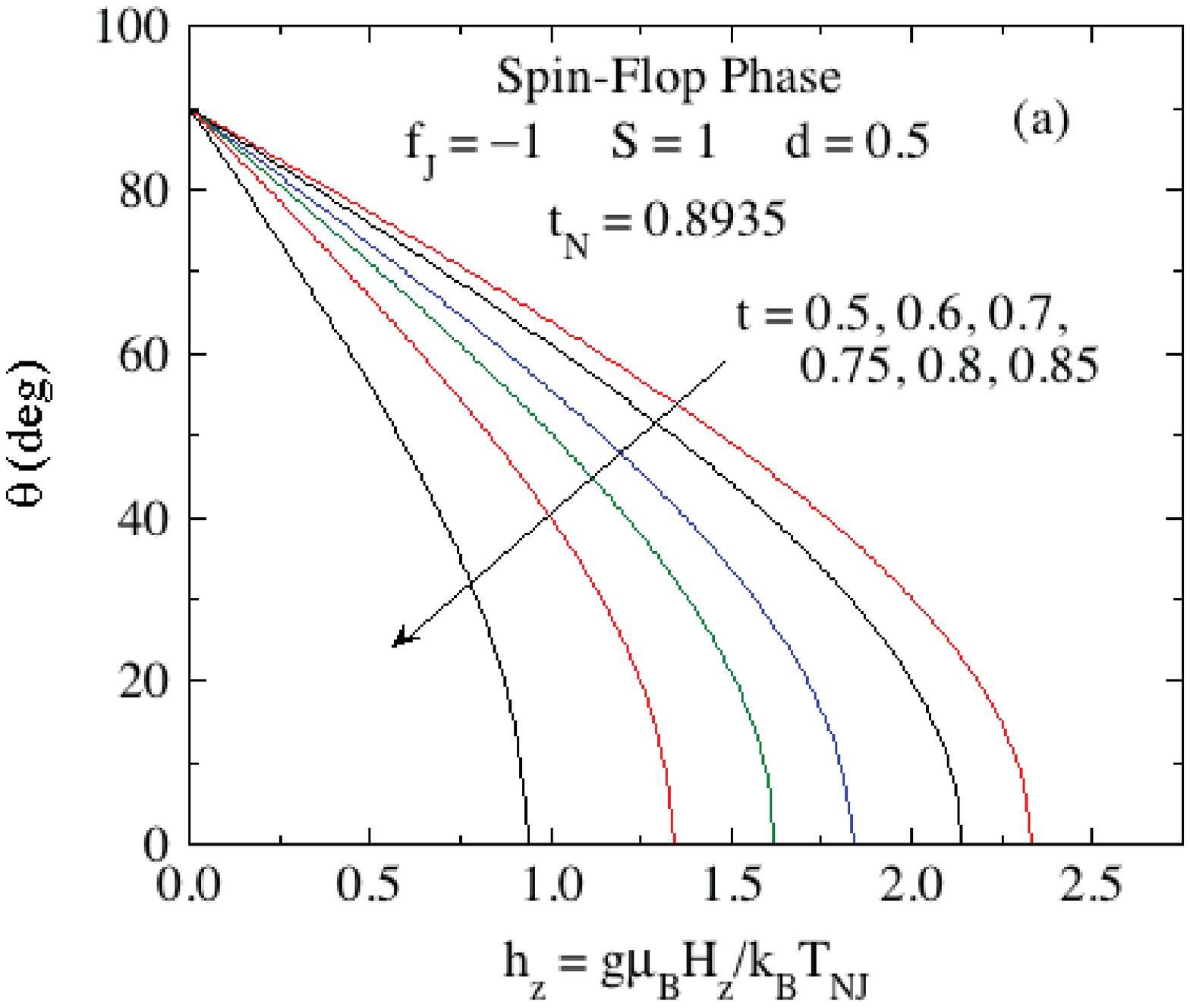}
\includegraphics [width=3.3in]{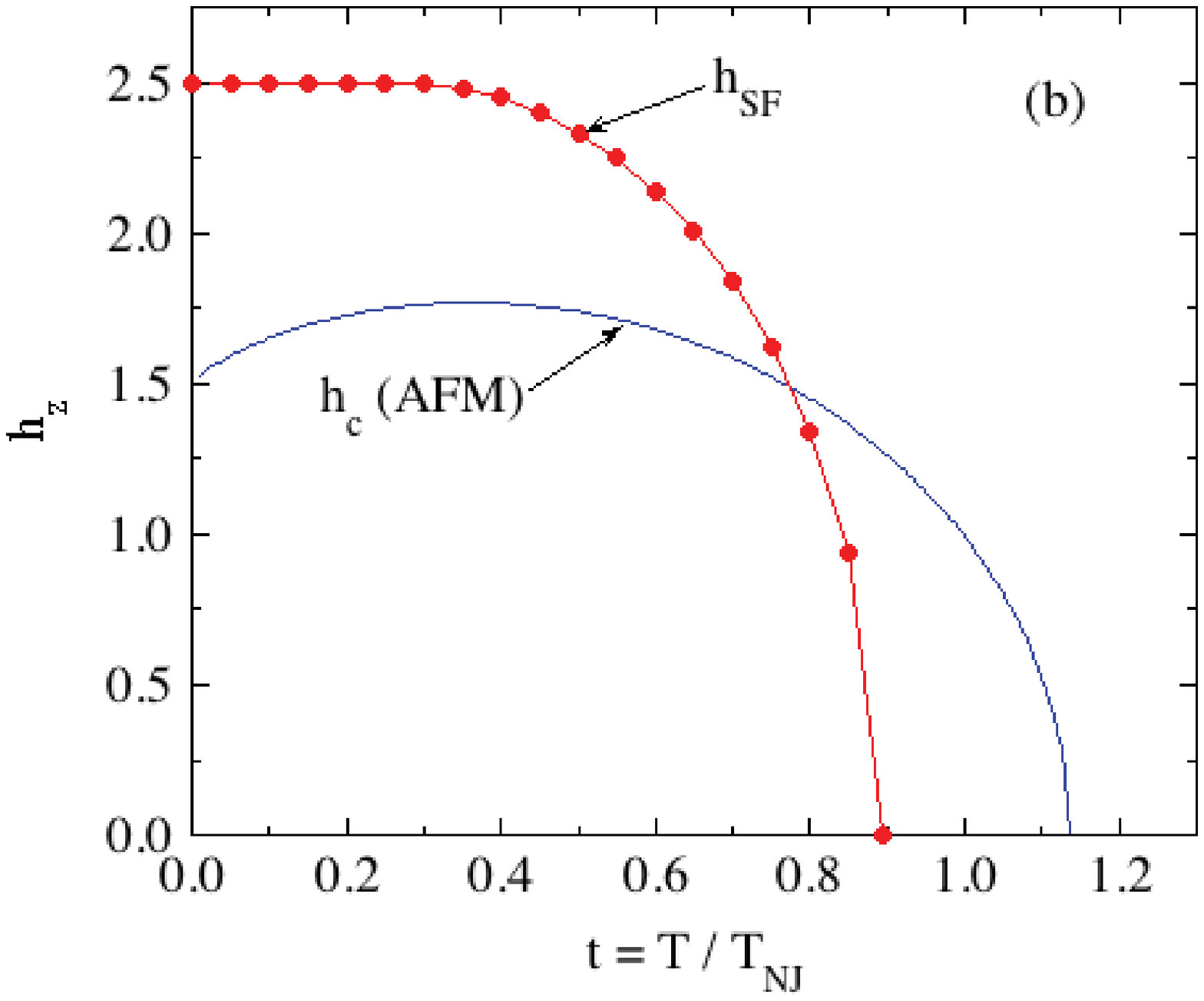}
\caption{(Color online) (a)~Angle~$\theta$ between an ordered moment in the spin-flop phase and the $z$~axis versus~$h_z$ for $S=1$, $f_J=-1$ and~$d=0.5$ for six values of the reduced temperature~$t$.  (b)~Spin-flop ransition field $h_{\rm SF}$ between the spin-flop (SF) and paramagnetic (PM) phases versus~$t$ for $S=1$, $f_J=-1$ and~$d=0.5$.  This transition field is the field at which $\theta\to0$ with increasing~$h_z$ such as obtained from the data in~(a). The data in (a) and~(b) were calculated using Eqs.~(\ref{Eqs:muxmuzSF}).  Also shown in (b) is the AFM critical field $h_{\rm c}$ versus~$t$ for the same parameters, obtained from Fig.~\ref{Fig:HcAFMS1fJm1d0To1}(a).}
\label{Fig:hzSFt0p1To1p2d50}
\end{figure}

\begin{figure}
\includegraphics [width=3.3in]{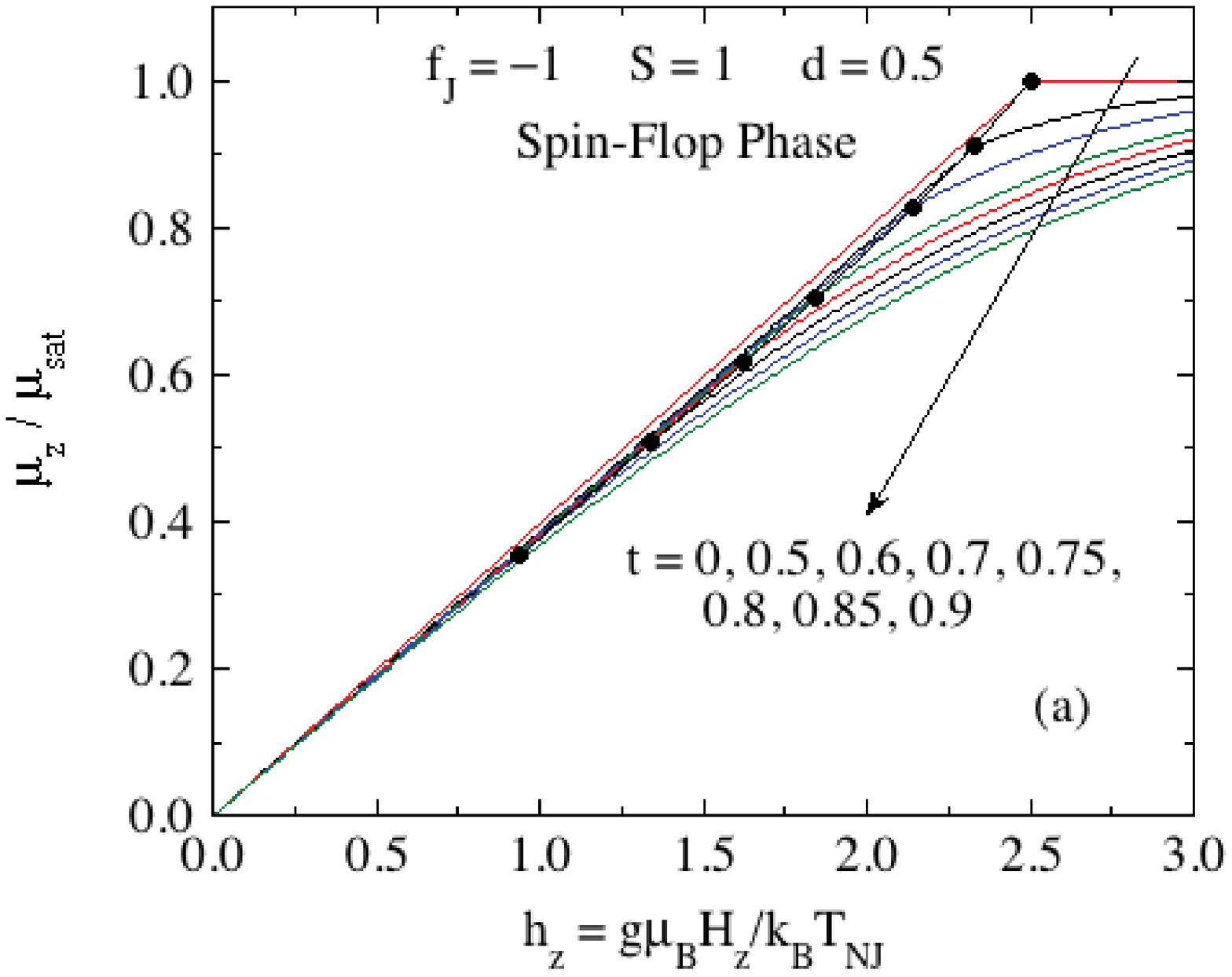}
\includegraphics [width=3.3in]{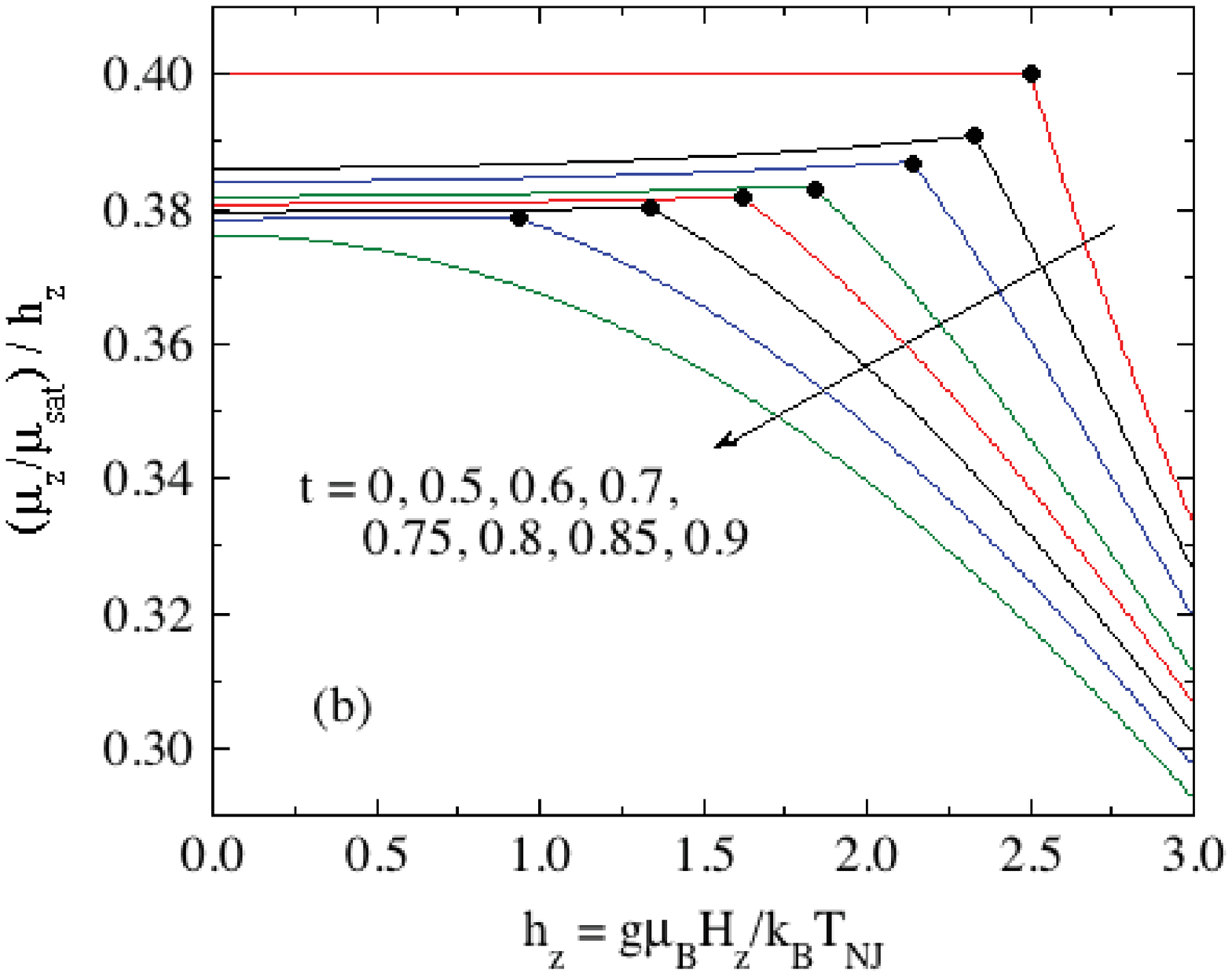}
\caption{(Color online) (a) Reduced ordered moment $\bar{\mu}_z$ of the SF phase versus reduced field~$h_z$ along the $z$~axis for $S=1$, $f_J=-1$ and~$d=0.5$ at the reduced temperatures $t$ indicated.  Also shown as filled black circles are the SF to PM transition fields~$h_{\rm SF}$ for the respective $t$ values from Fig.~\ref{Fig:hzSFt0p1To1p2d50}. For $t\geq t_{\rm N}(h_z=0) = 0.8935$ the system is in the PM state for all~$h_z$. (b)~Chordal slope $\bar{\mu}_z/h_z$ versus $h_z$ obtained from the data in~(a).  The SF to PM phase transition at each~$t$ is characterized by a discontinuity in $\bar{\mu}_z/h_z$ versus~$t$, again marked by a filled black circle for each~$t$ shown.  The temperature $t=0.9$ is slightly above $t_{\rm N}(h_z=0)$ so there is no transition versus $h_z$  for this~$t$. }
\label{Fig:hzSFt0p1To1p2d50muzbar}
\end{figure}

Shown in Fig.~\ref{Fig:hzSFt0p1To1p2d50}(a) are plots of $\theta = \arctan(\bar{\mu}_x/\bar{\mu}_z)$ in Fig.~\ref{Fig:mui_muj} versus reduced field $h_z$ calculated using Eqs.~(\ref{Eqs:muxmuzSF}) for different reduced temperatures~$t$ with $S=1,\ f_J=-1$ and $d=0.5$.  For each~$t$ one sees a second-order transition at which $\theta(t)\to0$ at the reduced spin-flop field $h_z \equiv h_{\rm SF}(t)$.  The $h_{\rm SF}$ for $S=1,\ f_J=-1$ and $d=0.5$ is plotted versus~$t$ in Fig.~\ref{Fig:hzSFt0p1To1p2d50}(b).  Also shown in Fig.~\ref{Fig:hzSFt0p1To1p2d50}(b) is the AFM critical field $h_{\rm c}$ versus~$t$ for the same parameters, obtained from the data in Fig.~\ref{Fig:HcAFMS1fJm1d0To1}(a). The crossover between these two curves in Fig.~\ref{Fig:hzSFt0p1To1p2d50}(b) occurs in part because a given value of $d>0$ suppresses the $t_{\rm N}$ of the SF phase below unity whereas it increases the $t_{\rm N}$ of the AFM phase above unity.

The normalized thermal-average moment $\bar{\mu}_z\equiv \mu_z/\mu_{\rm sat}$ for the SF phase calculated using Eqs.~(\ref{Eqs:muxmuzSF}) is plotted versus $h_z$ in Fig.~\ref{Fig:hzSFt0p1To1p2d50muzbar}(a) for $S=1$, $f_J=-1$ and~$d=0.5$ at the reduced temperatures~$t$ indicated.  The slopes of $\bar{\mu}(h_z)$ in the SF state for given values of $f_J$, $S$ and~$d$ at $t < t_{\rm N}$ are seen to be field and temperature dependent. The black filled circles are the SF to PM transition fields~$h_{\rm SF}$ for the respective temperatures.  At these values of $h_z$, there are discontinuities in the slopes of $\bar{\mu}_z$ versus~$t$, indicative of the second-order nature of the SF--PM transition as shown more clearly in the chordal slope $\bar{\mu}_z/h_z$ versus~$t$ data in Fig.~\ref{Fig:hzSFt0p1To1p2d50muzbar}(b).  

\section{\label{Sec:PhaseDiagram} Magnetic Fields Applied along the Uniaxial Easy Axis: Phase Diagrams}

Which of the AFM, SF and PM phases at a given temperature and field is more stable is determined by which phase has the lowest free energy.  Here we calculate the reduced free energies~$f_{\rm mag}$ versus reduced $z$-axis field~$h_z$ at a number of reduced temperatures~$t$ for each of these phases for the same parameters $S=1$, $d=0.5$ and $f_J=-1$.  The free energy of the PM phase appears as part of the calculations of those of the AFM and SF phases versus $t$ and~$h_z$.

In order to calculate the partition function $Z_S$ for the AFM phase one must first calculate the $t$-dependent energy eigenvalues using the $t$-dependent values of $\bar{\mu}_{1z}$ and $\bar{\mu}_{2z}$ from Eqs.~(\ref{Eqs:mu1zmu2z}) such as those plotted in Fig.~\ref{Fig:HzAFMS1fJm1d50mu}.  The reduced energy eigenvalues of the two sublattices~1 and~2 versus the respective spin magnetic quantum numbers $m_{S1}$ and $m_{S2}$ of sublattices~1 and~2 are 
\be
\epsilon(m_{S1},m_{S2}) = (h_{{\rm exch}1z} + h_z)m_{S1} + (h_{{\rm exch}2z} + h_z)m_{S2}
\ee
where the reduced exchange fields are given in Eqs.~(\ref{Eqs:Hexchiz12}).  Since $m_{S1}$ and $m_{S2}$ are independent of each other, the energy of a pair of spins with one spin on each sublattice is
\bea
\epsilon(m_{S1},m_{S2}) &=& \epsilon_1(m_{S1}) + \epsilon_2(m_{S2}),\\*
Z_S(t,h_z) &=& Z_{S1}(t,h_z)Z_{S2}(t,h_z).\nonumber
\eea
The average free energy per spin is then obtained from Eq.~(\ref{Eq:fmagFromZ}) as
\be
f_{\rm mag}(t,h_z) = -\frac{1}{2}t\ln Z_S(t,h_z).
\ee
For the SF phase, the reduced Hamiltonian is given in Eq.~(\ref{Eqs:HSF0}), where $\bar{\mu}_x(h_z,t)$ and~$\bar{\mu}_z(h_z,t)$ are determined by solving Eqs~(\ref{Eqs:muxmuzSF}) such as shown for $\bar{\mu}_z(h_z,t)$ in Fig.~\ref{Fig:hzSFt0p1To1p2d50muzbar}(a).  One inserts these values into Eq.~(\ref{Eqs:HSF0}) and diagonalizes the Hamiltonian to obtain the $t$- and $h_z$-dependent energy eigenvalues.  Using these, one then calculates the partition function and then $f_{\rm mag}(h_z,t)$.  

\begin{figure}[t]
\includegraphics [width=3.5in]{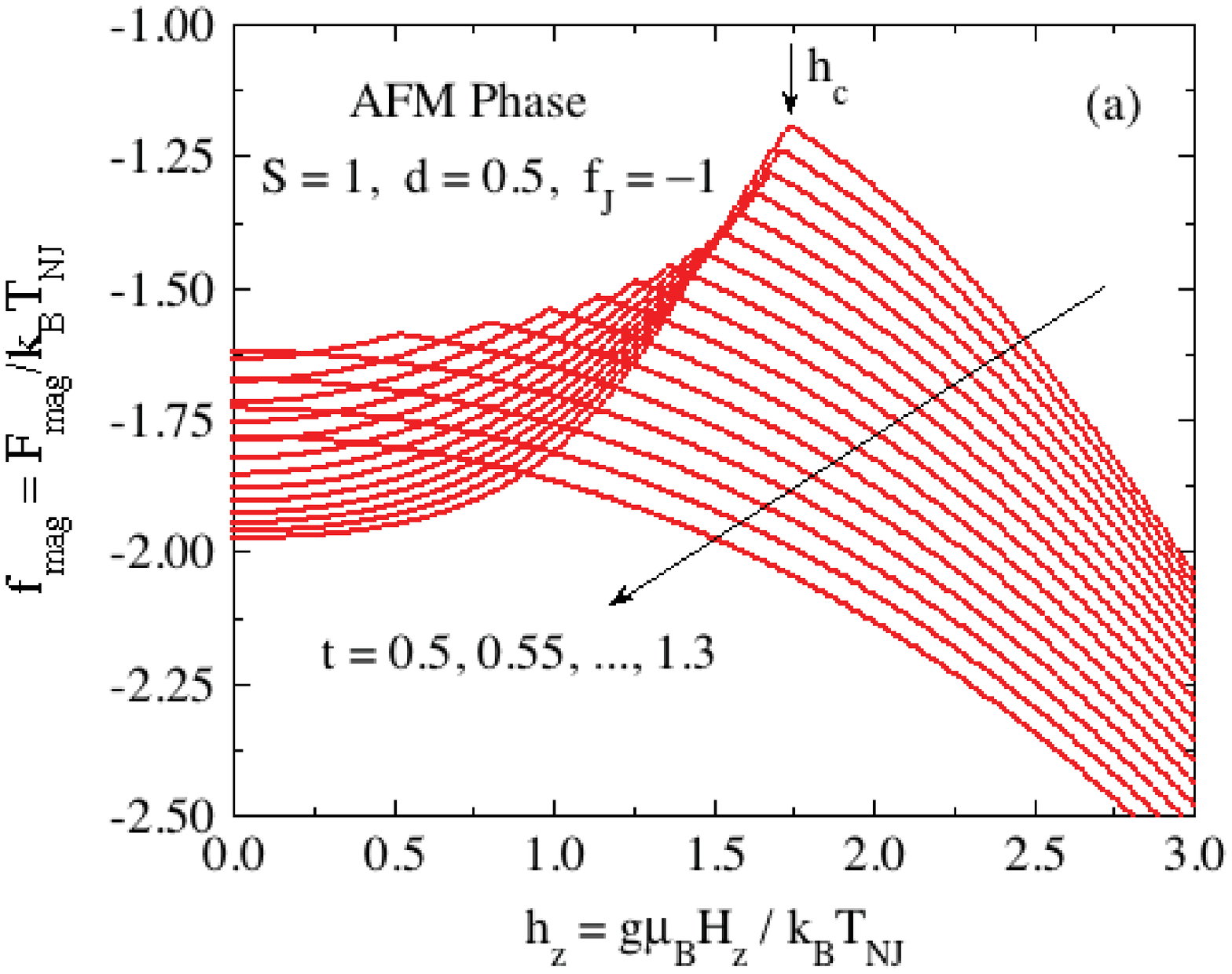}
\includegraphics [width=3.5in]{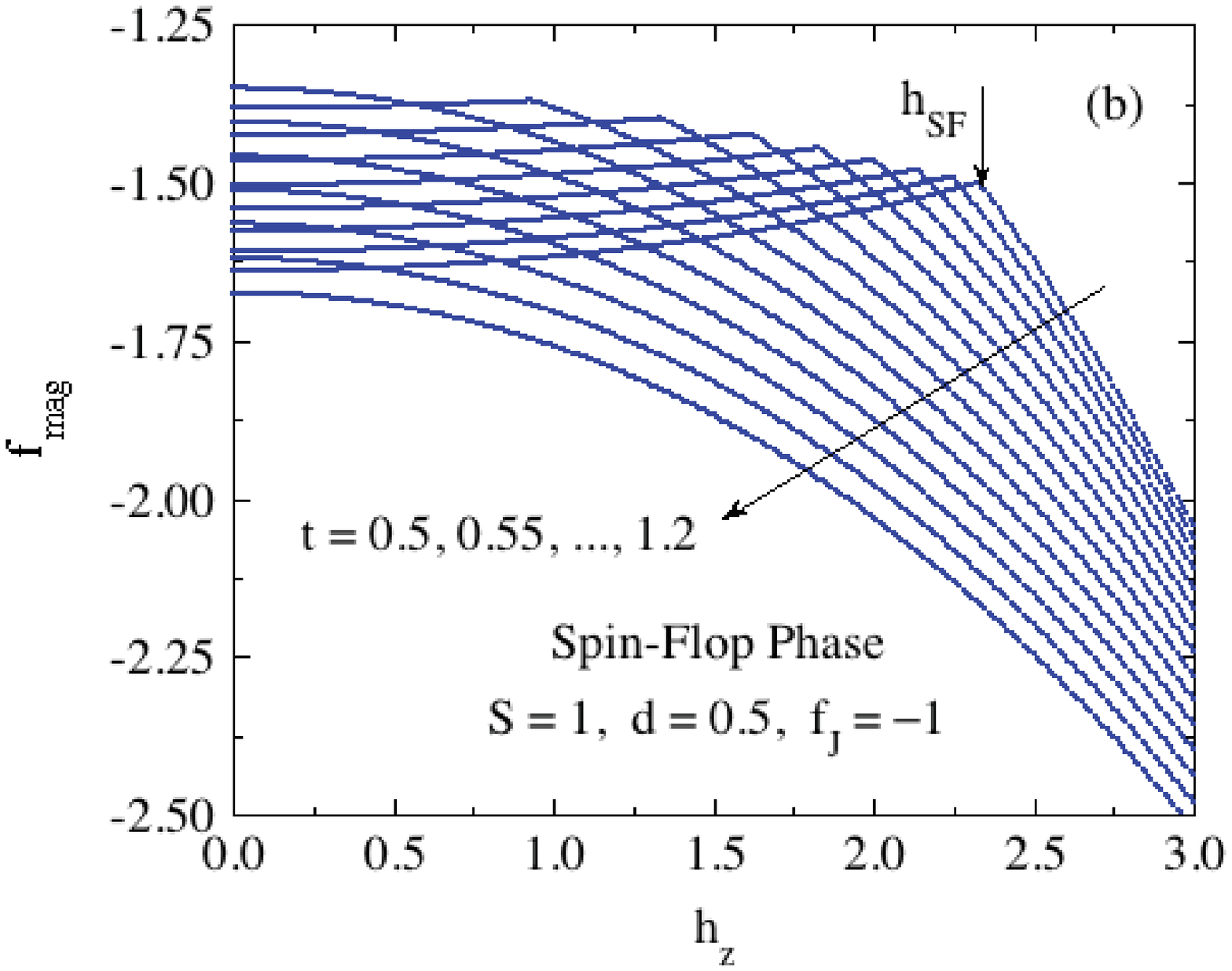}
\caption{(Color online) Reduced magnetic free energy~$f_{\rm mag}$ versus reduced applied field~$h_z$ at the reduced teperatures~$t$ indicated for the (a)~the antiferromagnetic phase and (b)~spin-flop phase with $S=1$, $d=0.5$ and $f_J=-1$.  The cusps in the data in~(a) occur at the critical fields $h_{\rm c}$ and those in~(b) occur at the spin-flop transition fields $h_{\rm SF}$ in Fig.~\ref{Fig:hzSFt0p1To1p2d50}, as indicated in the figures for $t=0.5$, respectively.}
\label{Fig:FmagSFS1fJm1d50}
\end{figure}

\begin{figure}
\includegraphics [width=3.3in]{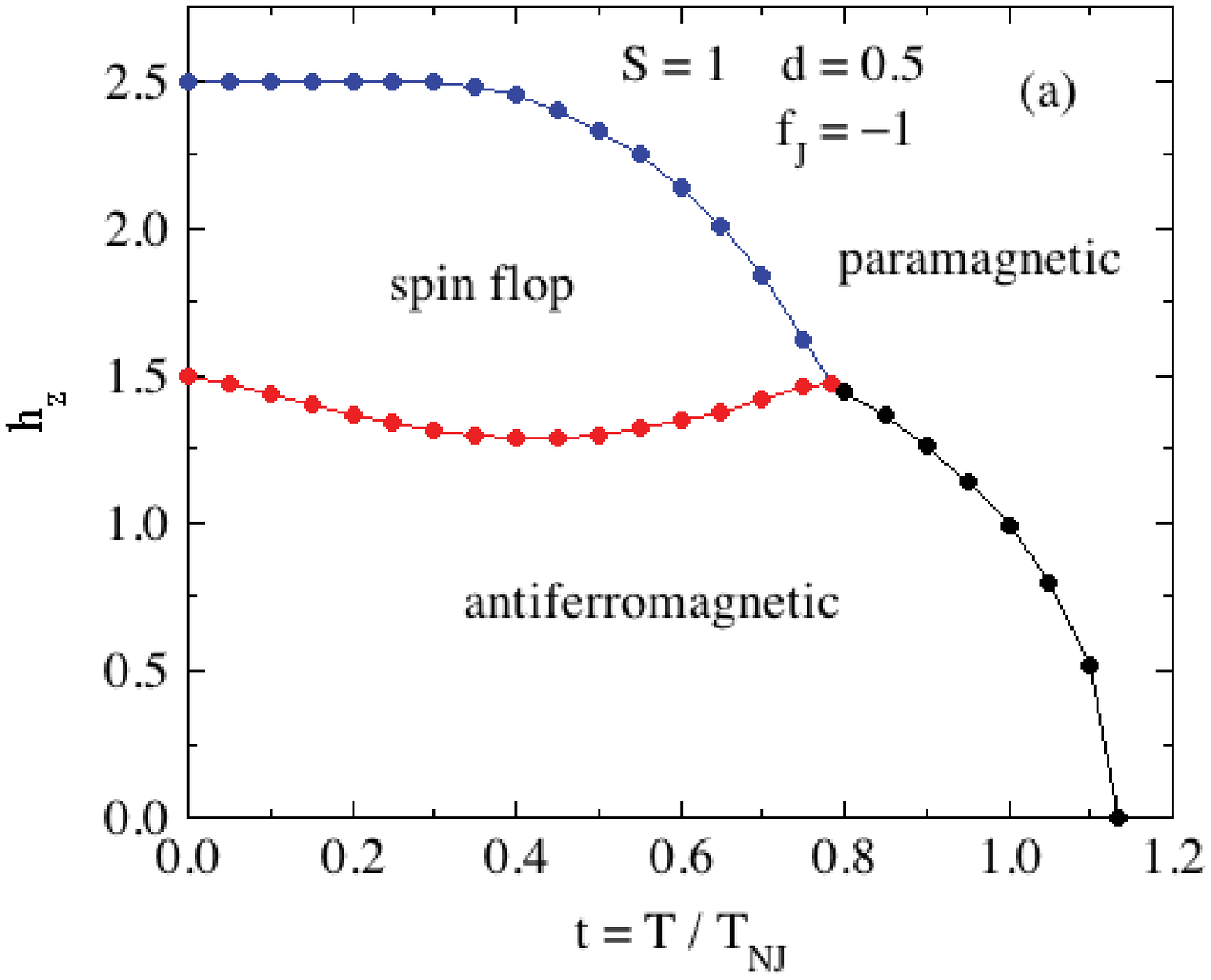}\vspace{-0.1in}
\includegraphics [width=3.3in]{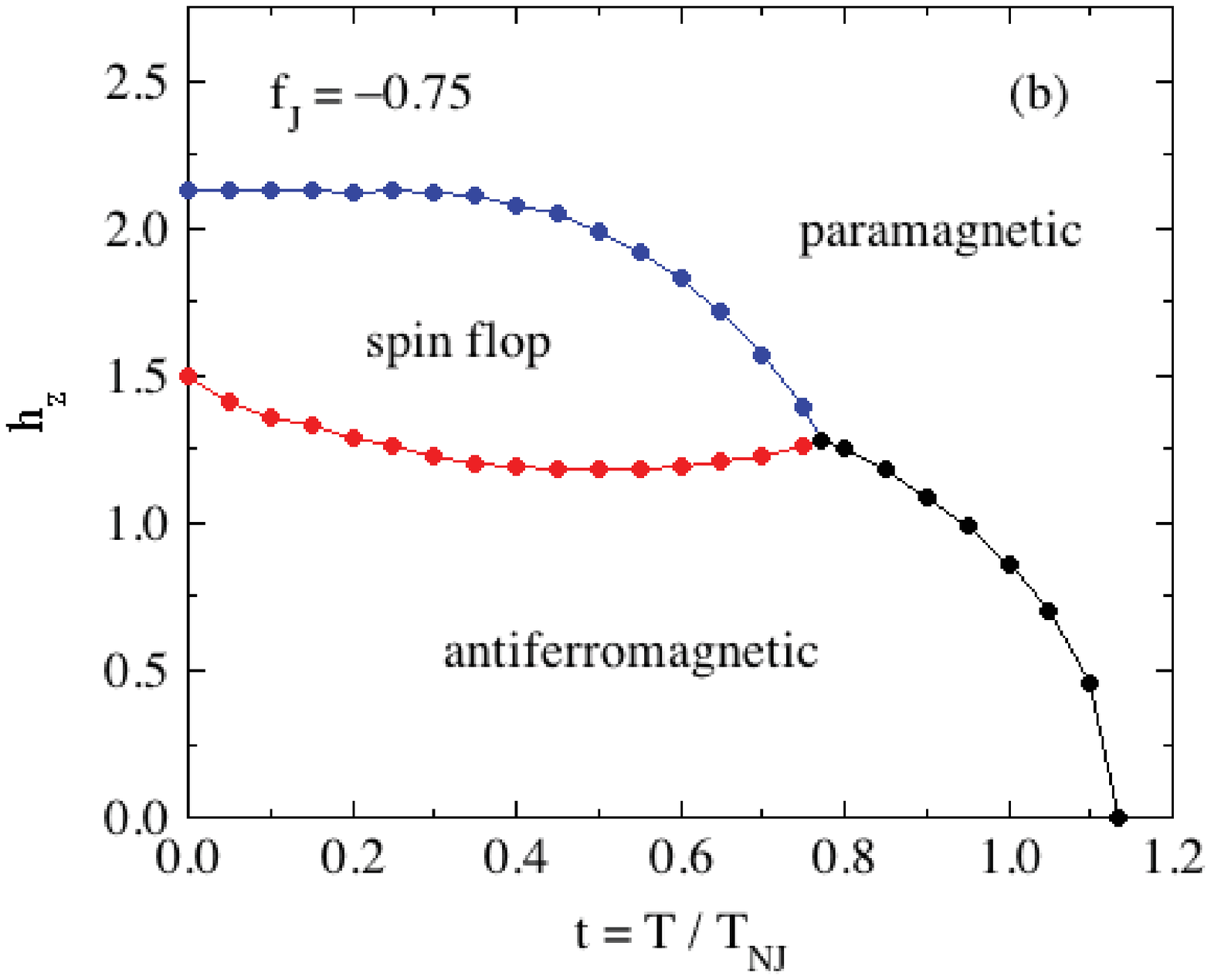}\vspace{-0.1in}
\includegraphics [width=3.3in]{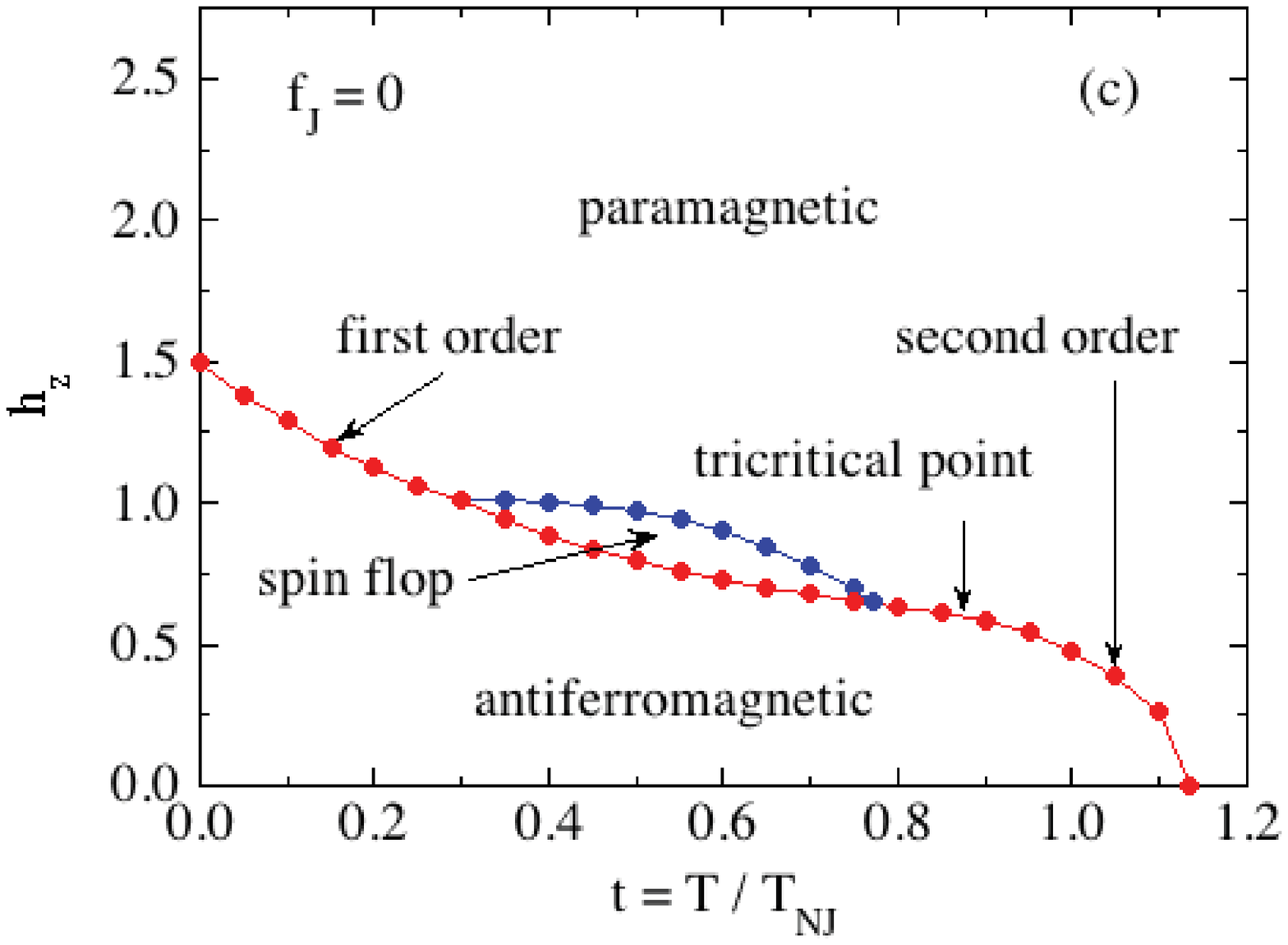}\vspace{-0.1in}
\caption{(Color online) Phase diagram for collinear antiferromagnetic ordering along the $z$~axis versus reduced temperature~$t$ and applied magnetic field~$h_z$ for spins~$S=1$ with reduced anisotropy parameter $d = D/k_{\rm B}T_{{\rm N}J}=0.5$ and parameter $f_J = \theta_{{\rm p}J}/T_{{\rm N}J}$ values of (a)~$-1$, (b)~$-075$ and (c)~0.  The phases in competition are the antiferromagnetic (AFM) phase, the paramagnetic (PM) phase and the spin-flop~(SF) phase.  The AFM/SF transition is intrinsically first order and the SF/PM and AFM/PM transitions are both second order for $f_J=-1$ and~$-0.75$.  For $f_J=0$ one sees a reentrant spin-flop phase bubble and a tricritical point in the high-field region of the AFM/PM transition curve (cf.~Fig.~\ref{Fig:HcAFMS1d50fJ0T0m1}).}
\label{Fig:PhaseDiagramS1d50}
\end{figure}

The $f_{\rm mag}$ for the AFM and SF phases versus $h_z$ were calculated for $S=1$, $d=0.5$ and $f_J=-1$ at various reduced temperatures~$t$ as described above.  Some of the results are shown for the AFM and SF phases in Figs.~\ref{Fig:FmagSFS1fJm1d50}(a) and~\ref{Fig:FmagSFS1fJm1d50}(b), respectively.  By finding which of the AFM, SF or PM phases is stable versus $h_z$ and~$t$ the phase diagram was constructed as shown in Fig.~\ref{Fig:PhaseDiagramS1d50}(a).  The upper boundary of the SF phase is part of the $h_{\rm SF}(t)$ curve in Fig.~\ref{Fig:hzSFt0p1To1p2d50}(b) and the phase boundary to the right of the AFM phase region is part of the $h_{\rm c}(t)$ curve in the same figure.  The AFM/PM and SF/PM transitions are inferred from our calculations to be thermodynamically of second-order because the free energy difference between them changes continuously on crossing the respective phase transition curve versus~$h_z$ at fixed~$t$.  On the other hand, the intrinsic first-order nature of the AFM/SF transition is manifested by a discontinuous change in the free energy on traversing the transition curve versus field.  The phase diagram is qualitatively similar to phase diagrams from the literature for fields applied parallel to the easy axis of a collinear Heisenberg antiferromagnet with uniaxial anisotropy where no first-order phase transitions occur between the AFM and PM phases \cite{Gorter1956, Stryjewski1977, Carlin1980}.  The XXZ model with uniaxial anistropy in spin space shows similar phase diagrams \cite{Selke2010, Selke2011}.

We also calculated the phase diagrams for $S=1$, $d=0.5$ and two values of $f_J>-1$ in the same manner as for $f_J=-1$.  This increase in $f_J = \theta_{{\rm p}J}/T_{{\rm N}J}$ from~$-1$ corresponds to including ferromagnetic interactions between a representive spin and its neighbors.  The phase diagram for $f_J=-0.75$ shown in Fig.~\ref{Fig:PhaseDiagramS1d50}(b) is similar to that for $f_J=-1$ in Fig.~\ref{Fig:PhaseDiagramS1d50}(a) but with shifted transition curves.  On the other hand, the phase diagram for $f_J=0$ shown in Fig.~\ref{Fig:PhaseDiagramS1d50}(c) has new features.  First, the AFM/PM transition curve at fields above the SF phase region exhibits a tricritical point as already discussed with respect to Fig.~\ref{Fig:HcAFMS1d50fJ0T0m1}.  Second, the spin-flop phase is reentrant, appearing with decreasing field and then disappearing at a lower field, resulting in a topological change to a spin-flop bubble in the phase diagram.  The AFM/PM phase transitions are first order at all fields below the tricritical point including fields lower than the minimum field for stability of the SF phase.

\section{\label{Sec:HPerp} Magnetic Fields applied Perpendicular to the Easy Axis}
When a field is applied along the $x$~axis, perpendicular to the easy $z$~axis for $D>0$, in the AFM state below $T_{\rm N}(d)$ the ordered moments tilt towards the applied field as shown in Fig.~\ref{Fig:ChiPerp}.  According to Fig.~\ref{Fig:ChiPerp},
\bse
\label{Eqs:muimujperp}
\bea
\vec{\mu}_i &=& \mu[\sin(\theta)\hat{\bf i} + \cos(\theta)\hat{\bf k}],\label{Eq:vecmui}\\*
\vec{\mu}_j &=& \mu[\sin(\theta)\hat{\bf i} - \cos(\theta)\hat{\bf k}],
\eea
\ese
where $\mu$ is the thermal-average magnitude of both $\vec{\mu}_i$ and $\vec{\mu}_j$.  Inserting Eqs.~(\ref{Eqs:muimujperp}) into~(\ref{Eq:HexchHGen}) and using the definitions $\bar{\mu} = \mu/gS\mu_{\rm B}$ as in Eq.~(\ref{Eq:barmuDef}) gives
\be
{\bf H}_{{\rm exch}i} = \frac{3k_{\rm B}T_{{\rm N}J}\bar{\mu}}{g\mu_{\rm B}(S+1)}[f_J\sin(\theta)\hat{\bf i} + \cos(\theta)\hat{\bf k}].
\label{Eq:HexchHx2}
\ee

\subsection{\label{Sec:ChiPerp} Perpendicular Magnetic Susceptibility}

\begin{figure}[t]
\includegraphics [width=1in]{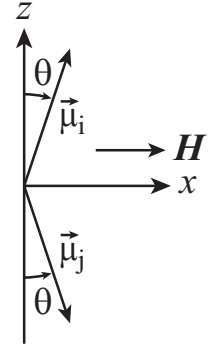}
\caption{(Color online) Geometry of two representative ordered moments~$\vec{\mu}_i$ and~$\vec{\mu}_j$ in the AFM phase with a field applied along the perpendicular $x$~axis.  Both moments have equal magnitudes and make equal angles~$\theta$ with respect to the easy $z$~axis.}
\label{Fig:ChiPerp}
\end{figure}

Here we consider infinitesimally small fields $H_x$ to calculate the perpendicular susceptibility $\chi_\perp\equiv \chi_x$ and we use second-order perturbation theory to obtain this quantity for arbitrary values of $d$, $f_J$, $S$ and~$t$.

For infinitesimal angle~$\theta$, to first order in $H_x$ and~$\theta$ the magnitude of each ordered moment is the value $\mu_0$ in zero field.  To first order in $\theta\propto\mu_x$, Eqs.~(\ref{Eqs:muimujperp}) and~(\ref{Eq:HexchHx2}) give
\bse
\bea
\vec{\mu}_i &=& \mu_0(\theta\hat{\bf i} + \hat{\bf k}),\\*
\vec{\mu}_j &=& \mu_0(\theta\hat{\bf i} - \hat{\bf k}),\\*
{\bf H}_{{\rm exch}i} &=& \frac{3k_{\rm B}T_{{\rm N}J}\bar{\mu}_0}{g\mu_{\rm B}(S+1)}\left(\theta f_J\hat{\bf i} + \hat{\bf k}\right),\label{Eq:HexchiPerp}
\eea
\ese
where $\bar{\mu}_0$ is the temperature-dependent reduced ordered moment in the AFM state at $H_x=0$ as discussed in Sec.~\ref{Sec:AFMmu0}.  We assume $\theta\ll1$ in Fig.~\ref{Fig:ChiPerp} since $h_x\ll1$.  Therefore
\be
\theta = \frac{\mu_x}{\mu_0} =  \frac{\bar{\mu}_x}{\bar{\mu}_0},
\ee
where $\mu_x$ is thermal average of the $x$~component of the magnetic moment of a spin and $\mu_0$ is unchanged to first order in~$\theta$ as noted above.  Substituting this into Eq.~(\ref{Eq:HexchiPerp}) gives
\be
{\bf H}_{{\rm exch}i} = \frac{3k_{\rm B}T_{{\rm N}J}}{g\mu_{\rm B}(S+1)}\left(f_J\bar{\mu}_x\hat{\bf i} + \bar{\mu}_0\hat{\bf k}\right).\label{Eq:HexchiPerp2}
\ee

The part of the Hamiltonian associated with the exchange field is then
\be
{\cal H}_{{\rm exch}i} = g\mu_{\rm B}{\bf S}\cdot{\bf H}_{{\rm exch}i} = \frac{3k_{\rm B}T_{{\rm N}J}}{S+1}\left(f_J\bar{\mu}_x S_x + \bar{\mu}_0S_z\right).
\ee
Normalizing the Hamiltonian by $k_{\rm B}T_{{\rm N}J}$ and including the anisotropy and applied field terms gives
\be
\frac{\cal H}{k_{\rm B}T_{{\rm N}J}} = \left(\frac{3f_J\bar{\mu}_x}{S+1} + h_x\right)S_x + \left(\frac{3\bar{\mu}_0}{S+1}\right)S_z-dS_z^2,
\label{Eq:HamHxDgtr0}
\ee
where $d$ is defined in Eq.~(\ref{Eq:dDef}) and according to Eq.~(\ref{Eq:bhDef}) the reduced applied field is
\be
h_x = \frac{g\mu_{\rm B}H_x}{k_{\rm B}T_{{\rm N}J}}.
\ee

To use second-order perturbation theory, we write Hamiltonian~(\ref{Eq:HamHxDgtr0}) as the sum of a diagonal unperturbed part~${\cal H}_0$ and a perturbed part~${\cal H}^\prime$:
\bse
\label{Eqs:HamMuPerp}
\bea
\frac{\cal H}{k_{\rm B}T_{{\rm N}J}} &=& \frac{{\cal H}_0}{k_{\rm B}T_{{\rm N}J}} +\frac{{\cal H}^\prime}{k_{\rm B}T_{{\rm N}J}},\\*
\frac{{\cal H}_0}{k_{\rm B}T_{{\rm N}J}} &=& b_zS_z - dS_z^2, \label{Eq:H0}\\*
\frac{{\cal H}^\prime}{k_{\rm B}T_{{\rm N}J}}  &=& b_xS_x, \\*
b_x &=& \frac{3f_J\bar{\mu}_x}{S+1} + h_x,\label{Eq:bxDef} \\*
b_z &=& \frac{3\bar{\mu}_0}{S+1},\label{Eq:bzPerpDef}
\eea
\ese
where $\bar{\mu}_0(t)$ is calculated using Eq.~(\ref{Eq:barmuFromGS}). 
The perpendicular magnetizations for both integer and half-integer~$S$ are calculated using Eqs.~(\ref{Eqs:muPerpIntSpins}) in Sec.~\ref{Sec:PerpPertThy}.  These equations hold for integer spins at all temperatures.  For the temperature range $t\geq t_{\rm N}$ in which the ordered moment $\bar{\mu}_0$ is zero, we set $b_z = 10^{-6}$ for half-integer spins, with negligible error in the derived perpendicular susceptibility. 

To first order in $b_x$, Eqs.~(\ref{Eqs:muPerpIntSpins}) yield

\be
\bar{\mu}_x =  b_xF_{x1},
\label{Eqs:muxbarSoln}
\ee
where the function $F_{x1}(d,b_z,t)$ is given in Eq.~(\ref{Eq:Fx1}).  Inserting Eq.~(\ref{Eq:bxDef}) for $b_x$ into~(\ref{Eqs:muxbarSoln}) and solving for $\bar{\mu}_x$ gives
\be
\bar{\mu}_x = \frac{\frac{S+1}{3}h_x}{\frac{S+1}{3F_{x1}}-f_J}.
\ee
Then using Eq.~(\ref{Eq:ChiBarDef}) gives the reduced perpendicular susceptibility $\bar{\chi}_\perp$ as
\be
\bar{\chi}_\perp \equiv \frac{\chi_\perp T_{{\rm N}J}}{C_1} = \frac{1}{\frac{S(S+1)}{3F_{x1}}-f_J}\quad{\rm (integer~spins).}
\label{Eq:barchiperp}
\ee
where the single-spin Curie constant $C_1$ is given in Eq.~(\ref{C1}).

We find $\bar{\chi}_\perp$ to be finite at $t=0$, given by
\be
\frac{1}{\bar{\chi}_\perp(t=0)} = 1-f_J+\frac{d}{3}(S+1)(2S-1).
\label{Eq:chiPerpT0}
\ee
Expanding Eq.~(\ref{Eq:barchiperp}) to second order in $1/t$ for the high-temperature behavior gives the Curie-Weiss law~(\ref{Eqs:CWLaw}) with reduced Weiss temperature
\bse
\be
\frac{\theta_{\rm p\perp}}{T_{{\rm N}J}} = f_J - d\left[\frac{(2S-1)(2S+3)}{30}\right].
\ee
Multiplying both sides of this equation by $T_{{\rm N}J}$ and using the definitions $f_J\equiv \theta_{{\rm p}J}/T_{{\rm N}J}$ and $d=D/k_{\rm B}T_{{\rm N}J}$ gives
\bea
\theta_{{\rm p}\perp} &=& \theta_{{\rm p}J} + \theta_{{\rm p}D\perp},\\*
\theta_{{\rm p}D\perp} &=& - \frac{D}{k_{\rm B}}\left[\frac{(2S-1)(2S+3)}{30}\right],
\eea
\ese
which is the sum of the contributions from the exchange interactions $\theta_{{\rm p}J}$ and the uniaxial anisotropy~$\theta_{{\rm p}D\perp}$.  The latter expression is identical to that found in Eq.~(\ref{Eq:thetaPerp}) in the presence of exchange interactions and in Eq.~(\ref{Eq:thetapx}) in the Appendix in the absence of these interactions. Thus the Weiss temperatures from different interactions are additive as noted previously.  

\begin{figure}
\includegraphics [width=3.3in]{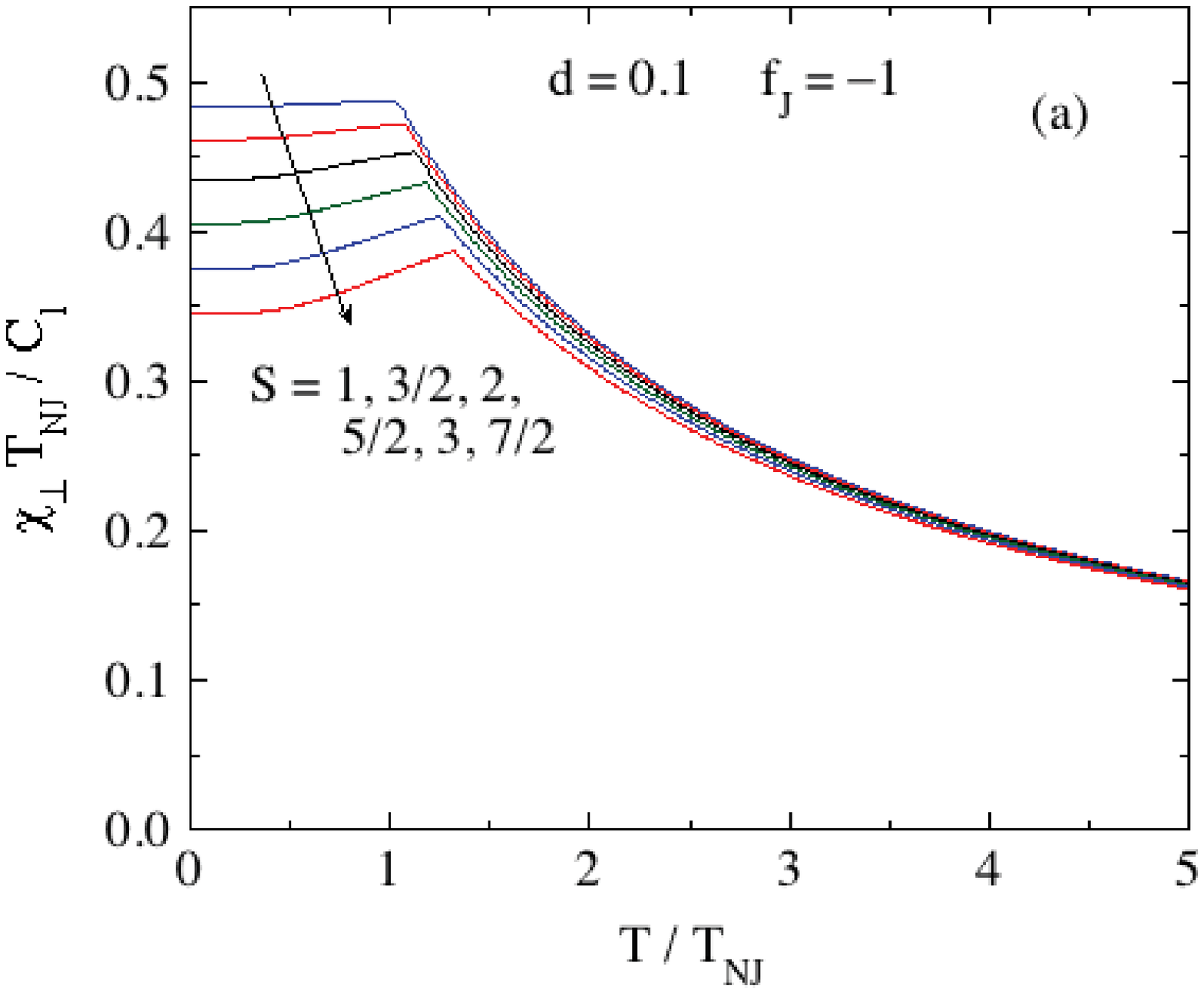}
\includegraphics [width=3.3in]{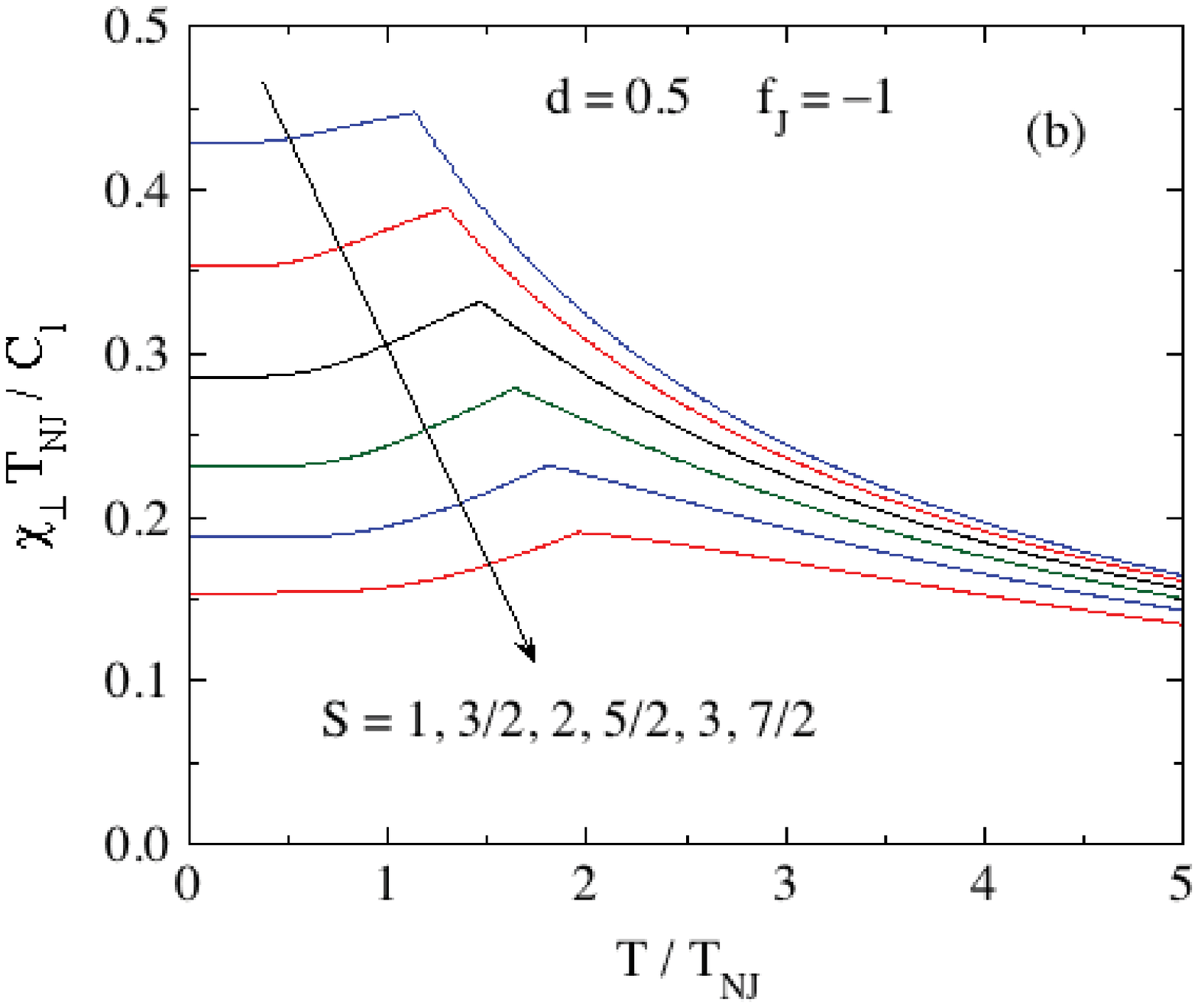}
\caption{(Color online) Normalized magnetic susceptibility $\bar{\chi}_\perp(T)\equiv \chi_\perp T_{{\rm N}J}/C_1$ versus $t = T/T_{{\rm N}J}$ obtained using Eq.~(\ref{Eq:barchiperp}) for spins $S= 1$, to~7/2 with $f_J = -1$ and reduced anisotropies (a)~$d=0.1$ and (b)~$d=0.5$. }
\label{Fig:ChixPerpd10S123fJm1}
\end{figure}

\begin{figure}
\includegraphics [width=3.3in]{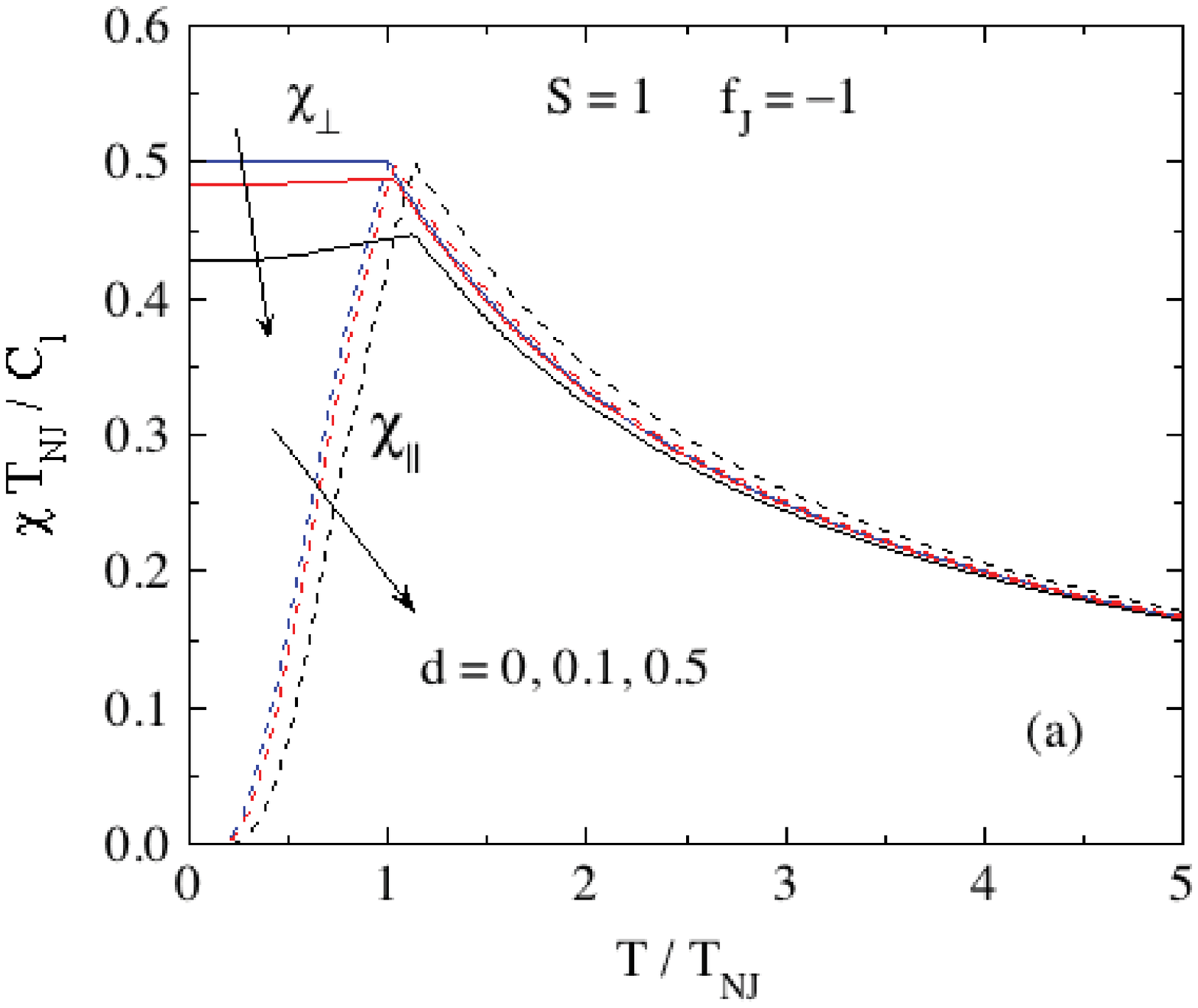}
\includegraphics [width=3.3in]{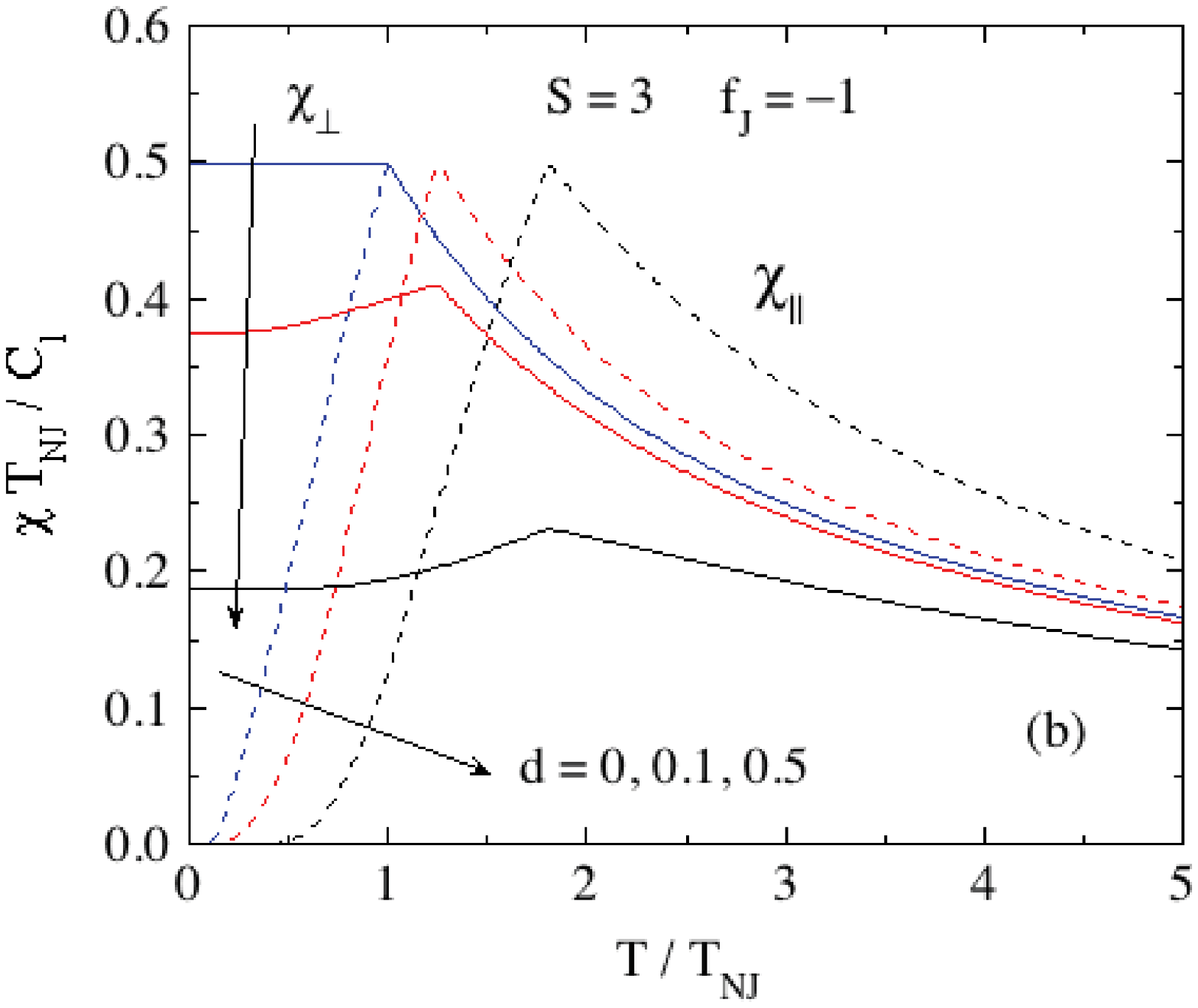}
\caption{(Color online) Normalized magnetic susceptibilities $\bar{\chi} \equiv \chi T_{{\rm N}J}/C_1$ where $\bar{\chi}=\bar{\chi}_\perp$ (solid curves) and $\bar{\chi}=\bar{\chi}_\parallel$ (dashed curves) versus $t = T/T_{{\rm N}J}$ obtained using Eqs.~(\ref{Eq:barchiperp}) and~(\ref{Eq:chibarz}), respectively, with $f_J = -1$ and $d=0$, 0.1 and~0.5 for spins (a)~$S=1$ and (b)~$S=3$. }
\label{Fig:ChixPerpd0To50S1fJm1}
\end{figure}

Shown in Fig.~\ref{Fig:ChixPerpd10S123fJm1} are plots of $\bar{\chi}_\perp$ versus~$t$ for fixed $f_J=-1$ and integer spins $S=1$ to~7/2 in increments of 1/2 with $d=0.1$ and $d=0.5$ obtained using Eq.~(\ref{Eq:barchiperp}).  Contrary to MFT predictions for the exchange interaction with or without a magnetic dipole anisotropy \cite{Johnston2016} or a generic anisotropy field where $\bar{\chi}_\perp$ is found to be independent of temperature for $T\leq T_{\rm N}$, here we find that a uniaxial anisotropy with $D>0$ causes $\bar{\chi}_\perp$ to decrease with decreasing temperature below~$T_{\rm N}$.  The $\bar{\chi}_\perp(t=0)$ values in Fig.~\ref{Fig:ChixPerpd10S123fJm1} are in agreement with the general expression~(\ref{Eq:chiPerpT0}).  A similar decrease in $\chi_\perp$ upon cooling below $T_{\rm N}$ was found in a MFT study for $S=2$ in the presence of single-ion anisotropy \cite{Honma1960}.  Figure~\ref{Fig:ChixPerpd0To50S1fJm1} shows plots of both $\chi_\perp$ and $\chi_\parallel$ versus~$t$ with $f_J = -1$ and $d = 0,$ 0.1 and~0.5 for spins~$S=1$ and~$S=3$.  One sees that $\chi_\perp(t>1)$ in the PM state is increasingly suppressed relative to $\chi_\parallel(t>1)$ with increasing~$d$, and that this effect is accentuated with increasing~$S$\@.

\subsection{Torque on an Integer-Spin Ordered Moment due to the Axial Anisotropy}

In Fig.~\ref{Fig:ChiPerp} above is shown a representative thermal-average magnetic moment $\vec{\mu}_i$ that makes a polar angle~$\theta$ with respect to the uniaxial $z$~axis.  Intuitively, the $DS_z^2$ term in the spin Hamiltonian with $D>0$ may lead to a torque $\vec{\tau}_D$ on $\vec{\mu}_i$ that tends to align $\vec{\mu}_i$ with the $+z$~axis.  Here we show that this is the case and calculate $\vec{\tau}_D$ using a simple strategy.  In equilibrium, the sum of the torques due to the axial anisotropy $\vec{\tau}_D$, the applied field $\vec{\tau}_H$ and the exchange field $\vec{\tau}_{{\rm exch}i}$ on the thermal-average moment $\vec{\mu}_i$ must be zero.  We know how to calculate the latter two torques.  Hence we calculate $\vec{\tau}_D$ from
\be
\vec{\tau}_D = -(\vec{\tau}_H + \vec{\tau}_{{\rm exch}i}).
\label{Eq:tauD}
\ee
From that we calculate the lowest-order anisotropy energy
\be
E_i = K_1\sin^2\theta\approx K_1\theta^2\qquad(\theta\ll1)
\label{Eq:AnisEnergy}
\ee
and the corresponding anistropy constant $K_1$.  Although it has been stated that this is not a useful approach for calculating $K_1$ \cite{Kanamori1962}, our approach gives the same expression for $K_1$ at $T=0$ as they obtain by a different route.  The temperature dependence of $K_1$ is also calculated and found to be proportional to the square of the ordered moment in the AFM state and therefore vanishes for $T \geq T_{\rm N}$.

Here we calculate the torques on $\vec{\mu}_i$ using the same construct as used above to calculate $\chi_\perp$ with $D>0$.  We thus calculate the torques only to first order in $\theta$.  From Eqs.~(\ref{Eq:vecmui}) and~(\ref{Eq:HexchiPerp}), one obtains
\be
\vec{\tau}_{{\rm exch}i} = \frac{3k_{\rm B}T_{{\rm N}J}S\bar{\mu}_0^2}{S+1}\theta (f_J - 1)\,\hat{\bf j}.
\ee
The torque on $\vec{\mu}_i$ due to ${\bf H} =H_x\,\hat{\bf i}$ is
\be
\vec{\tau}_H = \mu_0 H_x\,\hat{\bf j}.
\ee
Referring to Fig.~\ref{Fig:ChiPerp}, these torques both tend to rotate $\vec{\mu}_i$ away from the $+z$~axis.  From Eq.~(\ref{Eq:tauD}) and the definitions of the reduced variables we thus obtain
\be
\frac{\vec{\tau}_D}{k_{\rm B}T_{{\rm N}J}} = -\left[S\bar{\mu_0}h_x + \frac{3S\bar{\mu}_0^2(f_J-1)}{S+1}\theta\right]\,\hat{\bf j}.
\label{Eq:taud0}
\ee
The direction of this torque tends to align $\vec{\mu}_i$ parallel to the applied field in the $\hat{\bf k}$ direction.

In order to solve for $K_1$ in Eq.~(\ref{Eq:AnisEnergy}) one needs to write $h_z$ in Eq.~(\ref{Eq:taud0}) in terms of~$\theta$.  We first express $\bar{\mu}_x$ in terms of $h_x$.  Using Eq.~(\ref{Eq:ChiBarDef}) one obtains
\be
\bar{\mu}_x = \frac{S+1}{3}\bar{\chi}_\perp h_x.
\label{Eq:muxVShx}
\ee
From Fig.~\ref{Fig:ChiPerp} and using $\theta\ll1$ one has
\be
\theta=\frac{\mu_x}{\mu_0} = \frac{\bar{\mu}_x}{\bar{\mu}_0}.
\label{Eq:thetaVSbarmux}
\ee
Inserting Eq.~(\ref{Eq:muxVShx}) into~(\ref{Eq:thetaVSbarmux}) and solving for $h_x$ gives
\be
h_x = \frac{\bar{\mu}_0\theta}{\frac{S+1}{3}\bar{\chi}_\perp}.
\label{Eq:hxFromTheta}
\ee
Inserting this expression into Eq.~(\ref{Eq:taud0}) gives the torque from the axial anisotropy for $\theta\ll1$ as
\be
\frac{\vec{\tau}_D(t)}{k_{\rm B}T_{{\rm N}J}} = -\left[\frac{3S\bar{\mu}_0^2(t)}{S+1}\right]\left[\frac{1}{\bar{\chi}_\perp(t)} + f_J - 1\right]\theta\,\hat{\bf j}.
\label{Eq:taud1}
\ee
Finally, the anisotropy energy is obtained from $\tau_D$ as
\bea
\frac{E_i}{k_{\rm B}T_{{\rm N}J}} &=& \int_0^\theta \frac{\tau_D(\theta)}{k_{\rm B}T_{{\rm N}J}}\,d\theta\\*
&=& \left[\frac{3S\bar{\mu}_0^2(t)}{S+1}\right]\left[\frac{1}{\bar{\chi}_\perp(t)} + f_J - 1\right]\frac{\theta^2}{2},\nonumber
\eea
and hence the anisotropy constant in Eq.~(\ref{Eq:AnisEnergy}) is 
\be
\frac{K_1(t)}{k_{\rm B}T_{{\rm N}J}} = \left[\frac{3S\bar{\mu}_0^2(t)}{2(S+1)}\right]\left[\frac{1}{\bar{\chi}_\perp(t)} + f_J - 1\right].
\label{Eq:K1Soln}
\ee
Since $\bar{\mu}_0\to0$ as $T\to T_{\rm N}$, so does $K_1$.  From Eq.~(\ref{Eq:K1Soln}), in general $K_1$ is proportional to $T_{{\rm N}J}$ and hence on the exchange interactions.  However, as shown in the following section, for $t\to0$ one finds, perhaps nonintuitively, that $K_1$ only depends on $S$ and $D$ and not on the exchange interactions explicitly.

\begin{figure}
\includegraphics [width=3.3in]{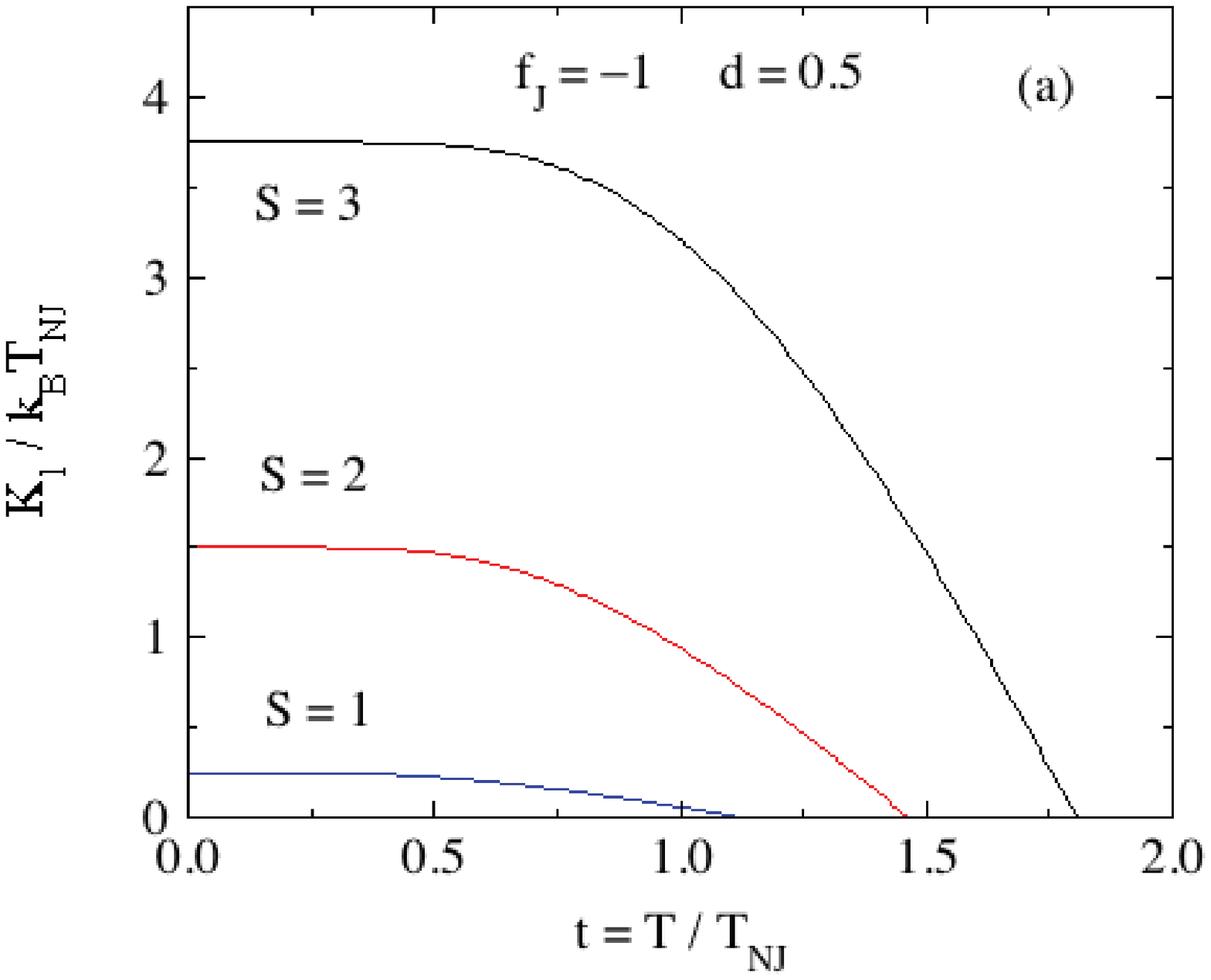}
\includegraphics [width=3.3in]{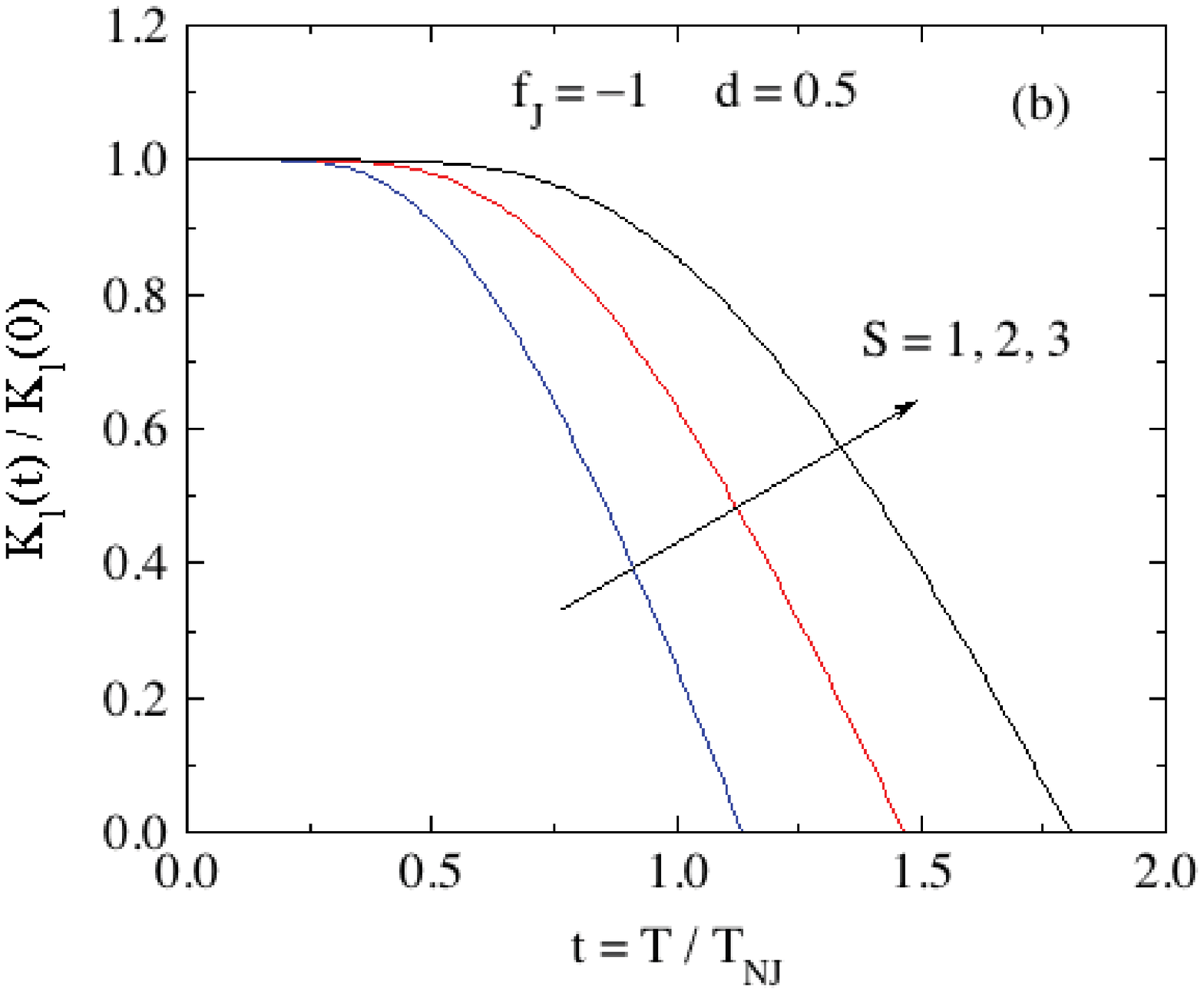}
\caption{(Color online) Normalized anisotropy constant (a)~$K_1(t)/k_{\rm B}T_{{\rm N}J}$ and~(b)~$K_1(t)/K_1(0)$ for integer spins $S = 1$, 2 and~3 and $d=0.5$ versus reduced temperature~$t$ obtained using Eq.~(\ref{Eq:K1Soln}). }
\label{Fig:K1_fJm1d50S1to3}
\end{figure}

Plots of $K_1(t)/k_{\rm B}T_{{\rm N}J}$ and the normalized $K_1(t)/K_1(0)$ versus~$t$ are shown for integer spins $S=1,\ 2,\ 3$, $d=0.5$ and $f_J=-1$ in Figs.~\ref{Fig:K1_fJm1d50S1to3}(a) and~\ref{Fig:K1_fJm1d50S1to3}(b), respectively.  The shapes of the curves do not depend strongly on~$S$\@.  The curves all approach zero linearly as $T\to T_{\rm N}$ because $\bar{\mu}_0\sim\sqrt{1-t}$ on approaching $t_{\rm N}$ from below.  The curve in Fig.~\ref{Fig:K1_fJm1d50S1to3}(b) for $S=2$ is similar to those calculated from MFT for $S=2$ and two values of $d$ \cite{Ohlmann1961}.

\subsubsection*{Anisotropy Constant $K_1$ at $T=0$}

Inserting $\bar{\mu}_0(t=0)=1$ and $1/\bar{\chi}(t=0)$ in Eq.~(\ref{Eq:chiPerpT0}) into Eq.~(\ref{Eq:K1Soln}) gives
\be
\frac{K_1(t=0)}{k_{\rm B}T_{{\rm N}J}} = dS\left(S-\frac{1}{2}\right).
\label{Eq:K1onkTNt0Red}
\ee
Then using the definition of~$d$ in Eq.~(\ref{Eq:dDef}) gives
\be
K_1(t=0) = DS\left(S-\frac{1}{2}\right) \qquad (\theta\ll1).
\label{Eq:K1onkTNt0}
\ee 
The same result was given in Ref.~\cite{Kanamori1962} obtained using a different approach.  Here, $K_1$ is obtained as the $t=0$ limit of the $t$-dependent~$K_1$ in Eq.~(\ref{Eq:K1Soln}).  Indeed, the $t\to0$ limits of $K_1(t)/k_{\rm B}T_{\rm N}$ in Fig.~\ref{Fig:K1_fJm1d50S1to3}(a) are seen to agree with Eq.~(\ref{Eq:K1onkTNt0Red}).

\subsection{High-Field Perpendicular Magnetization and Perpendicular Critical Field}

For the high-field behavior, it is convenient to use the same axes as in Fig.~\ref{Fig:mui_muj}.  The only change to be made to calculate the ordered moments in the parallel and perpendicular directions compared to the solutions for the SF phase with field along the $z$~axis, is to change $-dS_z^2$ in the reduced spin Hamiltonian~(\ref{Eqs:HSF0}) to $-dS_x^2$.  The field direction is still ${\bf H} = H_z\hat{\bf k}$, which is perpendicular to the easy $x$~axis.  In order to avoid confusion with the earlier notation for the SF phase, here we will refer to the $z$~direction of the field as the $\perp$ direction, so the induced magnetization is then $\mu_\perp(H_\perp)$.  The method of solution is the same as given for the high-field magnetization of the SF phase in Sec.~\ref{Sec:MofHzSF}.

\begin{figure}
\includegraphics [width=3.3in]{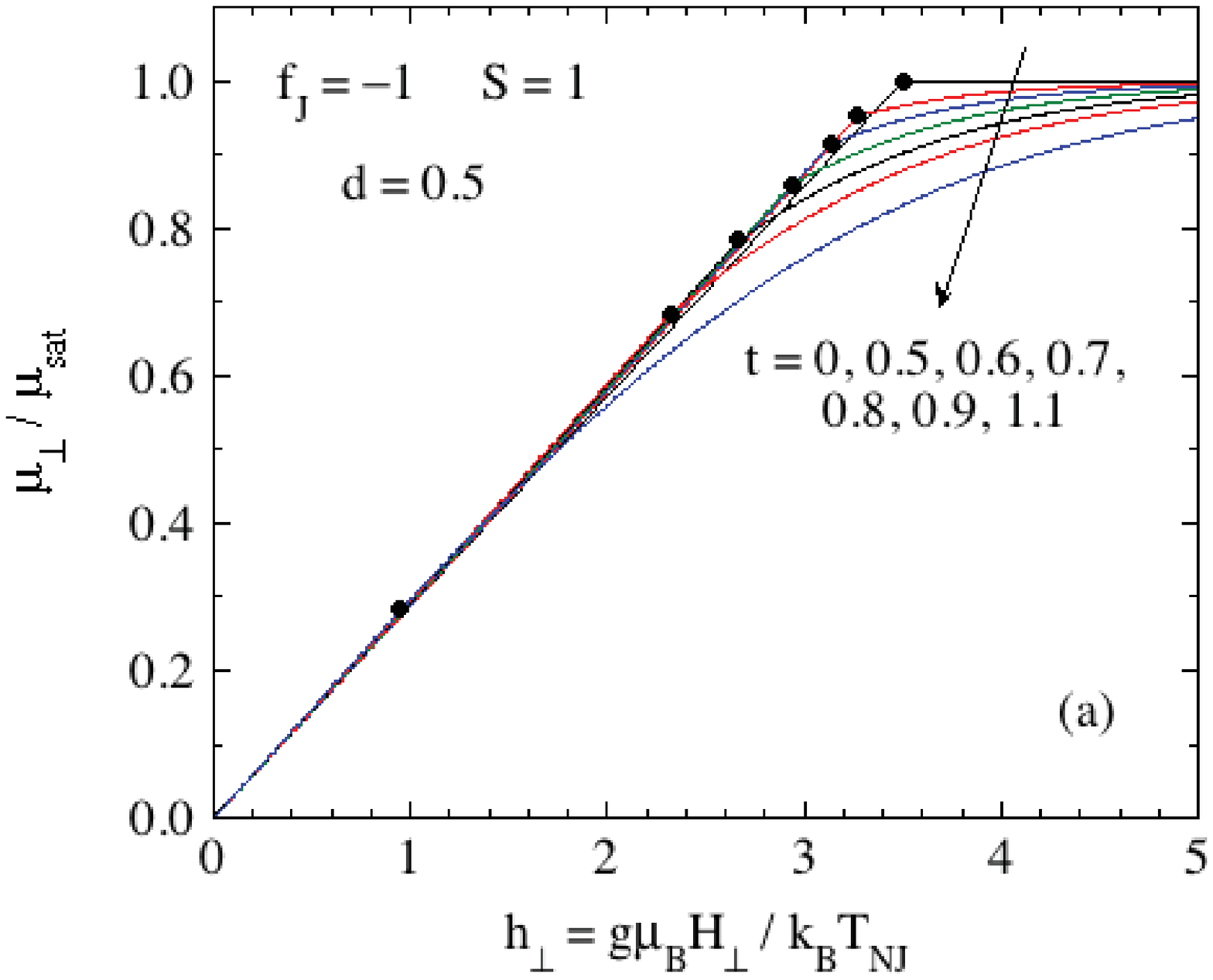}
\includegraphics [width=3.3in]{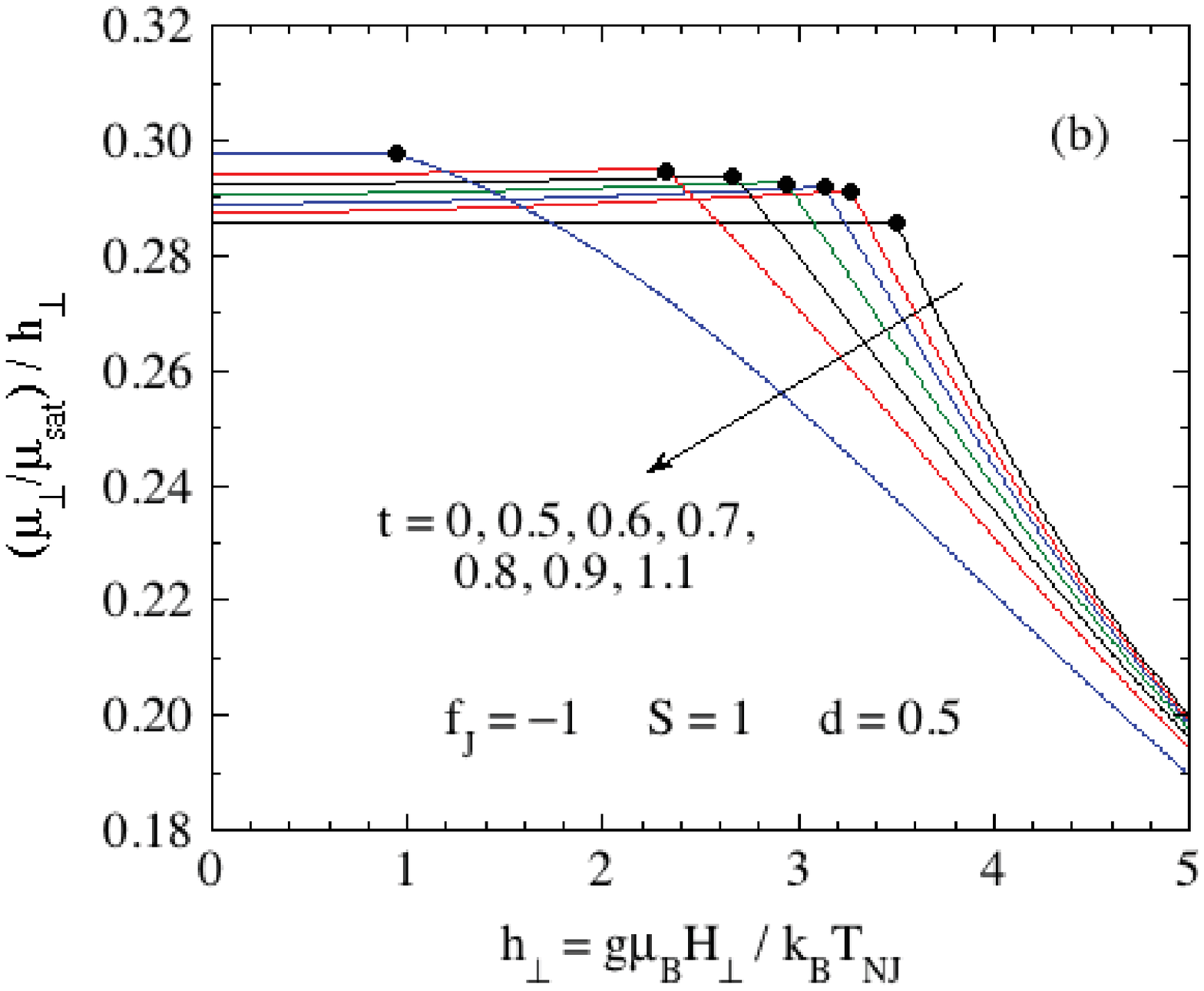}
\caption{(Color online) (a)~Reduced magntization $\bar{\mu}_\perp = \mu_\perp/\mu_{\rm sat}$ versus reduced field~$h_\perp$ for the listed values of reduced temperature~$t$ for $f_J=-1,\ d = 0.5$ and~$S=1$. (b)~Ratio of $\bar{\mu}_\perp/h_\perp$ versus $h_\perp$ for the same temperatures as in~(a).  The filled black circles in~(a) and~(b) denote the normalized perpendicular critical fields $h_{\rm c\perp}$.}
\label{Fig:muPerpVsHPerpd50AFMfJm1S1}
\end{figure}

\begin{figure}
\includegraphics [width=3.3in]{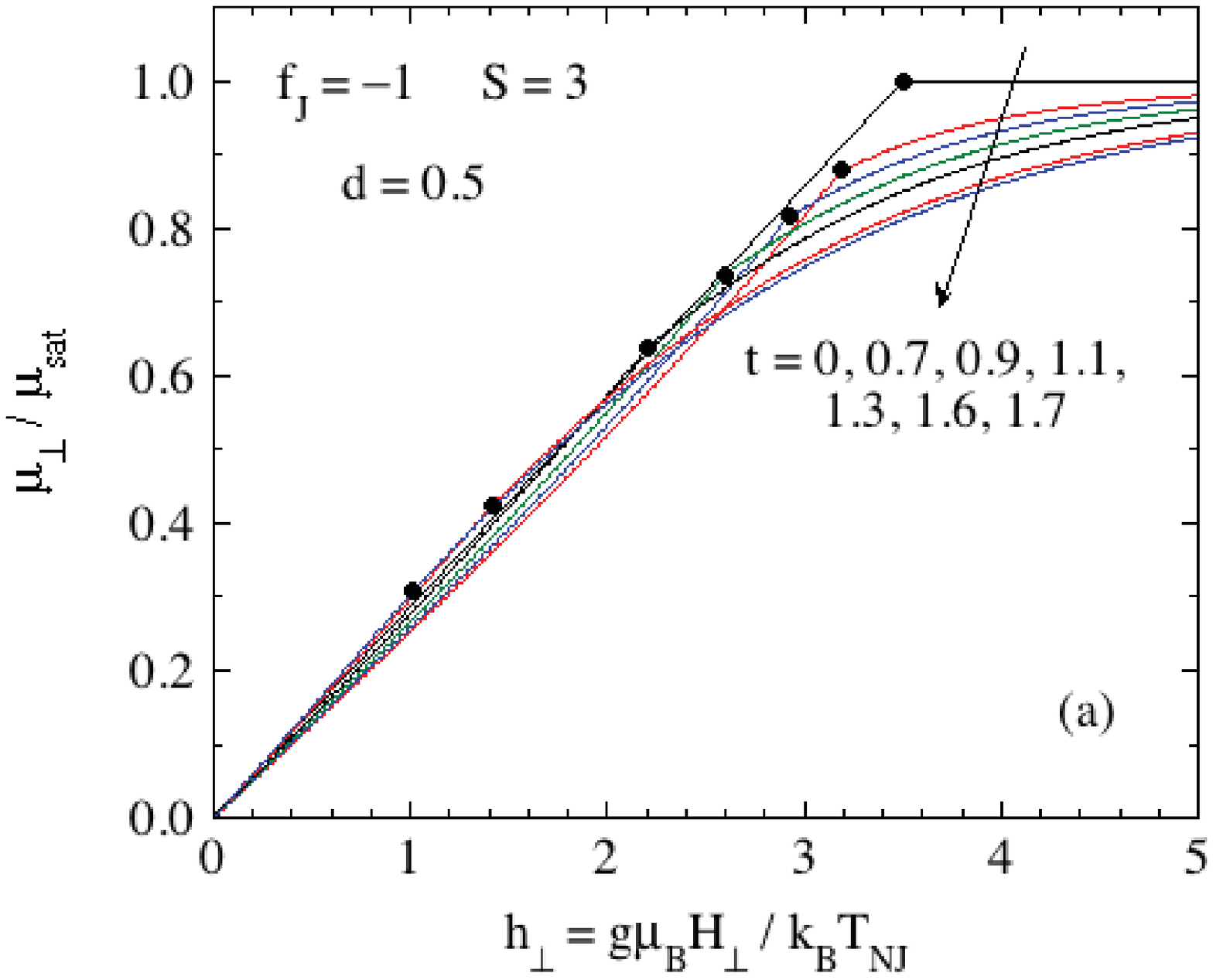}
\includegraphics [width=3.3in]{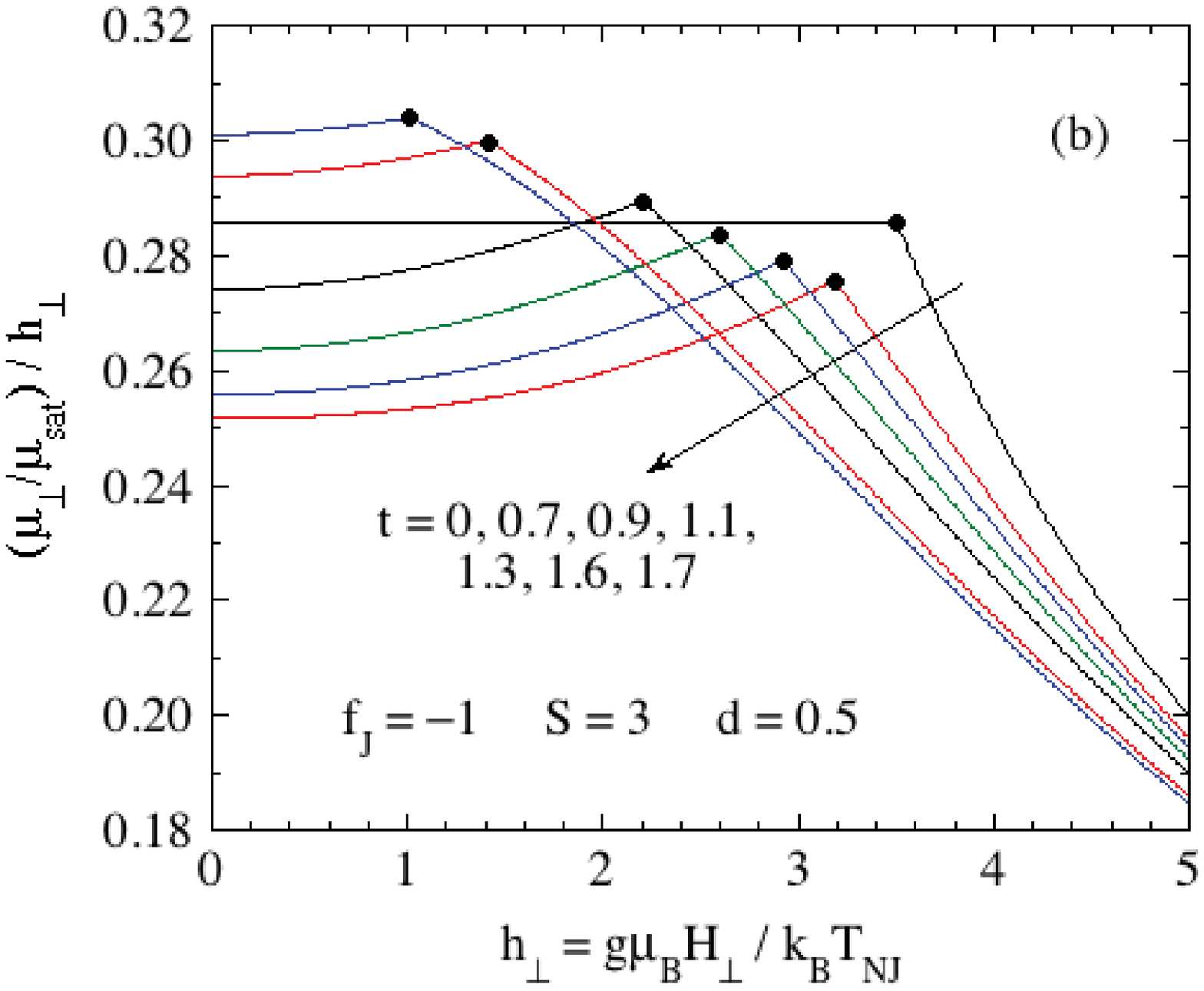}
\caption{(Color online) Same as Fig.~\ref{Fig:muPerpVsHPerpd50AFMfJm1S1} except with $S = 3$ and a different set of $t$ values.}
\label{Fig:muPerpVsHPerpd50AFMfJm1S3}
\end{figure}

\begin{figure}
\includegraphics [width=3.3in]{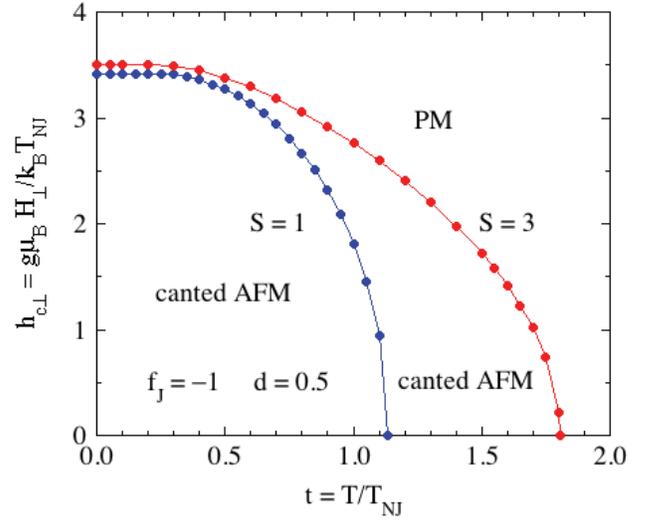}
\caption{(Color online) Reduced perpendicular critical field~$h_{\rm c\perp}$ versus reduced temperature~$t$ for spins $S = 1$ and $S = 3$ with reduced anisotropy parameter~$d=0.5$ and $f_J=-1$.  These data were obtained from data such as in Figs.~\ref{Fig:muPerpVsHPerpd50AFMfJm1S1} and~\ref{Fig:muPerpVsHPerpd50AFMfJm1S3}.}
\label{Fig:hcPerpd50S1S3}
\end{figure}

\begin{figure}
\includegraphics [width=3.3in]{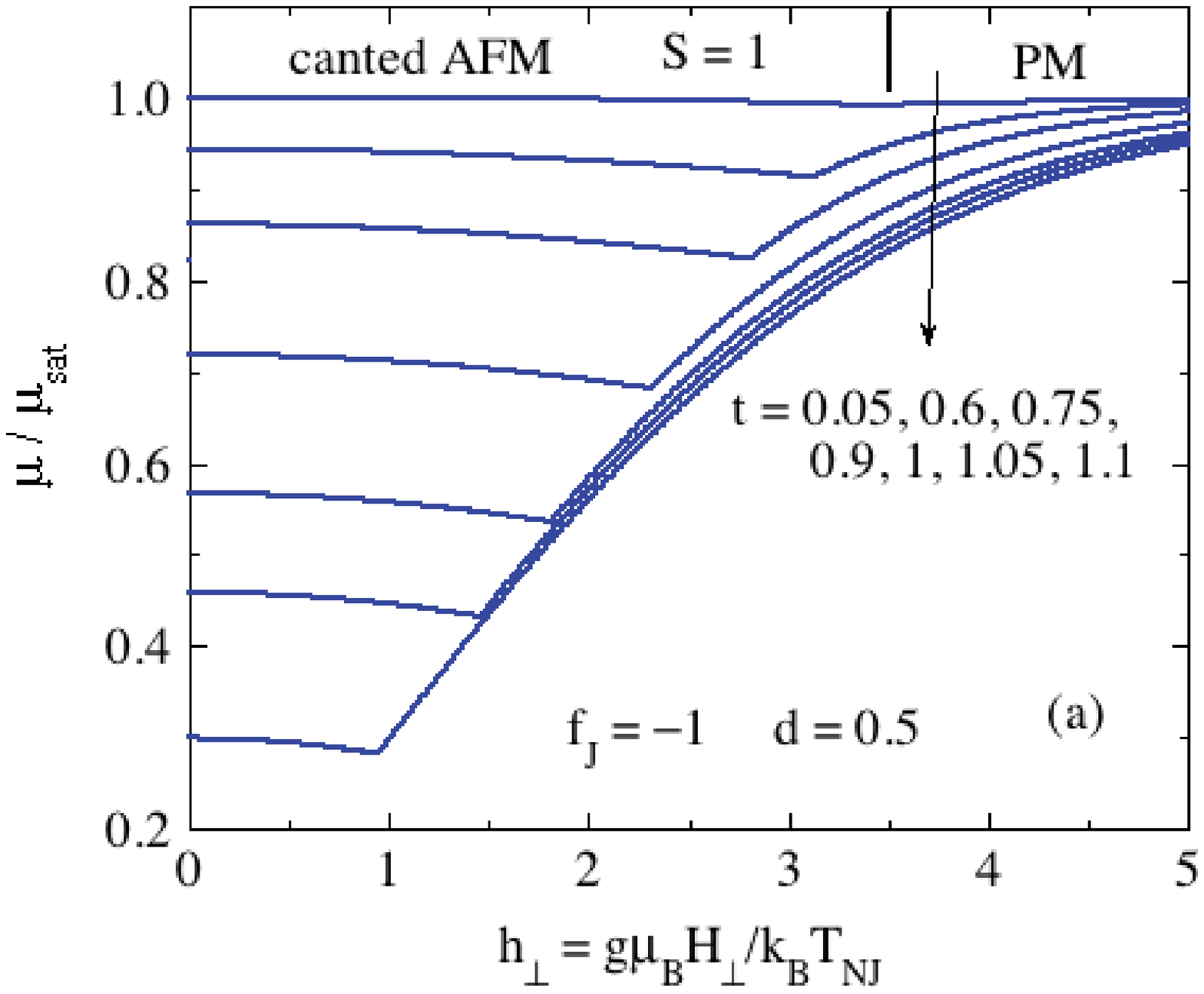}
\includegraphics [width=3.3in]{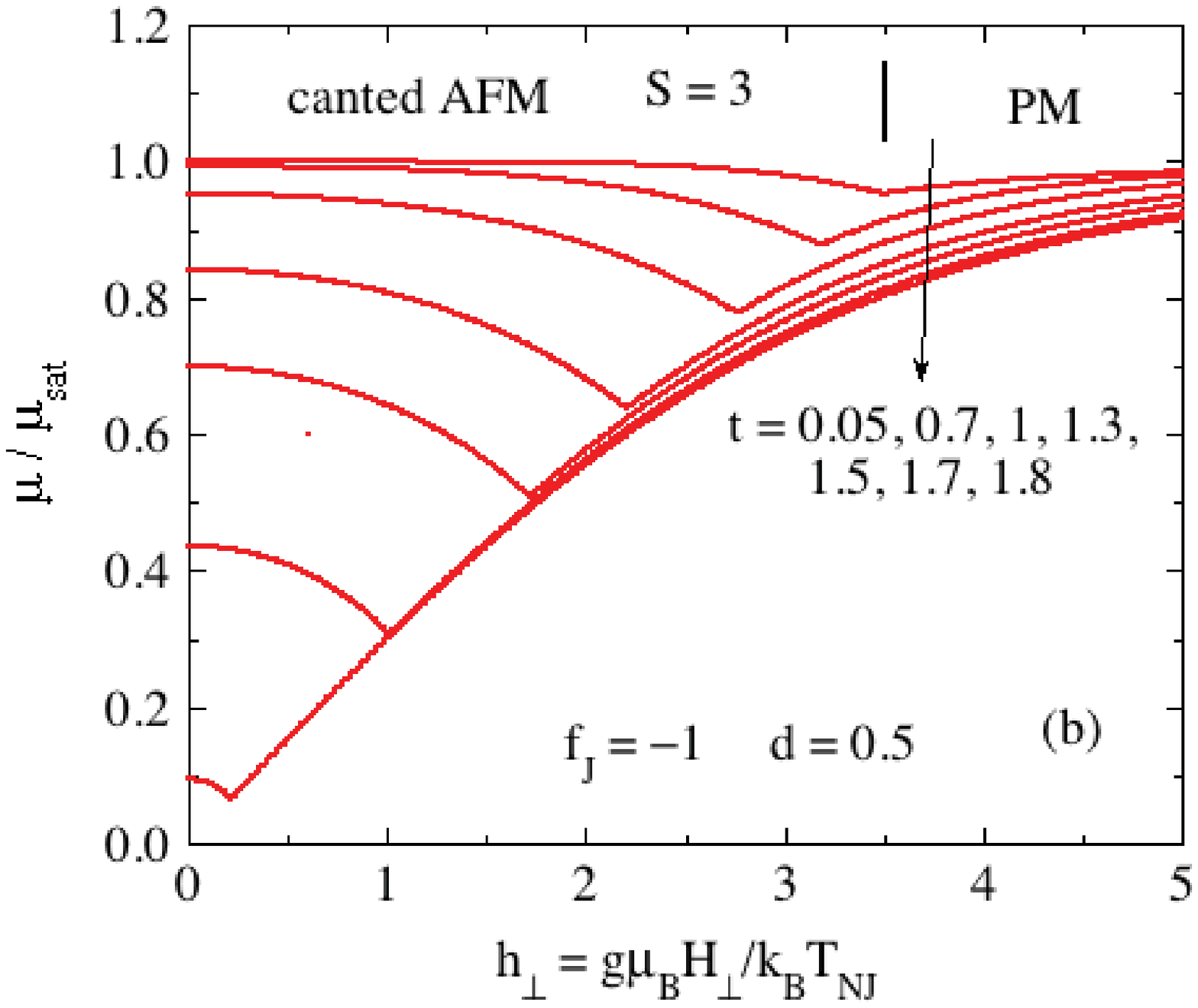}
\caption{(Color online) Dependence of the magnitude $\bar{\mu}$ on the reduced perpendicular field $h_\perp$ at the listed temperatatures for spins (a)~$S=1$ and (b)~$S=3$.  The cusps in the data occur at the transition field between the canted AFM state at the lower fields and the PM state at higher fields, as indicated for $t=0.05$ in each panel.}
\label{Fig:mubar_vs_hperp_d50}
\end{figure}

The dependence of $\bar{\mu}_\perp$ on $h_\perp$ for $d=0.5,\ f_J = -1$ and $S=1$ is shown in Fig.~\ref{Fig:muPerpVsHPerpd50AFMfJm1S1}(a) for several temperatures below $t_{\rm N}$.  The critical fields~$h_{\rm c\perp}$ for the second-order transitions from the canted AFM state to the PM state are denoted by filled black circles.  The chordal slope $\bar{\mu}_\perp/h_\perp$ is plotted versus $h_\perp$ for the same temperatures in Fig.~\ref{Fig:muPerpVsHPerpd50AFMfJm1S1}(b).  The same type of plots for $S=3$ are shown in Fig.~\ref{Fig:muPerpVsHPerpd50AFMfJm1S3}.  These plots are qualitatively similar to the perpendicular magnetization curves of the spin-flop phase with $S=1,\ d=0.5$ and $f_J=-1$ in Fig.~\ref{Fig:hzSFt0p1To1p2d50muzbar}. 

For plots as in Figs.~\ref{Fig:muPerpVsHPerpd50AFMfJm1S1}(b) and~\ref{Fig:muPerpVsHPerpd50AFMfJm1S3}(b), one defines the unreduced susceptibility as $\chi_\perp = \lim_{H_\perp\to0}(\mu_\perp/H_\perp)$.  The reduced susceptibility $\bar{\chi}_\perp$ is defined as in Eq.~(\ref{Eq:barchiperp}) and can be written in terms of $\bar{\mu}_\perp$ and $h_\perp$ as
\be
\bar{\chi}_\perp(t) \equiv \frac{\chi_\perp(t) T_{{\rm N}J}}{C_1} = \lim_{h_\perp\to0}\left(\frac{3}{S+1}\right)\frac{\bar{\mu}_\perp(t)}{h_\perp}.
\label{Eq:barchiperp2}
\ee
This relation is seen to be satisfied by comparing the low-field data in Figs.~\ref{Fig:muPerpVsHPerpd50AFMfJm1S1}(b) and~\ref{Fig:muPerpVsHPerpd50AFMfJm1S3}(b) with the corresponding data in  Figs.~\ref{Fig:ChixPerpd10S123fJm1} and~\ref{Fig:ChixPerpd0To50S1fJm1}.

The reduced perpendicular critical field $h_{\rm c\perp}$ at each temperature is defined as the second-order transition field between the canted AFM and the PM states.  The $h_{\rm c\perp}(t)$ is plotted versus~$t$ in Fig.~\ref{Fig:hcPerpd50S1S3}, obtained from data as in Figs.~\ref{Fig:muPerpVsHPerpd50AFMfJm1S1} and~\ref{Fig:muPerpVsHPerpd50AFMfJm1S3}.  For each $S$, the $h_{\rm c\perp}(t)$ curve separates the canted AFM state from the PM state, as indicated in Fig.~\ref{Fig:hcPerpd50S1S3}.

In contrast to the case for $d=0$ \cite{Johnston2015}, the magnitude of the ordered moment in the canted AFM state
\be
\bar{\mu} = \sqrt{\bar{\mu}_\parallel^2 + \bar{\mu}_\perp^2}
\ee
depends on the applied field, as shown in Fig.~\ref{Fig:mubar_vs_hperp_d50} for spins $S=1$ and $S=3$.

%\clearpage

\section{\label{Sec:In-PlaneOrdDless0} In-Plane Collinear AFM Ordering with $D<0$}

When the axial anisotropy parameter $D < 0$, AFM ordering with ordered-moment alignments along the $x$~axis, perpendicular to the $z$~axis, is favored over $z$-axis AFM ordering because then the lowest-energy states have minimum values of $\langle S_z^2\rangle$.  In zero field, the magnetic induction ${\bf B}_i$ seen by our central ordered moment~$\vec{\mu}_i = \mu_x\hat{\bf i}$, assumed to be aligned in the $+x$ direction, consists only of the exchange field that is also aligned in the $+x$ direction and is given by Eq.~(\ref{Eq:Hexch1}) as
\bse
\label{Eqs:TNxCalcs}
\be
B_x = H_{{\rm exch}0} = \frac{3k_{\rm B}T_{{\rm N}J}}{g\mu_{\rm B}(S+1)}\bar{\mu}_0.
\label{Eq:HexchPerp}
\ee
We use the definitions
\be
\bar{\mu}_x \equiv \bar{\mu}_0 = \frac{\mu_x}{gS\mu_{\rm B}}, \quad d = \frac{D}{k_{\rm B}T_{{\rm N}J}}, \quad t = \frac{T}{T_{{\rm N}J}},
\label{Eqs:PerpDefs}
\ee
\ese
and utilize the second-order perturbation theory results for the moment $\mu_x$ induced by a magnetic induction~$B_x$ described generically in Sec.~\ref{Sec:PerpPertThy}.  As explained in that section, different expressions are obtained for integer and half-integer spins.  Hence we expect and find the same dichotomy for the N\'eel temperatures.

\subsection{\label{Sec:TNxAxis} N\'eel Temperature}

For integer spins, substituting Eq.~(\ref{Eq:HexchPerp}) for $B_x$ into Eq.~(\ref{Eq:muxPerpInt}) and using the above definitions gives
\bea
\bar{\mu}_0 &\equiv& \frac{3\bar{\mu}_0}{S(S+1)d}F_{x1}\quad{\rm (integer}\ S),\label{Eq:mu0PerpInt}\\*
F_{x1} &=& \frac{1}{Z_S}\sum_{m_S=-S}^S \left[\frac{S(S+1)+m_S^2}{4m_S^2 -1}\right] e^{dm_S^2/t},\nonumber
\eea
where the partition function is
\be
Z_S=\sum_{m_S=-S}^S e^{dm_S^2/t}.
\label{Eq:ZSx}
\ee
For $t\to t_{\rm N}^-$, one can divide out $\bar{\mu}_0$ on both sides of Eq.~(\ref{Eq:mu0PerpInt}) and abtain an equation from which to numerically solve for the reduced ordering temperature $t_{\rm N} = T_{\rm N}/T_{{\rm N}J}$ versus~$d$, given by
\bse
\label{Eqs:TNx}
\bea
1 &=& \frac{3}{Z_SS(S+1)d}\sum_{m_S=-S}^S \frac{S(S+1)+m_S^2}{4m_S^2-1}\exp\left(\frac{dm_S^2}{t_{\rm N}}\right)\nonumber\\*
&& \hspace{1in} ({\rm integer}\ S).
\eea
For half-integer spins, using Eq.~(\ref{Eq:muxBxHalfInt}) one obtains a different expression for $t_{\rm N}$ given by
\bea
1 &=& \frac{3}{Z_SS(S+1)d}\qquad ({\rm half\ integer}\ S) \hspace{0.3in}\\*
&& \times \Bigg\{\frac{1}{2}\bigg[\frac{S(S+1)+1/4}{t_{\rm N}/d}-S(S+1)+\frac{3}{4}\bigg]\exp\left(\frac{d}{4t_{\rm N}}\right)\nonumber\\*
&&\ +\ 2\sum_{m_S=3/2}^S \frac{S(S+1)+m_S^2}{4m_S^2-1}\exp\left(\frac{dm_S^2}{t_{\rm N}}\right)\Bigg\}.\nonumber
\eea
\ese

For $|d| \ll 1$, one obtains
\bse
\bea
t_{\rm N} &=& 1 - \frac{dS(S+1)}{3}\quad (d<0,\ |d|\ll1,\ {\rm integer}\ S),\hspace{0.35in} \\*
t_{\rm N} &=& 1 - \frac{d(2S-1)\left(16S^3+40S^2+36S+9\right)}{96S(S+1)}\\*
&&\hspace{0.8in} (d<0,\ |d|\ll1,\ {\rm half\ integer}\ S),\nonumber
\eea
\ese
which both yield $t_{\rm N} = 1$ if $S=1/2$ as required.  The expression for integer~$S$ is quite different from that in Eq.~(\ref{tNdGTR0dll1}) for $c$-axis ordering with integer~$S$ and $d>0$.  For both integer and half-integer spins, one sees that a positive~$d$ suppresses~$t_{\rm N}$ whereas a negative~$d$ enhances it, consistent with expectation for $x$-axis ordering.

The variations of $t_{\rm N}$ versus (negative) $d$ for $S=1$ to $S=7/2$ are shown above in Figs.~\ref{Fig:tNVSd}(b) and~\ref{Fig:tNVSd}(c) for integer and half-integer spins, respectively.  One sees that with increasingly negative values of $d$, $t_{\rm N}$ initially increases for all values of $S$, reaches a maximum at $d\sim -1$ and then decreases.  For integer spins, $t_{\rm N}$ crashes to zero at $d=-3$.  The reason is that the anisotropy energy is $-dS_z^2$ and for integer spins a negative $d$ means the ground state has $S_z=0$ and is hence nonmagnetic.  For half-integer spins as in Figs.~\ref{Fig:tNVSd}(c), the same situation leads to the ground state having $S_z=1/2$ even though $S\geq 3/2$; hence the spin value is effectively diluted for large negative $d$ but in this case $t_{\rm N}$ approaches a constant value for large negative values of~$d$.  In the limit of large negative $d$, for half-integer spins we obtain
\be
t_{\rm N}(d\to-\infty)=\frac{3}{4}\left[1+\frac{1}{4S(S+1)}\right].
\ee

\subsection{Ordered Moment versus Temperature}

For $h_z=\bar{\mu}_z=0$, the ordered moments are aligned along the $x$~axis and the reduced Hamiltonian for in-plane AFM ordering is given by Eq.~(\ref{Eq:H0}). Then using Eq.~(\ref{Eq:barmualpha}) with $b_x=h_{\rm exch0} = 3\bar{\mu}_0/(S+1)$ from Eq.~(\ref{Eq:hexch0def}), the thermal-average ordered moment $\bar{\mu}_0(t)$ at each~$t$ is obtained by solving
\bse
\label{Eqs:mubarH0D<0}
\bea
\bar{\mu}_0(t) &=& -\frac{S+1}{3SZ_S}\sum_{n=1}^{2S+1}\frac{\partial\epsilon_n}{\partial \bar{\mu}_0}e^{-\epsilon_n/t},\\*
Z_S &=& \sum_{n=1}^{2S+1} e^{-\epsilon_n/t}.
\eea
\ese
These equations are valid for both integer and half-integer spins.

\begin{figure}
\includegraphics [width=3.3in]{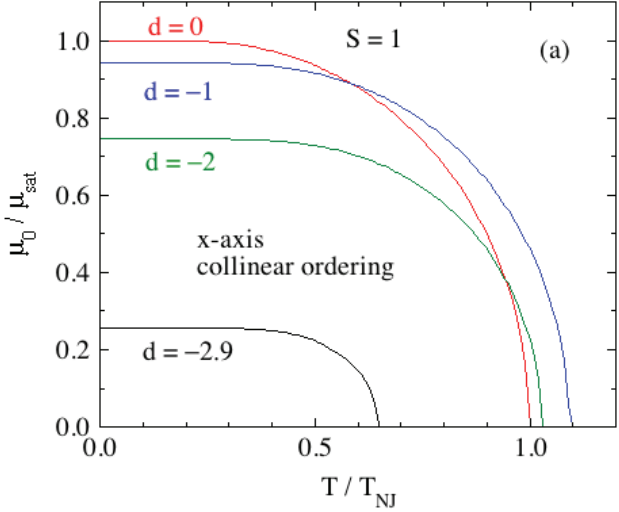}
\includegraphics [width=3.3in]{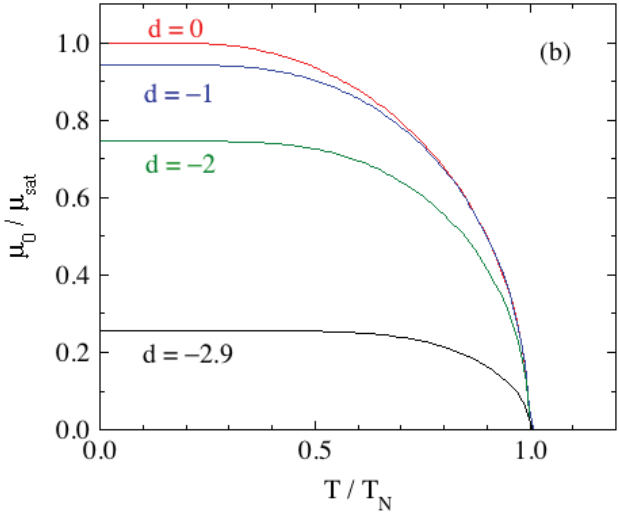}
\caption{(Color online) Reduced ordered moment $\bar{\mu}_0 = \mu_0/\mu_{\rm sat}$ versus reduced temperatures (a) $T/T_{{\rm N}J}$ and (b) $T/T_{\rm N}$ calculated using Eqs.~(\ref{Eqs:mubarH0D<0}) for spins $S=1$ with $x$-axis collinear AFM ordering with reduced anisotropy parameters $d=D/k_{\rm B}T_{{\rm N}J} = 0$, $-1$, $-2$ and~$-2.9$.}
\label{Fig:mu0xS1}
\end{figure}

\begin{figure}
\includegraphics [width=3.3in]{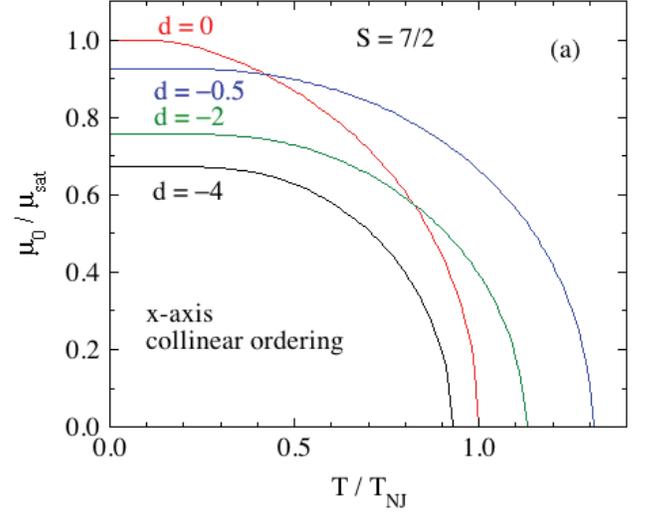}
\includegraphics [width=3.3in]{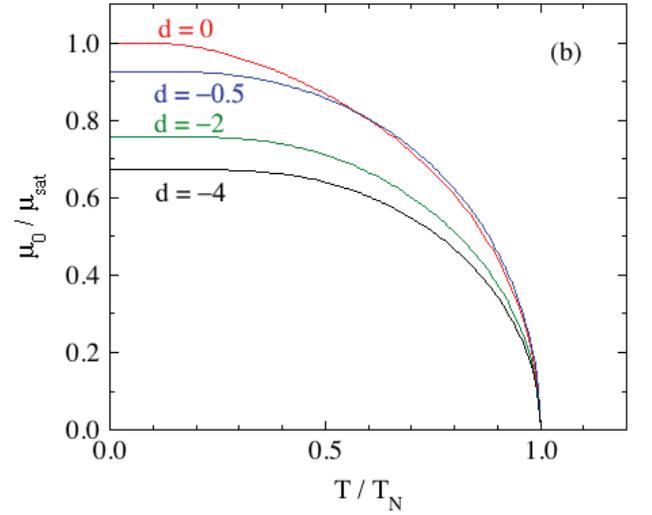}
\caption{(Color online) Same as Fig.~\ref{Fig:mu0xS1} except for $S=7/2$ and $d=D/k_{\rm B}T_{{\rm N}J} = 0,\ -0.5$, $-2$ and~$-4$.}
\label{Fig:mu0xS72}
\end{figure}

Plots of $\bar{\mu}_0$ versus~$t$ and versus $T/T_{\rm N}$ for $S=1$ and $S=7/2$ are shown in Figs.~\ref{Fig:mu0xS1} and~\ref{Fig:mu0xS72}, respectively, each with reduced anisotropy parameters $d=0,\ -0.5,\ -2$ and~$-4$.  For this in-plane orientiation of the easy axis, the normalized saturation moment does not go to unity at $T\to0$ for $d<0$, contrary to the case of $z$-axis ordering with $d>0$.  On the other hand, with $d>0$ a suppression of the ordered moment at $T=0$ was found for the spin-flop phase in Fig.~\ref{Fig:mu0barSFS1_vs_t_d}, as with $x$-axis ordering in Figs.~\ref{Fig:mu0xS1} and~\ref{Fig:mu0xS72}.

From Fig.~\ref{Fig:mu0xS1}(a), one sees that with increasingly negative values of~$d$, $t_{\rm N}$ for $S=1$ first increases, then decreases, and then stongly decreases for $d\to-3$, consistent with the explicit calculation of $t_{\rm N}(d)$ for $S=1$ in Fig.~\ref{Fig:tNVSd}(b) above.  On the other hand, with increasingly negative~$d$,  one sees from Fig.~\ref{Fig:mu0xS72}(a) that $T_{\rm N}(d)$ for $S=7/2$ initially increases but asymptotes to a constant value somehwat less than unity, consistent with $t_{\rm N}(d)$ for $S=7/2$ in Fig.~\ref{Fig:tNVSd}(c). 

\section{\label{Sec:Summary} Summary}

Theory was presented to calculate the magnetic and thermal properties of Heisenberg antiferromagnets with quantum uniaxial anisotropy of $-DS_z^2$ type.  The uniaxial anisotropy was included exactly and the Heisenberg interactions were treated within the unified molecular field theory in which the various parameters are expressed in terms of measurable properties.  This feature facilitates comparison of the theoretical predictions with experimental results compared to previous treatments in which the magnetic properties were expressed in terms of the Heisenberg exchange interactions themselves in addition to $D$\@.

Once the basic theory was formulated in Sec.~\ref{MFTBckgrnd}, it was applied to calculate many properties of these spin systems.  Of greatest interest are likely those associated with $D>0$ for which collinear AFM occurs along the $z$~axis.  The zero-field properties calculated include the N\'eel temperature~$T_{\rm N}$ versus~$D$, the ordered moment versus $D$ and temperature~$T$, and the magnetic entropy, internal energy, heat capacity and free energy versus~$D$ and~$T$\@.  In the absence of an ordered moment above~$T_{\rm N}$, the heat capacity is a Schottky anomaly arising from the zero-field splittings of the energy levels.  In addition to calculating the parallel susceptibility, we also obtained the perpendicular susceptibility using second-order perturbation theory.  The high-field uniform magnetization along the $z$~axis was calculated versus~$D$ and~$T$, together with the average staggered magnetization per spin (the ordered moment) which is the AFM order parameter.  A complete treatment of the magnetic properties of the spin-flop (SF) phase was also presented in which the applied field was along the $z$~axis.  We also considered the influence of a perpendicular field along the $x$~axis on the magnetization and presented the perpendicular critical field versus~$D$ and~$T$ for the resulting second-order AFM/PM transition.

Together with the results for the paramagnetic (PM) and SF phases, these results were used to construct phase diagrams in the $H_z-T$ plane for spin $S=1$, a particular value of~$D$, and for three different values of $f_J\equiv \theta_{{\rm p}J}/T_{{\rm N}J}$.  The value $f_J=-1$ is obtained, e.g., for a bipartite AFM spin lattice with equal nearest-neighbor AFM exchange interactions and no further-neighbor interactions.  Upon algebraically increasing $f_J$, as occurs if ferromagnetic interactions are present, the phase diagrams evolve.  For $f_J=-1$ and~$-0.7$ the phase diagrams are similar to previous calculations.  However, for $f_J=0$ we find a topologically distinct phase diagram in which the SF phase exists as a bubble at finite $H_z$ and~$T$\@.  It would be very interesting to extend the present work to a detailed study of how the phase diagram evolves with increasing $f_J$ at fixed~$D$\@.

We also studied the magnetic properties of systems with~$D<0$, which results in AFM ordering within the $xy$~plane.  We considered the case of collinear AFM ordering for which $T_{\rm N}(D)$ and the ordered moment versus $D$ and~$T$ were calculated.

It is interesting and useful to compare the magnetic and thermal results on the above systems with correponding results on noninteraction spin systems with quantum uniaxial anisotropy only.  For this purpose such calculations were carried out and plots of the results made, which are are included in the Appendix.

The main purpose of this work was to provide a convenient and detailed framework to quantitatively estimate the influence of uniaxial anisotropy on the measured thermal and magnetic properties of real Heisenberg antiferromagnets from measurements of the anisotropic properties of single crystals.  The influence of the magnetic dipole interaction in producing such anisotropies was previously considered in detail for a variety of spin lattices within the same unified MFT utilized here \cite{Johnston2016}.

\acknowledgments

This work was supported by the U.S. Department of Energy, Office of Basic Energy Sciences, Division of Materials Sciences and Engineering.  Ames Laboratory is operated for the U.S. Department of Energy by Iowa State University under Contract No.~DE-AC02-07CH11358.

%\clearpage

\appendix

\section{\label{NonintSpins} Noninteracting Spins}

Here we compute the magnetic behaviors of magnetically noninteracting spins because they illustrate the influence of the uniaxial anisotropy on the magnetic properties of a spin system without the complications of including Heisenberg exchange interactions.  In the absence of spin interactions, the Hamiltonian~(\ref{Eq:Ham1}) becomes
\be
{\cal H} = g\mu_{\rm B}{\bf S}_i\cdot {\bf H} - DS_z^2.
\label{Eq:Ham2}
\ee

\subsection{Longitudinal magnetic fields ${\bf H} = {H}_z\hat{\bf k}$}

\subsubsection{Zeeman Energy Levels}

From Eq.~(\ref{Eq:Ham2}) one obtains the single-spin eigenenergies
\bse
\be
E(m_S) = g\mu_{\rm B}H_z m_S - Dm_S^2.
\label{Eq:Ei}
\ee
We normalize energies by $|D|$ and define the reduced field
\be
h_z = \frac{g\mu_{\rm B}H_z}{|D|},
\label{Eq:hzDef}
\ee
yielding the reduced energy 
\be
\epsilon(m_S) \equiv\frac{E(m_S)}{|D|} = h_zm_S - {\rm sgn}(D)m_S^2.
\label{Eq:eps(m_S)}
\ee
Thus if $D>0$ the larger $|m_S|$ states have the lower energy and hence alignment of the spins along the $z$~axis is favored, whereas if $D<0$ the smaller $|m_S|$ states have the lower energy and hence alignment of the spins within the $xy$~plane is favored. Similarly, we define the reduced temperature as 
\be
t = \frac{k_{\rm B}T}{|D|}.
\ee
\ese

\begin{figure}
\includegraphics [width=3.3in]{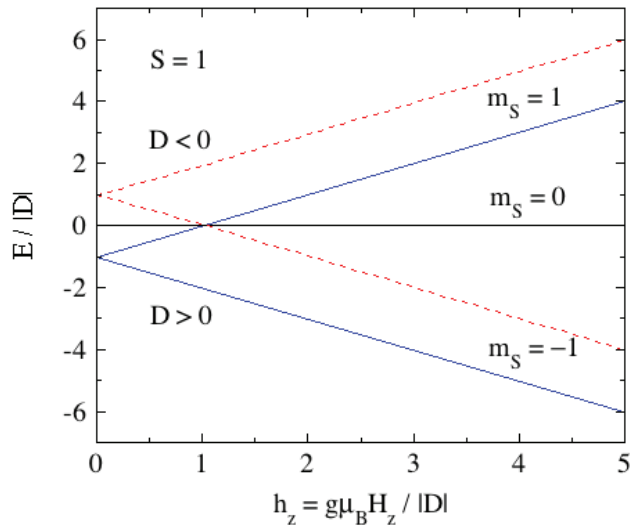}
\caption{(Color online) Reduced Zeeman energy levels $\epsilon(m_S) = E[m(S)]/|D|$ versus reduced field $h_z$ obtained for $S=1$ and both $D>0$ (solid lines) and $D<0$ (dashed lines) obtained from Eq.~(\ref{Eq:eps(m_S)}). The $m_S = 0$  Zeemann levels for the two signs of $D$ overlap.}
\label{Fig:EvsHzS1d1m1}
\end{figure}

The dependences of the Zeeman energy levels $E(m_S)$ versus $h_z$ for $S=1$ and both $D>0$ (solid lines) and $D < 0$ (dashed lines) are shown in Fig.~\ref{Fig:EvsHzS1d1m1}.  The $E(m_S = 0)$ levels are the same for $D>0$ and~$D<0$.  One notices that for $D>0$ the ground state is the $m_S=-1$ Zeeman level for all~$h_z$.  On the other hand, for $D<0$ the ground state is the $m_S=0$ level for $h_z<1$ whereas the $m_S=-1$ state is the ground state for $D>0$ and $h_z>1$.  Thus for $D<0$ a first-order transition occurs at $h_z=1$ between a nonmagnetic ground state and a magnetic one.  This results in a first-order transition at $T=0$ in the $\mu_z(h_z)$ isotherm [see Fig.~\ref{Fig:muzbarVsHzS1d1m1}(b) below].

\begin{figure}
\includegraphics [width=3.3in]{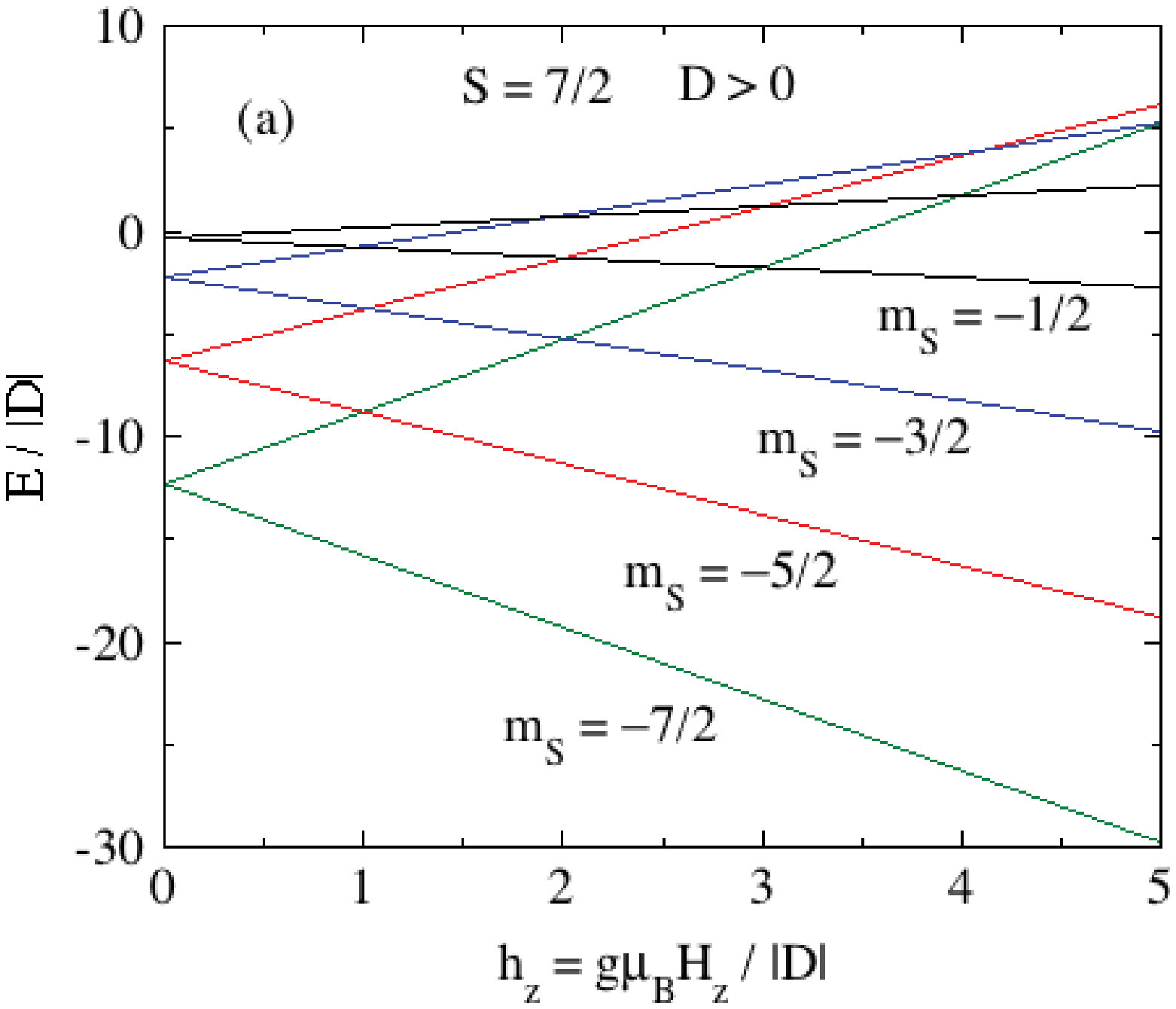}
\includegraphics [width=3.3in]{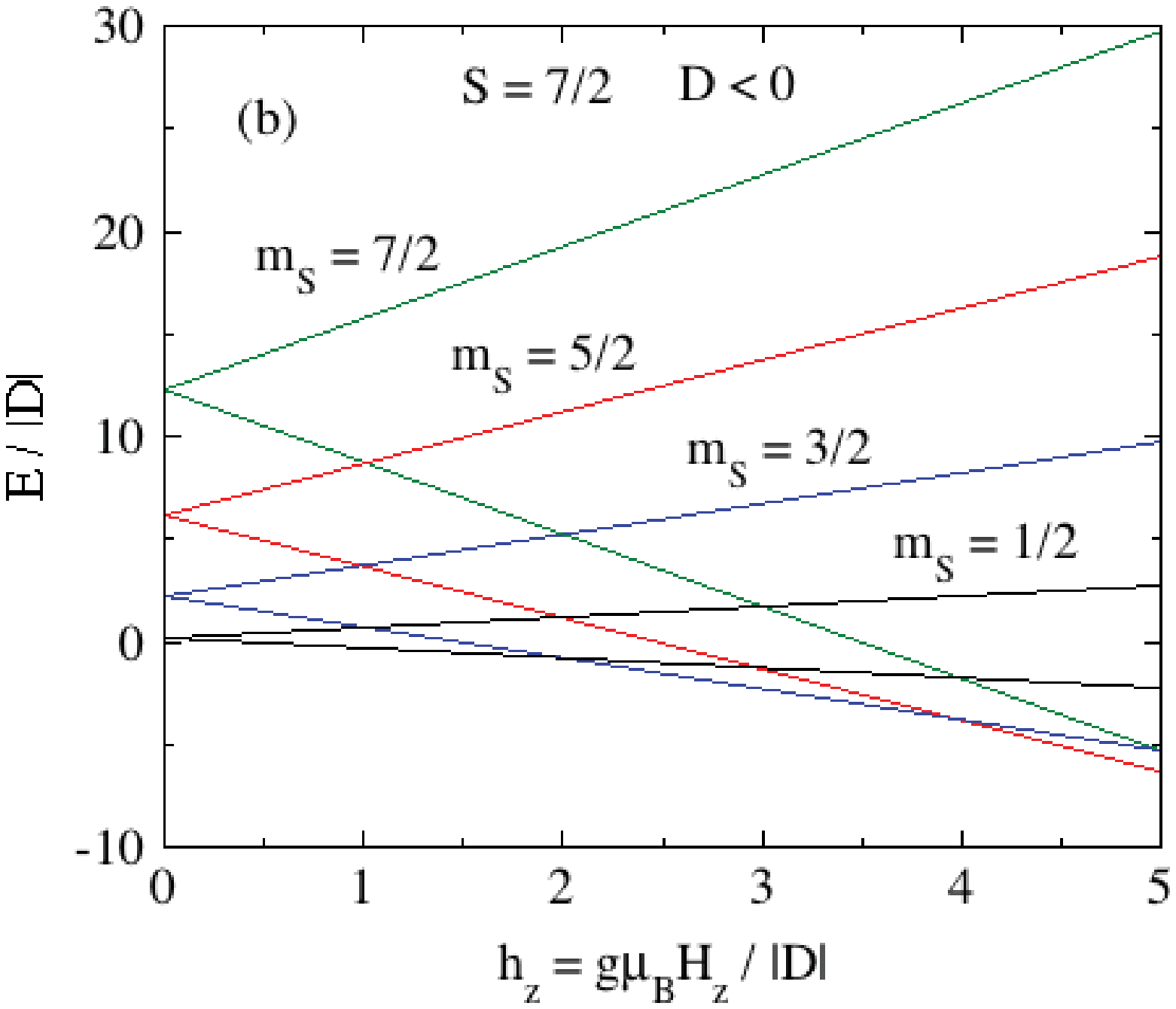}s
\caption{(Color online) Zeeman energy levels $\epsilon(m_S) = E[m(S)]/|D|$ versus reduced field $h_z$ obtained from Eq.~(\ref{Eq:eps(m_S)}) for $S=7/2$ and both (a)~$D>0$ and (b)~$D<0$.}
\label{Fig:EvsHzS72d1m1}
\end{figure}

The corresponding plots of $\epsilon(m_S)$ versus $h_z$ for $S=7/2$ with $D>0$ and $D<0$ are shown in Figs.~\ref{Fig:EvsHzS72d1m1}(a) and~\ref{Fig:EvsHzS72d1m1}(b), respectively.  For $D>0$ the ground state is always the $m_S=-7/2$ Zeeman level.  For $D<0$, Zeeman level crossings occur with increasing field where the ground state is the $m_S=-1/2$ level for $h_z< 2$, the $m_S=-3/2$ level for $2< h_z<4$, the $m_S=-5/2$ level for $4< h_z<6$ and the $m_S=-7/2$ level for $h_z>6$.  These result in three first-order transitions at $T=0$ in the $\mu_z(h_z)$ isotherm [see Fig.~\ref{Fig:muzbarVsHzS72d1m1}(b) below].

\subsubsection{Magnetic Entropy}

The magnetic entropy is calculated using Eqs.~(\ref{Eqs:SmagCmag}) and the eigenenergies in Eq.~(\ref{Eq:eps(m_S)}).  
For example, for $h_z=0$ the molar magnetic entropies for spins~1 and~3/2 are given for $D>0$ and $D<0$ by
\bse
\label{Eqs:SmagEqsS1S32}
\bea
\frac{S_{\rm mag}(S=1,D>0,t)}{R} &=& \\*
&&\hspace{-0.5in}\ln(1 + 2e^{1/t}) + \frac{1}{t(1+2e^{1/t})}- \frac{1}{t},\nonumber\\*
\frac{S_{\rm mag}(S=1,D<0,t)}{R} &=& \\*
&&\hspace{-0.5in}\ln(2 + e^{1/t}) - \frac{1}{t(1+2e^{-1/t})},\nonumber\\*
\frac{S_{\rm mag}(S=3/2,t)}{R} &=& \\*
&&\hspace{-0.6in}\ln[2(1 + e^{2/t})] - \frac{1+\tanh(1/t)}{t},\nonumber
\eea
\ese
where the expressions for $S=3/2$ are the same for $D>0$ and $D<0$, just as for the $C_{\rm mag}(t)$ expressions in the following section.

From Figs.~\ref{Fig:EvsHzS1d1m1} and~\ref{Fig:EvsHzS72d1m1}, for integer~$S$ the degeneracy of the ground state is~2 for $D>0$ and $h_z = 0$; 1~for $D>0$ and $h_z > 0$; and~1 for $D < 0$, yielding
\bse
\label{Eqs:SmagIsoSpins232}
\bea
\frac{S_{\rm mag}}{R}(h_z = 0,t=0) &=& \ln2 \quad (D>0),\\*
\frac{S_{\rm mag}}{R}(h_z > 0,t=0) &=& 0 \quad(D>0),\qquad({\rm integer}~S)\nonumber\\*
\frac{S_{\rm mag}}{R}(h_z \geq 0,t=0) &=& 0 \quad(D<0).\nonumber
\eea
On the other hand, for half-integer~$S$ with either $D>0$ or $D<0$ the degeneracy of the ground state is either 2~$(h_z = 0)$ or 1~$(h_z > 0)$, so one obtains
\bea
\frac{S_{\rm mag}}{R}(h_z = 0,t=0) &=& \ln2 \ ,\\*
\frac{S_{\rm mag}}{R}(h_z > 0,t=0) &=& 0 \quad ({\rm half~integer~}S).\nonumber
\eea
\ese
Results obtained from Eqs.~(\ref{Eqs:SmagCmag}) are consistent with these requirements at $t\to0$, and also give the correct entropy per spin $S_{\rm mag}/k_{\rm B} = \ln(2S+1)$ for arbitrary spin~$S$ at $t\to\infty$.

\subsubsection{Magnetic Heat Capacity}

\begin{figure}
\includegraphics [width=3.3in]{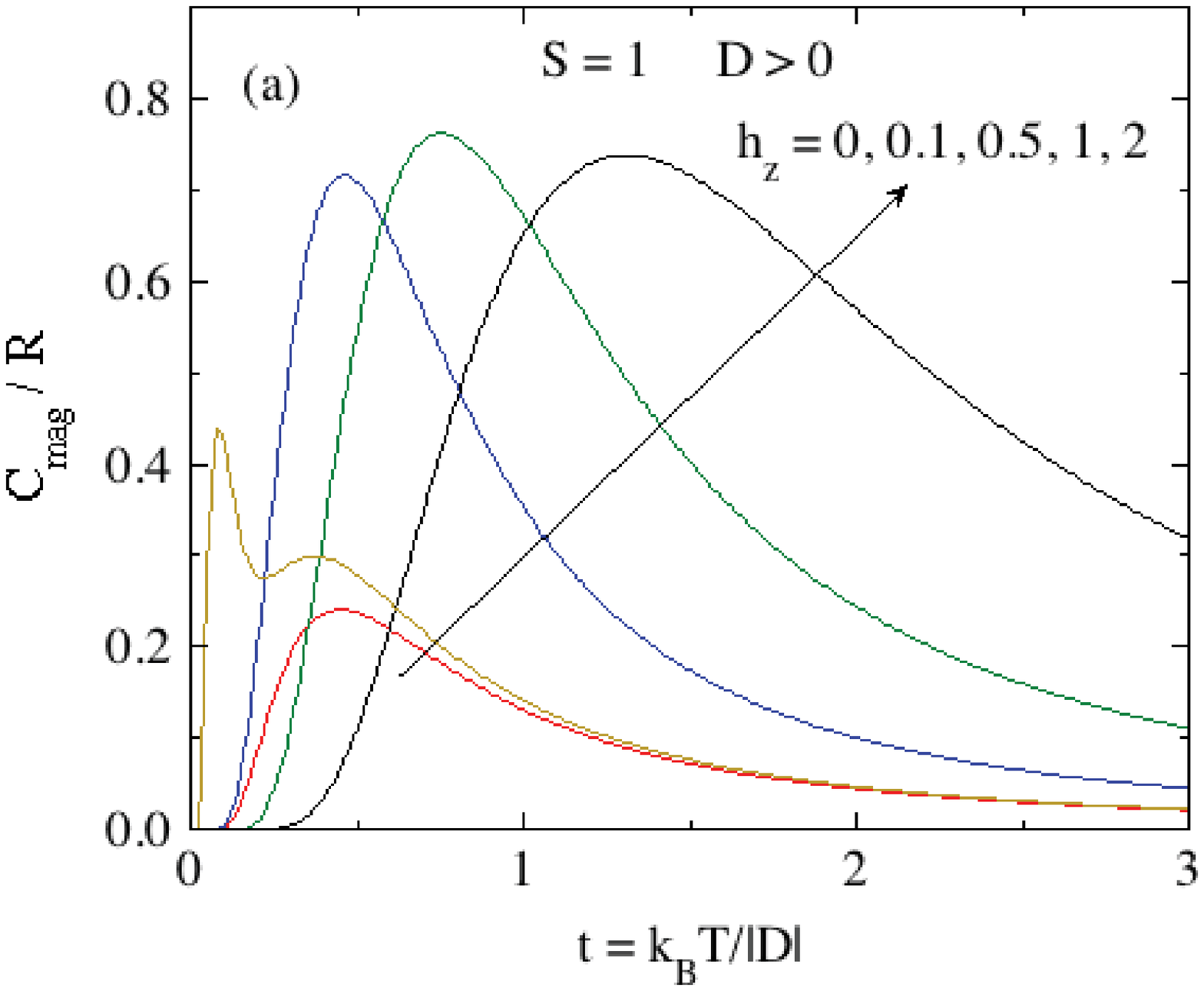}
\includegraphics [width=3.3in]{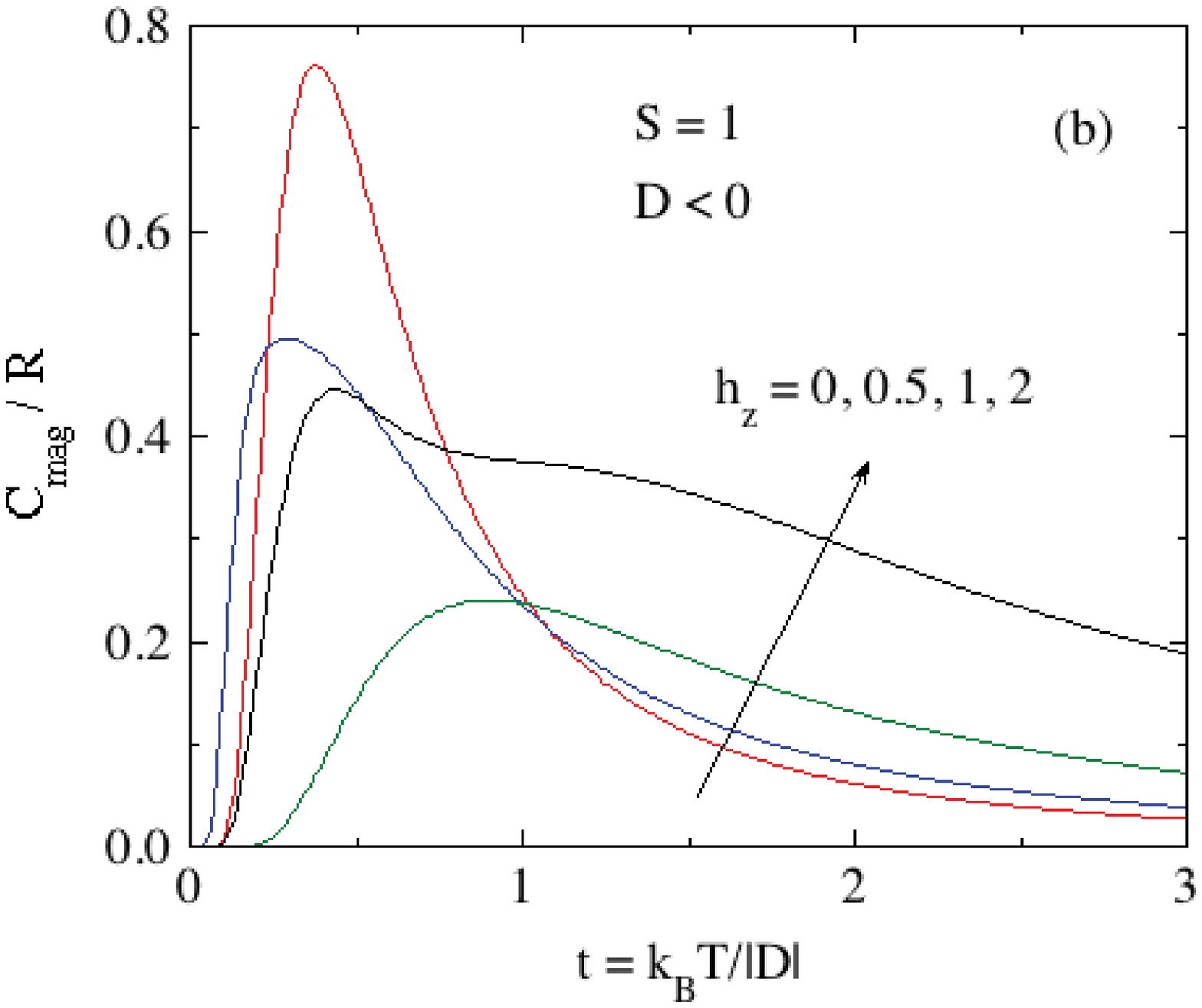}
\caption{(Color online) Molar magnetic heat capacity $C_{\rm mag}/R$ versus reduced field temperature $t$ obtained for $S=1$ and (a)~$D>0$ and (b)~$D<0$ from Eqs.~(\ref{Eqs:umagCmag}).}
\label{Fig:CmagS1d1m1Iso}
\end{figure}

\begin{figure}
\includegraphics [width=3.3in]{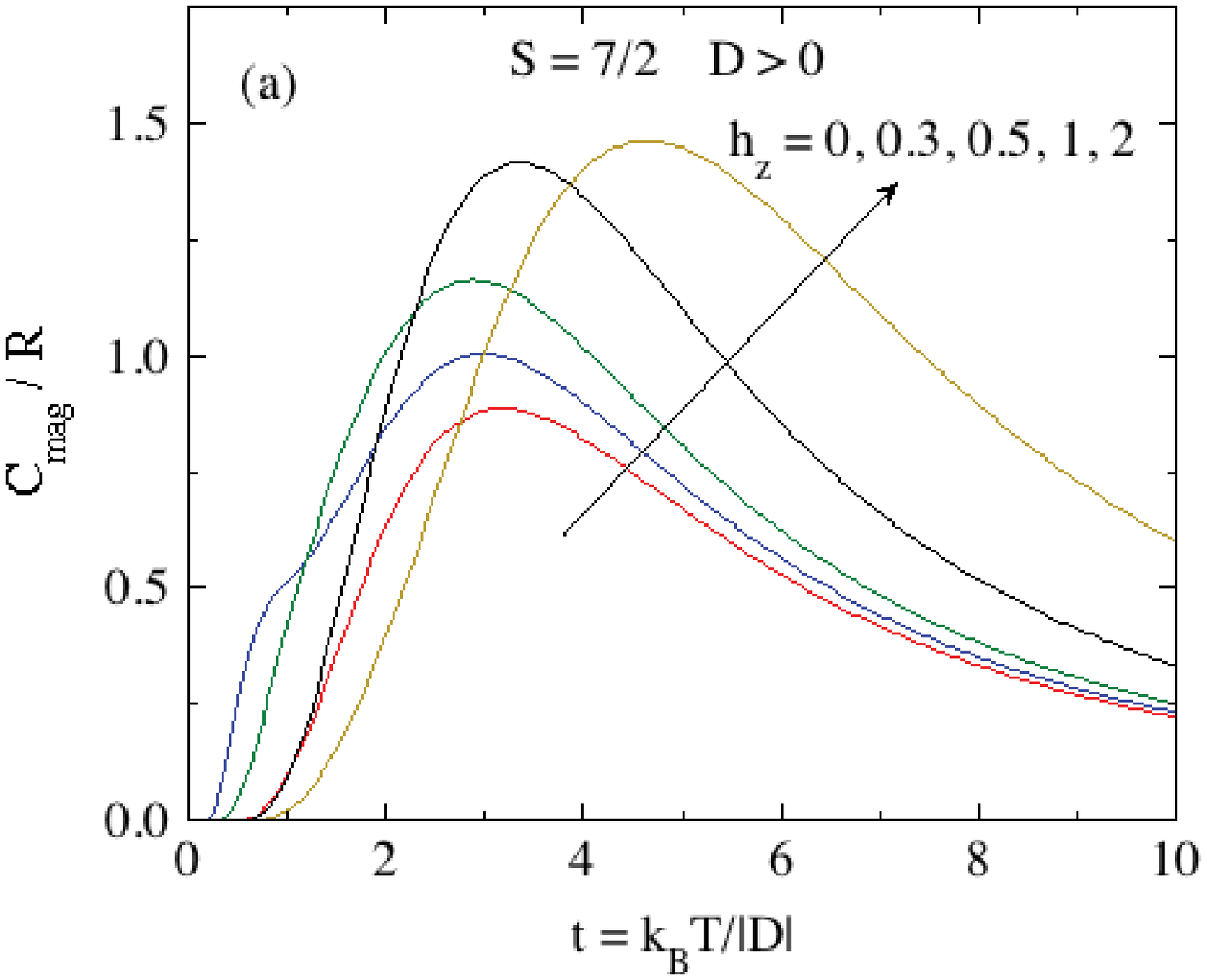}
\includegraphics [width=3.3in]{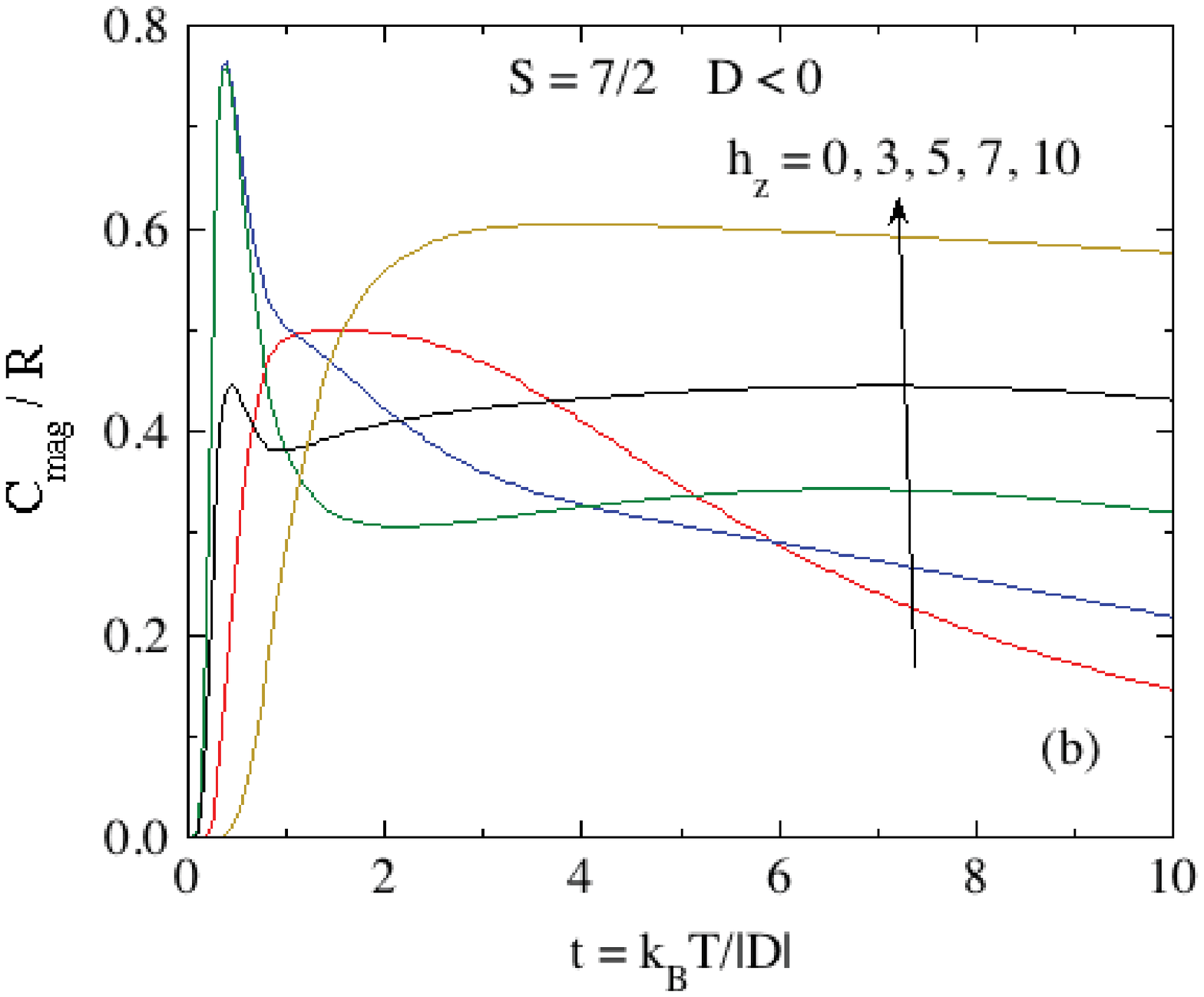}
\caption{(Color online) Same as Fig.~\ref{Fig:CmagS1d1m1Iso} but with $S=7/2$.}
\label{Fig:CmagS72d1m1Iso}
\end{figure}

The thermal-average reduced internal magnetic energy $u_{\rm mag}$ for spin~$S$ is given per spin in a reduced field $h_z$ by
\bse
\label{Eqs:umagCmag}
\be
u_{\rm mag} = \frac{U_{\rm mag}}{|D|} = \frac{1}{Z_S}\sum_{m_S=-S}^S \epsilon(m_S)e^{- \epsilon(m_S)/t},
\ee
where the reduced energy $\epsilon(m_S)$ is given in Eq.~(\ref{Eq:eps(m_S)}) and the partition function $Z_S$ is
\be
Z_S = \sum_{mS=-S}^S e^{- \epsilon(m_S)/t}.
\label{Eq:ZShz}
\ee
Then for a mole of spins the magnetic heat capacity $C_{\rm mag}$ is
\be
\frac{C_{\rm mag}}{R} = \frac{du_{\rm mag}}{dt},
\ee
\ese
where $R$ is the molar gas constant.  Figures~\ref{Fig:CmagS1d1m1Iso} and \ref{Fig:CmagS72d1m1Iso} show $C_{\rm mag}/R$ versus~$t$ for both $D>0$ and $D<0$ with $S=1$ and $S=7/2$, respectively.  One sees that the evolution in the shapes of $C_{\rm mag}/R$ versus~$t$ for $S=1$ and $S=7/2$ for each of $D>0$ and $D<0$ are similar, although for each spin value the $C_{\rm mag}/R$ versus~$t$ for $D>0$ and $D<0$ are quite different, as one might infer from the different organization of the Zeeman levels for $D>0$ and $D<0$ in Figs.~\ref{Fig:EvsHzS1d1m1} and~\ref{Fig:EvsHzS72d1m1} for the two spin values, respectively.

Analytic expressions for the molar $C_{\rm mag}(t)$ in $h_z=0$ are obtained from Eqs.~(\ref{Eqs:umagCmag}).  For $S=3/2$, the same expression is obtained for both positive and negative~$D$, but this does not occur generally.  We obtain
\bse
\label{CmagIsoSpins}
\bea
\frac{C_{\rm mag}(S=1,D>0,t)}{R} &=& \frac{2e^{1/t}}{t^2(1+2e^{1/t})^2},\\*
\frac{C_{\rm mag}(S=1,D<0,t)}{R} &=& \frac{2e^{1/t}}{t^2(2+e^{1/t})^2},\nonumber\\*
\frac{C_{\rm mag}(S=3/2,t)}{R} &=& \frac{4e^{2/t}}{t^2(1+e^{2/t})^2},\nonumber\\*
\frac{C_{\rm mag}(S=2,D>0,t)}{R} &=& \frac{2e^{1/t}(1+16e^{3/t}+18e^{4/t})}{t^2(1+2e^{1/t}+2e^{4/t})^2},\nonumber\\*
\frac{C_{\rm mag}(S=2,D<0,t)}{R} &=& \frac{2e^{3/t}(18+16e^{1/t}+e^{4/t})}{t^2(2+2e^{3/t}+e^{4/t})^2},\nonumber\\*
\frac{C_{\rm mag}(S=5/2,D>0,t)}{R} &=& \frac{4e^{2/t}(1+9e^{4/t}+4e^{6/t})}{t^2(1+e^{2/t}+e^{6/t})^2},\nonumber\\*
\frac{C_{\rm mag}(S=5/2,D<0,t)}{R} &=& \frac{4e^{4/t}(4+9e^{2/t}+e^{6/t})}{t^2(1+e^{4/t}+e^{6/t})^2},\nonumber\\
\frac{C_{\rm mag}(S=3,D>0,t)}{R} &=&\nonumber\\*
&& \hspace{-1.7in}\frac{2e^{1/t}(1+16e^{3/t}+18e^{4/t} + 81e^{8/t} + 128e^{9/t}+50e^{12/t})}{t^2(1+2e^{1/t}+2e^{4/t}+2e^{9/t})^2},\nonumber\\*
\frac{C_{\rm mag}(S=3,D<0,t)}{R} &=&\nonumber\\*
&& \hspace{-1.6in}\frac{2e^{5/t}(50+128e^{3/t}+81e^{4/t} + 18e^{8/t} + 16e^{9/t}+e^{12/t})}{t^2(2+2e^{5/t}+2e^{8/t}+e^{9/t})^2},\nonumber\\*
\frac{C_{\rm mag}(S=7/2,D>0,t)}{R} &=&\nonumber\\*
&& \hspace{-1.45in}\frac{4e^{2/t}(1+9e^{4/t}+4e^{6/t} + 36e^{10/t} + 25e^{12/t}+9e^{16/t})}{t^2(1+e^{2/t}+e^{6/t}+e^{12/t})^2},\nonumber\\*
\frac{C_{\rm mag}(S=7/2,D<0,t)}{R} &=&\nonumber\\*
&& \hspace{-1.45in}\frac{4e^{6/t}(9+25e^{4/t}+36e^{6/t} + 4e^{10/t} + 9e^{12/t}+e^{16/t})}{t^2(1+e^{6/t}+e^{10/t}+e^{12/t})^2}.\nonumber
\eea
\ese
Each of these results satisfy the $t\to\infty$ entropy requirement $S_{\rm mag}(t=0)/R + \int_0^\infty [C_{\rm mag}(t)/R]/t = \ln(2S+1)$ for $h_z=0$, where the $S_{\rm mag}(t=0)/R$ is given in Eqs.~(\ref{Eqs:SmagIsoSpins232}) for the respective $D>0$ or $D<0$ and integer or half-integer spin.

\subsubsection{Thermal-Average Magnetic Moment}

In the absence of exchange interactions and using the current reduced variables, the induced moment is described by an equation similar to Eq.~(\ref{Eq:barmuFromGS}),
\bse
\label{Eqs:muz}
\bea
\bar{\mu}_z &=& G_S(y),\label{Eq:GSDef}\\*
G_{S}(y) &=& -\frac{1}{SZ_S}\sum_{m_S = -S}^S m_S \exp\left[\frac{{\rm sgn}(D)m_S^2}{t}\right]e^{-ym_S},\nonumber\\*
\\*
Z_S &=& \sum_{mS=-S}^S \exp\left[\frac{{\rm sgn}(D)m_S^2}{t}\right]e^{-ym_S},\label{Eq:ZShz2}\\*
y &=& \frac{g\mu_{\rm B}H_z}{k_{\rm B}T} = \frac{h_z}{t}.
\eea
\ese

\subsubsection{Low-Field Magnetization and Magnetic Susceptibility}

Here we only keep terms in Eqs.~(\ref{Eq:GSDef}) to first order in $h_z$, yielding
\bea
\bar{\mu}_z &=& \frac{h_z}{t SZ_S}\sum_{m_S=-S}^S m_S^2\exp\left[\frac{{\rm sgn}(D)m_S^2}{t}\right],\label{Eq:muLowH}\\*
Z_S &=& \sum_{m_S=-S}^S \exp\left[\frac{{\rm sgn}(D)m_S^2}{t}\right].\nonumber
\eea
Since the magnetic susceptibility is $\chi_z = \mu_z/H_z$, using Eqs.~(\ref{Eq:hzDef}) and~(\ref{Eq:muLowH}) one obtains the reduced $z$-axis magnetic susceptibility $\bar{\chi}_z$ for $N$ spins as
\bea
\bar{\chi}_{z} &\equiv& \frac{\chi|D|}{Ng^2\mu_{\rm B}^2} = \frac{S\bar{\mu}_z}{h_z}\label{Eq:chiz} \\* 
&=& \frac{1}{t Z_S}\sum_{m_S=-S}^S m_S^2\exp\left[\frac{{\rm sgn}(D)m_S^2}{t}\right]\nonumber\\*
&=& \frac{1}{t}\langle S_z^2\rangle,\nonumber
\eea
where $\langle \cdots\rangle$ is the thermal average of the quantity between the angular brackets and the third equality is a well-known application of the fluctuation-dissipation theorem.  

At high temperatures one can expand Eq.~(\ref{Eq:chiz}) to second order in $1/t$ and thereby obtain the single-spin Curie-Weiss law~(\ref{Eqs:CWLaw}) with the Curie constant given in Eq.~(\ref{C1}) and with Weiss temperature
\be
\theta_{{\rm p}z} = \left(\frac{D}{k_{\rm B}}\right)\frac{(2S-1)(2S+3)}{15}.
\label{Eq:thetapz}
\ee
Thus for $S=1/2$ one obtains $\theta_{{\rm p}z} = 0$, i.e., there is no anisotropy since the anisotropy term is just a constant for $S=1/2$.  Because magnetic anisotropy tensors in the PM state with principal axis bases are traceless, one has
\be
\theta_{{\rm p}x} = \theta_{{\rm p}y} = -\frac{\theta_{{\rm p}z}}{2} = -\left(\frac{D}{k_{\rm B}}\right)\frac{(2S-1)(2S+3)}{30}.
\label{Eq:thetapx}
\ee
If $D>0$ then $\theta_{{\rm p}z}>0$ and $\theta_{{\rm p}x,y}<0$, favoring moment alignment along the $z$~axis as expected, whereas if $D>0$ then $\theta_{{\rm p}z}<0$ and $\theta_{{\rm p}x,y}>0$, favoring moment alignment within the $ab$~plane.  Also, if $D>0$ and hence $\theta_{{\rm p}z}>0$, the susceptibility diverges on cooling to $t=0$, irrespective of~$S$.  The Curie-Weiss law with the same Curie constant and Weiss temperatures can also be obtained from a perturbation theory calculation as described generically in Sec.~\ref{Sec:PerpPertThy}.  

\begin{figure}
\includegraphics [width=3.3in]{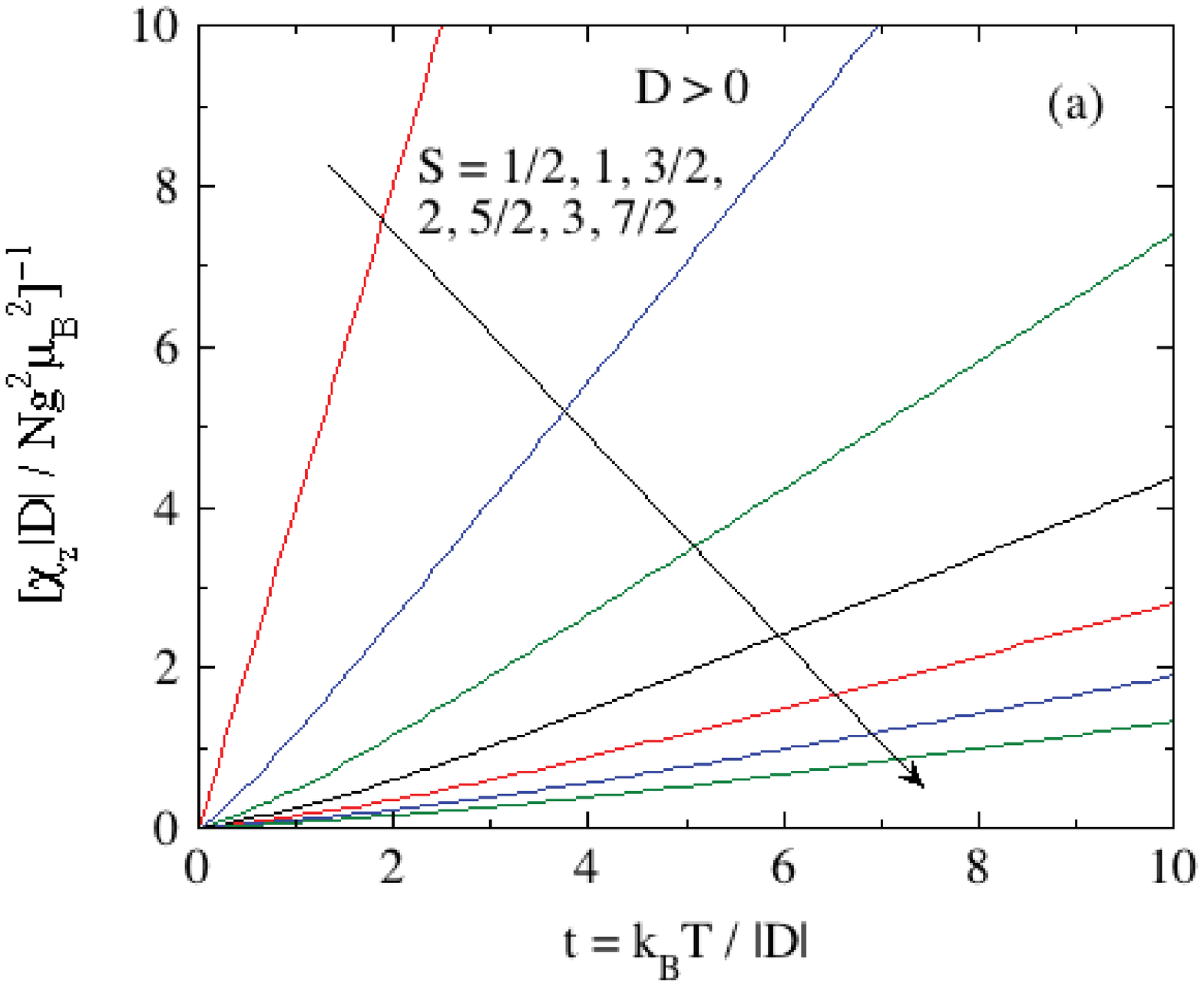}
\includegraphics [width=3.3in]{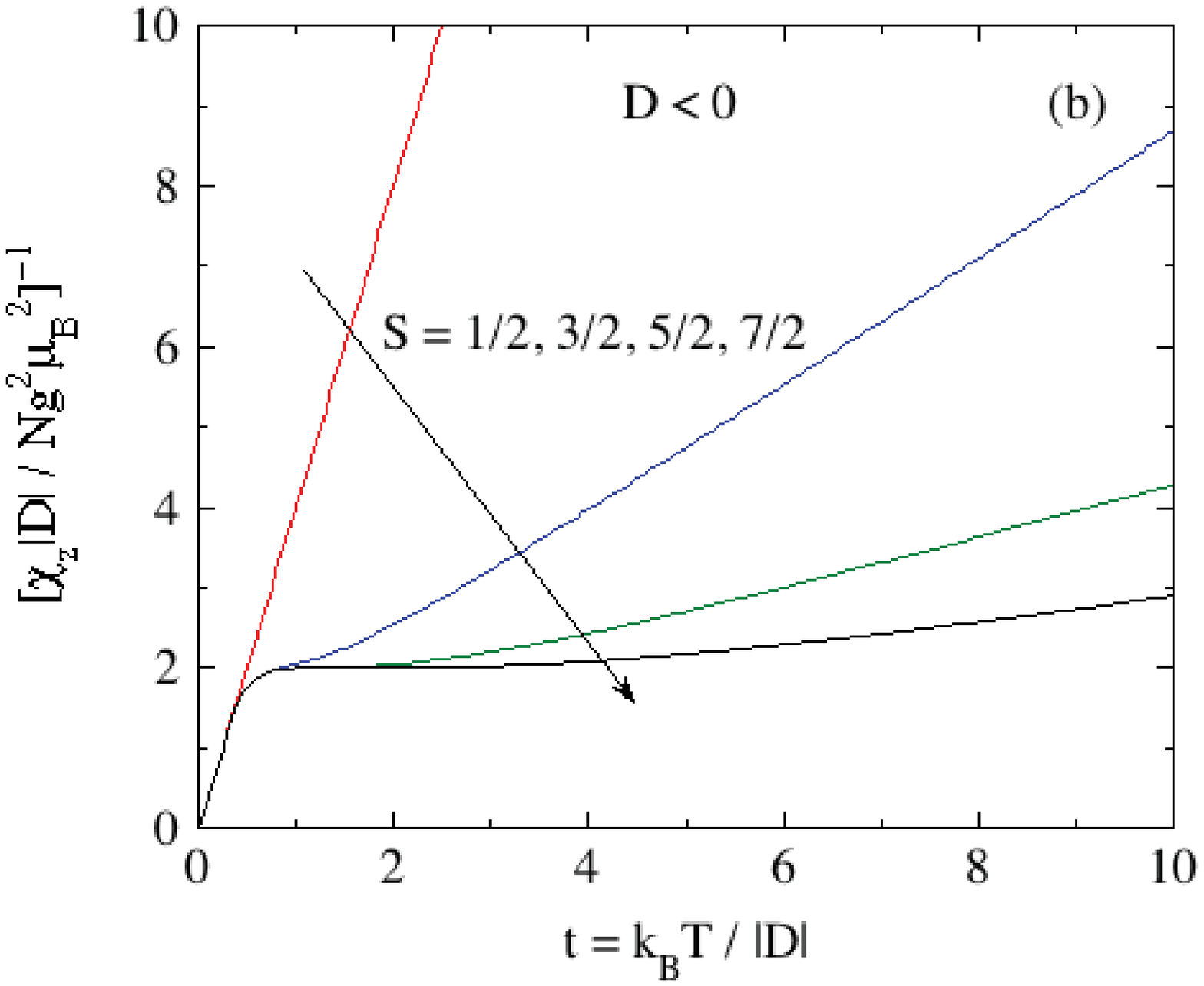}
\includegraphics [width=3.3in]{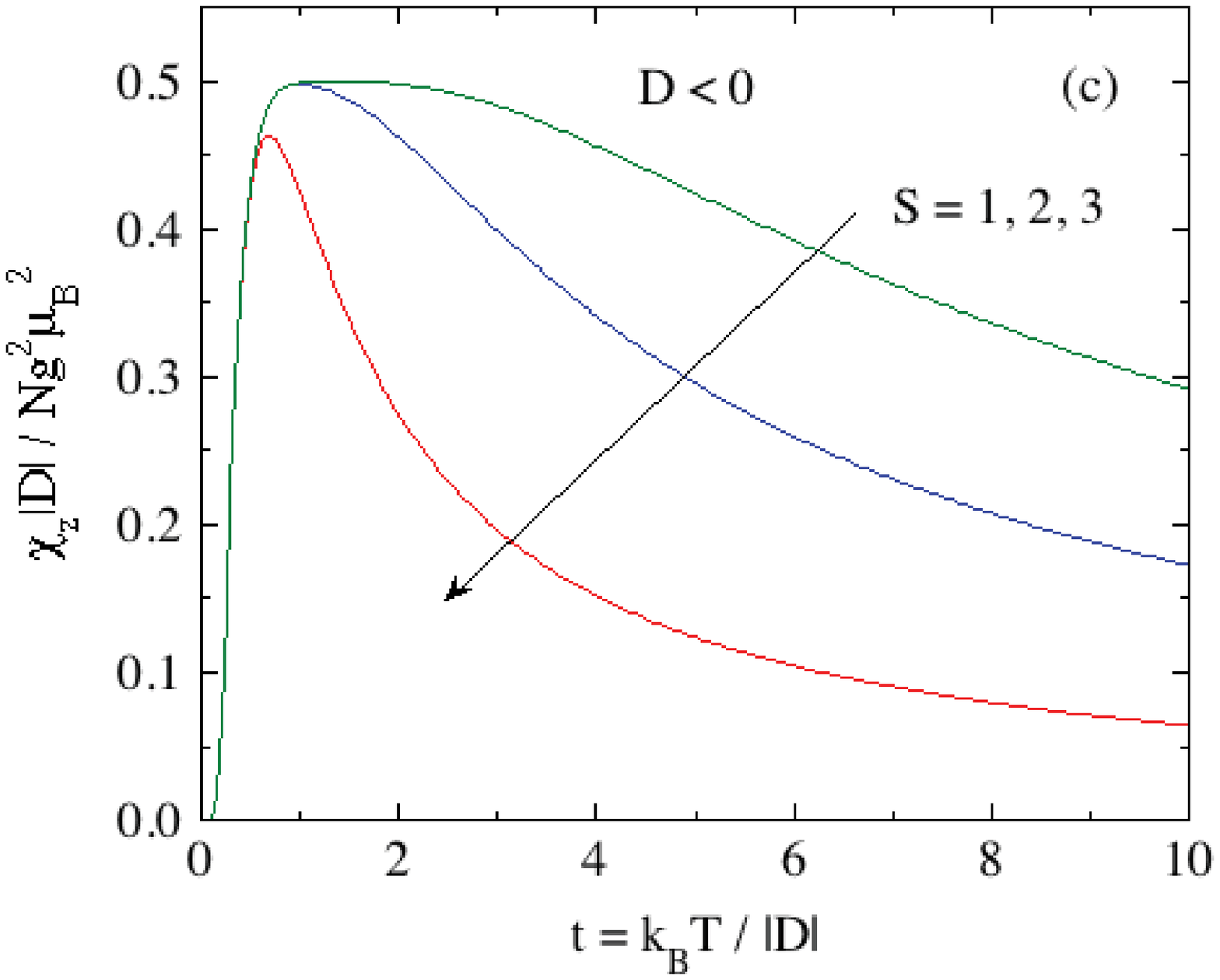}
\caption{(Color online) (a)~Inverse magnetic susceptibility $\bar{\chi}_z^{-1}\equiv (\chi_z|D|/Ng^2\mu_{\rm B}^2)^{-1}$ versus reduced temperature~$t$ for $S=1/2,\ldots,7/2$ with $D>0$. (b)~$\bar{\chi}_z^{-1}$ versus~$t$ for half-integer spins and $D<0$.  (c)~$\bar{\chi}_z$ versus~$t$ for integer spins $S=1$, 2, 3 and $D<0$.  These results were obtained using Eq.~(\ref{Eq:chiz}).}
\label{Fig:ChizS1S72d1dm1}
\end{figure}

\begin{table}
\caption{\label{Table:NoninteractingSpinData} Summary of properties of noninteracting spins with uniaxial single-ion anisotropy versus spin~$S$ of the magnetic ion.  Listed are the molar Curie constant~$C$ obtained using Eq.~(\ref{C1}), the Weiss temperature $\theta_{{\rm p}z}$ from Eq.~(\ref{Eq:thetapz}), the reduced temperature $t(\chi^{\rm max})$ at which the susceptibility reaches its maximum for integer spins if $D<0$. For $D>0$, the $\theta_{{\rm p}z}$ values are those listed (FM-like), whereas for $D<0$ the $\theta_{{\rm p}z}$ values are the negative of those listed (AFM-like). }
\begin{ruledtabular}
\begin{tabular}{cccc}
Spin~$S$ 	& $C$				& $\theta_{{\rm p}z}/(|D|/k_{\rm B})$	& $t(\chi^{\rm max})$	\\
		& $\left({\rm \frac{cm^3\,K}{mol}}\right)$	& (K)	& ($D<0$, K)					\\
\hline 
1/2  	& 0.375  	& 0  	& 	\\
1  		& 1.000  	& 1/3  	& 0.6835	\\
3/2  	& 1.876  	& 4/5 	& 	\\
2  		& 3.001  	& 7/5  	& 1.0017	\\
5/2  	& 4.377  	& 32/15	& 	\\
3  		& 6.002  	& 3  	& 1.3229\\
7/2  	& 7.878  	& 4  	& 	\\
\end{tabular}
\end{ruledtabular}
\end{table}

The $\bar{\chi}_z^{-1}(t)$ for $D>0$ calculated using Eq.~(\ref{Eq:chiz}) is shown for $S=1$ to $S=7/2$ in Fig.~\ref{Fig:ChizS1S72d1dm1}(a). Figure~\ref{Fig:ChizS1S72d1dm1}(b) shows $\bar{\chi}_z^{-1}(t)$ for half-integer~$S$ with $D<0$ and Fig.~\ref{Fig:ChizS1S72d1dm1}(c) shows $\bar{\chi}_z(t)$ for integer~$S$ for $D<0$.  In~(a), the $\chi(t)$ diverges at $t=0$ for all~$S$.  Figure~\ref{Fig:ChizS1S72d1dm1}(b) shows that $\bar{\chi}_z(t)$ crosses over on cooling to a $S=1/2$ behavior for half-integer spins at $t\lesssim0.3$ because of the $S=1/2$ doublet ground state in Fig.~\ref{Fig:EvsHzS72d1m1}(b).  On the other hand, for $D<0$ and integer spins, on cooling the $\chi_z(t)$ goes over a maximum at a temperature $t(\chi^{\rm max})$ and then approaches zero exponentially for $t\to0$.  This happens because of the singlet ground state for integer spins if $D<0$ as shown for $S=1$ in Fig.~\ref{Fig:EvsHzS1d1m1}.  The values of $C$, $\theta_{{\rm p}z}/(|D|/k_{\rm B})$ and $t(\chi^{\rm max})$ are listed for spins~1/2 to 7/2 in Table~\ref{Table:NoninteractingSpinData}.  The expressions for $\bar{\chi}_z(S,t)$ for $D>0$ and $D<0$ are given in Eqs.~(\ref{Eq:ChiParPerpList}) and~(\ref{Eqs:ChiParD<0List}), respectively, each for $S=1/2$ to~7/2.

%\newpage

The expressions for $\bar{\chi}_z(S,t)$ and $\bar{\chi}_x(S,t)$ (obtained later) with $D>0$ are
\bea
\bar{\chi}_z(S=1/2) &=& \frac{1}{4 t} \qquad (D>0),\label{Eq:ChiParPerpList}\\
\bar{\chi}_x(S=1/2) &=& \frac{1}{4 t},\nonumber\\
\bar{\chi}_z(S=1) &=& \frac{2 e^{1/t}}{(1+2 e^{1/t}) t }, \nonumber\\
\bar{\chi}_x(S=1) &=& \frac{2 \left(-1+e^{1/t}\right)}{1+2 e^{1/t}},\nonumber\\
\bar{\chi}_z(S=3/2) &=& \frac{1+9 e^{2/t}}{4 (1 + e^{2/t})t }, \nonumber\\
\bar{\chi}_x(S=3/2) &=& \frac{4-3 t(1 - e^{2/t})}{4 (1 + e^{2/t})t }\nonumber\\
\bar{\chi}_z(S=2) &=& \frac{2 e^{1/t}+8 e^{4/t}}{(1+2 e^{1/t} +2 e^{4/t}) t}, \nonumber\\
\bar{\chi}_x(S=2) &=& \frac{2 \left(-9+7 e^{1/t}+2 e^{4/t}\right)}{3 (1 +2 e^{1/t}+2 e^{4/t})},\nonumber\\
\bar{\chi}_z(S=5/2) &=& \frac{1+9 e^{2/t}+25 e^{6/t}}{4 \left(1+e^{2/t}+e^{6/t}\right) t}, \nonumber\\
\bar{\chi}_x(S=5/2) &=& \frac{18+\left(-16+11 e^{2/t}+5 e^{6/t}\right) t}{8 \left(1+e^{2/t}+e^{6/t}\right) t},\nonumber\\
\tilde{\chi}_z(S=3) &=& \frac{2 (e^{1/t}+4 e^{4/t}+9 e^{9/t})}{(1+2 e^{1/t} +2 e^{4/t} +2 e^{9/t} )t} ,\nonumber\\
\tilde{\chi}_x(S=3) &=& \frac{2 \left(-90+65 e^{1/t}+16 e^{4/t}+9 e^{9/t}\right)}{15 \left(1+2 e^{1/t}+2 e^{4/t}+2 e^{9/t}\right)},\nonumber\\
\tilde{\chi}_z(S=7/2) &=& \frac{1+9 e^{2/t}+25 e^{6/t}+49 e^{12/t}}{4 \left(1+e^{2/t}+e^{6/t}+e^{12/t}\right) t}, \nonumber\\
\tilde{\chi}_x(S=7/2) &=& \frac{48+\left(-45+27 e^{2/t}+11 e^{6/t}+7 e^{12/t}\right) t}{12 \left(1+e^{2/t}+e^{6/t}+e^{12/t}\right) t}.\nonumber
\eea
Our expressions for $\chi_z(T)$ and $\chi_x(T)$ for $S=3/2$ in Eqs.~(\ref{Eq:ChiParPerpList}) agree with the respective expressions in Refs.~\cite{Carlin1985, Merabet1990}.

The expressions for $\bar{\chi}_z(S,t)$ {\rm with}\ $D<0$ are
\bea
\bar{\chi}_z(S=1/2) &=& \frac{1}{4 t}\qquad (D<0), \label{Eqs:ChiParD<0List}\\
\bar{\chi}_z(S=1) &=& \frac{2}{(2 + e^{1/t}) t }, \nonumber\\
\bar{\chi}_z(S=3/2) &=& \frac{9+ e^{2/t}}{6(1 + e^{2/t})t }, \nonumber\\
\bar{\chi}_z(S=2) &=& \frac{4+ e^{3/t}}{(2+2 e^{3/t} + e^{4/t}) t}, \nonumber\\
\bar{\chi}_z(S=5/2) &=& \frac{25+9 e^{4/t}+ e^{6/t}}{10 (1+e^{4/t}+e^{6/t}) t}, \nonumber\\
\tilde{\chi}_z(S=3) &=& \frac{2 (9 + 4 e^{5/t}+ e^{8/t})}{3(2+2 e^{5/t} +2 e^{8/t} + e^{9/t} )t}, \nonumber\\
\tilde{\chi}_z(S=7/2) &=& \frac{49+25 e^{6/t}+9 e^{10/t}+ e^{12/t}}{14 (1+e^{6/t}+e^{10/t}+e^{12/t}) t}. \nonumber
\eea

\subsubsection{High-Field Magnetization versus Field Isotherms with Fields along the $z$~axis}

\begin{figure}
\includegraphics [width=3.3in]{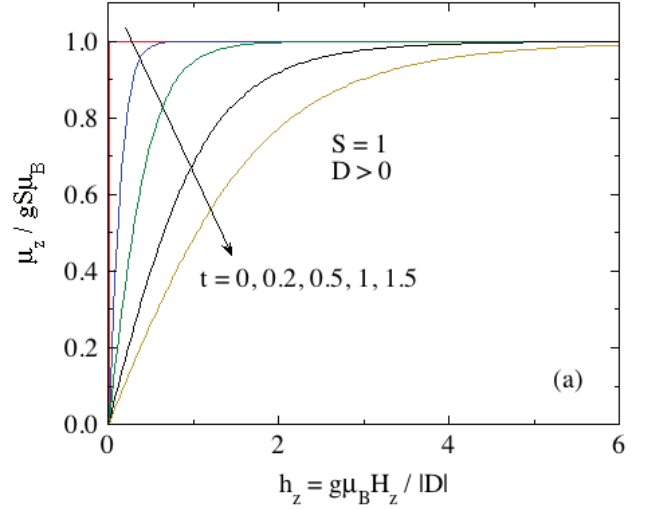}
\includegraphics [width=3.3in]{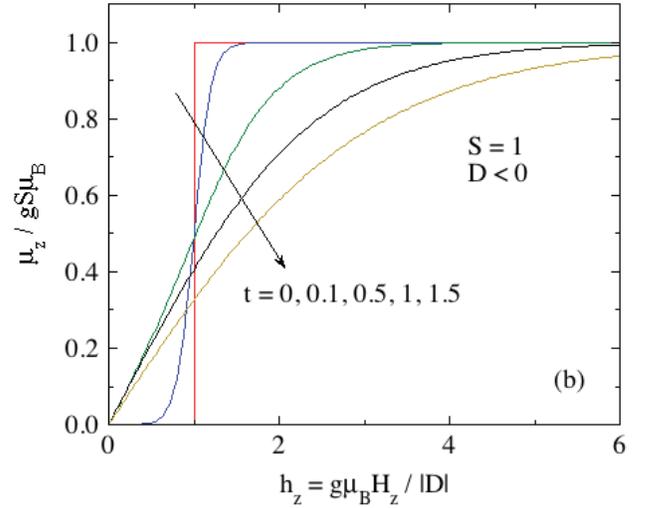}
\caption{(Color online) Normalized magnetic moment along the $z$~axis, $\bar{\mu}_z = \mu_z/gS\mu_{\rm B}$, versus reduced $z$-axis magnetic field $h_z$ obtained using Eq.~(\ref{Eq:GSDef}) for $S=1$ with (a) $D>0$ and~(b)~$D<0$.}
\label{Fig:muzbarVsHzS1d1m1}
\end{figure}

\begin{figure}
\includegraphics [width=3.3in]{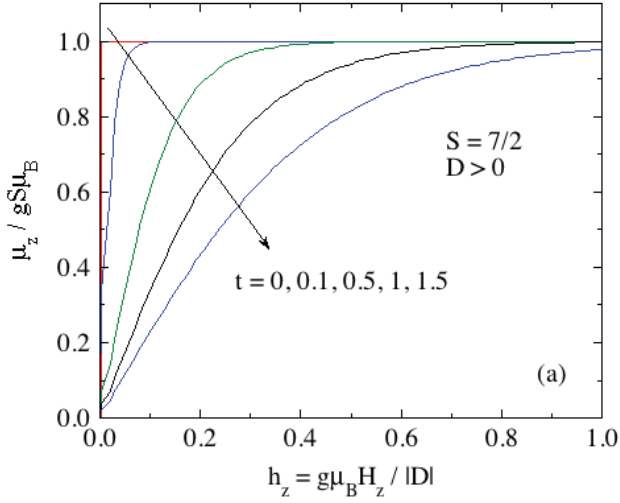}
\includegraphics [width=3.3in]{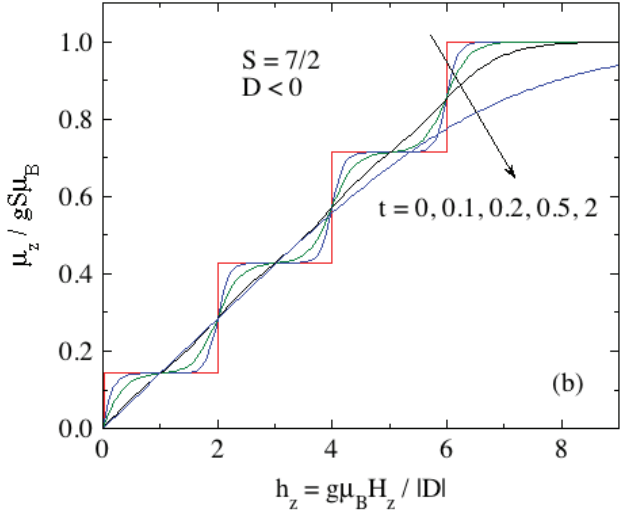}
\caption{(Color online) Same as Fig.~\ref{Fig:muzbarVsHzS1d1m1} but with $S = 7/2$.}
\label{Fig:muzbarVsHzS72d1m1}
\end{figure}

The reduced magnetic moment $\bar{\mu}_z = \mu_z/\mu_{\rm sat}$ versus reduced magnetic field $h_z$ obtained using Eq.~(\ref{Eq:GSDef}) is plotted for both $D>0$ and $D<0$ for spins $S=1$ and $S=7/2$ in Figs.~\ref{Fig:muzbarVsHzS1d1m1} and \ref{Fig:muzbarVsHzS72d1m1}, respectively.  For $D>0$, the magnetization curves are typical for a paramagnetic species.  However, for $D<0$ one sees steps in the data at low~$t$ and nonzero $h_z$ which occur for $S=1$ at $h_z=1$ and for $S=7/2$ at $h_z =2$, 4 and~6 arising from crossings of the ground-state Zeeman level by zero-field excited states  shown in Figs.~\ref{Fig:EvsHzS1d1m1} and~\ref{Fig:EvsHzS72d1m1}(b) for $S=1$ and $S=7/2$, respectively.  These crossings result in discontinuous increases in the ground state moment with increasing $h_z$ for $t\to0$.  The discontinuities become washed out with increasing temperature. 

\subsection{\label{Sec:XverseField} Perpendicular Magnetic Field ${\bf H} = H_x\hat{\bf i},\ D>0$}

\begin{figure}
\includegraphics [width=3.3in]{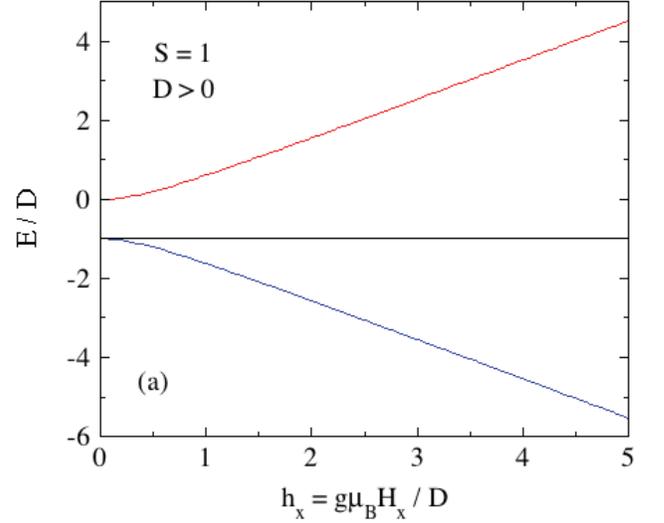}
\includegraphics [width=3.3in]{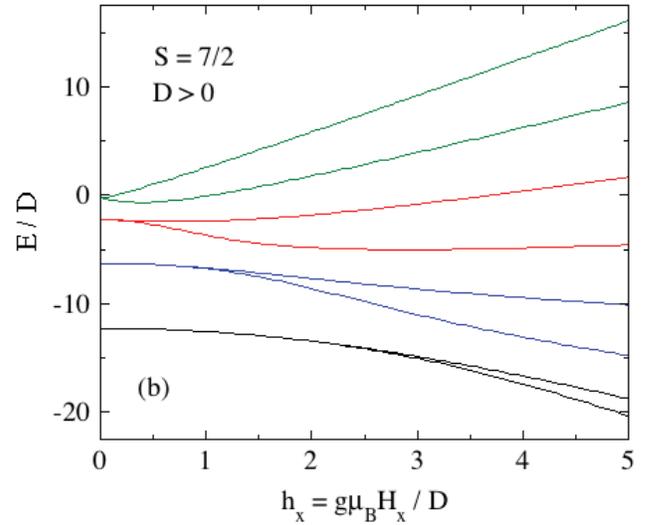}
\caption{(Color online) Normalized eigenvalue energies $E/D$ with $D>0$ versus reduced $x$-axis magnetic field $h_x$ for  (a)~$S=1$ and (b)~$S=7/2$.}
\label{Fig:EofHxS1S72d1}
\end{figure}

Here we choose a representative applied field direction ($x$) perpendicular to the uniaxial $z$ direction.  Then the single-spin  Hamiltonian is 
\be
{\cal H}_i = g\mu_{\rm B}H_xS_x  - DS_z^2,
\label{Eq:HxHamil}
\ee
where $S_x = (S_+ + S_-)/2$.  As above, we normalize all energies by $D$, which here is $>0$.  Thus the Hamiltonian becomes
\bse
\be
\frac{{\cal H}_i}{D} = h_xS_x - S_z^2, 
\label{Eq:HamSxSz2}
\ee
where we define the reduced $x$-axis field as
\be
h_x \equiv \frac{g\mu_{\rm B}H_x}{D}.
\ee
\ese
Plots of the reduced eigenenergies $\epsilon_n\equiv E_n/D\ (n=1\ {\rm to}\ 2S+1$) are given for $S=1$ and $S=7/2$ in Figs.~\ref{Fig:EofHxS1S72d1}(a) and~\ref{Fig:EofHxS1S72d1}(b), respectively.  For integer~$S$, at $h_x=0$ there are $S$ doublets with different energies with the highest-energy state being nondegenerate.  For half-integer spins there are $2S$ doublets at different energies.  In contrast to fields applied along the $z$~axis, the energy versus field relationships are in general nonlinear.  For example, for $S=1$ the three energy eigenvalues are
\be
\frac{E(S=1)}{D} = -1,\qquad -\frac{1}{2}\left(1 \pm \sqrt{1+4h_x^2}\right).
\label{Eq:ES=1}
\ee 

For $H_x=0$ the eigenenergies are $Dm_S^2$ with $m_S=-S,\ -S+1, \ldots,\ S\ (2S+1$ values).  As seen in Eq.~(\ref{Eq:ES=1}) and Fig.~\ref{Fig:EofHxS1S72d1}, the lowest-order field dependences of the eigenenergies are $H_x$ for half-integer spins and $H_x^2$ for integer spins.  The former result is consistent with Kramers' degeneracy theorem for a spin system containing an odd number of fermions which states that the ground state of the system in the absence of a magnetic field is at least doubly degenerate.

\begin{figure}
\includegraphics [width=3.3in]{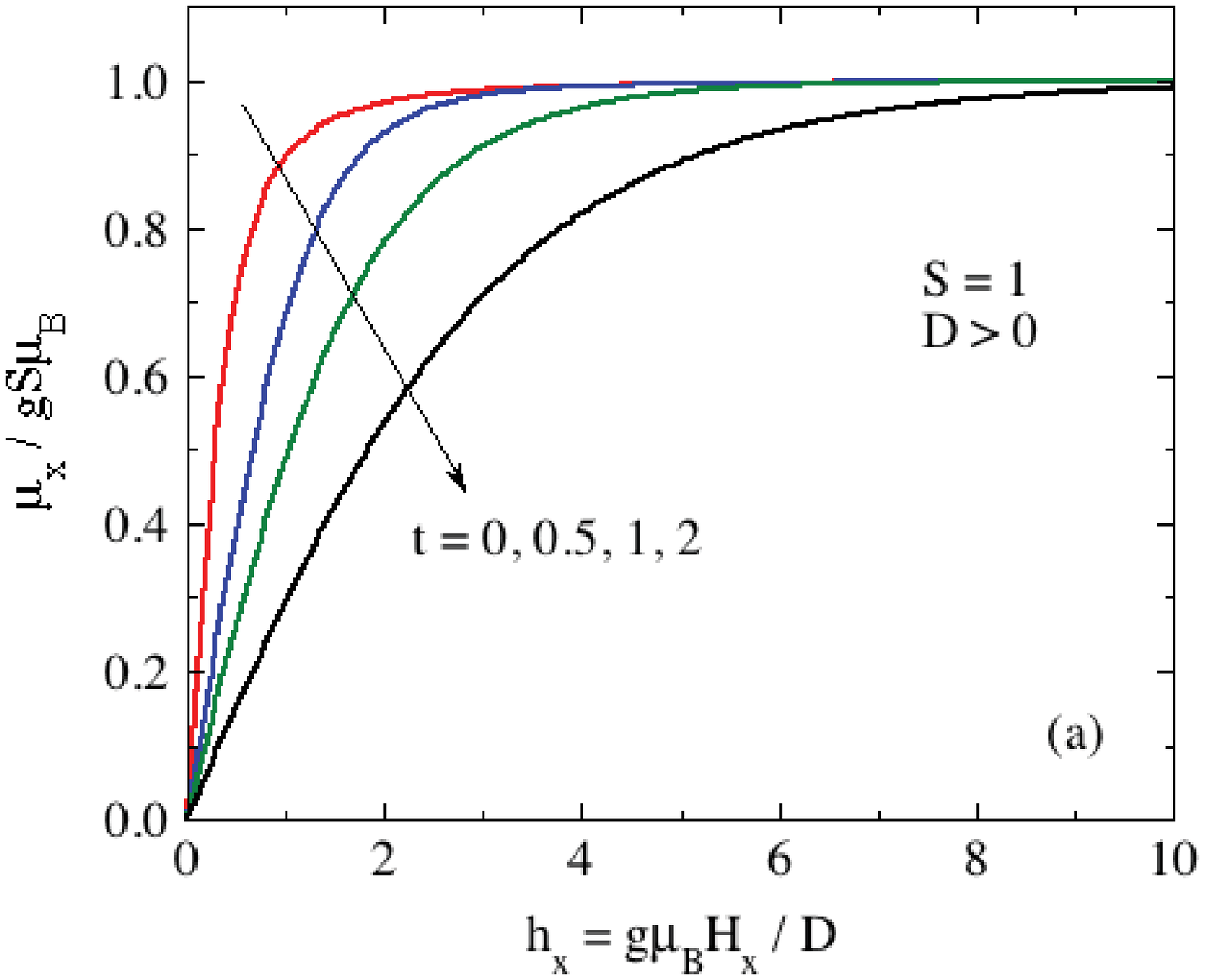}
\includegraphics [width=3.3in]{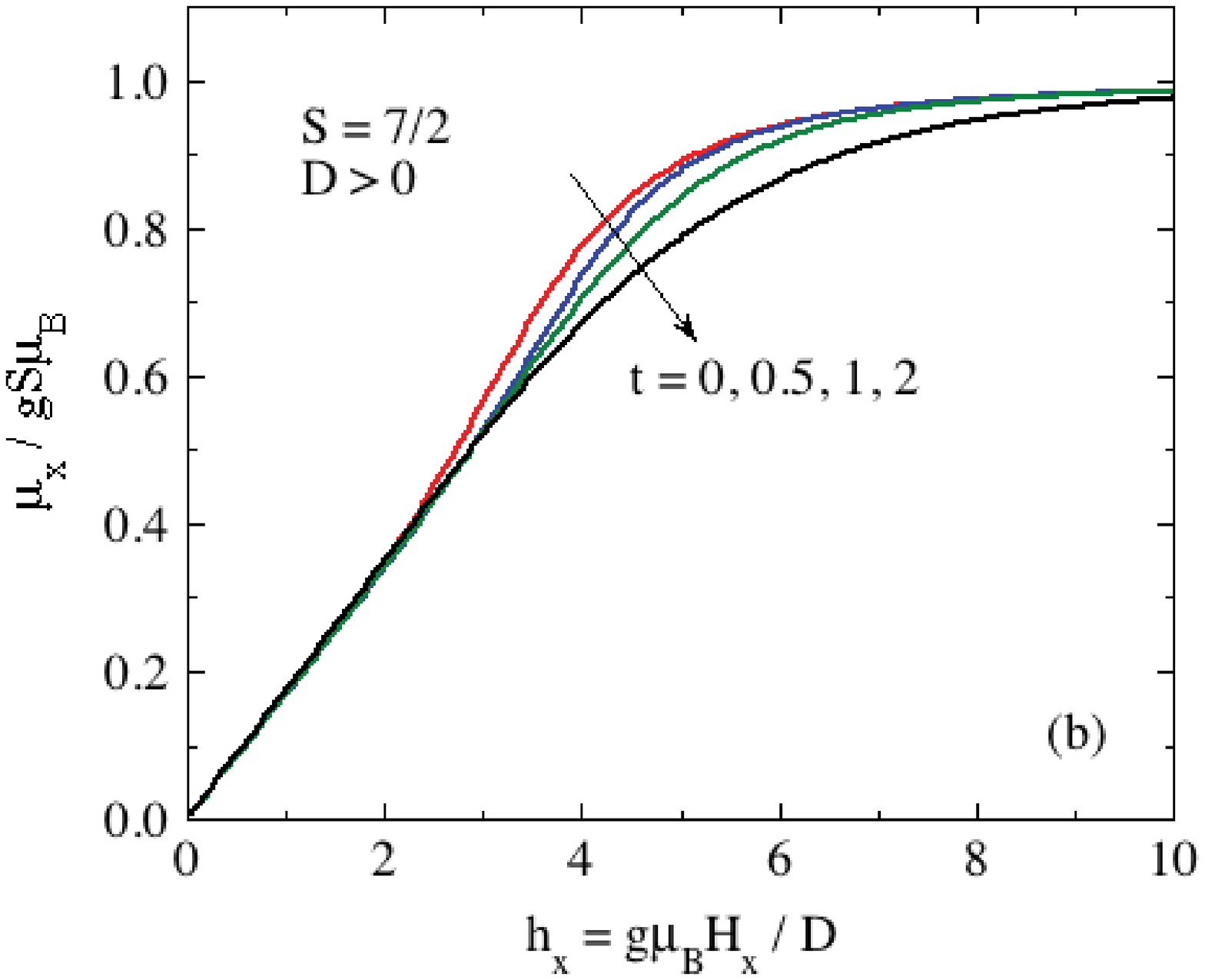}
\caption{(Color online) High-field transverse magnetization $\bar{\mu}_x = \mu_x/\mu_{\rm sat}$ for $D>0$ versus reduced transverse field~$h_x$ for (a)~$S=1$ and (b)~$S=7/2$ obtained from Eqs.~(\ref{Eqs:muixAve}).}
\label{Fig:MHxS1S72d1}
\end{figure}

\begin{figure}
\includegraphics [width=3.3in]{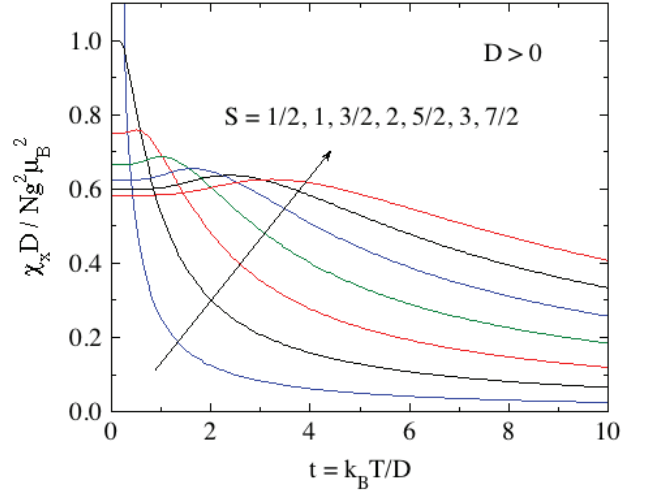}
\caption{(Color online) Normalized magnetic susceptibility along the $x$~axis $\bar{\chi}_x = \chi_xD/Ng^2\mu_{\rm B}^2$ versus reduced temperature~$t$ obtained from Eqs.~(\ref{Eq:ChiParPerpList}).}
\label{Fig:ChixS12S72d1}
\end{figure}

For $D>0$ the $x$-axis magnetization is obtained from
\bse
\label{Eqs:muixAve}
\be
\bar{\mu}_{x}(h_x,t) = -\frac{1}{SZ_S}\sum_{n = 1}^{2S+1} \frac{\partial\epsilon_n(h_x,t)}{\partial h_x} \exp\left[-\frac{\epsilon_n(h_x,t)}{t}\right],
\ee
where the partition function is
\be
Z_S = \sum_{n = 1}^{2S+1} \exp\left[-\frac{\epsilon_n(h_x,t)}{t}\right].
\ee
\ese
The results for the high-field transverse magnetization for $S=1$ and $S=7/2$ obtained from Eqs.~(\ref{Eqs:muixAve}) are shown in Figs.~\ref{Fig:MHxS1S72d1}(a) and~\ref{Fig:MHxS1S72d1}(b), respectively.  One sees that the plots, especially for $S=7/2$, are quite different from $\bar{\mu}_z(h_z,t)$ in corresponding Figs.~\ref{Fig:muzbarVsHzS1d1m1}(a) and~\ref{Fig:muzbarVsHzS72d1m1}(a).

\subsubsection{Perpendicular Susceptibility from Exact Diagonalization}

After expanding $\bar{\mu}_{x}(h_x,t)$ in Eqs.~(\ref{Eqs:muixAve}) to first order in $h_x$, the reduced perpendicular molar susceptibility
\be
\bar{\chi}_x(t)\equiv \frac{\chi_x(t)D}{Ng^2\mu_{\rm B}^2} = \frac{S\bar{\mu}_x}{h_x}
\ee
is obtained, as listed in Eqs.~(\ref{Eq:ChiParPerpList}) for spins 1/2 to 7/2 and plotted in Fig.~\ref{Fig:ChixS12S72d1}. 

\subsubsection{Perpendicular Susceptibility of Integer Spins from Perturbation Theory}

Defining the reduced temperature
\be
t = \frac{k_{\rm B}T}{D},
\label{Eq:tDDef}
\ee
the normalized perpendicular susceptibility per spin is obtained from Eq.~(\ref{Eq:muxPerpInt}) as
\bse
\label{Eqs:chiPerpD}
\be
\frac{\chi_\perp D}{g^2\mu_{\rm B}^2} = \frac{1}{Z_S}\sum_{m_S=-S}^S \left[\frac{S(S+1)+m_S^2}{4m_S^2 -1}\right] e^{m_S^2/t},
\ee
where the partition function is
\be
Z_S=\sum_{m_S=-S}^S e^{m_S^2/t}.
\label{Eq:ZSx2}
\ee
\ese

In the limit $t\ll1$, only the $m_S=\pm S$ terms in Eqs.~(\ref{Eqs:chiPerpD}) survive, giving the finite value
\be
\frac{\chi_\perp D}{g^2\mu_{\rm B}^2} = \frac{S}{2S-1}  \hspace{0.1in} \qquad (t \ll 1).
\label{Eq:chi0PerpT0Int}
\ee
In the limit $t\gg 1$, expanding the exponentials in Eqs.~(\ref{Eqs:chiPerpD}) in Taylor series to second order in $1/t$ gives a Curie-Weiss law
\bse
\label{Eqs:CWPerp}
\bea
\chi_\perp &=& \frac{C_1}{T-\theta_{{\rm p}\perp}} \quad (t \gg 1,\ {\rm integer}\ S),\\*
\theta_{{\rm p}\perp} &=& -\frac{D}{k_{\rm B}}\left[\frac{4S(S+1) -3}{30}\right],\label{Eq:CWPerpTheta}
\eea
\ese
where $C_1$ is the single-spin Curie constant in Eq.~(\ref{C1}).  Thus $\theta_{{\rm p}x}$ is negative if $D>0$ and is in agreement with Eq.~(\ref{Eq:thetapx}).  Equation~(\ref{Eq:CWPerpTheta}) yields $\theta_{{\rm p}x}=0$ for $S=1/2$ as required.

\subsubsection{Perpendicular Susceptibility of Half-Integer Spins from Perturbation Theory}

Equations~(\ref{Eq:muxBxHalfInt}) give the perpendicular susceptibility per spin $\chi_\perp = \mu_x/H_x$ after normalization as
\bse
\label{Eqs:ChiPerpDHalfInt}
\bea
\frac{\chi_\perp D}{g^2\mu_{\rm B}^2} &=& \frac{1}{Z_S} \bigg\{\frac{e^{1/4t}}{2}\bigg[\frac{S(S+1)+1/4}{t}-\ S(S+1)+3/4\bigg]\nonumber\\*
&&+\ 2\sum_{m_S=3/2}^S \left[\frac{S(S+1)+m_S^2}{4m_S^2 -1}\right] e^{m_S^2/t}\bigg\},\\
Z_S &=&\sum_{m_S=-S}^S e^{m_S^2/t}.
\eea
\ese
where we used the definition of the reduced temperature~$t$ in Eq.~(\ref{Eq:tDDef}).

For $k_{\rm B}T/|D|\ll1$, one obtains the same expression for $\chi_\perp$ as for integer spins in Eq.~(\ref{Eq:chi0PerpT0Int}).  For $k_{\rm B}T/|D|\gg1$, one obtains the Curie-Weiss law~(\ref{Eqs:CWLaw}) with the same Weiss temperature as for integer spins in Eq.~(\ref{Eq:CWPerpTheta}).  For either integer or half-integer spins, the powder average of the Weiss temperature from Eqs.~(\ref{Eq:thetapz}) and~(\ref{Eq:CWPerpTheta}) is the expected value
\be
\theta_{\rm p\ ave} = \frac{1}{3}(\theta_\parallel + 2\theta_\perp) = 0.
\ee

%\clearpage

\end{document}